\documentclass[preprintnumbers,article,amsmath,amssymb,floatfix,10pt,prd,superscriptaddress,nofootinbib,twocolumn]{revtex4-2}

\usepackage{doi}
\usepackage{hyperref}
\hypersetup{
  colorlinks=true,        
  linkcolor=blue,         
  citecolor=cyan,         
}

\usepackage{graphicx}
\usepackage{dcolumn}
\usepackage{bm}
\usepackage{color}
\usepackage{enumitem}

\newcommand{\beq}{\begin{equation}}
\newcommand{\eeq}{\end{equation}}
\newcommand{\bea}{\begin{eqnarray}}
\newcommand{\eea}{\end{eqnarray}}

\usepackage{bbm}
\usepackage{amsfonts}
\usepackage{mathrsfs}
\usepackage{latexsym}
\usepackage{epsfig}
\usepackage{epstopdf}
\usepackage{dcolumn}
\usepackage{bm}
\usepackage{color}
\usepackage{comment}
\usepackage{xcolor}
\usepackage{orcidlink}

\begin{document}

\title{Flip-Flop Shocks and Tori in Bondi–Hoyle–Lyttleton Accretion onto Rotating Kerr–Weyl Black Holes and Their Quasi-Periodic Variability}

\author{Orhan~Donmez}
\email{orhan.donmez@aum.edu.kw}
\affiliation{College of Engineering and Technology, American University of the Middle East, Egaila 54200, Kuwait}

\author{Imtiaz Khan}
\email{ikhanphys1993@gmail.com}
\affiliation{Department of Physics, Zhejiang Normal University, Jinhua, Zhejiang 321004, China}

\author{G. Mustafa}
\email{gmustafa3828@gmail.com}
\affiliation{Department of Physics, Zhejiang Normal University, Jinhua 321004, People's Republic of China}

\begin{abstract}
We present GRHD simulations of BHL accretion onto a rotating black hole in Weyl conformal gravity, described by the Kerr--Weyl metric. The flow is evolved on the equatorial plane for spins $a=0.9M$ and $a=0.5M$ and for a range of Weyl coefficients, which enter the dynamics through an effective curvature parameter $k$. All models are compared with Kerr runs that share the same numerical setup and asymptotic flow. For the values of $k$ considered, the horizon, the innermost stable circular orbit and the orbital and epicyclic frequencies of test particles differ from their Kerr values by at most $\sim2\%$. The curvature term, however, grows as $kr^{2}$ and acts at the accretion radius $R_{\rm acc}$, where the gas is captured. We show that the flow is controlled by a single quantity, the ratio $\eta=|k|R_{\rm acc}^{3}/M$ of the curvature-induced to the Newtonian acceleration at $R_{\rm acc}$. For $\eta\lesssim0.03$, a stable shock cone forms that is indistinguishable from the Kerr case. For $\eta\simeq0.2$--$0.3$, the anti-de Sitter-like term traps shocked gas behind the black hole, widens the cone and drives persistent flip-flop oscillations. For $\eta\gtrsim1.6$, the cone is destroyed and an oscillating torus forms. The flow depends on the Weyl coefficients only through $k$, while the spin, through frame dragging, modifies the inner few gravitational radii without changing the flow regime. In the strongly deviating models, the mean accretion rate is $1.7$--$3$ times the Kerr value and varies by $20$--$60\%$ rms, compared with $0.1$--$1\%$ in Kerr-like flows. The power spectra reveal global QPOs with fundamentals at $2.9$--$4.8$\,Hz for a $10\,M_\odot$ black hole and ladders of harmonics, whose $3{:}2$ and $2{:}1$ ratios arise from a single non-sinusoidal oscillation rather than from resonances. These frequencies lie in the low-frequency QPO band of X-ray binaries and are unrelated to the test-particle frequencies at the same radii, $\approx70$--$200$\,Hz. They scale with the Keplerian frequency at $R_{\rm acc}$, $f_0=\xi f_{\rm K}(R_{\rm acc})$, with $\xi\simeq0.7$--$1.1$ for stable cones, as in Newtonian simulations, and $\xi\simeq0.23$--$0.38$, decreasing with $\eta$, for the flip-flop and torus regimes. For supermassive black holes, the same modes correspond to periods of days for Sgr~A$^*$ and years for M87$^*$. For a bright Galactic source, photon statistics would allow the strong oscillations to be detected within a few seconds, whereas Kerr-like cone oscillations would require $10^{4}$--$10^{8}$\,s. Since the required $|k|$ exceeds the values inferred from galactic rotation curves by $18$--$36$ orders of magnitude, a detected shock-driven QPO, combined with the wind parameters of the source, would measure a local deviation from the Kerr geometry, while its absence would constrain $|k|$.\\
\end{abstract}

\maketitle

\date{\today}


\section{Introduction}\label{S1}
General relativity (GR) predicts that sufficiently massive compact objects cannot be supported against their own gravity~\cite{BHS6,BHS7} and must collapse into black holes, whose exterior geometry is described by a small family of exact vacuum solutions of Einstein's equations~\cite{BHS1,BHS2,BHS3,BHS4,BHS5,Ker1}. Among them, the Kerr metric~\cite{metric} plays a privileged role. By the uniqueness theorems, an isolated, stationary and uncharged black hole in GR is completely characterized by its mass $M$ and spin parameter $a$~\cite{Ker2,Ker3,Ker4}, and this statement remains approximately valid for black holes embedded in realistic astrophysical environments~\cite{Ker5}. Soon after the Kerr solution was found, accretion of gas onto such objects was proposed as the energy source of quasars~\cite{BHS8,BHS9}, because it converts rest-mass energy into radiation far more efficiently than nuclear burning. Accretion flows have since become our main window onto the strong-field region. In the last decade, the Kerr paradigm has been confronted with observations of unprecedented directness: the gravitational waves emitted by merging binary black holes~\cite{GW150914,Ker15}, the horizon-scale images of M87$^*$ and Sgr~A$^*$~\cite{EHT2019I,Ker18,EHT2019VI,EHT2022SgrA}, and the stellar orbits around the Galactic Center~\cite{GRAVITY2022} are all consistent with the Kerr geometry.

Consistency, however, is not proof. Present observations still allow sizable deviations from the Kerr metric close to the horizon, and a broad program aims to test the Kerr hypothesis with electromagnetic and gravitational-wave observations~\cite{Psaltis2008,Ker7,Ker9,Ker10,Ker8,Ker13,Ker14}. Parametrized and theory-specific non-Kerr metrics have been confronted with X-ray reflection spectroscopy~\cite{Ker11,Ker12}, black hole images and shadows~\cite{Ker16,Ker17,QPOs15,Mizuno2018}, and the ringdown of merger remnants~\cite{Ker19,Ker26}. Because the Kerr metric is also a solution of many alternative theories~\cite{Ker24}, tests based on the metric alone can be degenerate, and the dynamics of matter in the spacetime provides complementary information. The theoretical motivation for such tests is strong. GR has passed all Solar-System and binary-pulsar tests~\cite{Ker23}, yet the accelerated expansion of the Universe~\cite{Riess1998,Perlmutter1999}, the cosmological-constant problem~\cite{Weinberg1989}, the missing mass in galaxies, and the non-renormalizability of the Einstein--Hilbert action have motivated a large variety of modified theories of gravity~\cite{Ker20,Ker21,Ker22,Ker25}.

Weyl conformal gravity is one of the most economical of these alternatives. The theory, which goes back to the work of Weyl and Bach
~\cite{Weyl1918,Bach1921}, replaces the Einstein--Hilbert action with the square of the Weyl tensor, $I_W\propto\int d^4x\sqrt{-g}\,C_{\lambda\mu\nu\kappa}C^{\lambda\mu\nu\kappa}$, which is invariant under local conformal transformations of the metric, $g_{\mu\nu}\to\Omega^2(x)g_{\mu\nu}$. Conformal invariance forbids a fundamental cosmological constant and makes the theory power-counting renormalizable~\cite{Stelle1977,Mannheim2006}. The ghost problem associated with its fourth-order field equations has been addressed through a non-Hermitian, $\mathcal{PT}$-symmetric quantization~\cite{BenderMannheim2008}, and Einstein gravity can be recovered from conformal gravity with suitable boundary conditions~\cite{Maldacena2011}. The static, spherically symmetric vacuum solution of the theory was found by Mannheim and Kazanas~\cite{Mannheim1989} and shown to be unique~\cite{Riegert1984}. In addition to a Schwarzschild-like term, it contains a potential that grows linearly with radius, proportional to a coefficient $\gamma$, and a quadratic, cosmological-like term proportional to $\kappa$. With these two terms, conformal gravity has been used to fit the rotation curves of large samples of galaxies without dark matter~\cite{MannheimOBrien2011,MannheimOBrien2012}. The viability of the theory remains debated, in particular regarding light deflection and its weak-field limit~\cite{EderyParanjape1998,Flanagan2006}. This makes independent tests in different physical regimes, and especially in the strong-field regime around black holes, particularly valuable.

Such tests have recently become possible for rotating black holes. Asuncion et al.~\cite{Asuncion:2025cxw} constructed and analyzed the stationary, rotating generalization of the Mannheim--Kazanas solution, the Kerr--Weyl black hole. In this metric, the Weyl coefficients $(\beta,\gamma,\kappa)$ enter only through an effective mass parameter and an effective curvature parameter $k$. The latter acts as a cosmological term, de Sitter-like for $k>0$ and anti-de Sitter-like for $k<0$, although the solution differs in detail from the Kerr--(anti-)de Sitter metric of GR~\cite{Carter1968}. The motion of test particles in this spacetime, including its innermost stable circular orbit (ISCO), its epicyclic frequencies and its periastron and Lense--Thirring precession frequencies, has been analyzed together with planar accretion onto the same black hole~\cite{ZULQARNAIN2026117449}. A key property of the Kerr--Weyl correction is that the curvature term grows as $kr^2$ with radius. For the small values of $|k|$ that preserve the black hole structure, it leaves the horizon, the ISCO and the local orbital frequencies practically unchanged, so near-horizon probes are only weakly sensitive to it. Its influence is instead expected to be largest at tens of gravitational radii, precisely where an accretion flow captures gas from its surroundings and forms large-scale structures.

Quasi-periodic oscillations (QPOs) in the X-ray flux of accreting black holes are among the most precisely measured observables associated with the inner accretion flow~\cite{RemillardMcClintock2006,Motta2016,IngramMotta2019}. In black hole X-ray binaries, low-frequency QPOs (LFQPOs), at $\sim0.1$--$30$\,Hz, are classified into types A, B and C, whose properties correlate with the spectral state of the source and with other timing features~\cite{Belloni2002,Casella2005,QPOs31}. High-frequency QPOs (HFQPOs), at $\sim40$--$450$\,Hz, are rarer, but some sources show pairs in a $3{:}2$ frequency ratio, such as the $300$ and $450$\,Hz QPOs of GRO~J1655$-$40~\cite{QPOs18,Strohmayer2001,QPOs20}. HFQPOs have also been detected in XTE~J1550$-$564 and XTE~J1650$-$500~\cite{QPOs20,QPOs5}. The microquasar GRS~1915$+$105 shows a particularly rich variety of QPOs, whose evolution with the accretion rate has been followed in detail~\cite{QPOs19,Liu:2020jig,QPOs25}, and QPOs have also been found in active galactic nuclei~\cite{QPOs26}.

Many models have been proposed to explain these frequencies. Kinematic models identify them with the orbital, epicyclic and precession frequencies of test particles, as in the relativistic precession model~\cite{StellaVietri1998,Stella1999,Motta2014} and in nonlinear resonance models, which naturally explain the $3{:}2$ ratio~\cite{KluzniakAbramowicz2001,AbramowiczKluzniak2001,Torok2005,QPOs10,QPOs11,QPOs12,QPOs17,QPOs29}. Type-C LFQPOs are often attributed to Lense--Thirring precession of a hot inner flow~\cite{Ingram2009}. Other models invoke oscillation modes trapped in the disk~\cite{QPOs33} or the global modes of pressure-supported tori, whose eigenfrequencies approximately follow sequences of small integers~\cite{QPOs9,Zanotti2003,QPOs38,DonmezMNRAS2014}. Because the characteristic frequencies of geodesic motion depend directly on the metric, QPOs have become a widely used tool to test the Kerr hypothesis~\cite{JohannsenPsaltis2011,QPOs27,QPOs21}. The same approach has been applied to a broad variety of non-Kerr and modified-gravity black holes~\cite{QPOs50,QPOs1,QPOs2,QPOs3,QPOs4,QPOs6,QPOs7,QPOs13,QPOs14,QPOs16,QPOs23,QPOs28,QPOs34,QPOs35,QPOs41,QPOs42,QPOs43,QPOs44,QPOs45,QPOs46,QPOs48,B1,B2,B6,MustafaJcap2025}. These studies, however, rely on the dynamics of test particles. A real accretion flow is shaped by pressure, shocks and hydrodynamic instabilities, which can generate oscillations whose frequencies are not those of geodesic motion. Capturing these effects requires the numerical solution of the general relativistic hydrodynamic (GRHD) equations~\cite{Font2008,RezzollaZanotti2013}.

Bondi--Hoyle--Lyttleton (BHL) accretion~\cite{Hoyle1939,Bondi1944,Edgar2004} provides a natural framework for such studies. It describes a compact object moving through, or fed by, a supersonic flow of ambient gas, as in wind-fed X-ray binaries, black holes moving through the interstellar medium, and black holes embedded in dense gaseous environments~\cite{Edgar2004}. Gravity focuses the incoming gas behind the black hole, where the streams collide and form a shock cone. Newtonian simulations showed that this cone can become unstable and swing from side to side, the so-called flip-flop instability~\cite{FryxellTaam1988,Foglizzo2005,Blondin2009}. Relativistic BHL accretion has been studied since the pioneering work of Petrich et al.~\cite{Petrich1989} and of Font and collaborators~\cite{FontIbanez1998,Font1999}, and has since been extended to magnetized flows~\cite{Penner2011} and to three dimensions~\cite{GraciaLinares2015}. These studies showed that the shock cone vibrates and can trap oscillation modes~\cite{LoraClavijo2013}. These modes modulate the mass accretion rate and produce QPOs whose frequencies and commensurate ratios 
resemble the observed ones~\cite{DonmezMNRAS2011,DonmezMNRAS2012,KoyuncuMpla2014,QPOs24}. When the BHL inflow onto a Kerr black hole is halted, the shock cone turns into a plasma structure with spiral shock waves, which excites both low- and HFQPOs~\cite{DonmezNPB2026}. Since the amplitude of flip-flop oscillations can also depend on the coordinates and on the numerical treatment of the inner boundary~\cite{CruzOsorio2012}, careful control of numerical effects is required.

In recent years, the same approach has been applied to black holes in alternative theories of gravity, including hairy black holes in Horndeski gravity~\cite{DonmezJheap2024}, quantum-corrected black holes~\cite{MustafaJcap2025} and other modified spacetimes~\cite{Donmezpdu2024,DonmezEpjc2025}. These studies showed that deviations from the Kerr metric can change the morphology of the shock cone, the onset of instabilities and the resulting QPO spectra. For the Kerr--Weyl black hole, accretion has so far been studied only for planar injection~\cite{ZULQARNAIN2026117449}. BHL accretion is particularly well suited for probing this spacetime. The region in which the gas is captured is set by the accretion radius, $R_{\rm acc}=2M/(v_\infty^2+c_s^2)$, where $v_\infty$ and $c_s$ are the asymptotic velocity and sound speed of the gas. This radius lies far from the horizon, and it is there that the curvature term of the Kerr--Weyl metric can become comparable to the Newtonian potential. A BHL flow can therefore respond strongly to deviations that leave the near-horizon geometry and the test-particle frequencies practically unchanged. Whether, and how, this response appears in the morphology and variability of the flow has not been investigated so far.

In this paper, we address this question with two-dimensional GRHD simulations of BHL accretion onto the rotating Kerr--Weyl black hole on the equatorial plane, for the spins $a=0.9M$ and $a=0.5M$. We consider aggressive, mild-deviation and near-Kerr sets of Weyl coefficients and compare them with Kerr runs that share the same numerical setup and asymptotic flow. We follow the formation and evolution of the shock cone, the mass accretion rate close to the horizon and in the inner flow, and the azimuthal motion of the shock. We then compute the power spectra of the accretion rate to identify the QPO frequencies and their commensurate ratios, and compare them with the characteristic frequencies of test particles in the same spacetime. We show that the flow is controlled by a single dimensionless quantity, the ratio of the curvature-induced to the Newtonian acceleration at the accretion radius. Depending on its value, the flow forms a stable shock cone, a flip-flop oscillating shock, or an oscillating torus around the black hole, each with a distinct QPO signature. Finally, we translate these results into observable predictions for stellar-mass and supermassive black holes and discuss how they can be used to constrain the Kerr--Weyl correction. The paper is organized as follows. In Sec.~\ref{S2}, we present the Kerr--Weyl metric, the GRHD equations, the numerical method, and the initial and boundary conditions. In Sec.~\ref{NumSim}, we explain the choice of model parameters and present the numerical results, namely the morphology of the flow, the mass accretion rates, the azimuthal density profiles and the power spectra. In Sec.~\ref{obser_signature}, we discuss the astrophysical signatures and observational prospects of the resulting QPOs. Finally, Sec.~\ref{S4} summarizes our conclusions.

\section{Kerr--Weyl Spacetime and Numerical Framework}\label{S2}

In this section, we describe the spacetime in which the accretion flow evolves, the GRHD equations that govern the gas, the numerical method used to solve them, and the initial and boundary conditions of the simulations. Throughout, we use geometrized units ($G=c=1$), and lengths and times are measured in units of the black hole mass $M$.

\subsection{The rotating black hole in Weyl conformal gravity}\label{S2_metric}

Weyl conformal gravity replaces the Einstein--Hilbert action by the square of the Weyl tensor, $I_W\propto\int d^{4}x\sqrt{-g}\,C_{\lambda\mu\nu\kappa}C^{\lambda\mu\nu\kappa}$, which is invariant under local conformal transformations of the metric, $g_{\mu\nu}\to\Omega^{2}(x)g_{\mu\nu}$. The resulting field equations are of fourth order. Their static, spherically symmetric vacuum solution, found by Mannheim and Kazanas~\cite{Mannheim1989}, contains three integration constants, $\beta$, $\gamma$ and $\kappa$. In addition to the Newtonian-like term, it includes a potential that grows linearly with radius, proportional to $\gamma$, and a quadratic one, proportional to $\kappa$. These terms have been used to describe galactic rotation curves without dark matter~\cite{MannheimOBrien2012}. The stationary, rotating generalization of this solution was constructed and analyzed by Asuncion et al.~\cite{Asuncion:2025cxw}. In Boyer--Lindquist-like coordinates $(t,r,\theta,\phi)$, it reads
\begin{eqnarray}
\mathrm{d} s^2 &=&-\left(1-\frac{2 \widetilde{M} r}{\rho^2}-k\left(r^2-a^2 \cos ^2 \theta\right)\right) \mathrm{d} t^2 +2k_1 \mathrm{d} t \mathrm{~d} \phi \nonumber \\
&+&\left(\frac{\rho^2}{\Delta^{\mathrm{H}}}\right) \mathrm{d} r^2+\left(\frac{\rho^2}{\Delta^\theta}\right) \mathrm{d} \theta^2+\left(\frac{\Sigma^2 \sin ^2 \theta}{\rho^2}\right) \mathrm{d} \phi^2 ,
\label{metric}
\end{eqnarray}
where
\begin{eqnarray*}
k_1&\equiv &\frac{-2 \widetilde{M} r a \sin ^2 \theta+k a\left(a^2\left(r^2+a^2\right) \cos ^4 \theta-r^4 \sin ^2 \theta\right)}{\rho^2},\\
 k &\equiv & \kappa+\frac{\gamma^2(1-\beta \gamma)}{(2-3 \beta \gamma)^2}, \\
 \rho^2 &\equiv & r^2+a^2 \cos ^2 \theta, \\
 \Delta^{\mathrm{H}} &\equiv &-k r^4+r^2-2 \widetilde{M} r+a^2, \\
 \Delta^\theta &\equiv & 1-k a^2 \cos ^2 \theta \cot ^2 \theta, \\
\Sigma^2 &\equiv &\Delta^\theta\left(r^2+a^2\right)^2-a^2 \Delta^{\mathrm{H}} \sin ^2 \theta .
\end{eqnarray*}
Here $a$ is the spin parameter of the black hole, and $\widetilde{M}$ is the effective mass parameter of the solution. $\widetilde{M}$ reduces to the black hole mass $M$ in the Kerr limit and differs from it by terms of order $\beta\gamma\lesssim10^{-2}$ for the models considered here. The Weyl coefficients enter the rotating metric only through $\widetilde{M}$ and the effective curvature parameter $k$. The function $\Delta^{\rm H}$ determines the horizons, $\Delta^\theta$ the angular structure, and $k_1=g_{t\phi}$ the frame dragging. For $k=0$ and $\widetilde{M}=M$, Eq.~(\ref{metric}) reduces to the Kerr metric. The term proportional to $k$ acts as a cosmological curvature term, de Sitter-like for $k>0$ and anti-de Sitter-like for $k<0$, although the rotating solution differs in detail from the Kerr--(A)dS metrics of GR~\cite{Asuncion:2025cxw}. The motion of test particles in this spacetime, including the effective potential, the ISCO, the radial, vertical and azimuthal epicyclic frequencies, and the periastron and Lense--Thirring precession frequencies, has been analyzed in Ref.~\cite{ZULQARNAIN2026117449}. That reference also compared these results with numerical simulations of planar accretion onto the same black hole.

Since the gas is evolved on the equatorial plane, we need the 3+1 decomposition of Eq.~(\ref{metric}) at $\theta=\pi/2$. Writing the line element as $ds^{2}=-(\alpha^{2}-\beta_i\beta^{i})dt^{2}+2\beta_i\,dx^{i}dt+\gamma_{ij}dx^{i}dx^{j}$, we obtain
\begin{eqnarray}
\alpha &=& \left(\frac{r^{2}\Delta^{\rm H}}{\Sigma^{2}}\right)^{1/2}, \qquad \beta^{r}=0, \nonumber\\
\beta^{\phi} &=& -\frac{a\left(2\widetilde{M}r+kr^{4}\right)}{\Sigma^{2}}, \nonumber\\
\gamma_{rr} &=& \frac{r^{2}}{\Delta^{\rm H}}, \qquad \gamma_{\phi\phi}=\frac{\Sigma^{2}}{r^{2}}, \nonumber\\
\Sigma^{2} &=& \left(r^{2}+a^{2}\right)^{2}-a^{2}\Delta^{\rm H},
\label{lapse_shift}
\end{eqnarray}
with $\sqrt{-g}=\alpha\sqrt{\gamma}=r^{2}$. The lapse $\alpha$ describes the gravitational redshift. The shift $\beta^{\phi}$ describes frame dragging, with angular velocity $\omega=-\beta^{\phi}$, and contains a curvature contribution $\propto kr^{4}$ in addition to the Kerr term. Both quantities enter the GRHD equations below, so the spin and the Weyl coefficients act on the gas only through Eq.~(\ref{lapse_shift}) and the corresponding Christoffel symbols.

\begin{figure*}[!ht]
  \vspace{1cm}
  \center
  \psfig{file=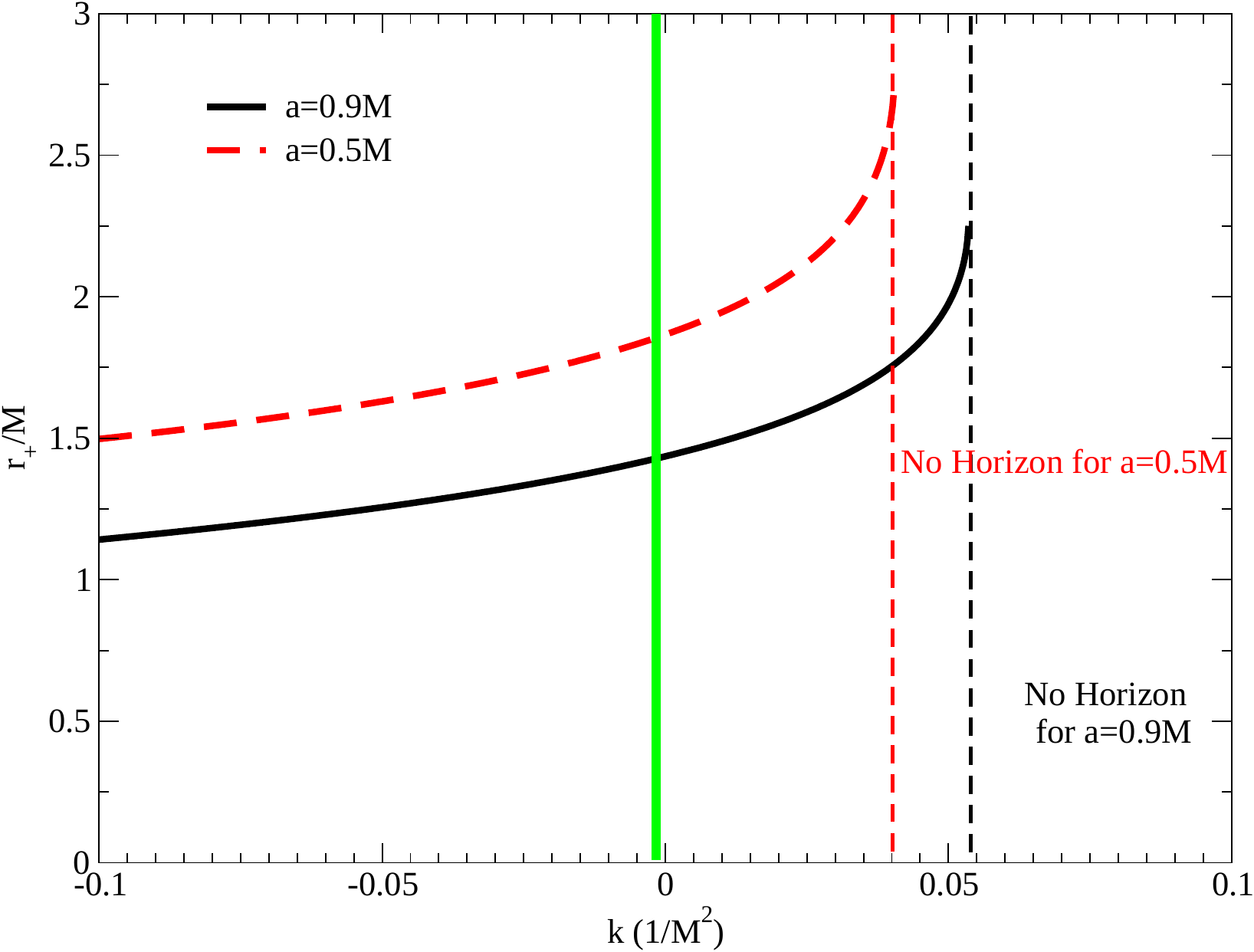,width=12.0cm,height=10cm}
     \caption{Outer horizon radius $r_+$ of the Kerr--Weyl black hole as a function of the curvature parameter $k$, for $a=0.9M$ (solid) and $a=0.5M$ (dashed). The horizon radius increases monotonically with $k$ and passes through the Kerr values, $1.436M$ and $1.866M$, at $k=0$. For $k>0$, the outer horizon approaches the cosmological horizon and merges with it at $k_c\simeq0.054\,M^{-2}$ ($a=0.9M$) and $k_c\simeq0.041\,M^{-2}$ ($a=0.5M$), indicated by the vertical dashed lines. This produces the steep rise of the curves. Beyond $k_c$, the spacetime no longer contains a black hole surrounded by a static region. The green band marks the range of $k$ used in the simulations (Table~\ref{Inital_Con_1}), in which $r_+$ is practically equal to its Kerr value.
}
\vspace{1cm}
\label{Weyl_horizon}
\end{figure*}

The lapse vanishes where $\Delta^{\rm H}=0$, which defines the horizons of the spacetime. For $k=0$, the roots are the inner and outer Kerr horizons, $r_\pm=\widetilde{M}\pm(\widetilde{M}^{2}-a^{2})^{1/2}$. For $k<0$, the term $-kr^{4}$ is positive and moves the outer horizon $r_+$ inward. For $k>0$, it moves $r_+$ outward and creates an additional cosmological horizon $r_c$, located at $r_c\simeq k^{-1/2}$ for small $k$. Fig.~\ref{Weyl_horizon} shows $r_+$ as a function of $k$ for $a=0.9M$ and $a=0.5M$. The outer horizon grows monotonically with $k$, from $r_+\simeq1.14M$ ($a=0.9M$) and $1.50M$ ($a=0.5M$) at $k=-0.1\,M^{-2}$ to the Kerr values, $1.436M$ and $1.866M$, at $k=0$. For $k>0$, the outer and cosmological horizons approach each other, which produces the steep rise of the curves. The two horizons merge at a critical value $k_c\simeq0.054\,M^{-2}$ for $a=0.9M$ and $k_c\simeq0.041\,M^{-2}$ for $a=0.5M$, shown by the vertical dashed lines. For $k>k_c$, only the inner root of $\Delta^{\rm H}$ remains, $\Delta^{\rm H}<0$ outside it, and the spacetime no longer contains a black hole surrounded by a static exterior region~\cite{Asuncion:2025cxw}. The values used in our simulations, $-3\times10^{-5}\lesssim k\lesssim5\times10^{-10}\,M^{-2}$ (green band), lie more than three orders of magnitude below $|k_c|$. For these values, $r_+$ differs from its Kerr value by less than $2\times10^{-4}M$. There is no cosmological horizon for $k<0$, and for the only positive value, Model $C$, $r_c\simeq4.5\times10^{4}M$. The whole computational domain, $2.3M\le r\le100M$, therefore lies in the static region outside the black hole, where $\Delta^{\rm H}>0$ and $\alpha$ is real, and the inner boundary at $r=2.3M$ lies outside the outer horizon for both spins. Since the horizon and the near-horizon geometry are practically those of the Kerr black hole, any difference between the Kerr and Kerr--Weyl flows must arise farther out, where the $kr^{2}$ term grows. This point is developed in the next section.

\subsection{General relativistic hydrodynamic equations}\label{S2_grhd}

The accreting gas is modeled as an inviscid, non-magnetized perfect fluid moving on the fixed background of Eq.~(\ref{metric}). Its self-gravity is neglected, since the mass of the gas is much smaller than that of the black hole. The evolution of the fluid follows from the conservation of rest mass and of energy and momentum,
\begin{equation}
\nabla_\mu(\rho u^\mu)=0,\qquad \nabla_\mu T^{\mu\nu}=0,\qquad
T^{\mu\nu}=\rho h\,u^\mu u^\nu+P g^{\mu\nu},
\label{grhd_cov}
\end{equation}
where $\rho$ is the rest-mass density, $u^\mu$ the four-velocity, $P$ the pressure, and $h=1+\epsilon+P/\rho$ the specific enthalpy, with $\epsilon$ the specific internal energy. The system is closed by the ideal-gas equation of state, $P=(\Gamma-1)\rho\epsilon$, for which the local sound speed is $c_s^{2}=\Gamma P/(\rho h)$.

To solve Eqs.~(\ref{grhd_cov}) with methods designed for hyperbolic conservation laws, we use the 3+1 conservative (Valencia) formulation~\cite{Banyuls1997,Font2008,RezzollaZanotti2013}. The fluid velocity measured by an Eulerian observer is $v^{i}=u^{i}/(\alpha u^{t})+\beta^{i}/\alpha$, and the Lorentz factor is $W=\alpha u^{t}=(1-\gamma_{ij}v^{i}v^{j})^{-1/2}$. On the equatorial plane, with $\partial_\theta=0$, the equations take the form
\begin{equation}
\partial_t\!\left(\sqrt{\gamma}\,\mathbf{U}\right)+\partial_r\!\left(\alpha\sqrt{\gamma}\,\mathbf{F}^{r}\right)+\partial_\phi\!\left(\alpha\sqrt{\gamma}\,\mathbf{F}^{\phi}\right)=\alpha\sqrt{\gamma}\,\mathbf{S}.
\label{grhd_cons}
\end{equation}
The vector of conserved variables is
\begin{equation}
\mathbf{U}=\begin{pmatrix} D\\ S_r\\ S_\phi\\ \tau\end{pmatrix}
=\begin{pmatrix} \rho W\\ \rho hW^{2}v_r\\ \rho hW^{2}v_\phi\\ \rho hW^{2}-P-\rho W\end{pmatrix},
\label{cons_var}
\end{equation}
where $D$, $S_j$ and $\tau$ are the rest-mass, momentum and energy densities measured by the Eulerian observer. The fluxes and sources are
\begin{eqnarray}
\mathbf{F}^{i}&=&\left(D\tilde v^{i},\ S_r\tilde v^{i}+P\delta^{i}_{r},\ S_\phi\tilde v^{i}+P\delta^{i}_{\phi},\ \tau\tilde v^{i}+Pv^{i}\right)^{T},\nonumber\\
\mathbf{S}&=&\Big(0,\ \tfrac{1}{2}T^{\mu\nu}\partial_r g_{\mu\nu},\ \tfrac{1}{2}T^{\mu\nu}\partial_\phi g_{\mu\nu},\nonumber\\
&&\ \ \alpha\left(T^{\mu t}\partial_\mu\ln\alpha-T^{\mu\nu}\Gamma^{t}_{\mu\nu}\right)\Big)^{T},
\label{flux_source}
\end{eqnarray}
with $\tilde v^{i}=v^{i}-\beta^{i}/\alpha$ and $\Gamma^{\lambda}_{\mu\nu}$ the Christoffel symbols of Eq.~(\ref{metric}). The momentum sources are written in the compact form $\tfrac{1}{2}T^{\mu\nu}\partial_j g_{\mu\nu}$, which is equivalent to $T^{\mu\nu}(\partial_\mu g_{\nu j}-\Gamma^{\lambda}_{\mu\nu}g_{\lambda j})$~\cite{Font2008}. The radial source contains the gravitational attraction, including the $kr^{2}$ contribution, together with the centrifugal and frame-dragging terms, and it is through this term that the Kerr--Weyl corrections act on the flow. Because the metric is stationary and axisymmetric, $\partial_\phi g_{\mu\nu}=0$ and the source of $S_\phi$ vanishes identically. The total angular momentum of the gas can therefore change only through fluxes across the boundaries of the domain. This property becomes important when the flow develops large-scale rotation, as in the flip-flop and torus regimes discussed in the next section. Finally, the flux Jacobians of Eqs.~(\ref{grhd_cons}) have real eigenvalues and a complete set of eigenvectors~\cite{Banyuls1997}, so the system is hyperbolic and can be solved with characteristic-based shock-capturing methods.

\subsection{Numerical scheme and computational grid}\label{S2_num}

Equations~(\ref{grhd_cons}) are solved with the GRHD code described in Ref.~\cite{QPOs37}, which has been used in our previous studies of BHL accretion onto Kerr black holes and onto black holes in alternative theories of gravity~\cite{DonmezMNRAS2011,DonmezMNRAS2012,DonmezJheap2024}. The code uses a finite-volume, high-resolution shock-capturing (HRSC) scheme. These methods are designed to resolve discontinuities sharply and without spurious oscillations, which is essential here, because the flow contains strong shocks: the shock cone, the swinging shock fronts of the flip-flop regime, and the spiral shocks of the torus. At each cell interface, the numerical fluxes are computed from the solution of local Riemann problems with Marquina's flux formula~\cite{DonatMarquina1996}, which uses the full spectral decomposition of the flux Jacobians~\cite{Banyuls1997}. The left and right states at each interface are obtained with a piecewise-linear MUSCL reconstruction of the primitive variables with a slope limiter~\cite{vanLeer1979}. This makes the scheme second-order accurate in smooth regions and monotonic near shocks. The equations are advanced in time with second-order accuracy, with a time step limited by the Courant--Friedrichs--Lewy condition. At each step, the primitive variables $(\rho,v_r,v_\phi,P)$ are recovered from the conserved ones by an iterative root-finding procedure.

The equations are solved on the equatorial plane on a polar grid of $N_r\times N_\phi=1024\times512$ zones, covering $2.3M\le r\le100M$ and $0\le\phi<2\pi$~\cite{DonmezJheap2024}. The azimuthal resolution is therefore $\Delta\phi=2\pi/512\simeq0.012$\,rad. The outer boundary lies at about $2.5$ times the accretion radius of the simulated flow, $R_{\rm acc}\simeq40M$ (see below), so the region in which the gas is captured, and in which the curvature term becomes important, is contained within the domain.

\subsection{Initial and boundary conditions}\label{S2_bc}

At $t=0$, the computational domain is filled with a uniform flow moving in the positive $x$ direction with asymptotic velocity $v_\infty$, which reproduces the BHL configuration of a black hole moving through a homogeneous medium. The velocity components are
\begin{equation}
v^{r}=\sqrt{\gamma^{rr}}\,v_\infty\cos\phi,\qquad
v^{\phi}=-\sqrt{\gamma^{\phi\phi}}\,v_\infty\sin\phi ,
\label{v_init}
\end{equation}
so that $\gamma_{ij}v^{i}v^{j}=v_\infty^{2}$ everywhere. With this choice, $\phi=\pi$ corresponds to the upstream direction, from which the gas arrives, and $\phi=0$ to the downstream axis, along which the shock cone forms. The rest-mass density is set to $\rho_\infty=1$. The pressure is fixed by the asymptotic sound speed $c_s$, through $c_s^{2}=\Gamma P/(\rho h)$ and the equation of state:
\begin{equation}
P_\infty=\frac{c_s^{2}\,\rho_\infty(\Gamma-1)}{\Gamma(\Gamma-1)-\Gamma c_s^{2}} .
\label{P_init}
\end{equation}
In all models, we use $v_\infty=0.2c$, $c_s=0.1c$ and $\Gamma=4/3$, which gives $P_\infty\simeq7.7\times10^{-3}\rho_\infty$. Because the gas does not affect the metric, Eqs.~(\ref{grhd_cons}) are invariant under $\rho\to\lambda\rho$, $P\to\lambda P$. All results therefore scale linearly with $\rho_\infty$, and accretion rates are given in arbitrary units. The uniform initial state is not an equilibrium of the curved spacetime. It relaxes during an initial transient lasting a few crossing times, $R_{\rm acc}/v_\infty\simeq200M$, after which the shock cone forms. The time series analyzed in the next section start at $t\simeq3\times10^{3}M$, well after this transient.

The boundary conditions are imposed through ghost zones. At the inner boundary, $r=2.3M$, we use outflow conditions, extrapolating the variables into the ghost zones, so that gas crossing this radius is treated as accreted by the black hole. The boundary must lie outside the horizon, because the Boyer--Lindquist-like coordinates of Eq.~(\ref{metric}) are singular at $\Delta^{\rm H}=0$, where $\alpha\to0$ and $\gamma_{rr}\to\infty$. As shown in Fig.~\ref{Weyl_horizon}, $r=2.3M$ lies outside $r_+$ for both spins and all values of $k$ considered. Since an inner boundary placed outside the horizon can affect the amplitude of shock-cone oscillations in relativistic simulations~\cite{CruzOsorio2012}, all models use the same inner boundary. The near-Kerr models, which differ from the aggressive ones only in $k$, thus serve as a control for this effect in the next section. At the outer boundary, $r=100M$, the treatment depends on the direction of the flow. On the upstream side ($\cos\phi<0$), the ghost zones are continuously filled with the injected gas, with $\rho_\infty$, $P_\infty$ and the velocities of Eq.~(\ref{v_init}), so that the black hole is supplied by a steady, uniform wind. On the downstream side ($\cos\phi\ge0$), outflow conditions let the gas leave the domain freely, without artificial reflections. Periodic conditions are applied at $\phi=0$ and $\phi=2\pi$.

The main diagnostic of the simulations is the mass accretion rate through a circle of radius $r$, denoted $dM/dt$ in the figures,
\begin{equation}
\dot M(r,t)=-\int_{0}^{2\pi}\sqrt{-g}\,\rho u^{r}\,d\phi=-\int_{0}^{2\pi}\alpha\sqrt{\gamma}\,\rho W v^{r}\,d\phi ,
\label{Mdot_def}
\end{equation}
where we have used $\beta^{r}=0$. The rate is positive for net inflow and negative for net outflow. We evaluate it at $r=2.3M$, close to the horizon, and at $r=6.11M$, in the inner part of the flow. Its power spectrum is then used to identify the QPO frequencies.

With this setup, all models share the same numerical method, grid, boundary conditions and asymptotic flow, and differ only in the spin $a$ and the Weyl coefficients $(\beta,\gamma,\kappa)$. Since Fig.~\ref{Weyl_horizon} shows that these coefficients leave the horizon and the near-horizon geometry practically unchanged, the question is how and where they affect the flow. The next section first answers this question by showing how the curvature term reshapes the gravitational potential at the scale of the accretion radius, and then presents the resulting evolution of the accretion flow.

\section{Numerical Simulation Results}\label{NumSim} 
In this section, we present the results of the simulations described in Sec.~\ref{S2}, which follow BHL accretion~\cite{Hoyle1939,Bondi1944,Edgar2004} onto the rotating Kerr--Weyl black hole. This approach extends earlier GRHD studies of BHL accretion onto Schwarzschild and Kerr black holes~\cite{FontIbanez1998,Font1999,DonmezMNRAS2011,DonmezMNRAS2012,KoyuncuMpla2014}, and onto black holes in alternative theories of gravity~\cite{Donmezpdu2024,DonmezJheap2024,DonmezEpjc2025,MustafaJcap2025}, as well as the study of planar accretion onto the same Kerr--Weyl black hole~\cite{ZULQARNAIN2026117449}.  Each model is evolved for $(2$--$3.5)\times10^{4}M$, long enough for the shock cone to form and for its long-term behavior, whether stable, oscillating or disrupted, to become clear.

We first explain the physical setup and the choice of model parameters. We then follow the global evolution of the rest-mass density on the equatorial plane to show how the shock cone forms, oscillates, or turns into a torus. Next, we examine the time evolution of the mass accretion rate at two radii: $r=2.3M$, close to the horizon, and $r=6.11M$, farther out in the inner flow. This shows how the variability imposed by the large-scale shock dynamics is transmitted to the inner flow. We also track the rest-mass density along the azimuthal direction at a fixed radius as a function of time, which directly reveals the side-to-side motion of the shock cone and the strength of the flip-flop instability. Finally, we compute the power spectral density (PSD) of the mass accretion rate. From the PSD, we identify the QPO frequencies and their commensurate ratios, and show how they depend on the curvature parameter $k$ and the black hole spin $a$.

\subsection{Physical Setup and Choice of Model Parameters}
\label{mot_inital_data} 

In BHL accretion, a compact object moves through a uniform gas or, equivalently, the gas streams past the object with asymptotic velocity $v_\infty$ and sound speed $c_s$. Gravity bends the trajectories of gas elements passing the black hole toward the axis behind it. Streams arriving from opposite sides collide in this downstream region, where they form a shock cone, and the shocked gas that has lost enough momentum falls into the black hole~\cite{Edgar2004}. The size of the region in which gas is captured is set by the accretion radius
\begin{equation}
R_{\rm acc}=\frac{2GM}{v_\infty^{2}+c_s^{2}},
\label{Racc}
\end{equation}
inside which the gravitational potential energy exceeds the kinetic and thermal energy of the gas.  For the asymptotic flow used in all simulations, $v_\infty=0.2c$ and $c_s=0.1c$, the incoming gas is supersonic, with Mach number $v_\infty/c_s=2$, and the accretion radius is $R_{\rm acc}\simeq40M$. 

The spacetime enters through the effective curvature parameter $k$ of Eq.~(\ref{metric}). For the models considered here, $\beta=1M$ and $|\beta\gamma|\leq10^{-2}$, so that $k\simeq\kappa+\gamma^{2}/4$ is set mainly by $\kappa$, with $\gamma$ adding only a small positive correction. The effect of $k$ is clearest in the temporal metric component on the equatorial plane,
\begin{equation}
g_{tt}=-\left(1-\frac{2\widetilde{M}}{r}-kr^{2}\right),
\label{gtt_eq}
\end{equation}
which does not depend on the spin $a$.

\begin{figure*}[!ht]
  \vspace{1cm}
  \center
  \psfig{file=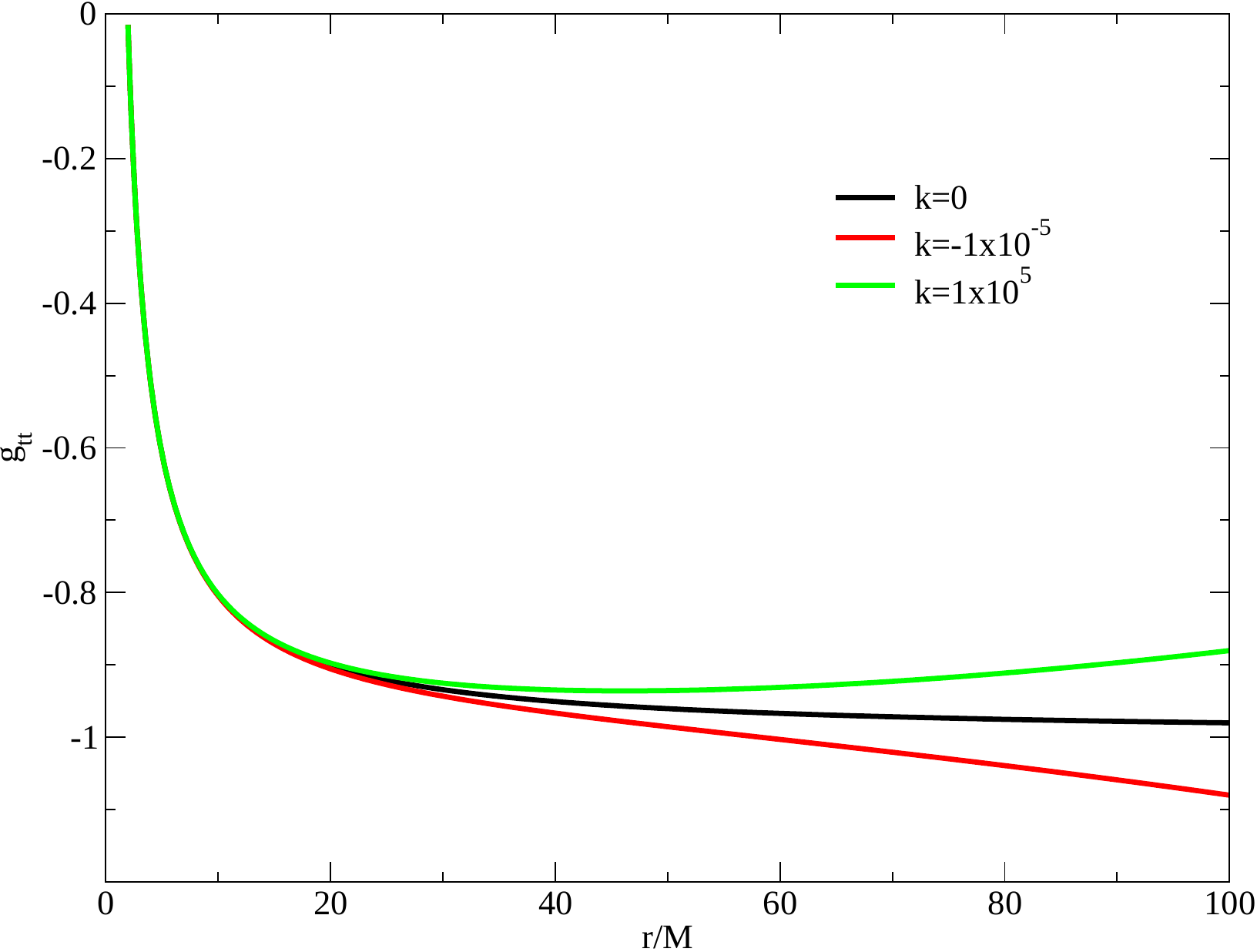,width=12.0cm,height=10cm}
     \caption{Temporal metric component $g_{tt}$ of the Kerr--Weyl spacetime on the equatorial plane as a function of the radial coordinate $r$, for $k=0$ (Kerr), $k=-10^{-5}\,M^{-2}$ (anti-de Sitter-like) and $k=+10^{-5}\,M^{-2}$ (de Sitter-like). The curves coincide near the black hole and separate at large radii, where the curvature term $kr^{2}$ becomes comparable to, and then larger than, the Newtonian term $2\widetilde{M}/r$.}
\vspace{1cm}
\label{gtt_r}
\end{figure*}

Fig.~\ref{gtt_r} shows $g_{tt}$ as a function of $r$ for $k=0$ (Kerr) and for $k=\pm10^{-5}\,M^{-2}$, a magnitude typical of our most strongly deviating models. Close to the black hole, the three curves coincide, because the Newtonian term $2\widetilde{M}/r$ dominates there. With increasing radius, the curvature term $kr^{2}$ grows and the curves separate. For $k<0$, the spacetime is anti-de Sitter-like: $g_{tt}$ becomes more negative, the potential well is deeper, and the additional attraction grows with distance. For $k>0$, the spacetime is de Sitter-like: $g_{tt}$ is less negative, and the additional repulsion weakens the inflow. At $r=100M$, for example, Eq.~(\ref{gtt_eq}) gives $g_{tt}\simeq-1.08$ for $k=-10^{-5}\,M^{-2}$ and $g_{tt}\simeq-0.88$ for $k=+10^{-5}\,M^{-2}$, compared with $g_{tt}\simeq-0.98$ in the Kerr case. The curvature term equals the Newtonian term at
\begin{equation}
r_k\simeq\left(\frac{2\widetilde{M}}{|k|}\right)^{1/3},
\label{rk}
\end{equation}
which is about $58M$ for $|k|=10^{-5}\,M^{-2}$ (taking $\widetilde{M}\simeq M$). This radius is comparable to $R_{\rm acc}$, so the curvature term acts on the very region where gas is captured and the shock cone forms. This is why values of $|k|$ as small as $10^{-6}$--$10^{-5}\,M^{-2}$ can change the flow so markedly. It is also why the formation, stability and possible destruction of the cone are controlled by the behavior of $k$ at large radii rather than near the horizon. For the same reason, no single critical value of $k$ separates the different flow regimes. What matters is the ratio $r_k/R_{\rm acc}$, which depends on the hydrodynamic parameters as well as on $k$. For $k\to0$, $r_k\to\infty$ and the Kerr solution is recovered.

The parameter sets used in the simulations are listed in Table~\ref{Inital_Con_1}. They fall into three groups, complemented by a pure Kerr run ($k=0$) at each spin as a reference. The aggressive models $A_1$--$A_5$ have $k$ between about $-3\times10^{-6}$ and $-3\times10^{-5}\,M^{-2}$, corresponding to a strongly AdS-like asymptotic structure. They are arranged in pairs of similar $|k|$ but different spin, which allows the effects of curvature and rotation to be separated. $A_1$ ($a=0.9M$) and $A_5$ ($a=0.5M$) both have $|k|\approx(2.4$--$3.0)\times10^{-5}\,M^{-2}$. $A_2$, $A_3$ ($a=0.9M$) and $A_4$ ($a=0.5M$) have $|k|\approx(3$--$5)\times10^{-6}\,M^{-2}$. $A_2$ and $A_3$ also share the same $\kappa$ but differ in the sign and magnitude of $\gamma$, which tests the sensitivity of the flow to $\gamma$ at almost fixed $k$. The mild-deviation models $B_1$--$B_3$ have $-5\times10^{-7}\lesssim k\lesssim-7\times10^{-9}\,M^{-2}$, so the geometry is close to Kerr while still containing a small Weyl-gravity correction. The near-Kerr baseline $C$ has a small positive value, $k=5\times10^{-10}\,M^{-2}$, and reproduces the Kerr solution within numerical accuracy.

With Eq.~(\ref{rk}), these parameter sets give $r_k\approx41$--$43M\approx R_{\rm acc}$ in $A_1$ and $A_5$, $r_k\approx75$--$87M\approx2R_{\rm acc}$ in $A_2$--$A_4$, and $r_k\gtrsim160M\approx4R_{\rm acc}$ in the $B$ and $C$ models. As the results below show, this ordering matches the three flow regimes found in the simulations. In the $B$, $C$ and Kerr models, a stable, standing shock cone forms in the downstream region. In $A_2$--$A_4$, the shock cone develops a widened opening angle and undergoes persistent flip-flop oscillations, an instability known from both Newtonian and relativistic BHL simulations~\cite{Blondin2009,DonmezMNRAS2012}. In $A_1$ and $A_5$, the cone is disrupted and an oscillating, quasi-steady toroidal structure forms around the black hole.

\begin{table*}[htbp]
\footnotesize
\caption{
Parameters of the simulated models: the black hole spin $a$, the Weyl-gravity coefficients $\gamma$, $\kappa$ and $\beta$, and the resulting effective curvature parameter $k$. The hydrodynamic parameters $v_\infty=0.2c$, $c_s=0.1c$ and $\Gamma=4/3$ are the same for all models, and a pure Kerr run ($k=0$) is also performed for each spin.
}
 \label{Inital_Con_1}
\begin{center}
  \begin{tabular}{ccccccc}
    \hline
    \hline
    Model & Model Name & \hspace{0.5cm}$a (M)\hspace{0.5cm}$ & \hspace{0.5cm}$\gamma\left(1/M\right)$\hspace{0.5cm} & \hspace{0.5cm}$\kappa\left(1/M^{2}\right)$ \hspace{0.5cm}&  \hspace{0.5cm}$\beta (M)$\hspace{0.5cm}  &  $k\left(1/M^{2}\right)$ \\    
   \hline       
        & $A_1$ & $0.9$ & $3 \times 10^{-4}$ & $-3 \times 10^{-5}$ & $1$ & $-2.9977\times 10^{-5}$\\
         & $A_2$  & $0.9$ & $-5 \times 10^{-5}$ & $-3 \times 10^{-6}$ & $1$ & $-2.9994\times 10^{-6}$\\
   A (Aggressive) & $A_3$ & $0.9$ & $3 \times 10^{-5}$ & $-3 \times 10^{-6}$ & $1$ & $-2.999775\times 10^{-6}$\\
 & $A_4$ & $0.5$ & $1 \times 10^{-3}$ & $-5 \times  10^{-6}$ & $1$ &  $-4.749\times 10^{-6}$\\ 
 & $A_5$ & $0.5$ & $1 \times 10^{-2}$ & $-5 \times  10^{-5}$ & $1$ &  $-2.449\times 10^{-5}$\\  
    \hline   
     & $B_1$ & $0.9$ & $3 \times 10^{-7}$ & $-1 \times 10^{-8}$ & $1$ &  $-9.999977\times 10^{-9}$\\
    B (Mild deviations) & $B_2$ & $0.5$ & $1 \times 10^{-4}$ & $-5 \times  10^{-7}$ & $1$ &  $-4.974995\times 10^{-7}$\\  
     & $B_3$ & $0.5$ & $1 \times 10^{-6}$ & $-7 \times 10^{-9}$ & $1$ & $-6.99975\times 10^{-9}$ \\
    \hline
     C (Near-Kerr baseline) & $C$ & $0.9$ & $-7 \times 10^{-8}$ & $5 \times 10^{-10}$ & $1$ & $5\times 10^{-10}$\\    
    \hline
    \hline
  \end{tabular}
\end{center}\label{tab1}
\end{table*}

\subsection{Global Flow Evolution and Shock Cone Morphology}\label{Sec_morphology} 

The previous subsection showed that the curvature parameter $k$ leaves the near-horizon geometry practically unchanged and acts on the flow through the $kr^{2}$ term in $g_{tt}$, whose importance at the scale of gas capture is measured by the ratio $r_k/R_{\rm acc}$, with $r_k$ given by Eq.~(\ref{rk}). A more direct dynamical measure follows from the weak-field limit of Eq.~(\ref{gtt_eq}), which is appropriate at $r\sim R_{\rm acc}\gg M$. Writing $g_{tt}\simeq-(1+2\Phi)$ gives the effective potential $\Phi\simeq-\widetilde{M}/r-kr^{2}/2$ and the radial acceleration
\begin{equation}
g_r=-\frac{d\Phi}{dr}\simeq-\frac{\widetilde{M}}{r^{2}}+kr .
\label{g_r}
\end{equation}
For $k<0$, the curvature term adds an inward acceleration $|k|r$ that grows linearly with distance, while the Newtonian pull decreases as $r^{-2}$. The two are equal at $r=2^{-1/3}r_k\simeq0.8\,r_k$, i.e., at about $32M$ in $A_1$ and $34M$ in $A_5$, inside the accretion radius. At $r=R_{\rm acc}$, their ratio is
\begin{equation}
\eta\equiv\frac{|k|R_{\rm acc}^{3}}{\widetilde{M}}=2\left(\frac{R_{\rm acc}}{r_k}\right)^{3}.
\label{eta}
\end{equation}
Taking $\widetilde{M}\simeq M$ and $R_{\rm acc}\simeq40M$, we obtain $\eta\simeq1.9$ and $1.6$ for $A_1$ and $A_5$, $\eta\simeq0.19$--$0.30$ for $A_2$--$A_4$, $\eta\simeq0.03$ for $B_2$, and $\eta<10^{-3}$ for $B_1$, $B_3$ and $C$. Hence, at the accretion radius, the curvature-induced pull dominates the Newtonian one in the first group, is a $20$--$30\%$ correction in the second, and is negligible in the remaining models.

The spin, by contrast, acts mainly close to the black hole. On the equatorial plane, the metric gives $g_{t\phi}=-a\,(2\widetilde{M}/r+kr^{2})$, and the frame-dragging angular velocity $\omega=-g_{t\phi}/g_{\phi\phi}$ decreases approximately as $2\widetilde{M}a/r^{3}$. The $k$-dependent part of $g_{t\phi}$ contributes at most $|k|a\lesssim3\times10^{-5}\,M^{-1}$ to $\omega$ and is negligible. For $a=0.9M$, $\omega$ is about $12\%$ of the Keplerian angular velocity $\Omega_{\rm K}=(\widetilde{M}/r^{3})^{1/2}$ at $r=6M$, but less than $1\%$ at $r=R_{\rm acc}$ (the orbital and precession frequencies of test particles in this spacetime are given in Ref.~\cite{ZULQARNAIN2026117449}). Spin and curvature therefore act on well-separated scales: $a$ shapes the inner flow within a few gravitational radii of the horizon, whereas $k$ shapes the region $r\sim R_{\rm acc}$ in which the shock cone forms. Since the hydrodynamic parameters are identical in all runs, and the Kerr runs with the same spins produce a stable cone, the differences between the morphologies described below can be attributed to $k$. The density maps of the $B$, $C$ and Kerr models are not shown, because they are nearly indistinguishable from one another. In all of them, a standing shock cone forms and remains stable, in agreement with earlier GRHD simulations of uniform BHL flows onto Kerr black holes~\cite{Font1999}.

\begin{figure*}[!ht]
  \vspace{1cm}
  \center
  \psfig{file=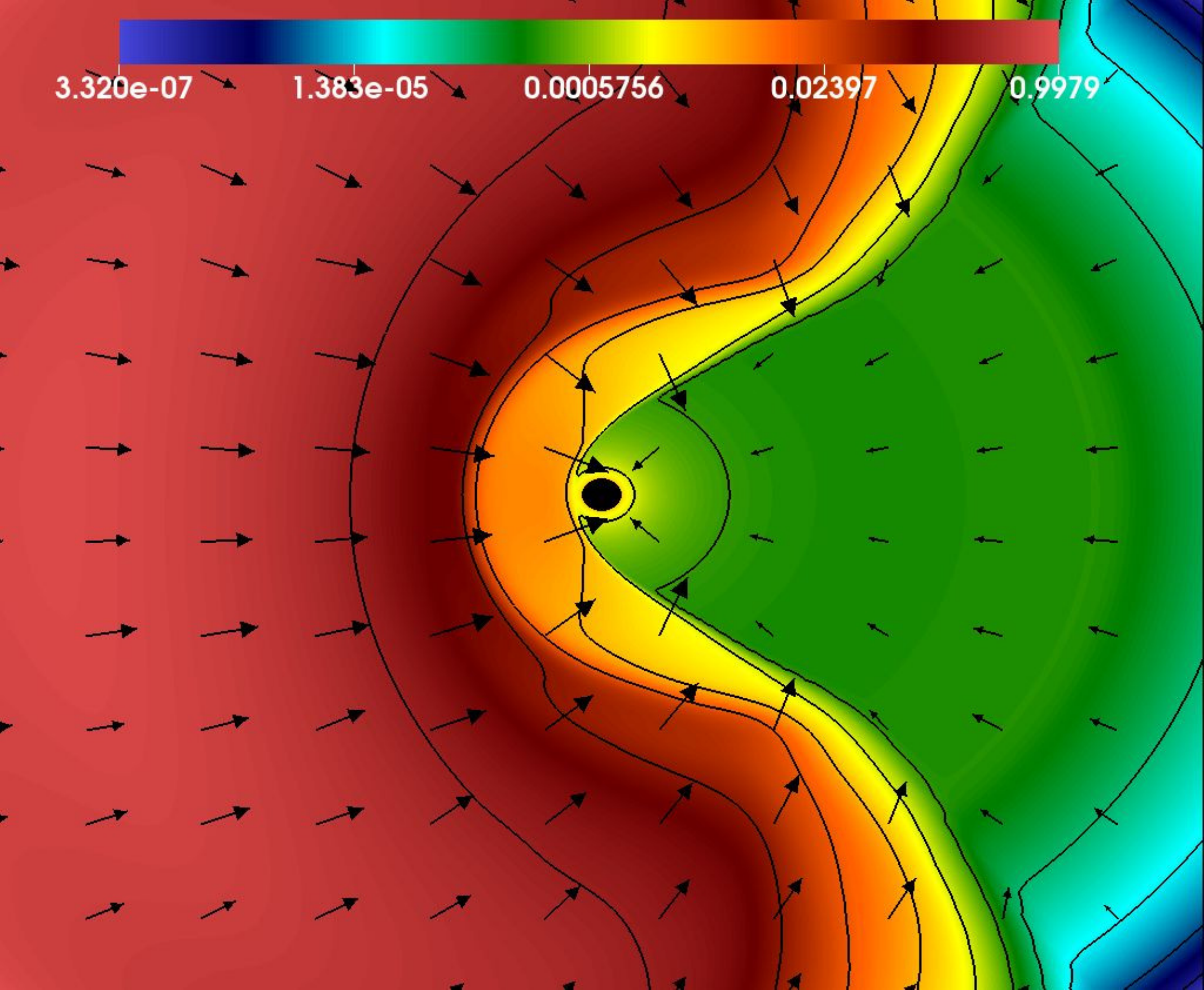,width=3.3cm,height=3.3cm} 
  \psfig{file=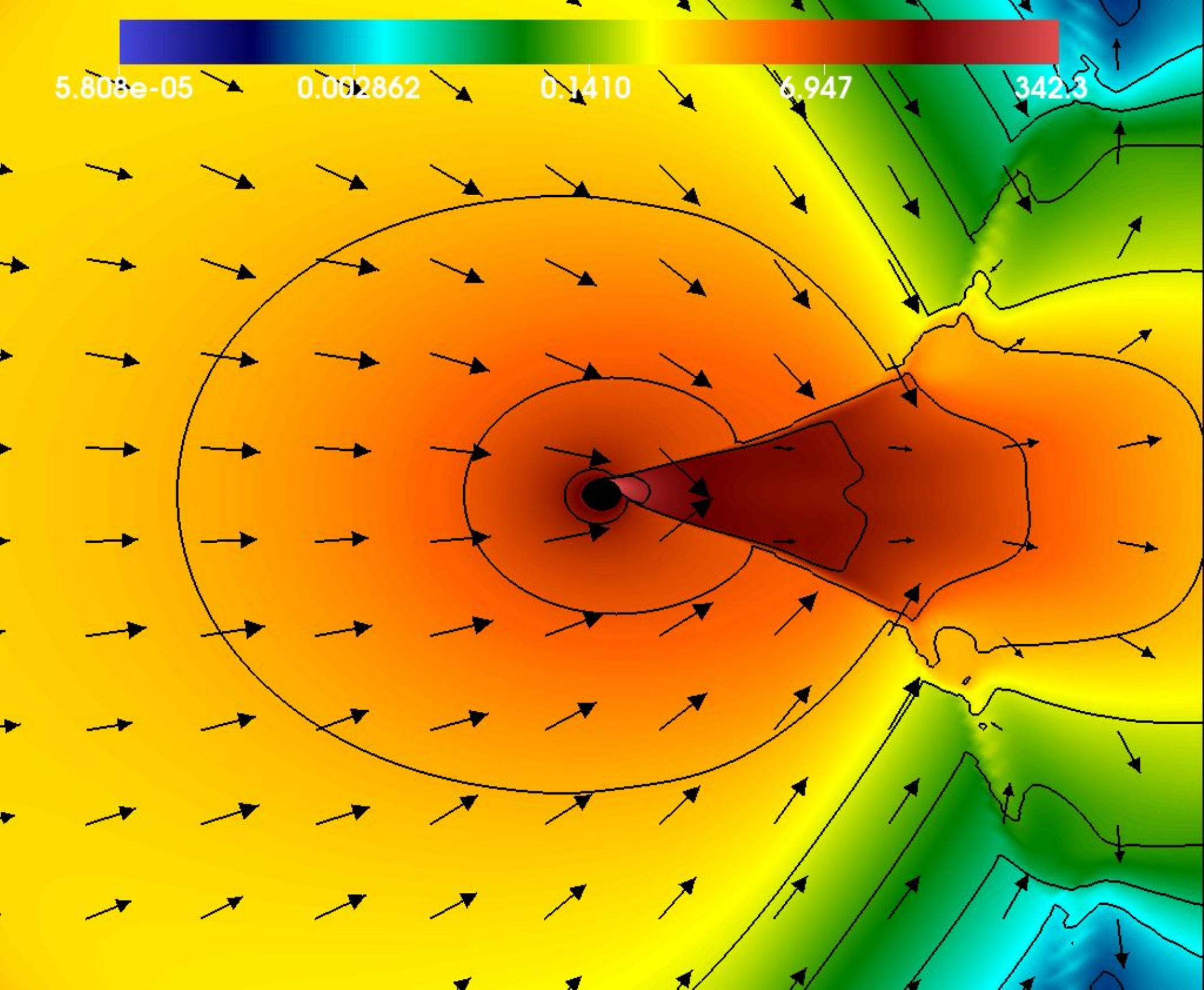,width=3.3cm,height=3.3cm}  
  \psfig{file=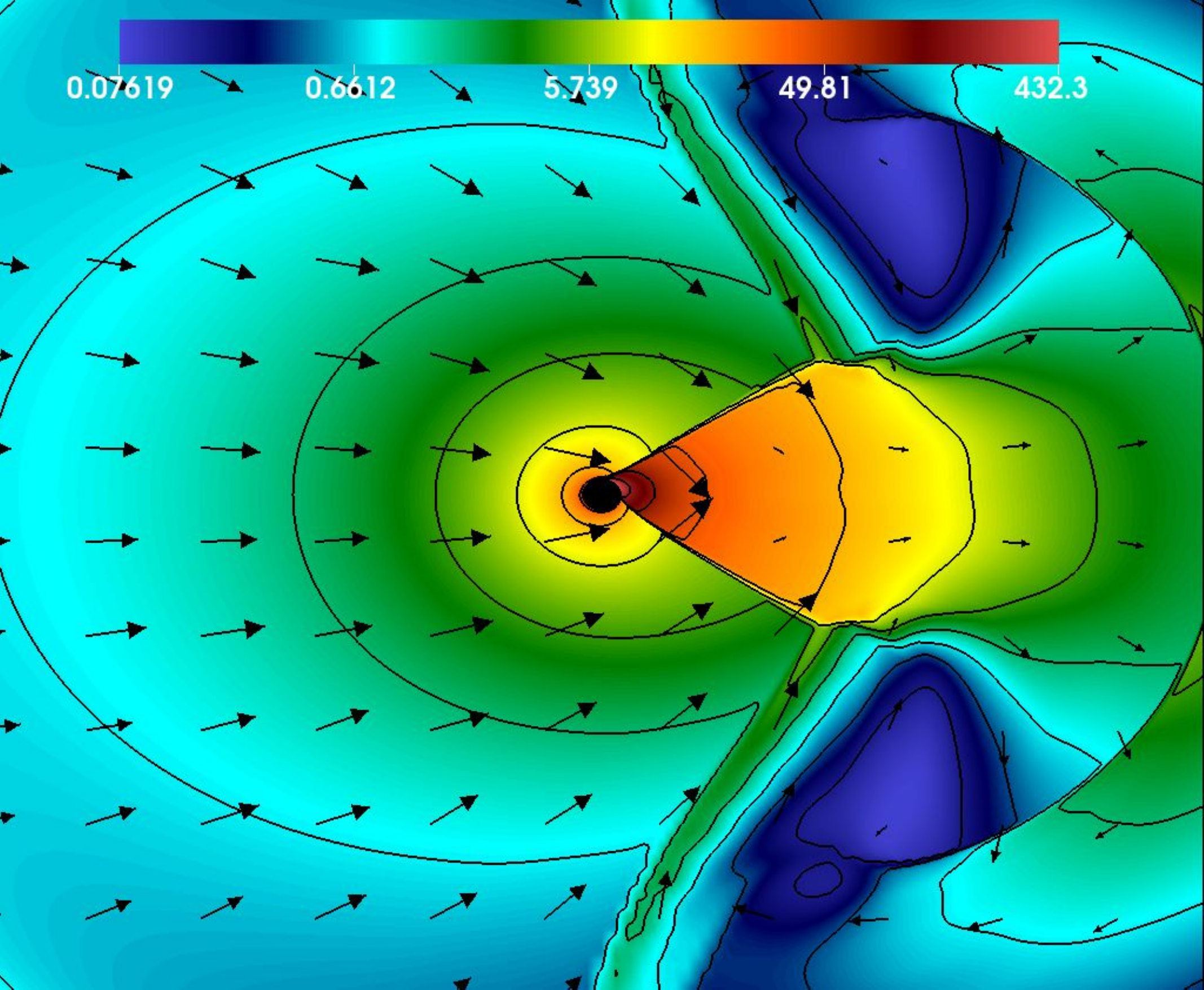,width=3.3cm,height=3.3cm} 
  \psfig{file=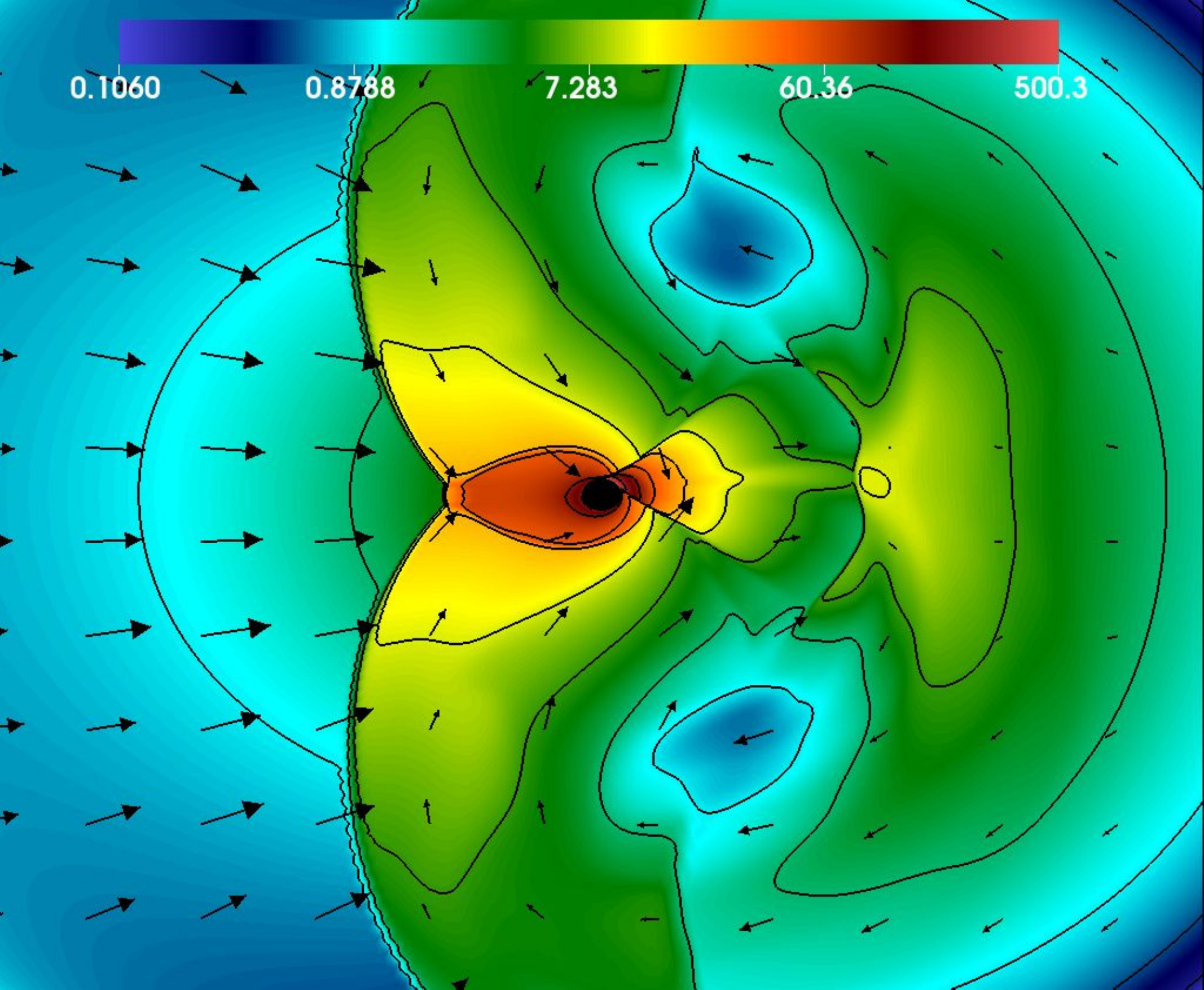,width=3.3cm,height=3.3cm}
  \psfig{file=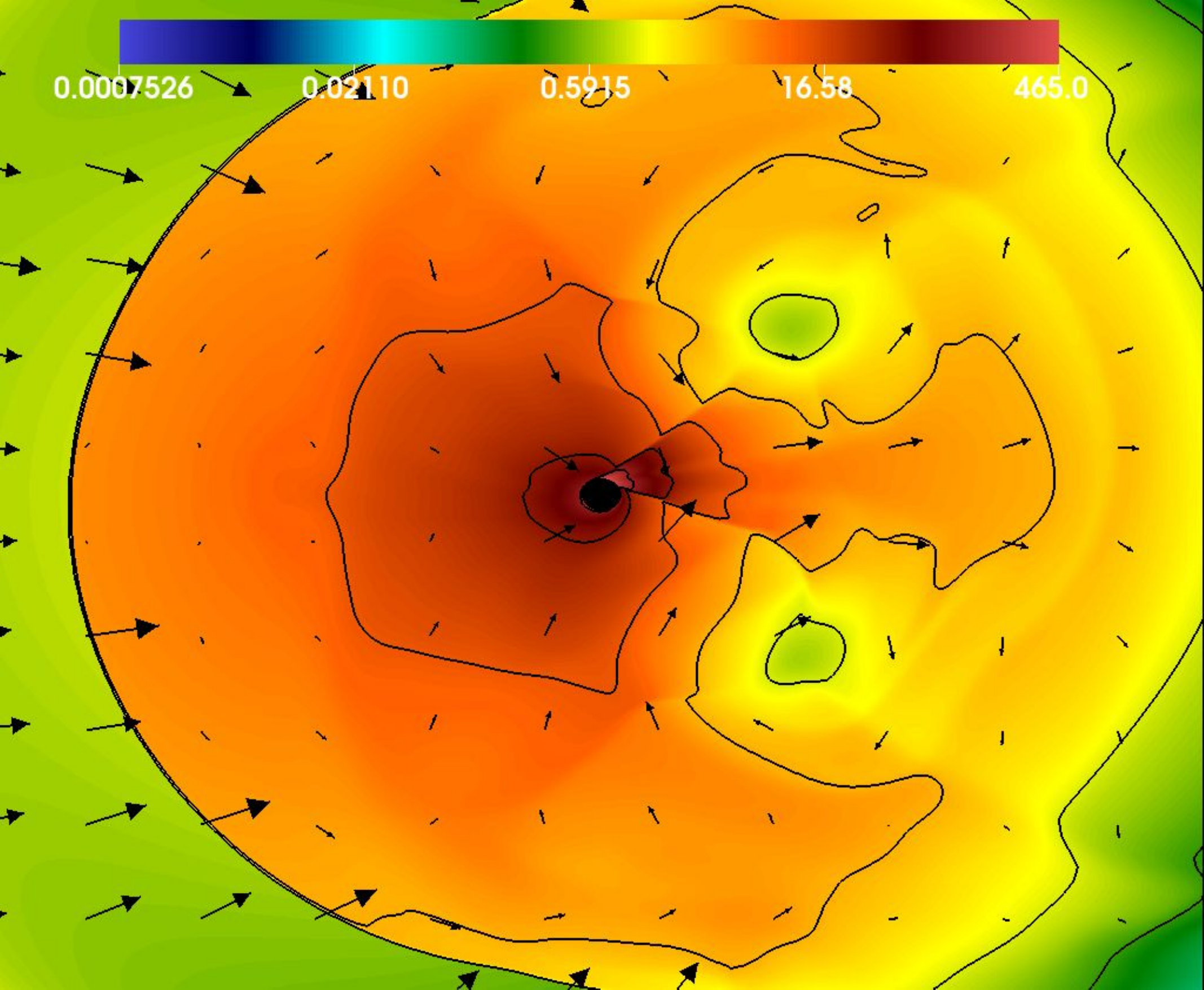,width=3.3cm,height=3.3cm}\\
  \psfig{file=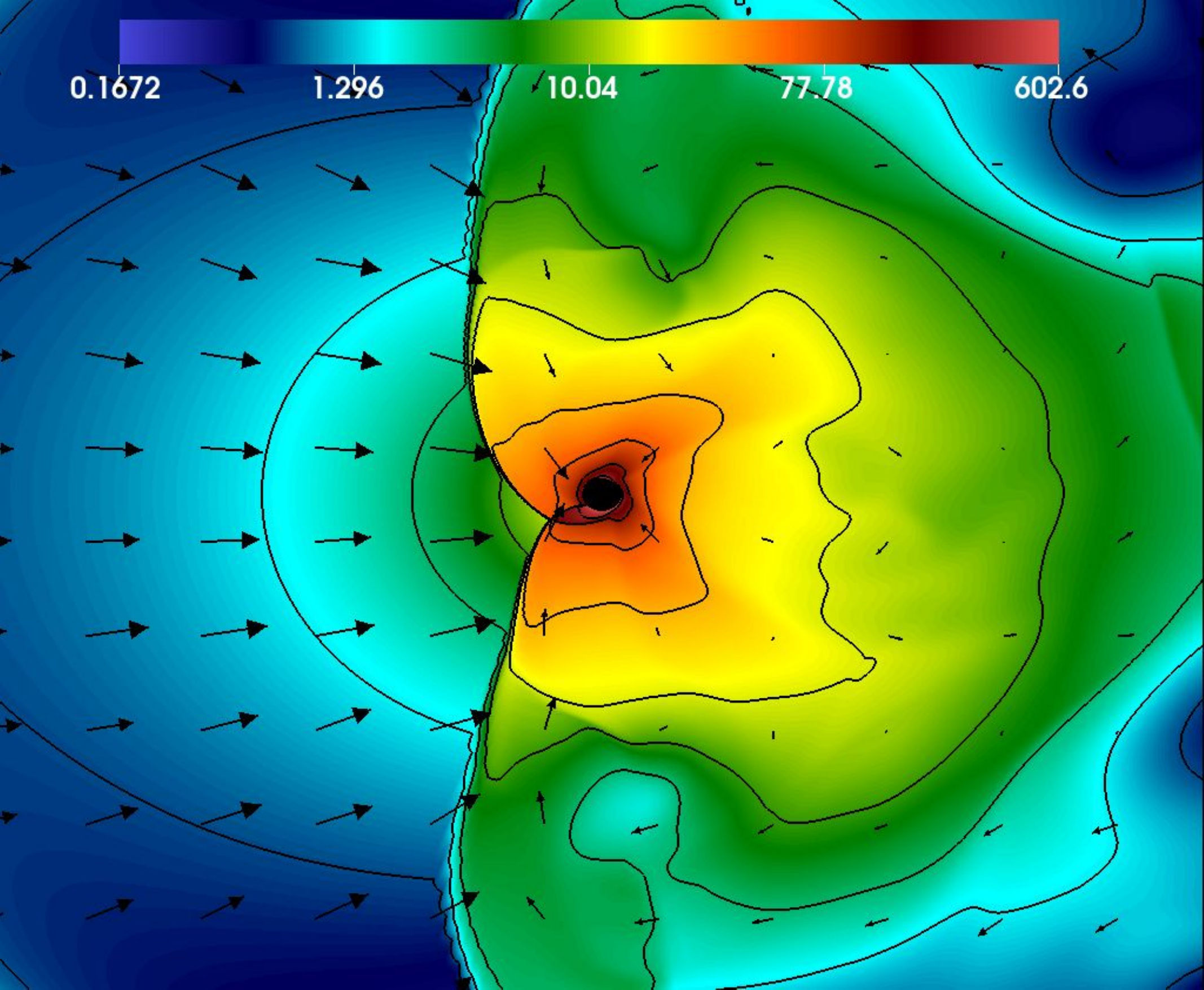,width=3.3cm,height=3.3cm}
  \psfig{file=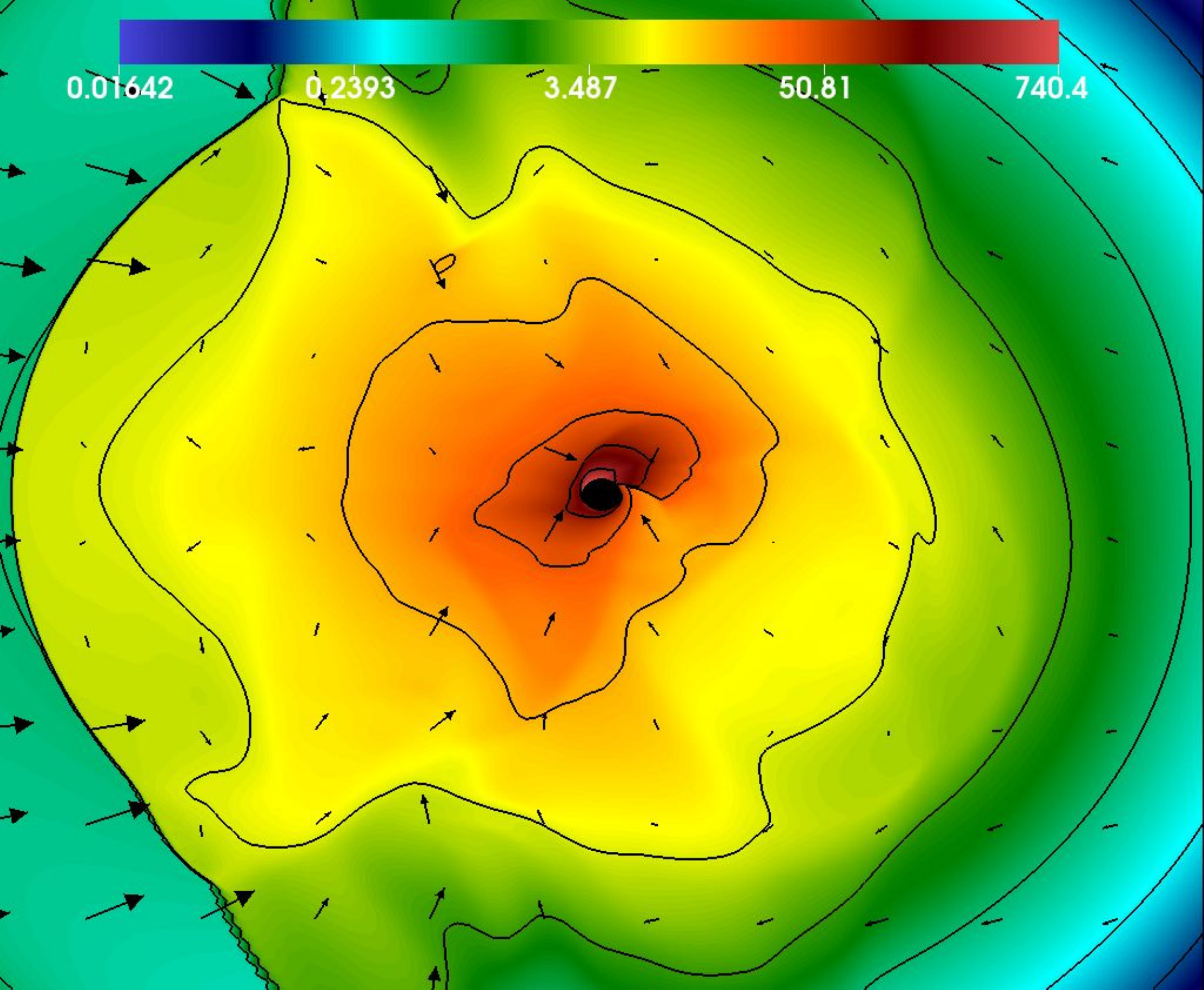,width=3.3cm,height=3.3cm}  
  \psfig{file=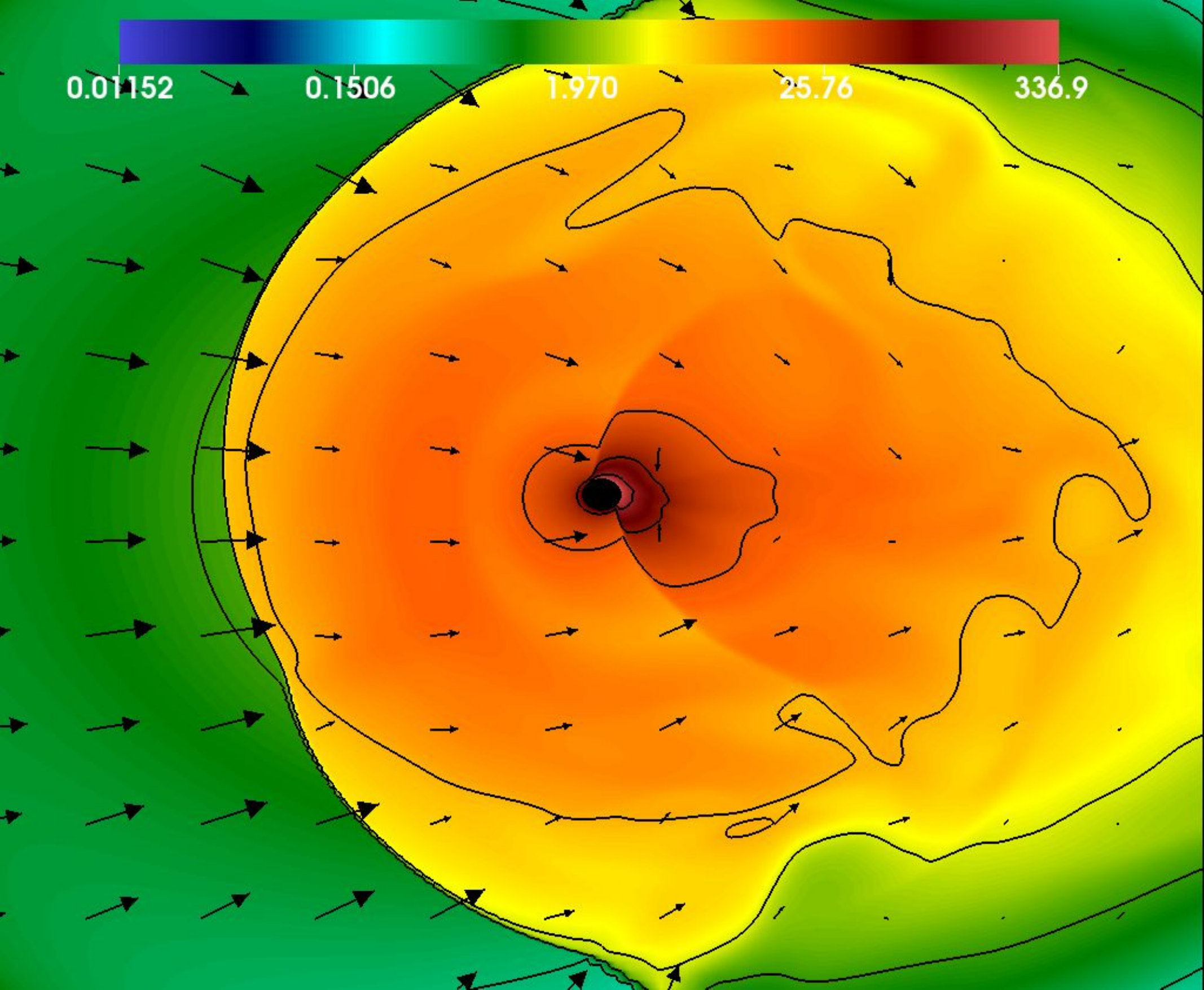,width=3.3cm,height=3.3cm} 
  \psfig{file=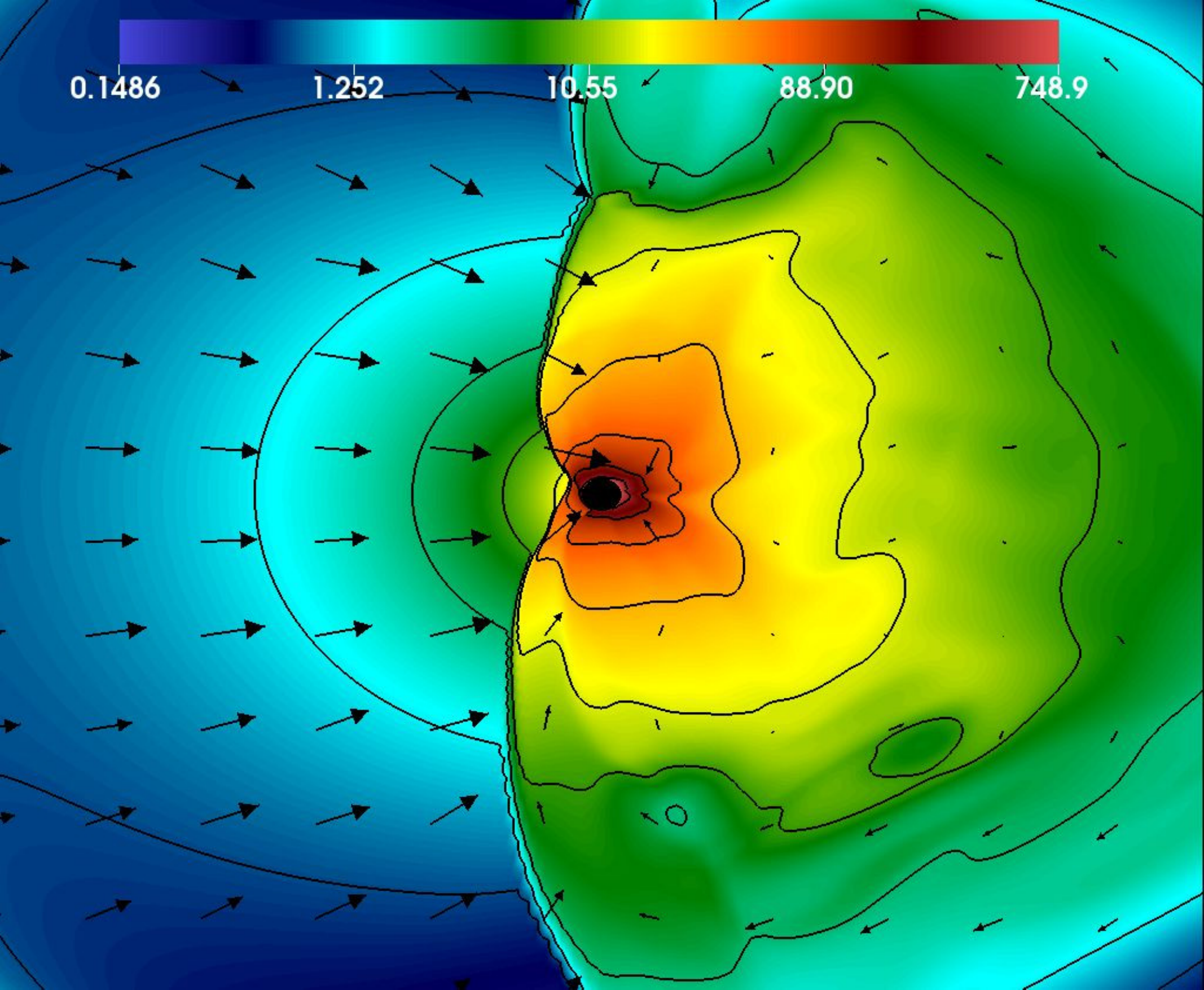,width=3.3cm,height=3.3cm}
  \psfig{file=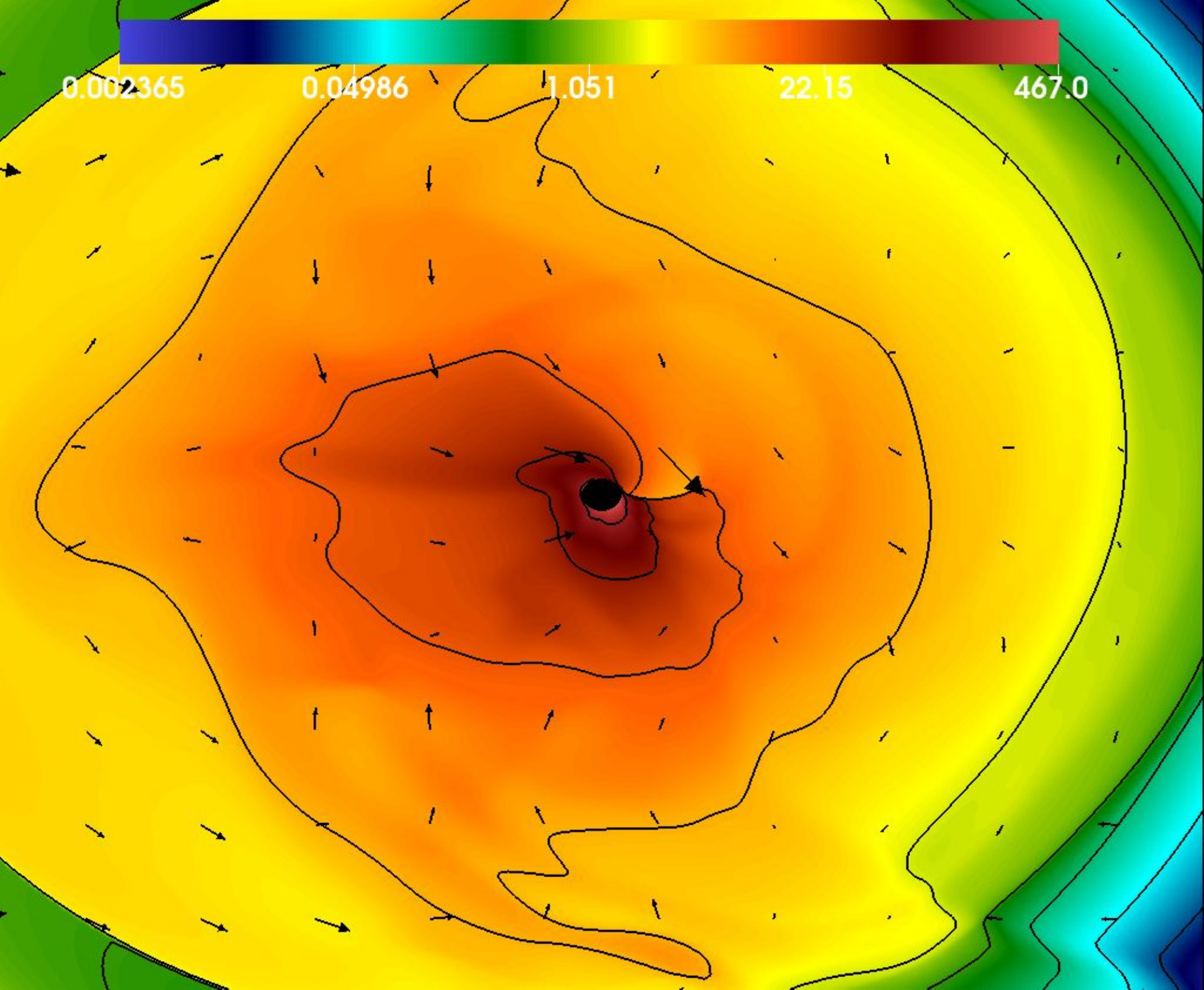,width=3.3cm,height=3.3cm}\\
  \psfig{file=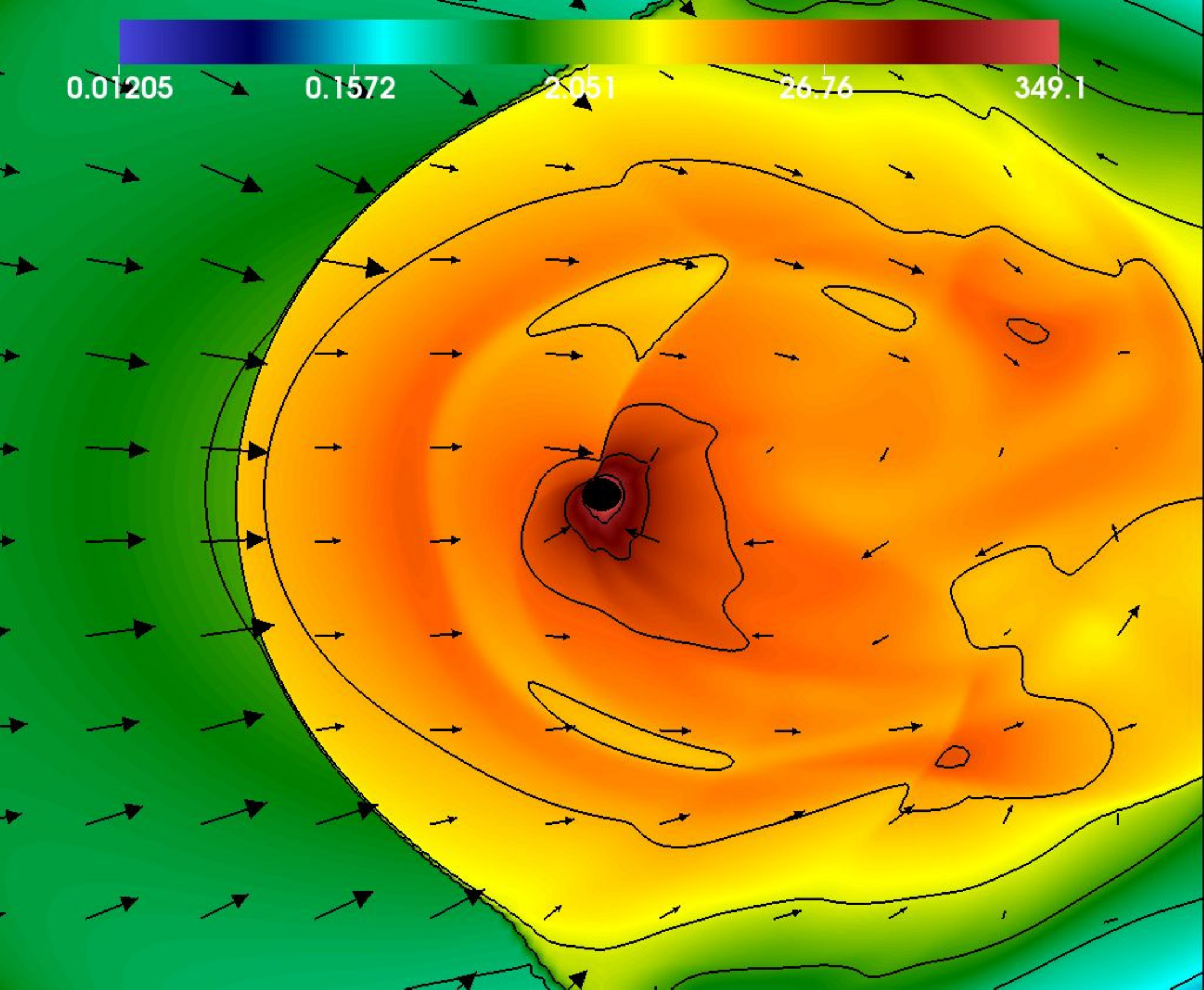,width=3.3cm,height=3.3cm}
  \psfig{file=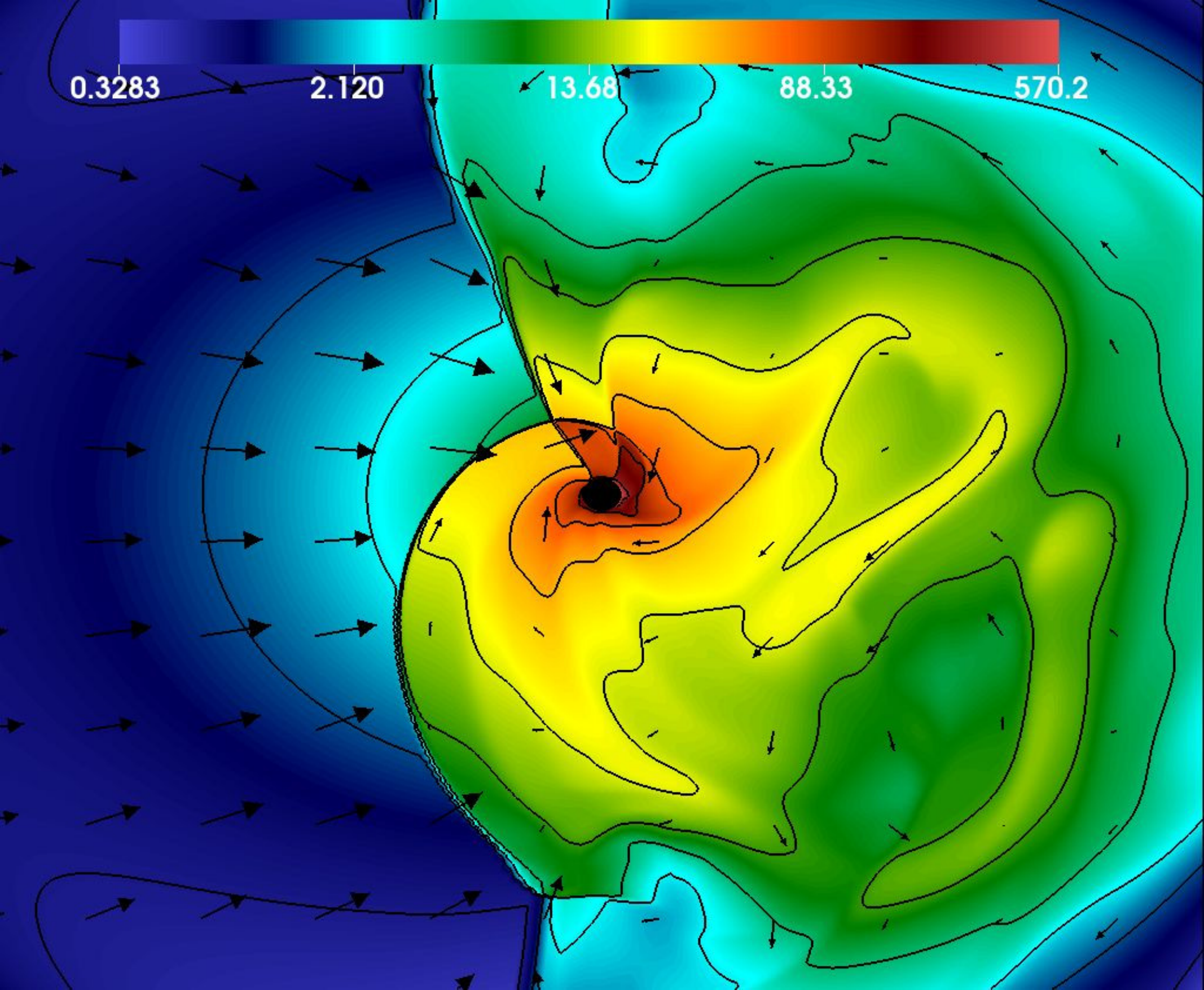,width=3.3cm,height=3.3cm}  
  \psfig{file=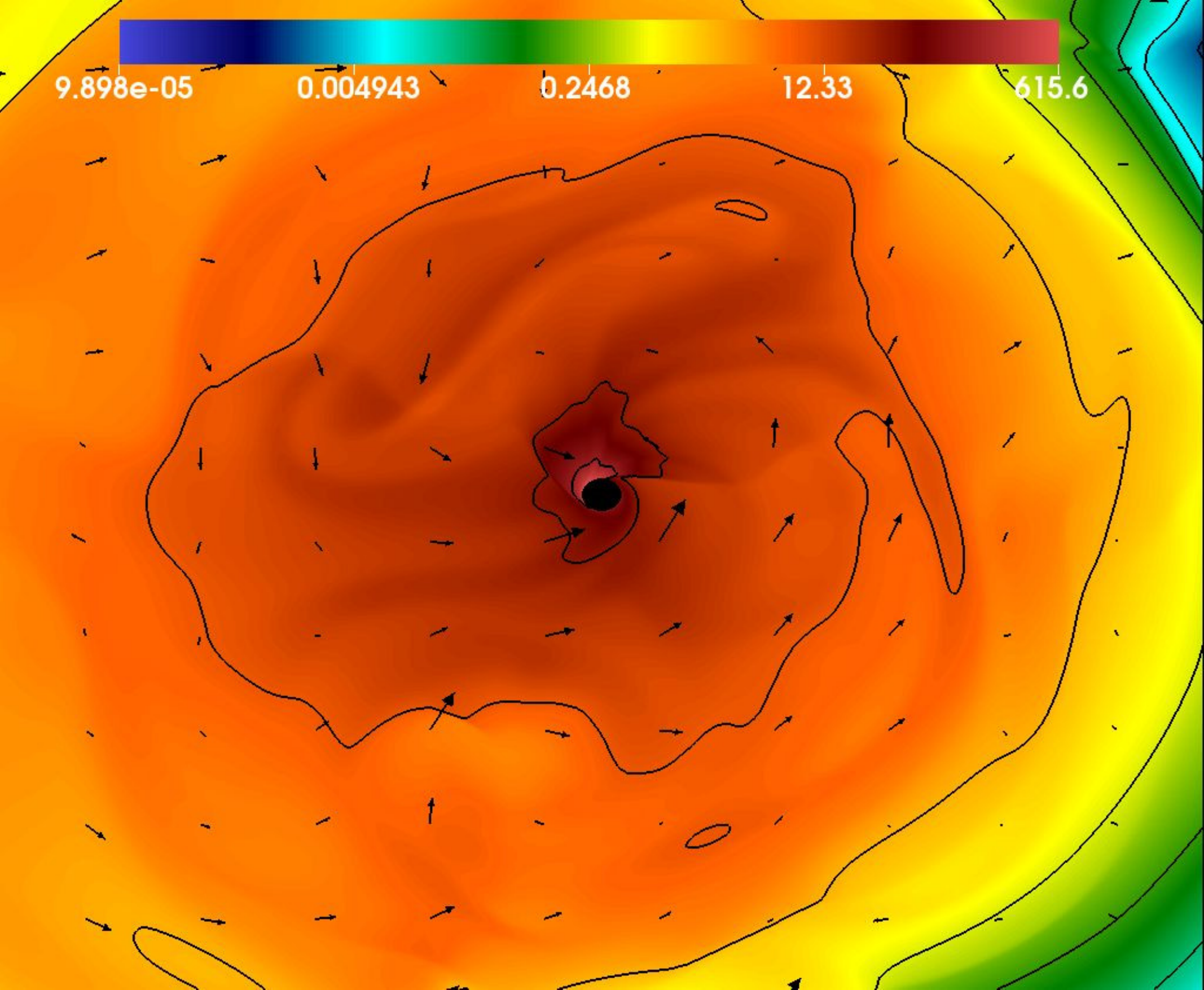,width=3.3cm,height=3.3cm} 
  \psfig{file=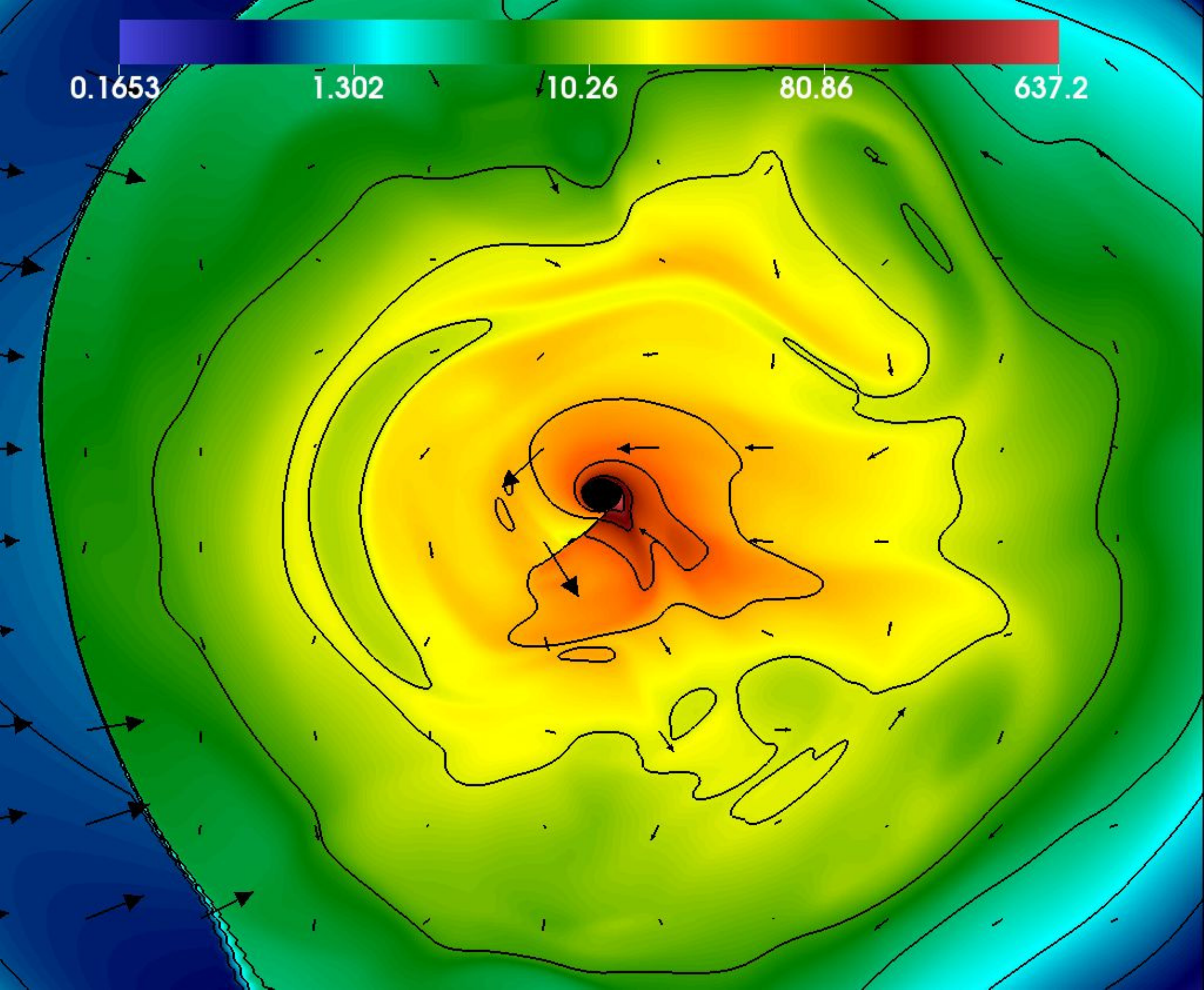,width=3.3cm,height=3.3cm}
  \psfig{file=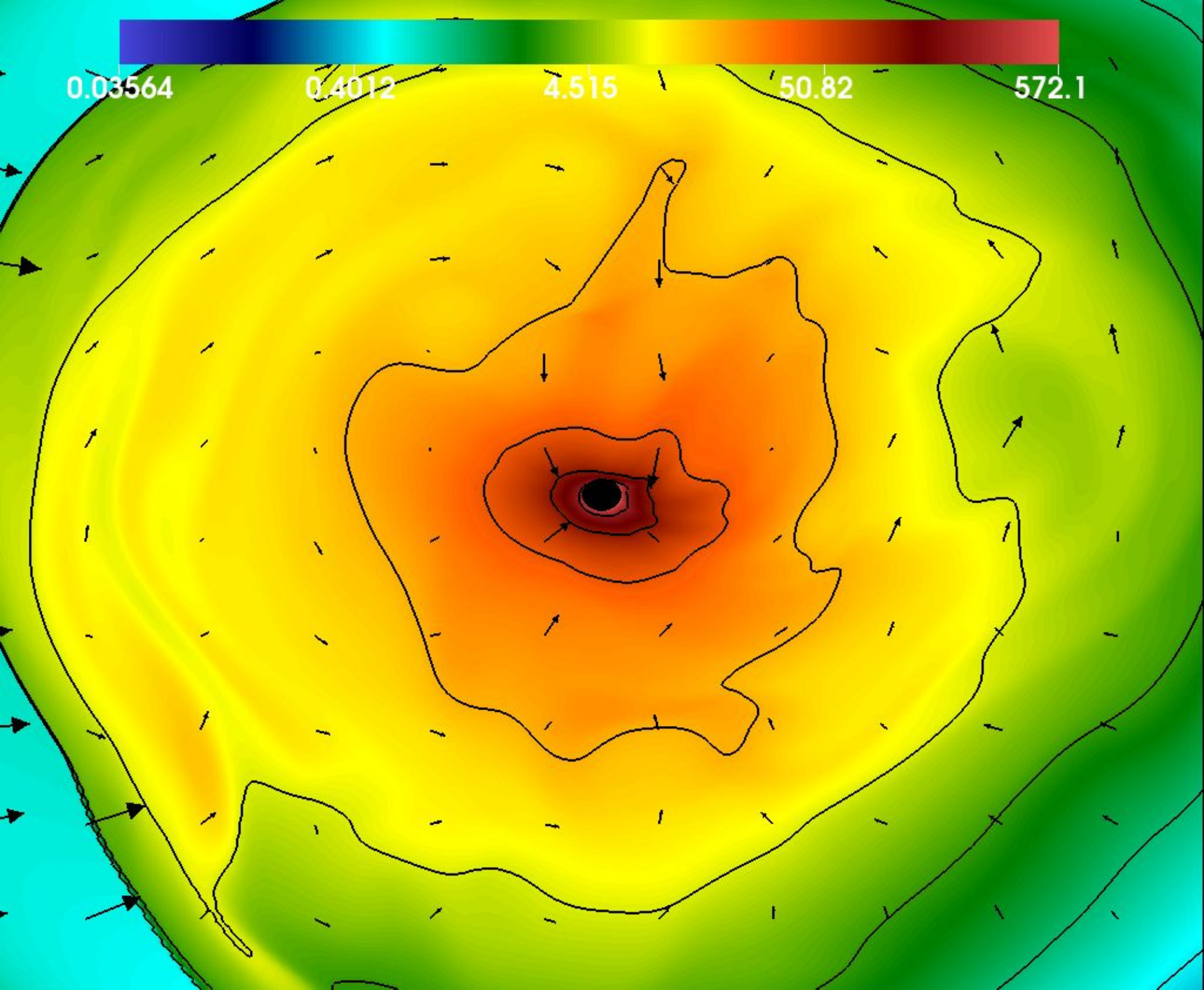,width=3.3cm,height=3.3cm}\\
  \psfig{file=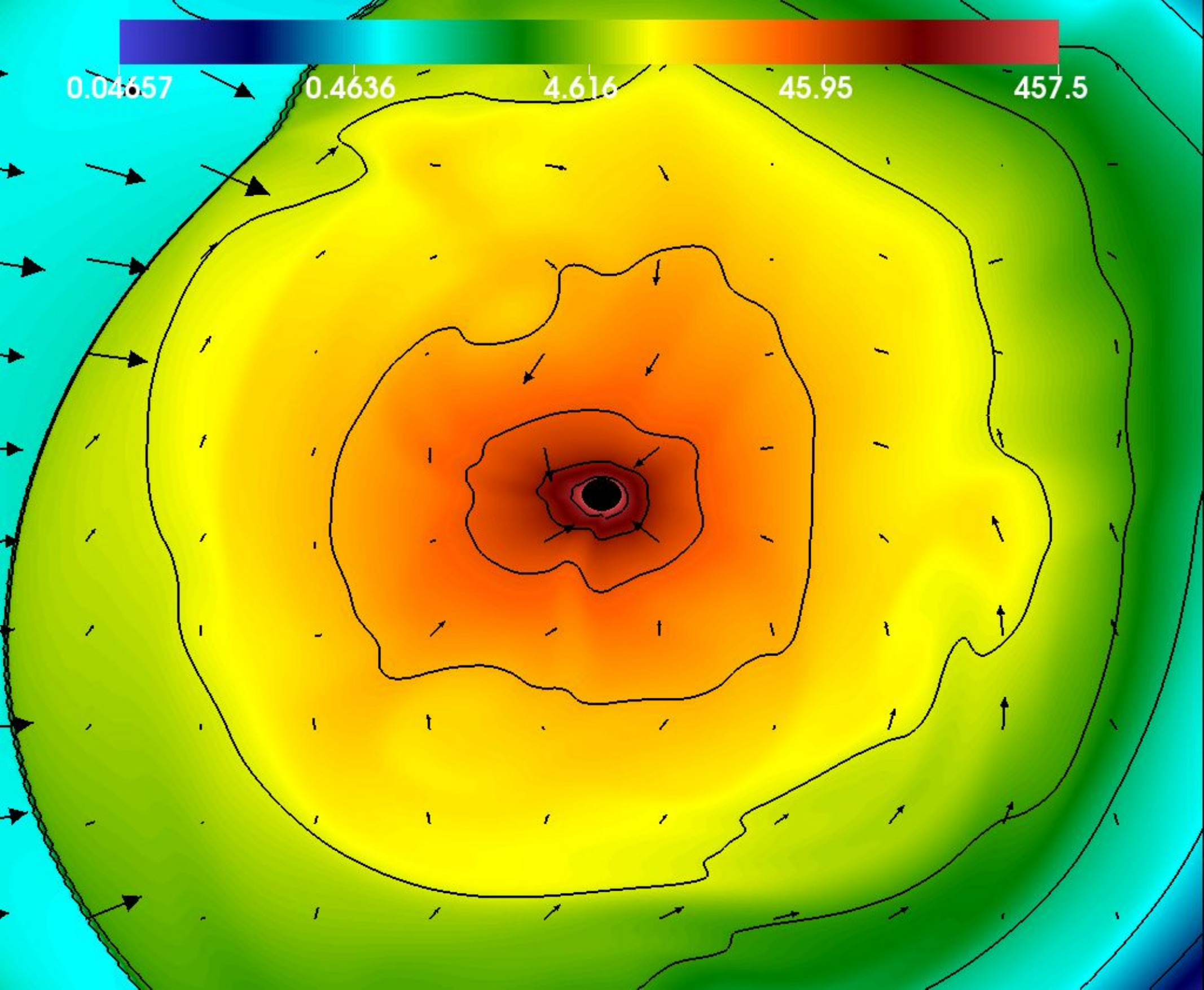,width=3.3cm,height=3.3cm}
  \psfig{file=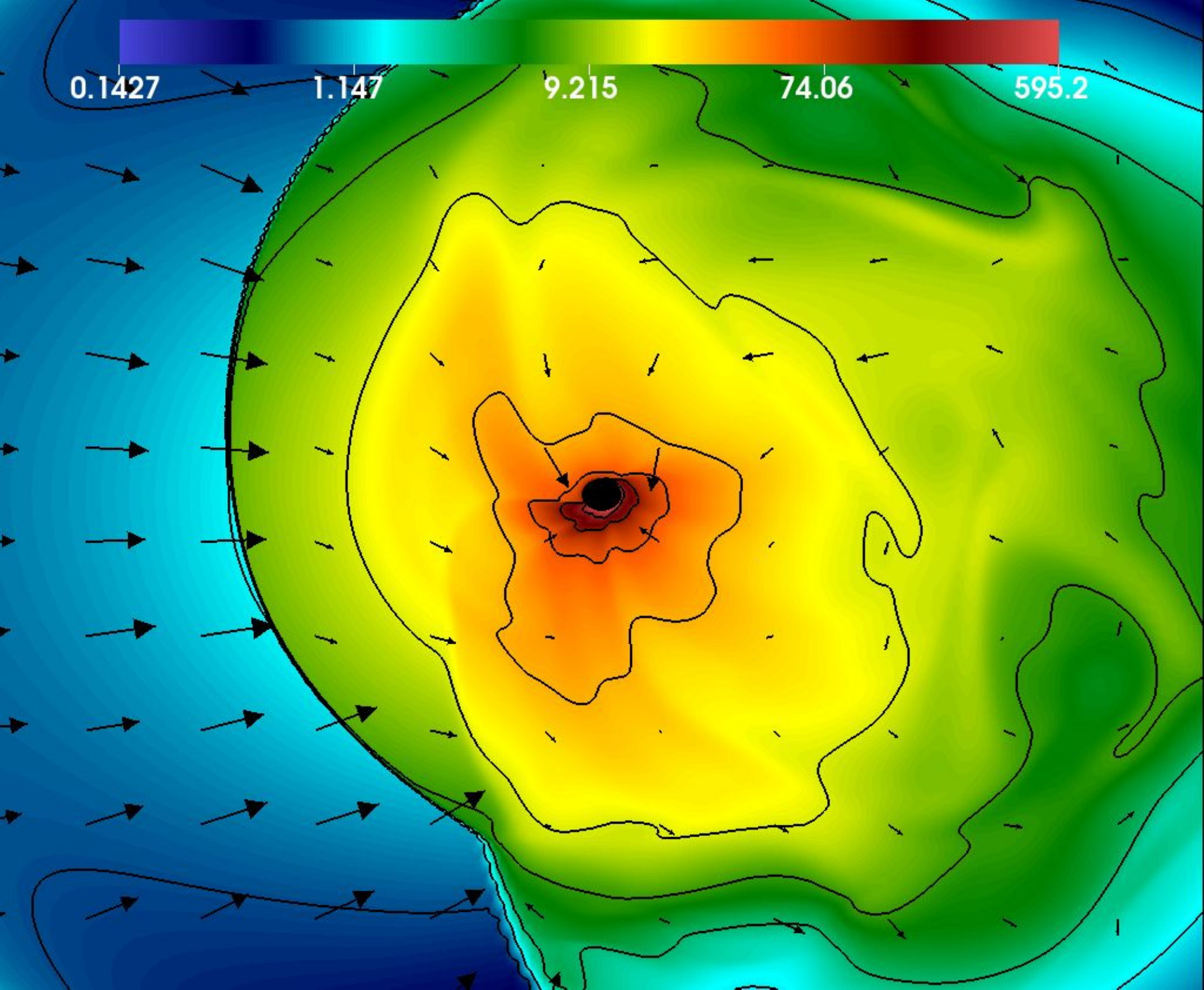,width=3.3cm,height=3.3cm}  
  \psfig{file=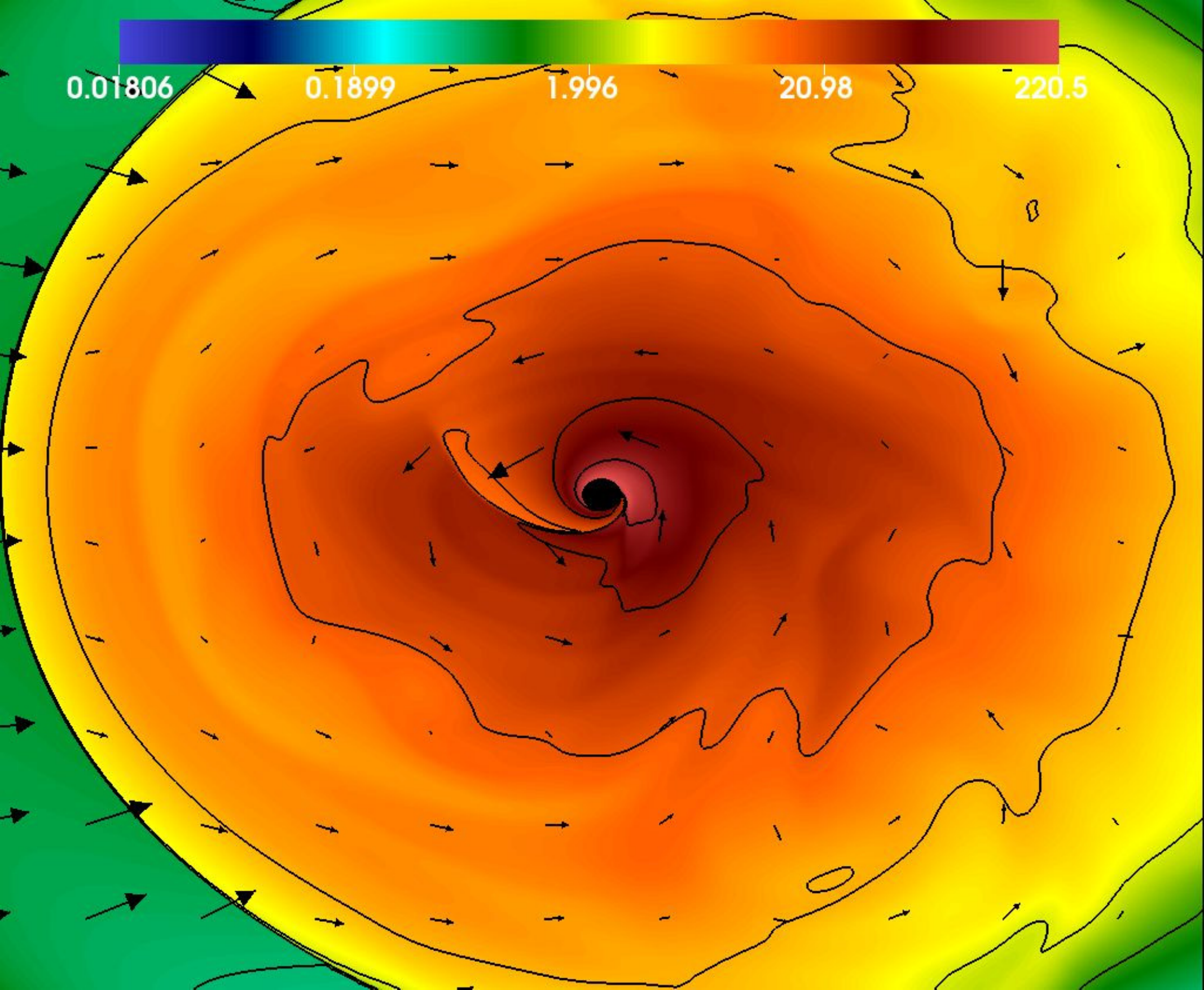,width=3.3cm,height=3.3cm} 
  \psfig{file=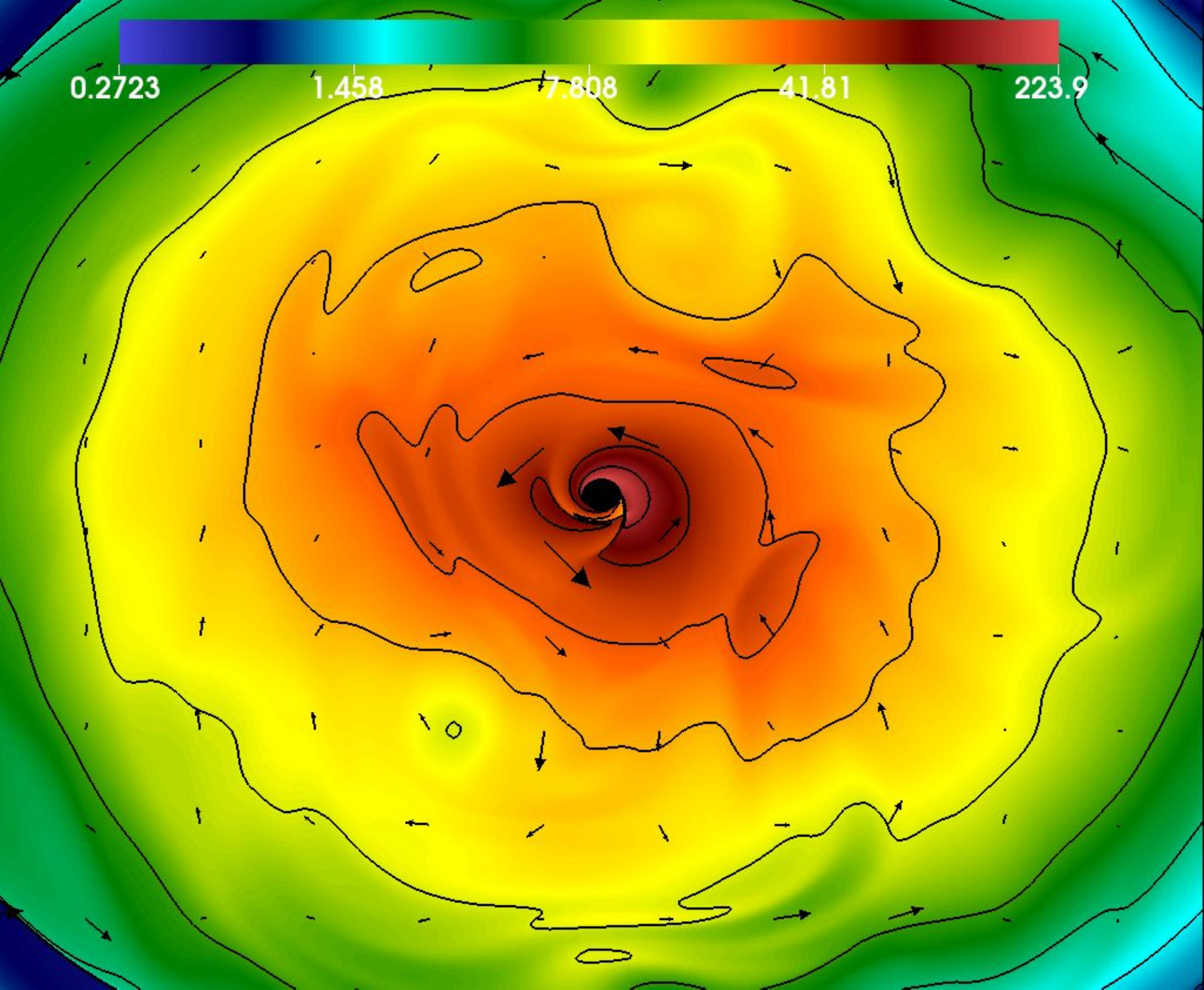,width=3.3cm,height=3.3cm}
  \psfig{file=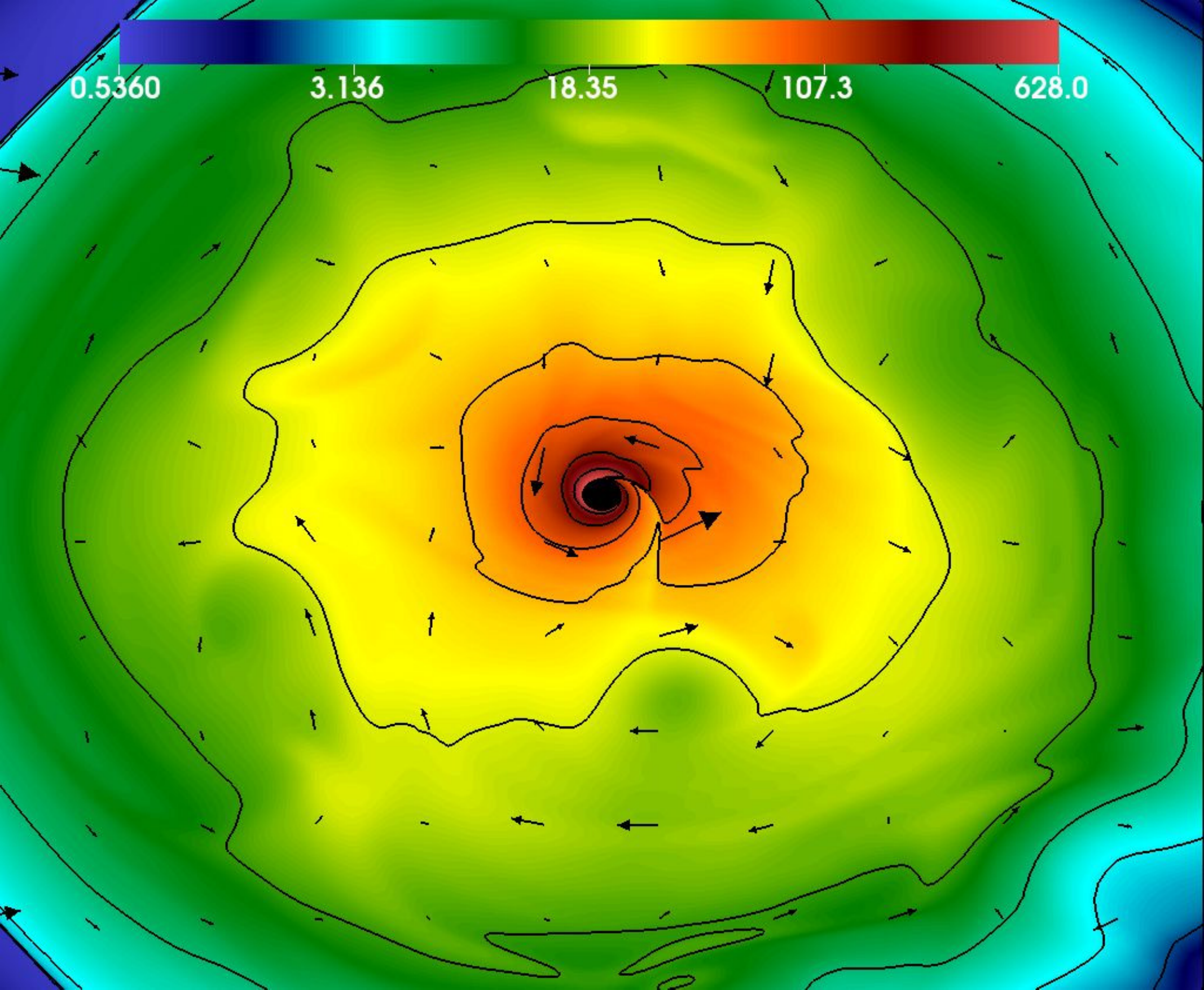,width=3.3cm,height=3.3cm}\\  
  \psfig{file=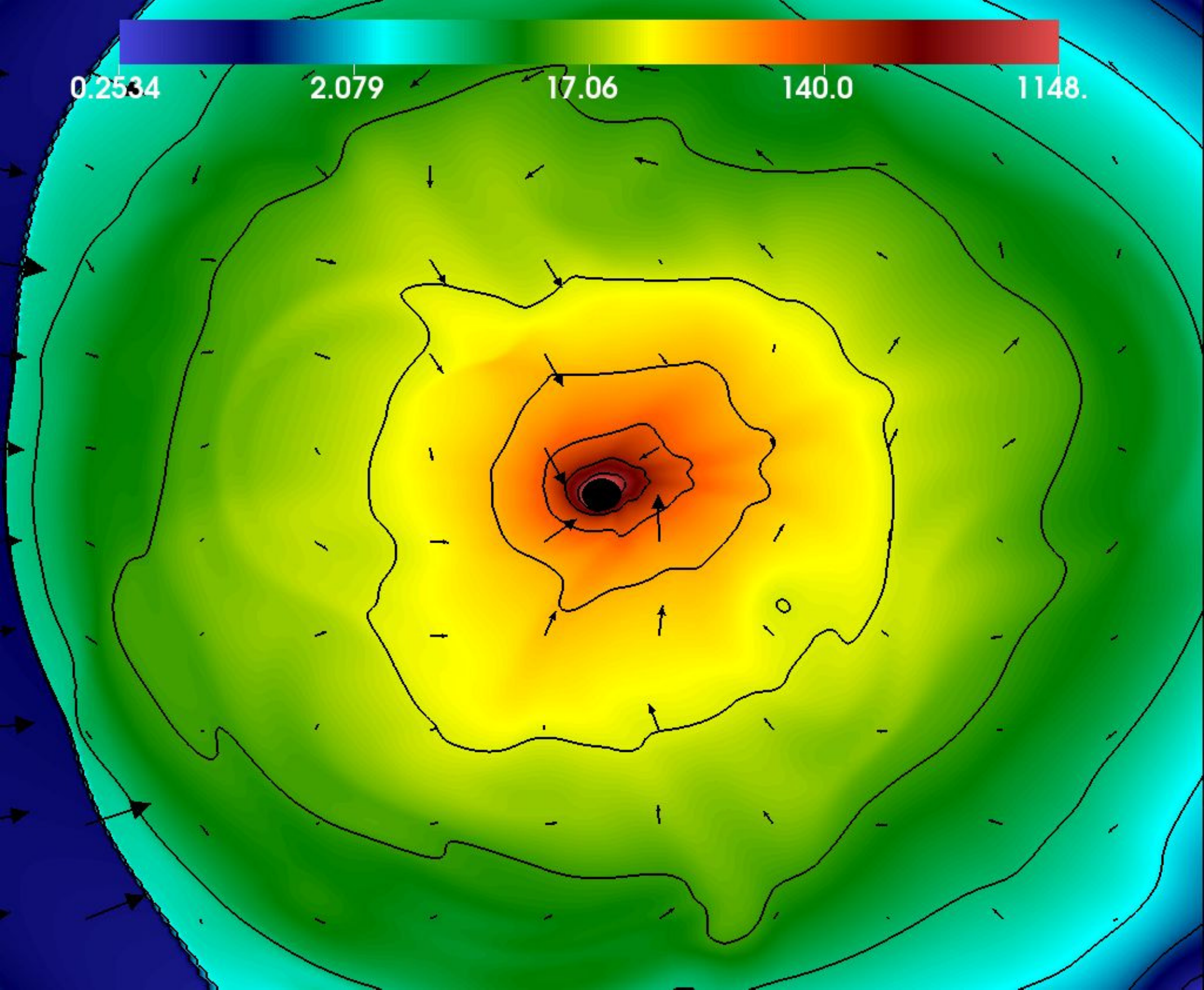,width=3.3cm,height=3.3cm}
  \psfig{file=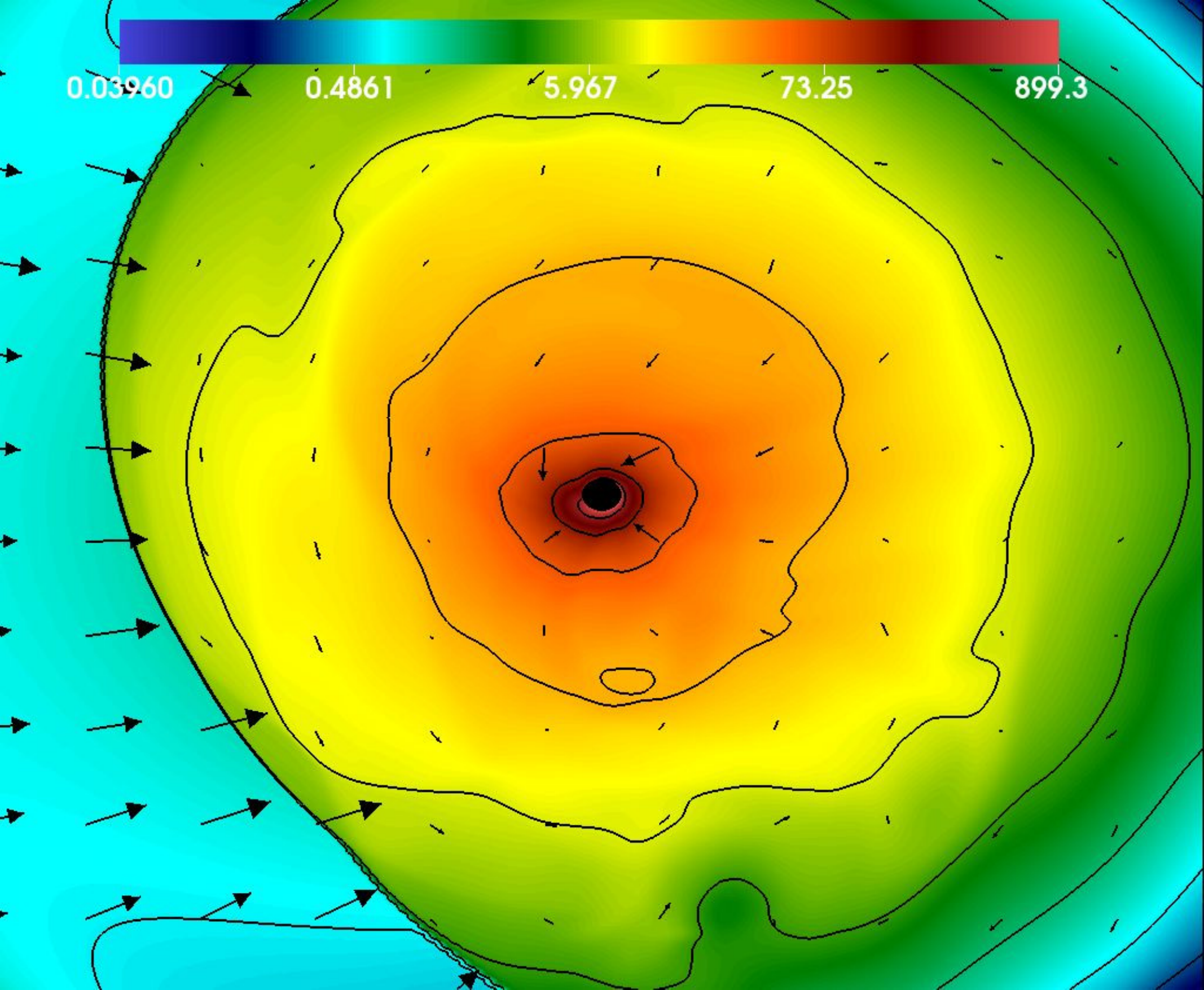,width=3.3cm,height=3.3cm}  
  \psfig{file=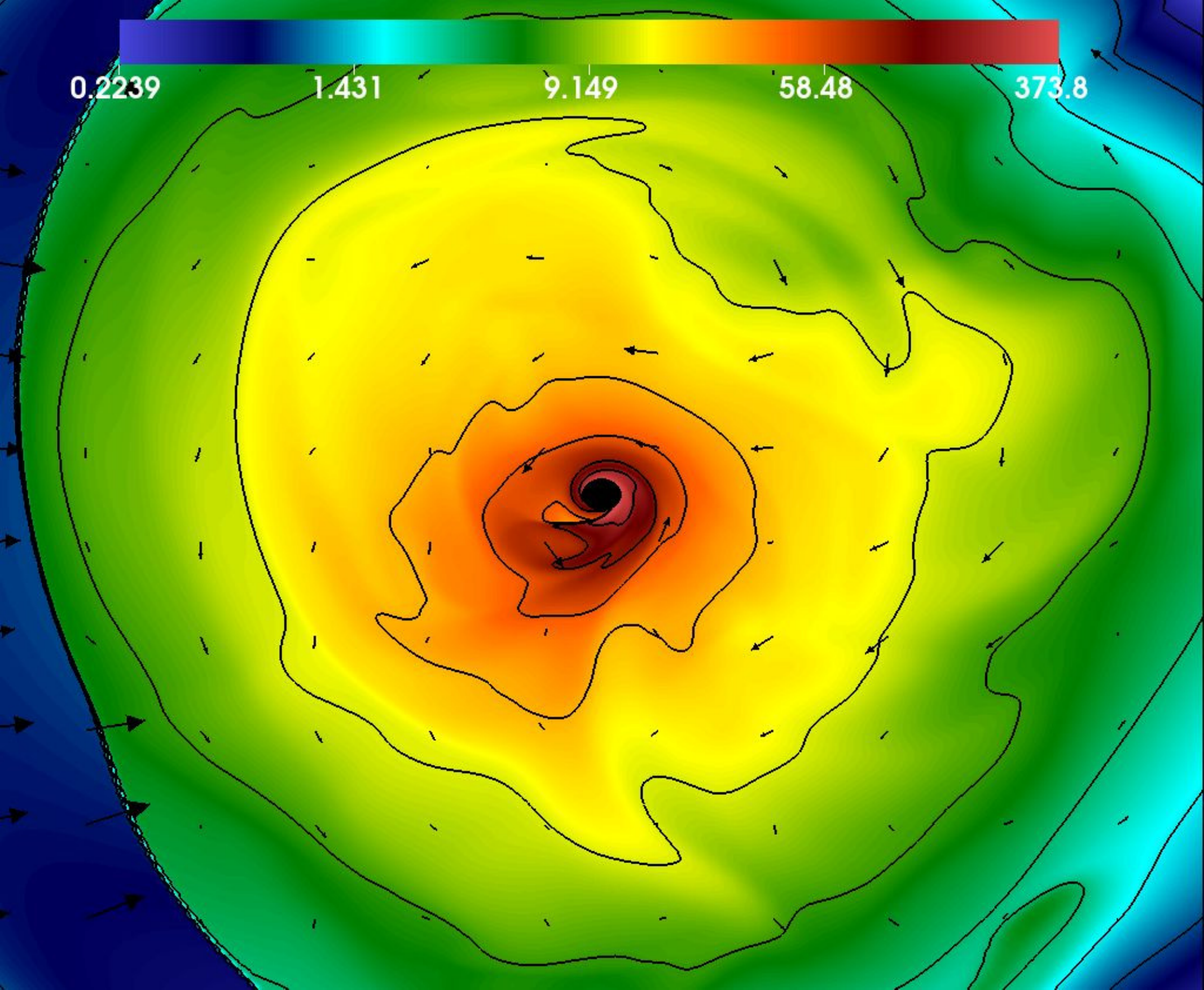,width=3.3cm,height=3.3cm} 
  \psfig{file=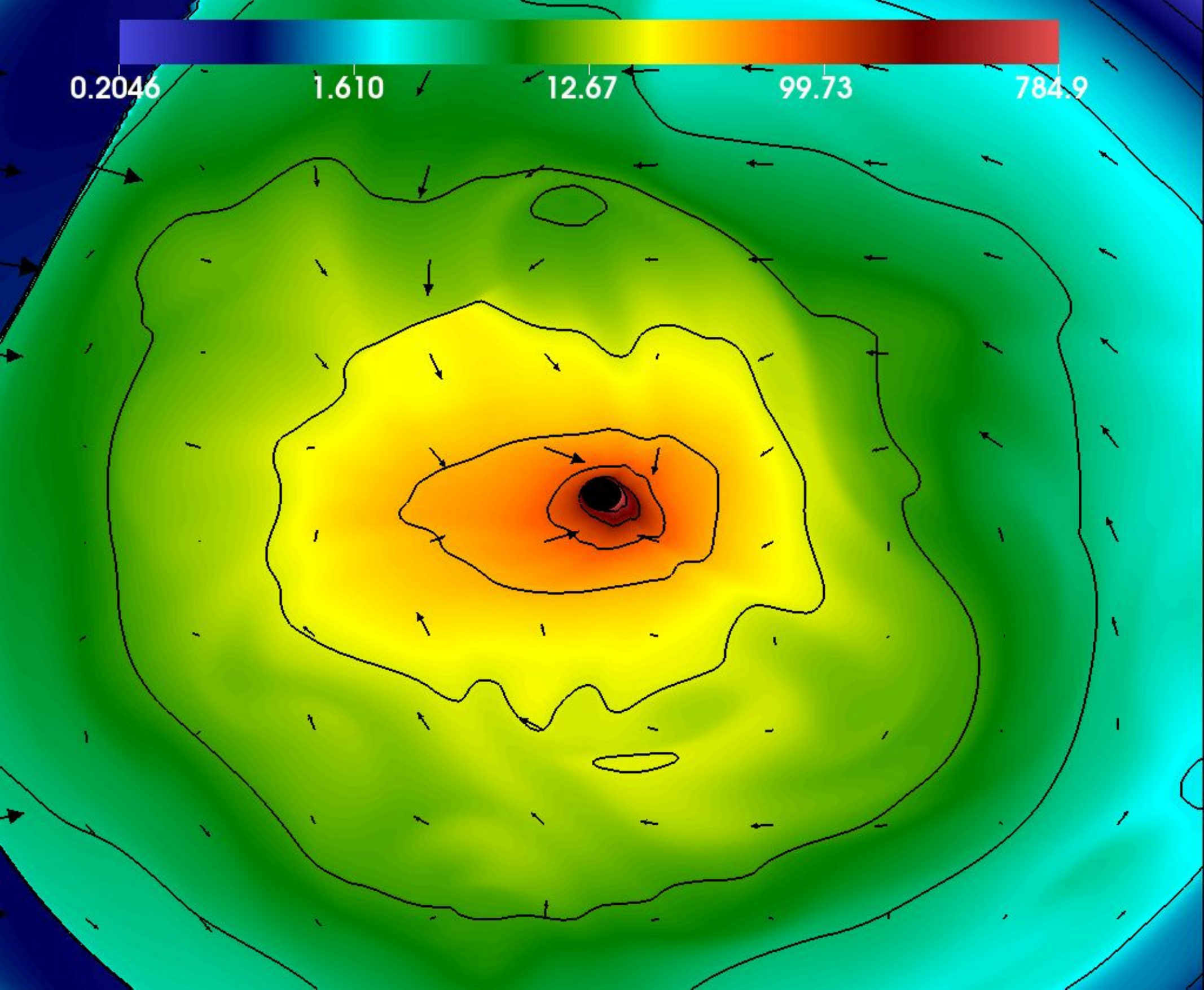,width=3.3cm,height=3.3cm}
  \psfig{file=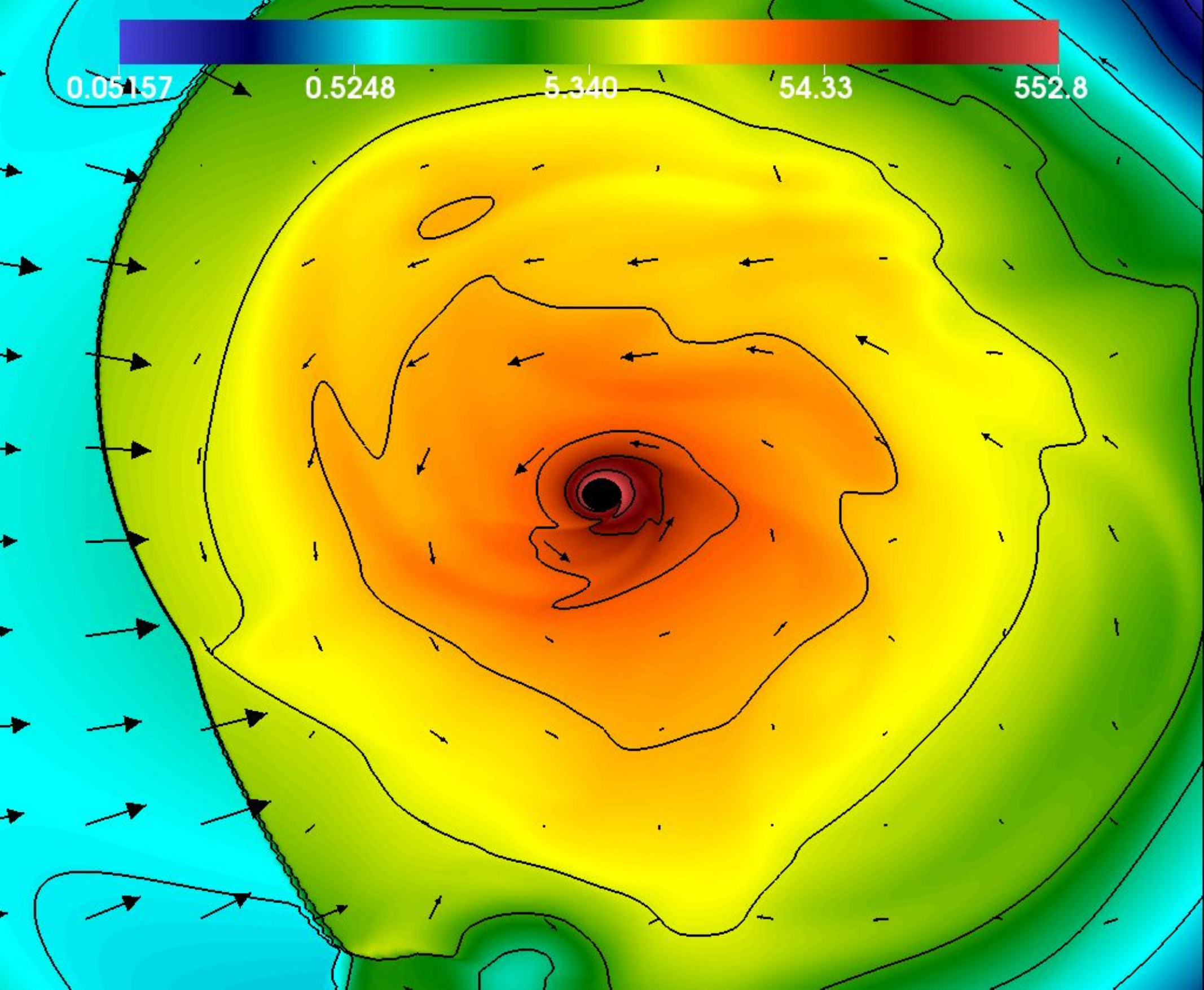,width=3.3cm,height=3.3cm}\\
  \psfig{file=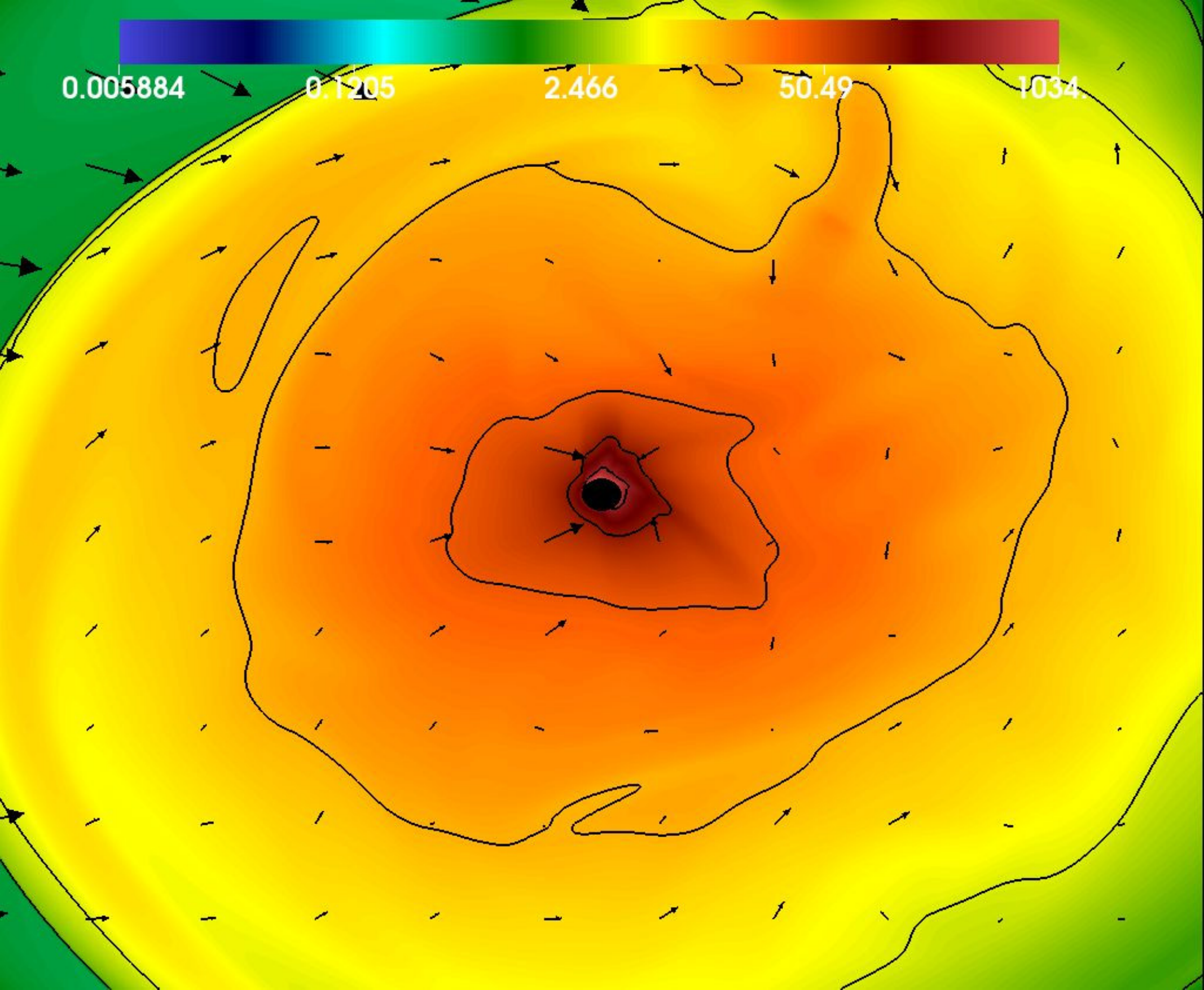,width=3.3cm,height=3.3cm}
  \psfig{file=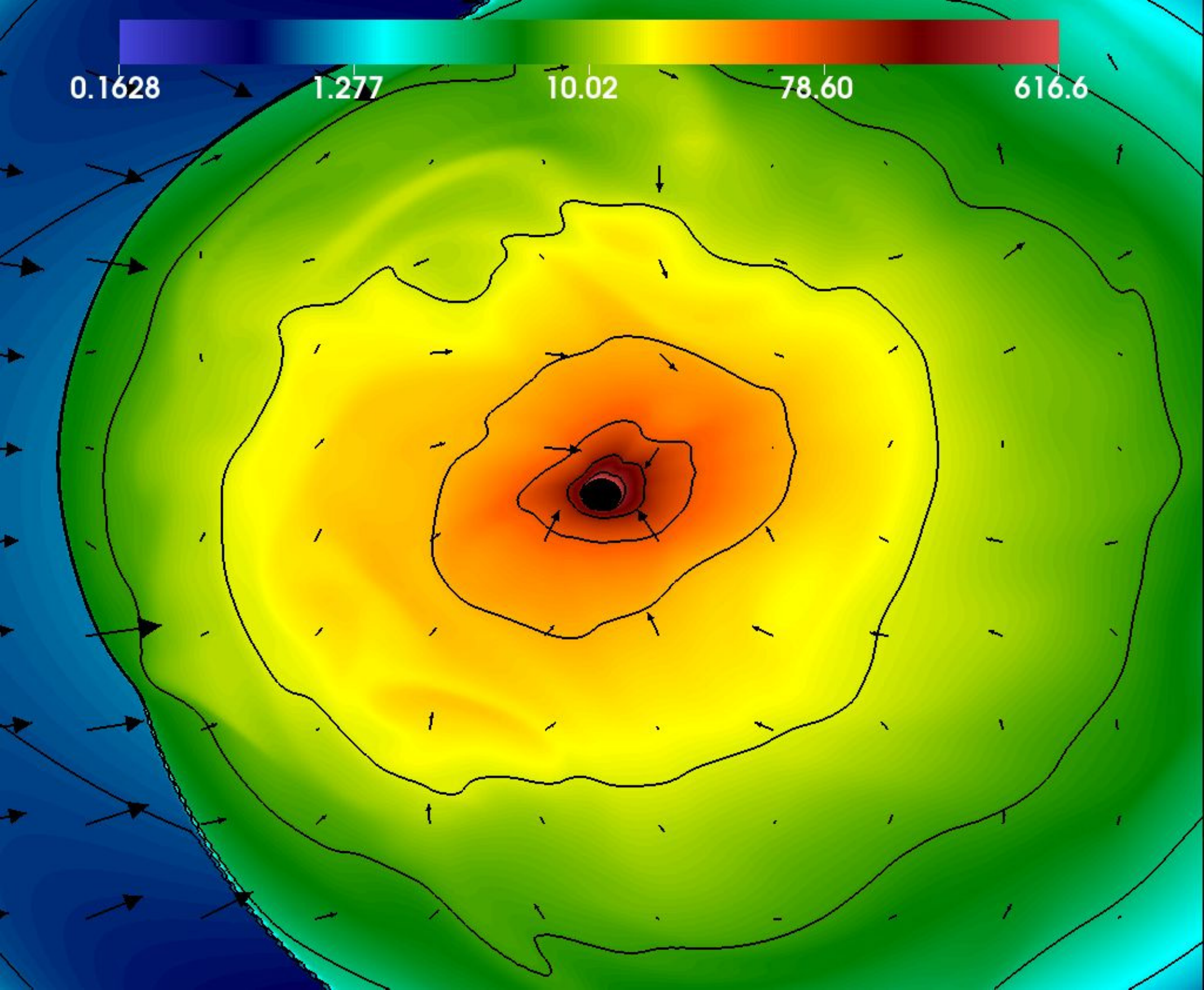,width=3.3cm,height=3.3cm}  
  \psfig{file=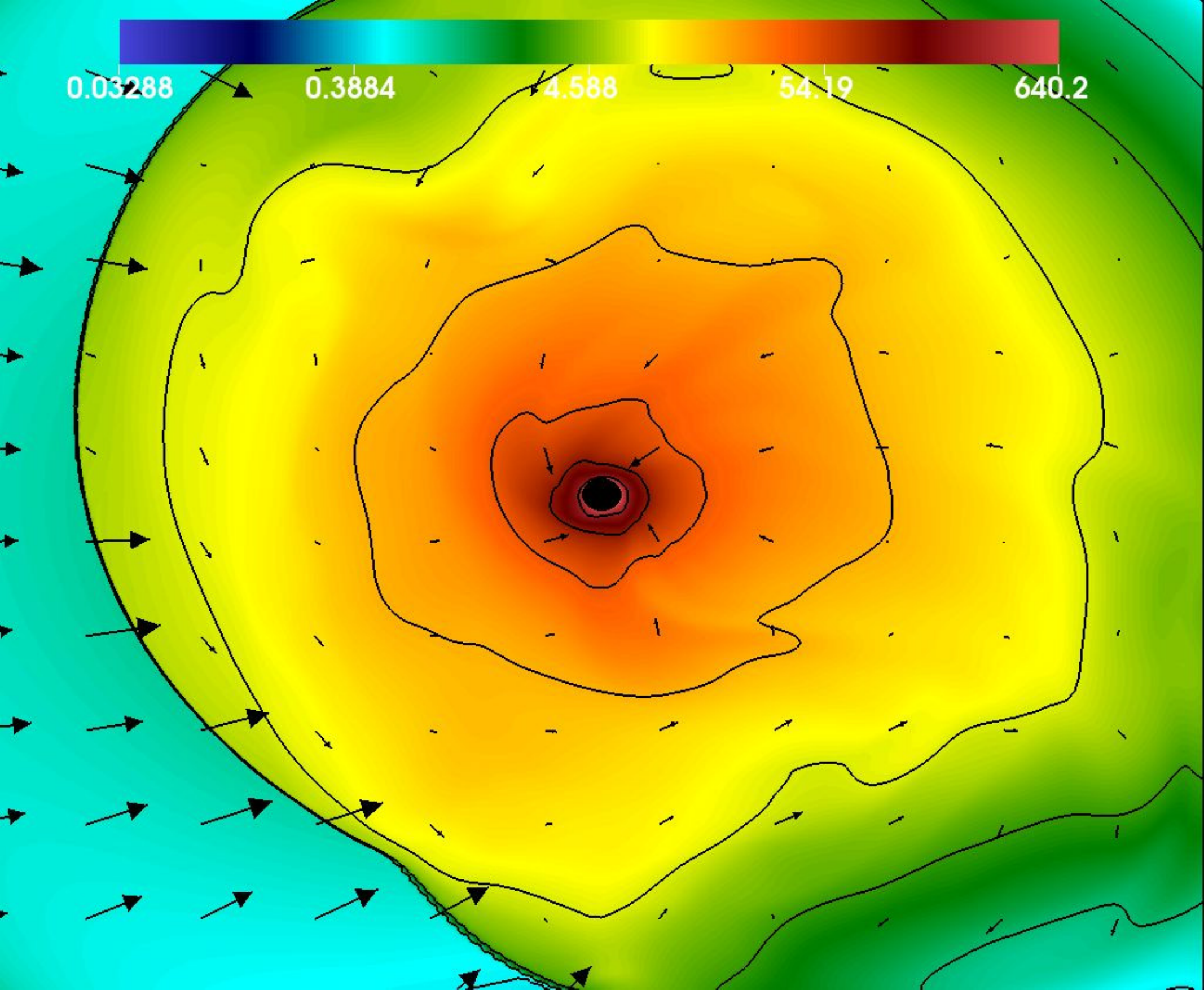,width=3.3cm,height=3.3cm} 
  \psfig{file=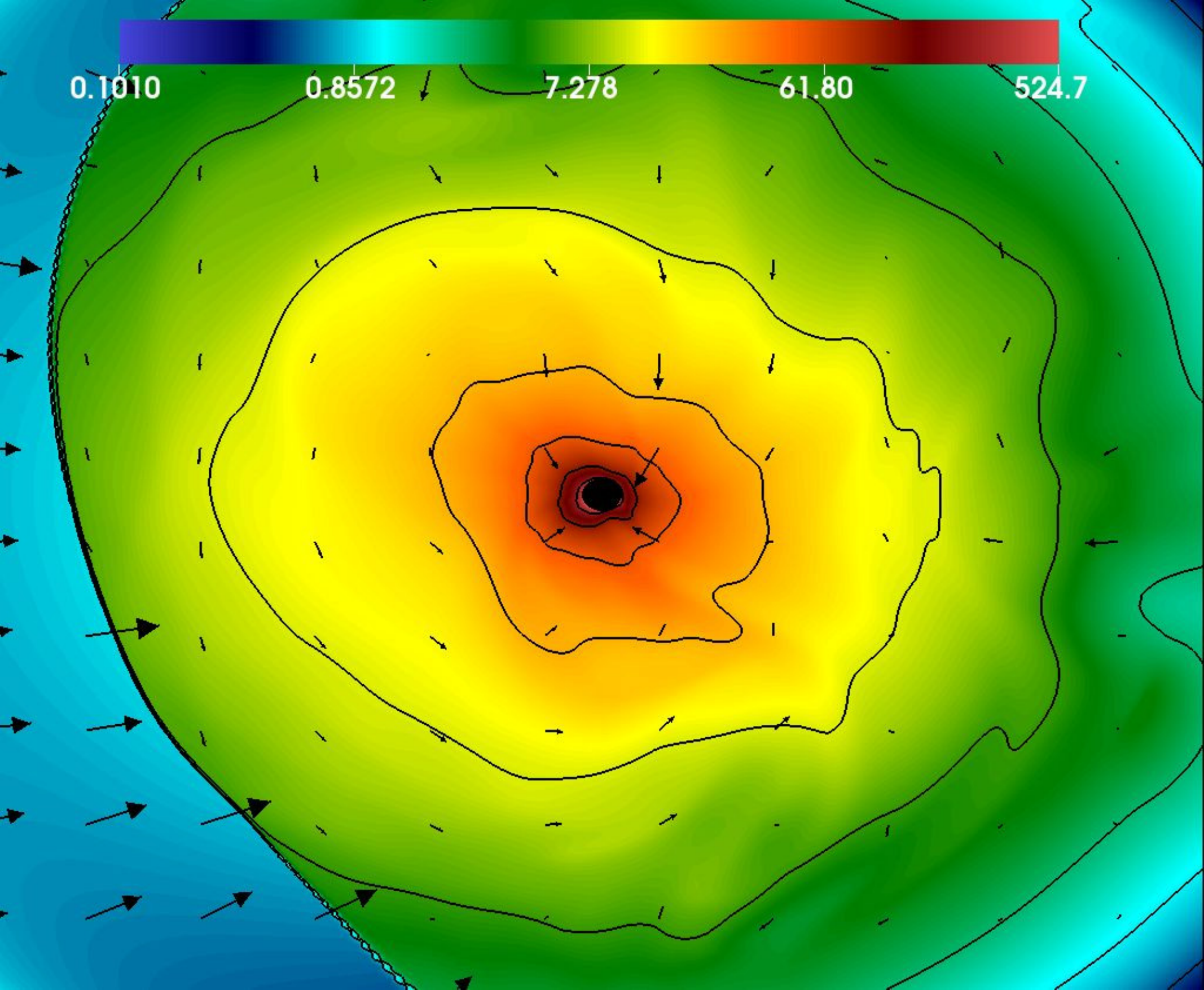,width=3.3cm,height=3.3cm}
  \psfig{file=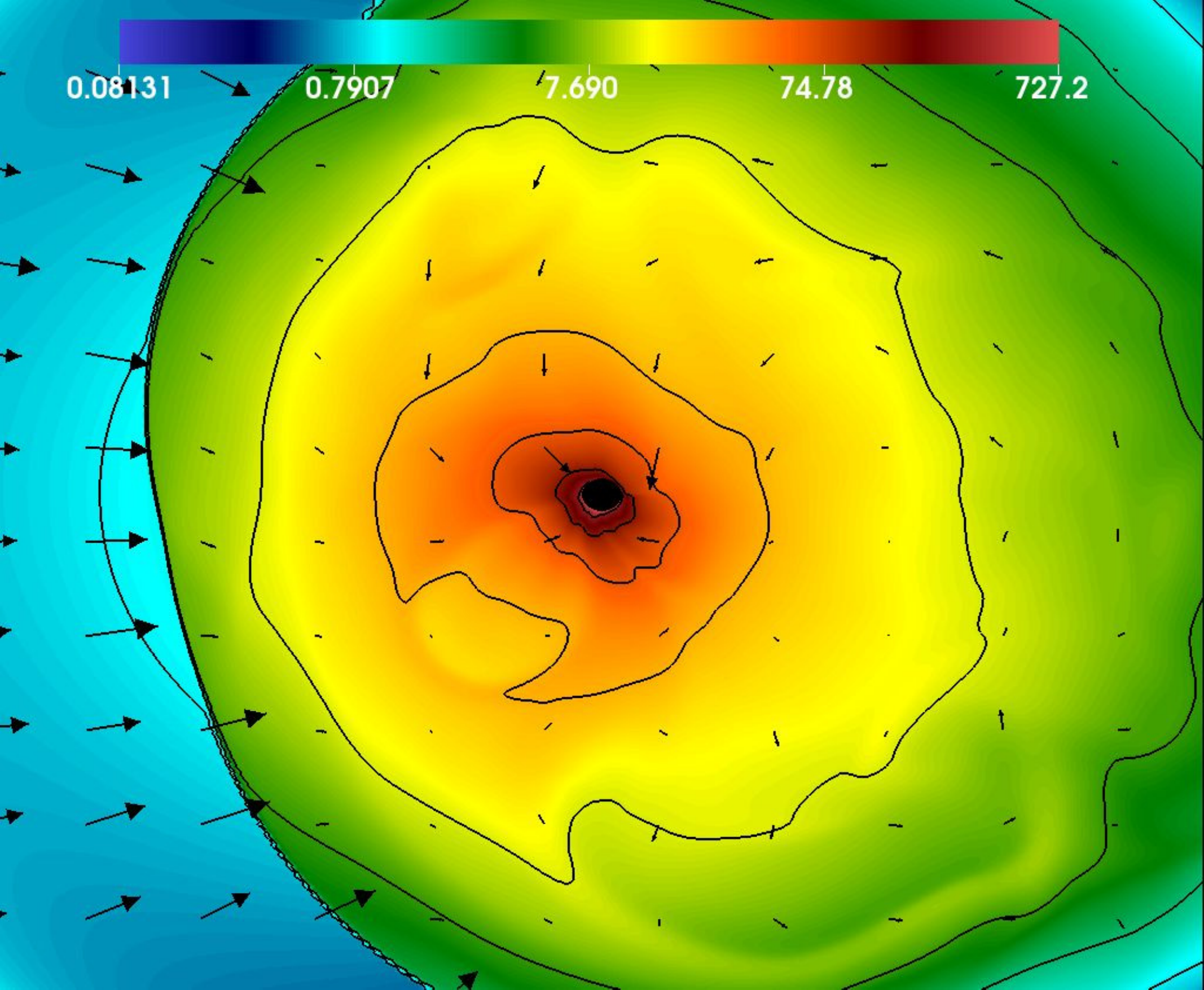,width=3.3cm,height=3.3cm}    
\caption{Visualization of the rest-mass density distribution on the equatorial plane around the rapidly rotating Kerr--Weyl black hole for Model $A_1$ ($a=0.9M$, $k\simeq-3\times10^{-5}\,M^{-2}$), as defined in Table~\ref{Inital_Con_1}. The figure combines vector fields with colored contour plots to illustrate the flow morphology. The top-left panel shows the early phase of the BHL accretion process, when matter from the upstream region begins to fall toward the black hole, and the bottom right panel corresponds to $t=3\times10^{4}M$. A well-defined shock cone first forms behind the black hole. It is then destabilized as shocked gas accumulates under the additional attraction of the AdS-like curvature term, and, up to the last snapshot in the third row, the flow exhibits flip-flop oscillations, with the accretion pattern continuously changing around the black hole. After this phase, the system evolves toward a quasi-steady toroidal structure with spiral shock arms surrounding the black hole.
}
\vspace{1cm}
\label{A1_animation}
\end{figure*}

A representative case of the strongest deviation, Model $A_1$ ($a=0.9M$, $k\simeq-3\times10^{-5}\,M^{-2}$, $\eta\simeq1.9$), is shown in Fig.~\ref{A1_animation}. The plasma surrounding the Kerr--Weyl black hole exhibits a rich and highly dynamical evolution. In the earliest stage, shown in the first row of Fig.~\ref{A1_animation}, matter falls supersonically toward the black hole through the BHL mechanism, is deflected toward the downstream axis, and forms a high-density shock cone in the downstream region. At this stage, the shocked region is still compact and lies where the Newtonian term dominates, so the cone is similar to that of the Kerr case. As the flow fills the region $r\sim R_{\rm acc}$, the AdS-like curvature term (Fig.~\ref{gtt_r}) makes the gravitational potential at the scale of the cone deeper than in the standard Kerr spacetime~\cite{DonmezMNRAS2011,KoyuncuMpla2014} or other alternative gravity models~\cite{Donmezpdu2024,DonmezJheap2024,DonmezEpjc2025,MustafaJcap2025}. The gas reaches the cone with a higher velocity, which leads to stronger focusing and a sharper density gradient near the axis, where the inflowing gas slows down and begins to accrete. A second effect is equally important. In the Kerr case, the stagnation point on the downstream axis separates the shocked gas that falls into the black hole from the gas that escapes downstream. For $k<0$, the escaping gas has to climb out of a potential that keeps deepening with radius, Eq.~(\ref{g_r}), so it is decelerated and partly pulled back. Shocked gas therefore accumulates behind the black hole, the post-shock pressure rises, and the shock fronts are pushed away from the axis, producing the first indications of a bow-shock structure.

At intermediate times, corresponding to the second and third rows of Fig.~\ref{A1_animation} (approximately $t=5\times10^{3}M$ to $1.5\times10^{4}M$), the shock cone becomes unstable. The flow alternates between states in which the shock stands upstream of the black hole and states in which the shocked gas surrounds it almost completely, while the cone swings from one side of the downstream axis to the other. It is important to note that the curvature term itself is smooth and monotonic in $r$ and introduces no oscillation into the potential. The oscillation is the flip-flop instability of the BHL wake, a hydrodynamic overstability found in both Newtonian and relativistic simulations, whose growth has been linked to an advective--acoustic feedback between the shock and the accretor~\cite{Foglizzo2005,Blondin2009,DonmezMNRAS2012}. The role of $k$ is to drive the background flow into the regime in which this instability grows, by widening the cone, raising the post-shock pressure, and moving the shock away from the black hole. Within the oscillating cone, vortices appear around the stagnation point as the flow becomes highly nonlinear, and the density distribution shows recurrent asymmetries with pronounced contrasts. The spin $a=0.9M$ breaks the left--right symmetry of the flow through frame dragging and therefore affects the details of these asymmetries. However, the same sequence is found for $a=0.5M$ in Model $A_5$ (see below), which shows that the instability is controlled by $k$ rather than by the spin.

In the later phases of the evolution, as seen in the remaining panels of Fig.~\ref{A1_animation}, the system gradually approaches a quasi-steady configuration. Because $\eta>1$, the curvature-induced pull exceeds the Newtonian one inside the capture region, and the shocked gas can no longer leave through the downstream wake. The gas accreted during the flip-flop phase carries angular momentum of alternating sign~\cite{Blondin2009}, and as this angular momentum is redistributed and energy is dissipated through the shocks, the violent oscillations are damped. The original shock cone is destroyed, and a dense toroidal structure with spiral shock arms (fourth row of Fig.~\ref{A1_animation}) forms and rotates around the black hole. This proto-disk configuration is maintained by the combined influence of gravitational attraction, centrifugal forces and the confining curvature term. Its morphology is clearly distinct from the stable and symmetric cone obtained in the Kerr run with the same spin and hydrodynamic parameters, so the transformation of the BHL shock cone into a toroidal plasma configuration is a consequence of the modification of the spacetime geometry induced by $k$. The negative $k$ thus has two effects: it enhances the early inflow toward the black hole and, at the same time, destabilizes the flow by suppressing the downstream escape of the shocked gas. Together, these effects are responsible for both the flip-flop motion of the shock cone and the subsequent evolution toward the quasi-steady state. Even a deviation as small as $|k|\sim10^{-5}\,M^{-2}$, which leaves the horizon and the near-horizon metric practically unchanged (Fig.~\ref{Weyl_horizon}), therefore produces a new physical regime around the black hole, because its effect is felt at $r\sim R_{\rm acc}$ rather than near the horizon.

The toroidal structure that emerges around the Kerr--Weyl black hole in the quasi-steady state remains dynamically active and gives rise to QPOs, which are analyzed below using the PSD of the mass accretion rate. The repeated compression and rarefaction of the gas near the black hole, together with the side-to-side motion of the cone during the flip-flop phase, produce periodic fluctuations in the density and accretion rate that appear as peaks in the PSD. Since the corresponding fluctuations in the Kerr runs are much weaker, the detection of such variability would provide a potential means of testing Kerr--Weyl gravity through astrophysical accretion phenomena.

\begin{figure*}[!ht]
  \vspace{1cm}
  \center
  \psfig{file=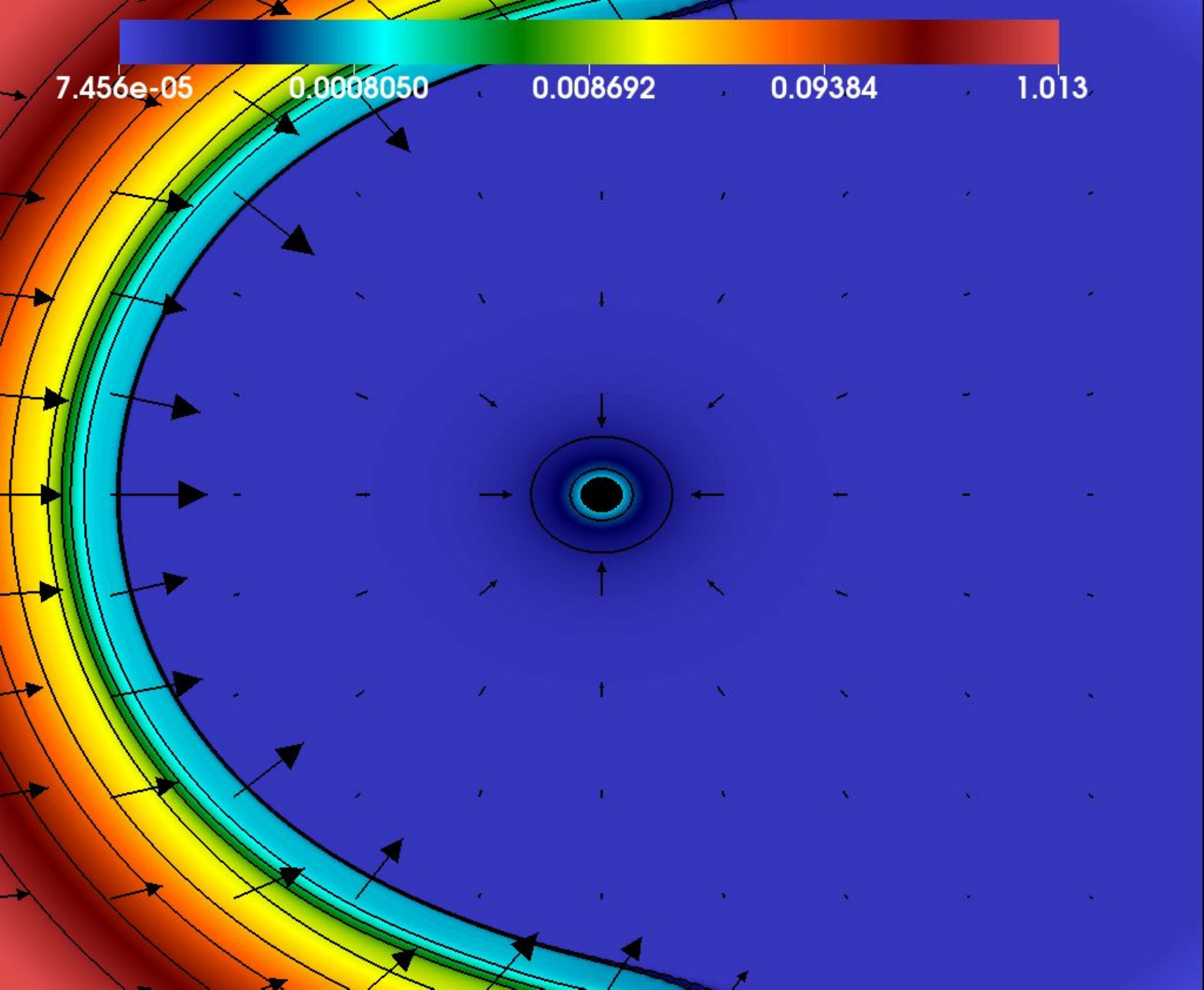,width=4.2cm,height=4.1cm}
  \psfig{file=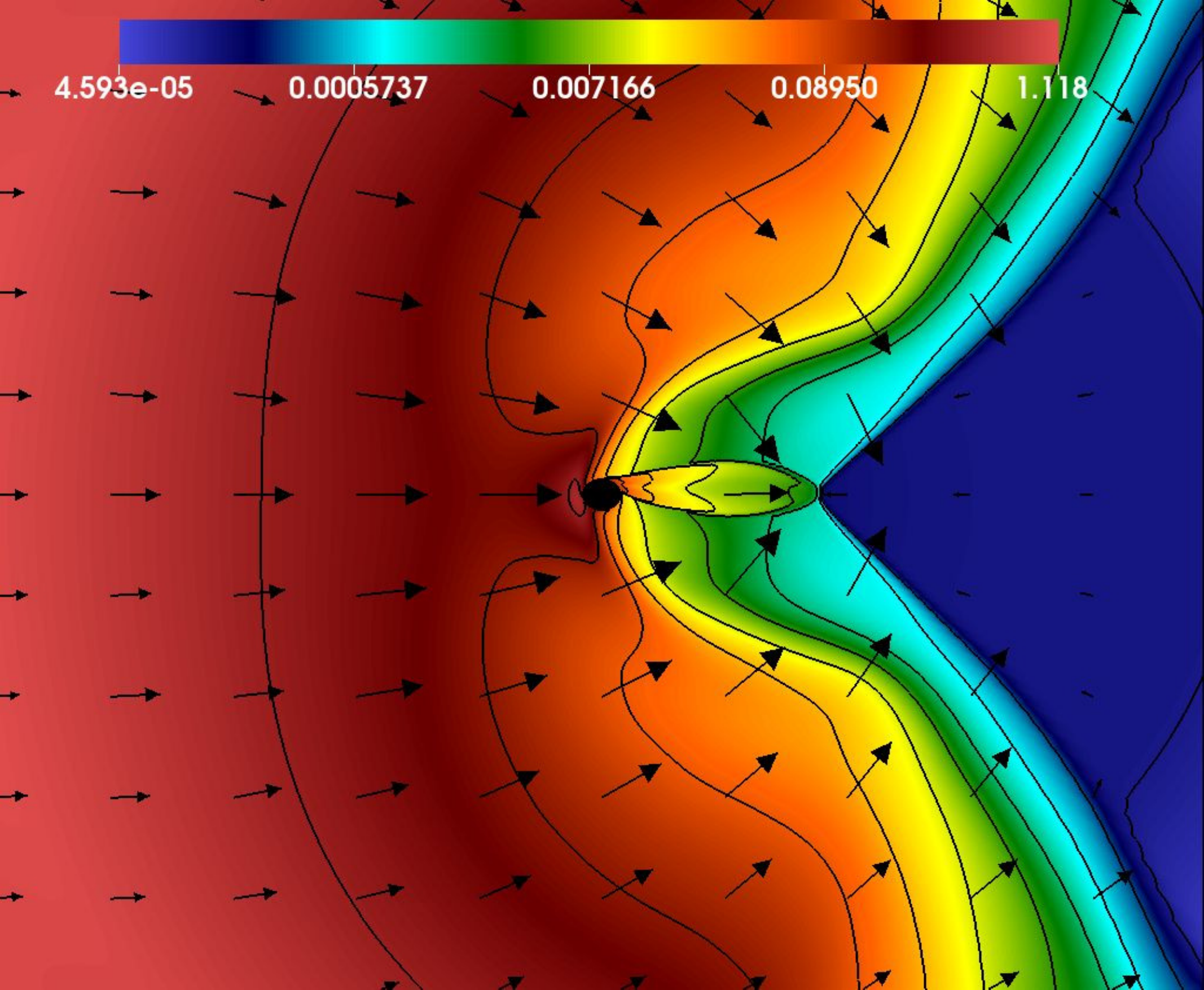,width=4.2cm,height=4.1cm}
  \psfig{file=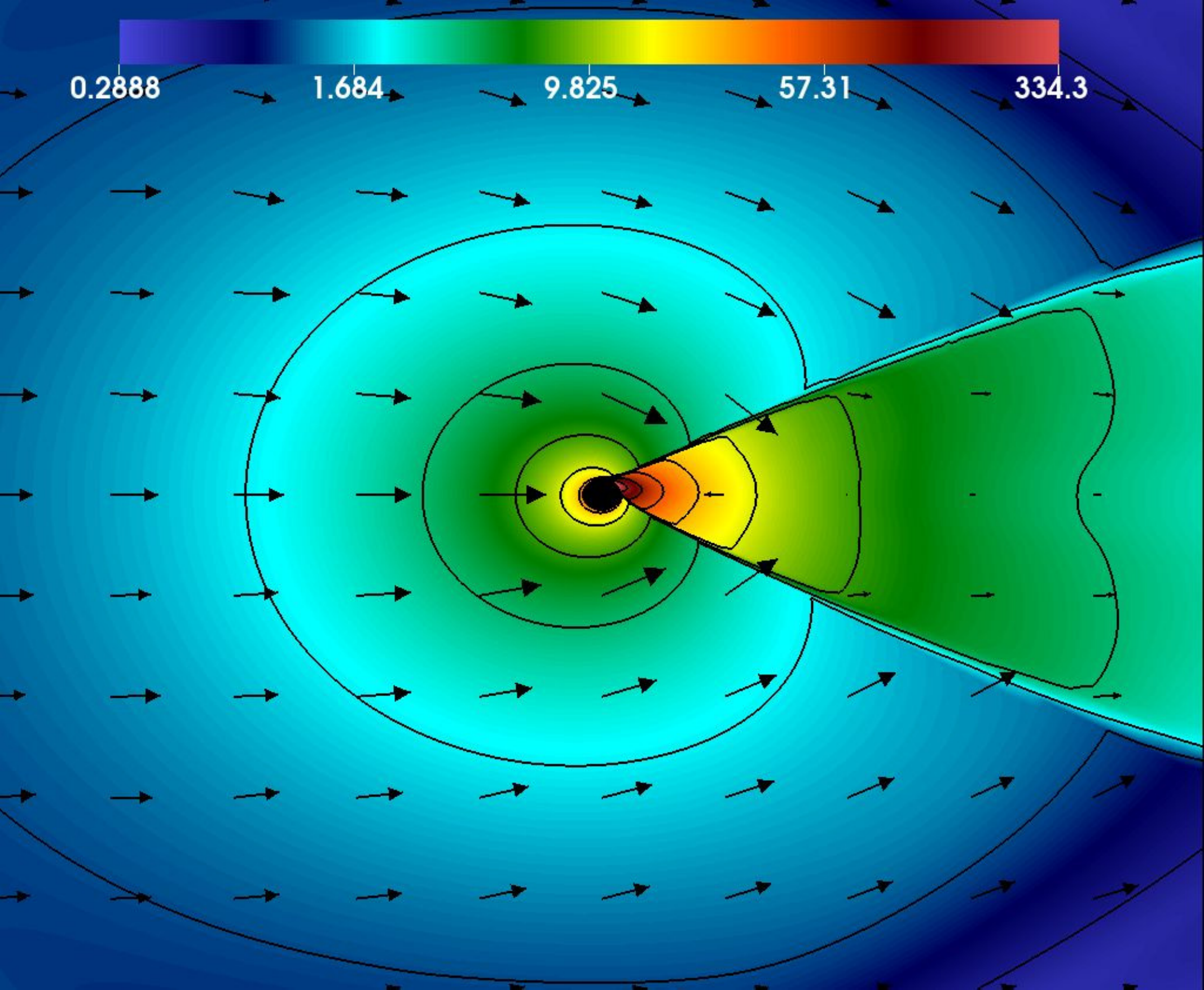,width=4.2cm,height=4.1cm}
  \psfig{file=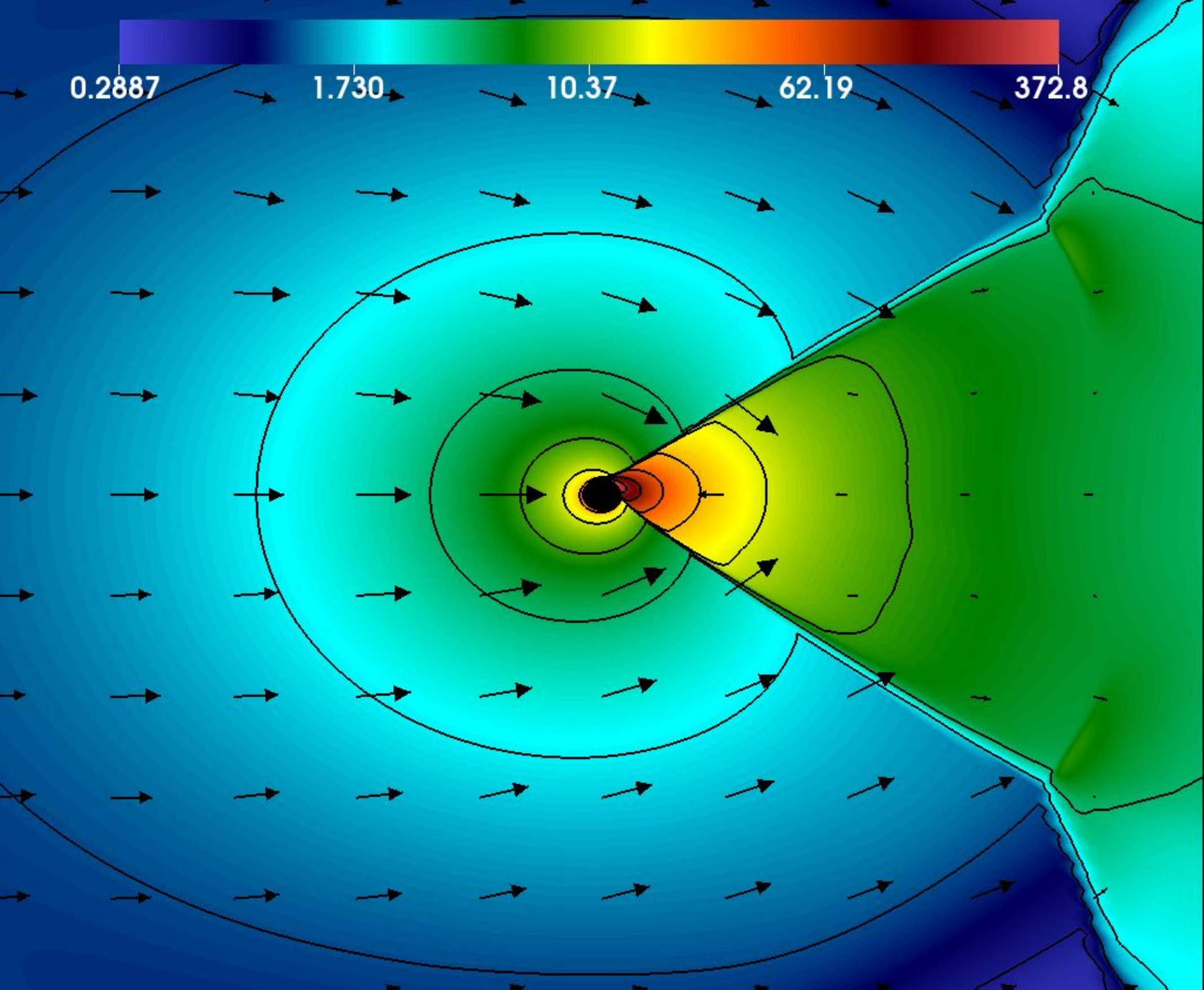,width=4.2cm,height=4.1cm}\\
  \psfig{file=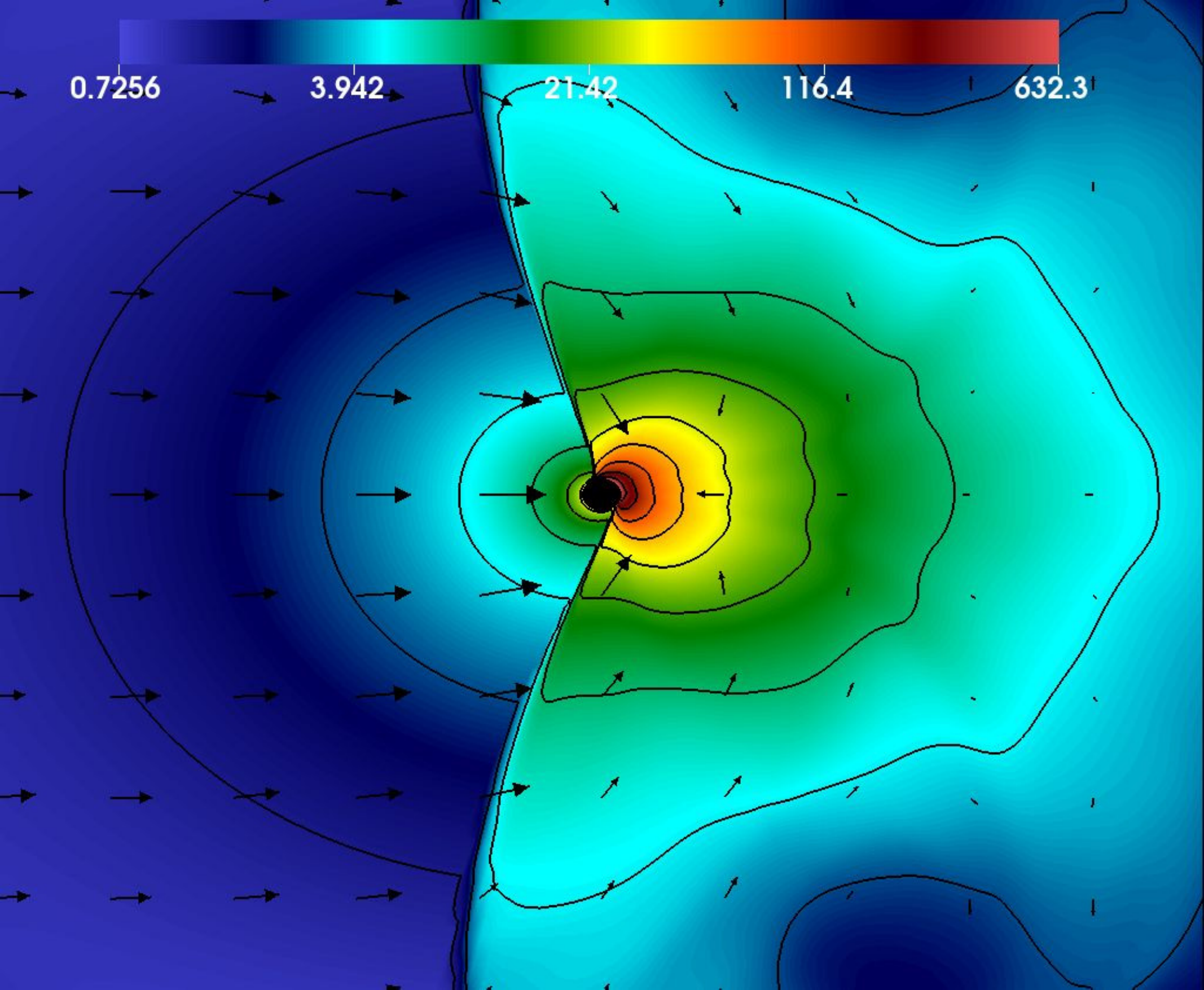,width=4.2cm,height=4.1cm}
  \psfig{file=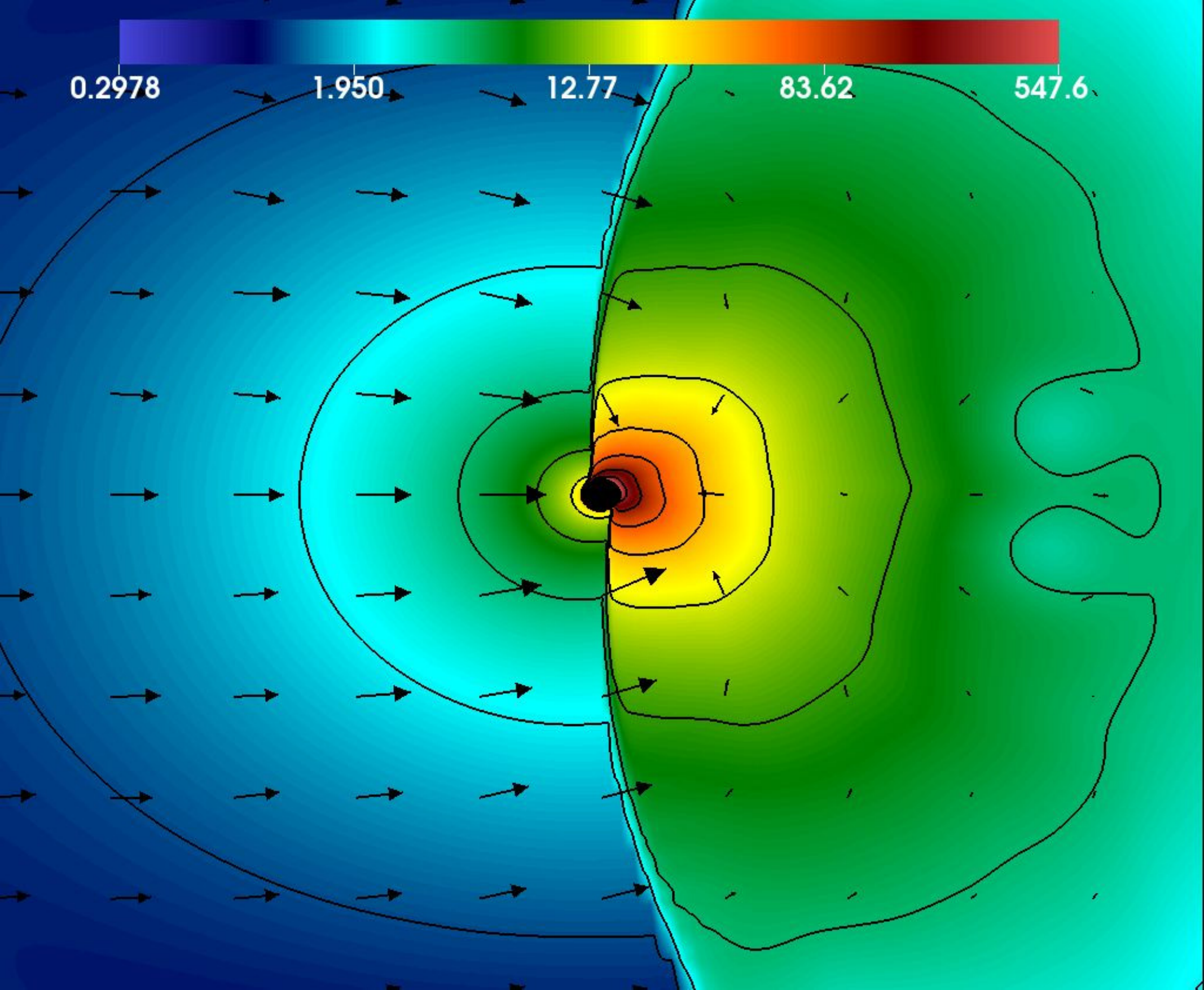,width=4.2cm,height=4.1cm}
  \psfig{file=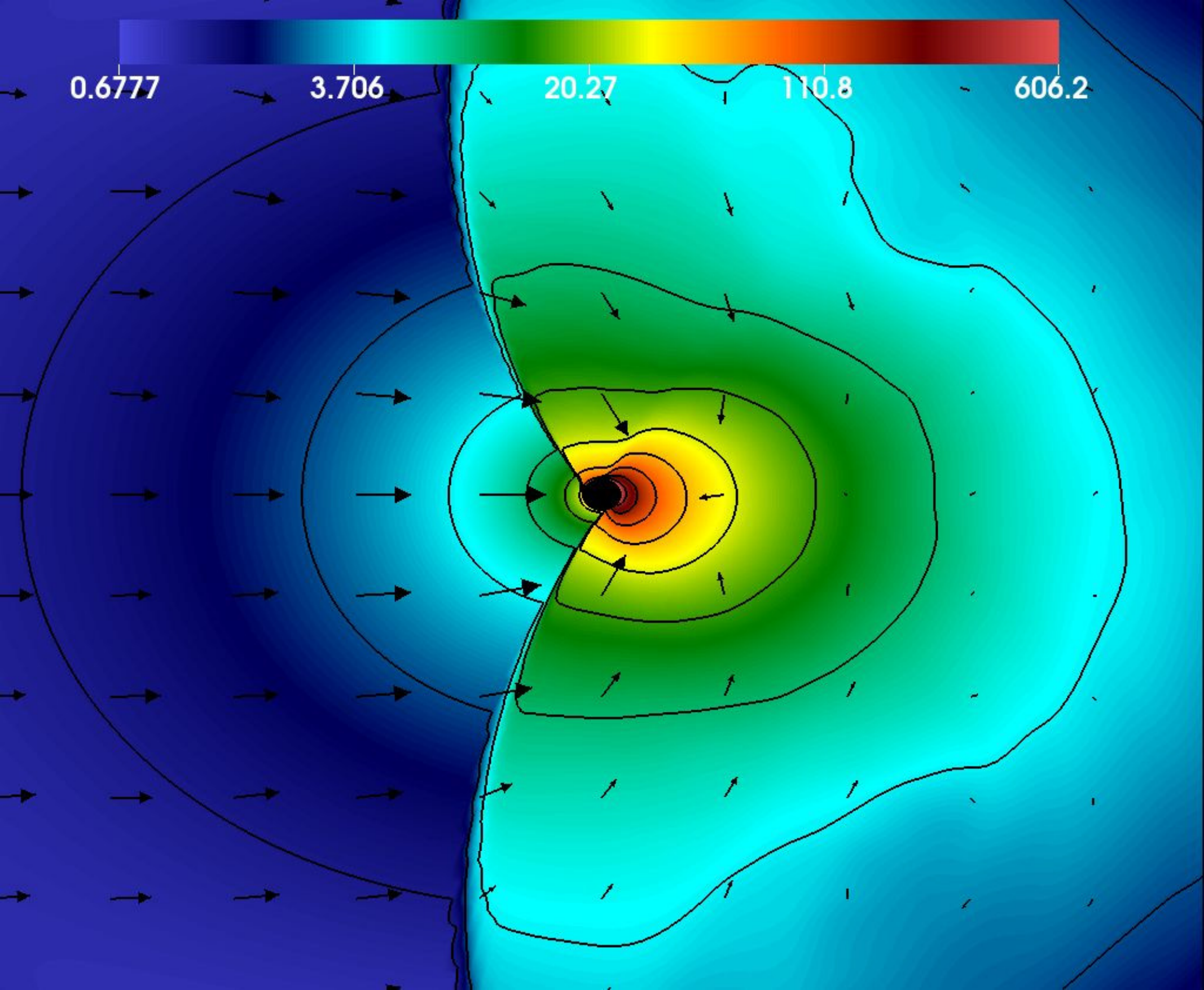,width=4.2cm,height=4.1cm}
  \psfig{file=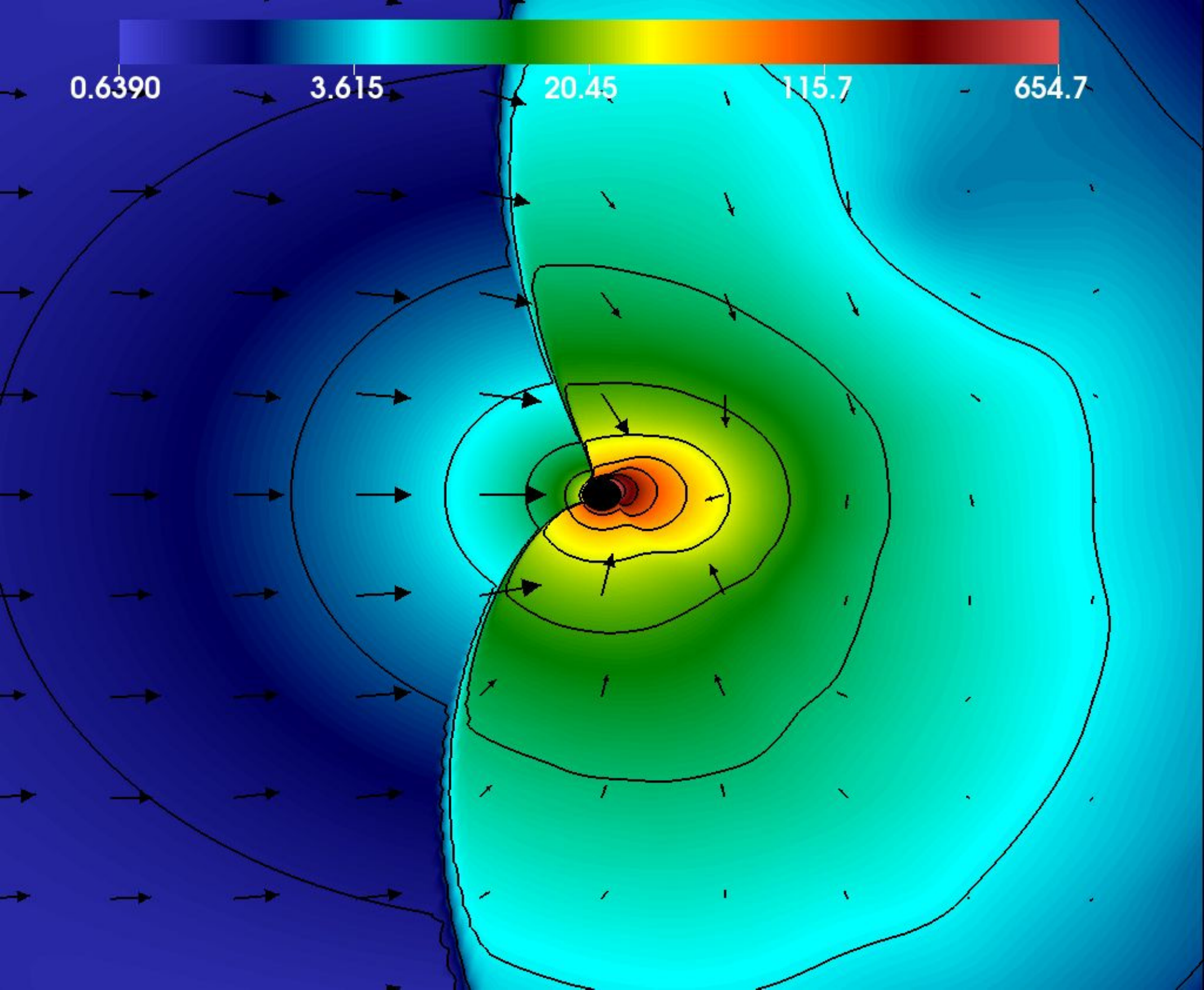,width=4.2cm,height=4.1cm} \\ 
  \psfig{file=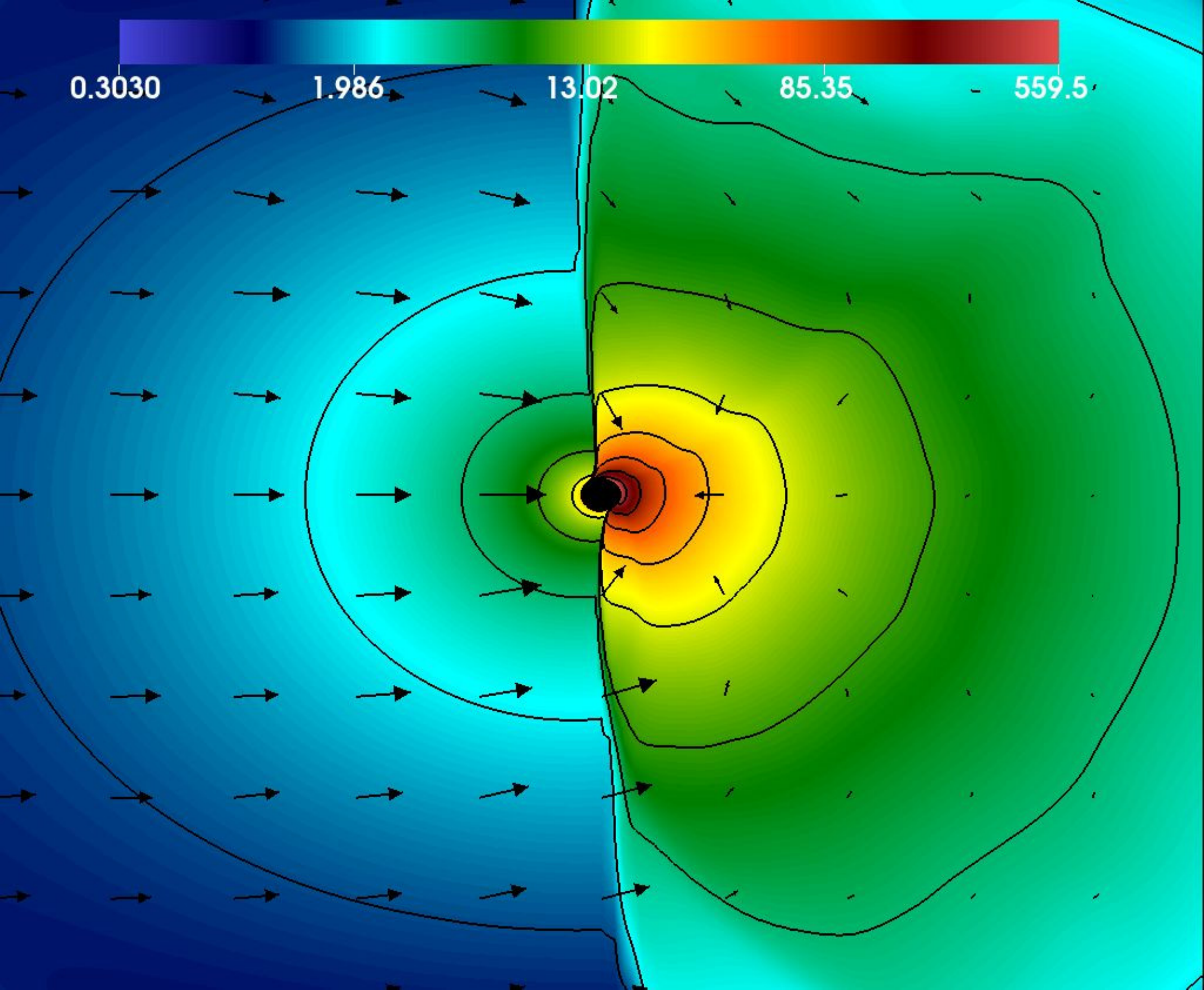,width=4.2cm,height=4.1cm}
  \psfig{file=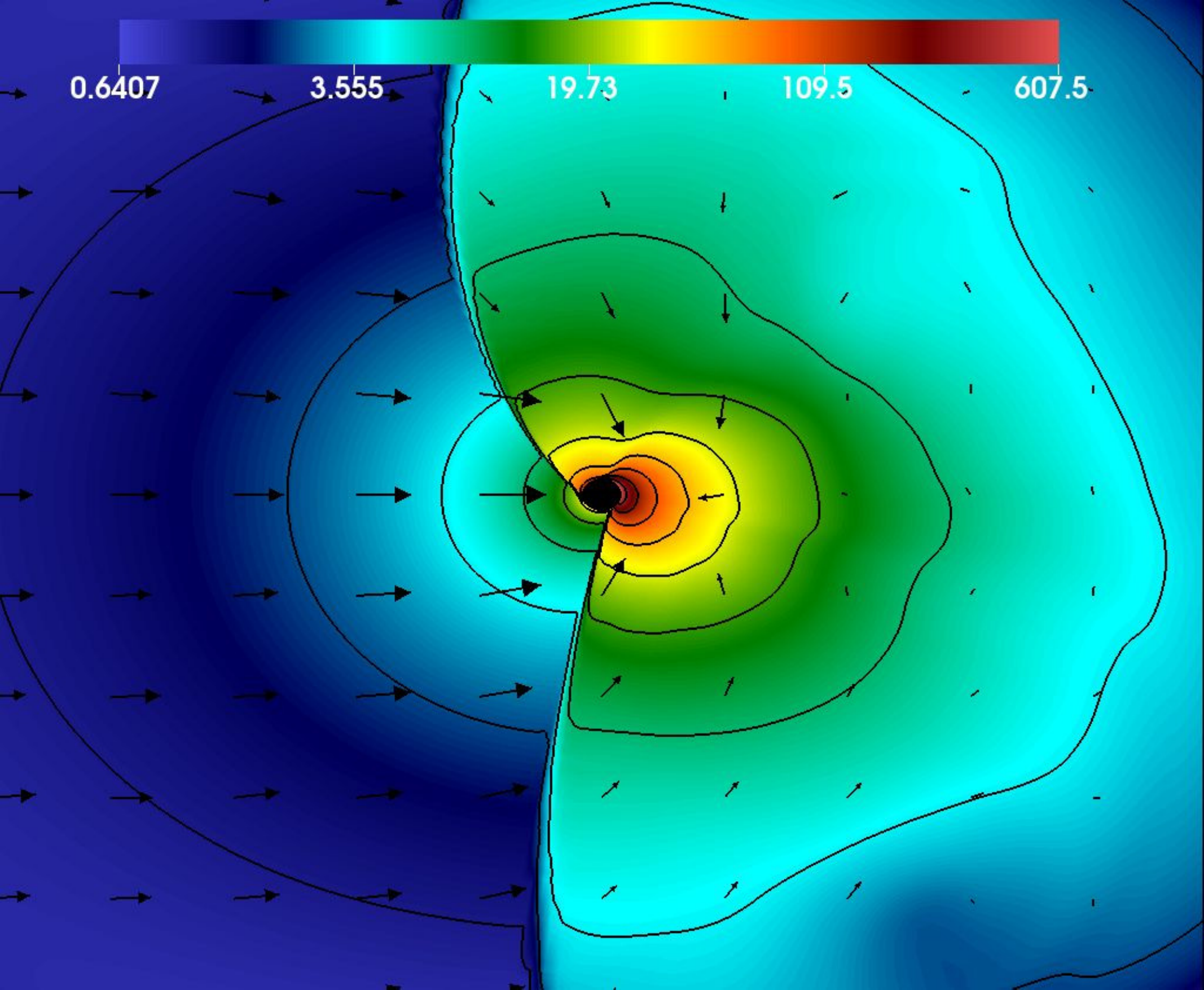,width=4.2cm,height=4.1cm}
  \psfig{file=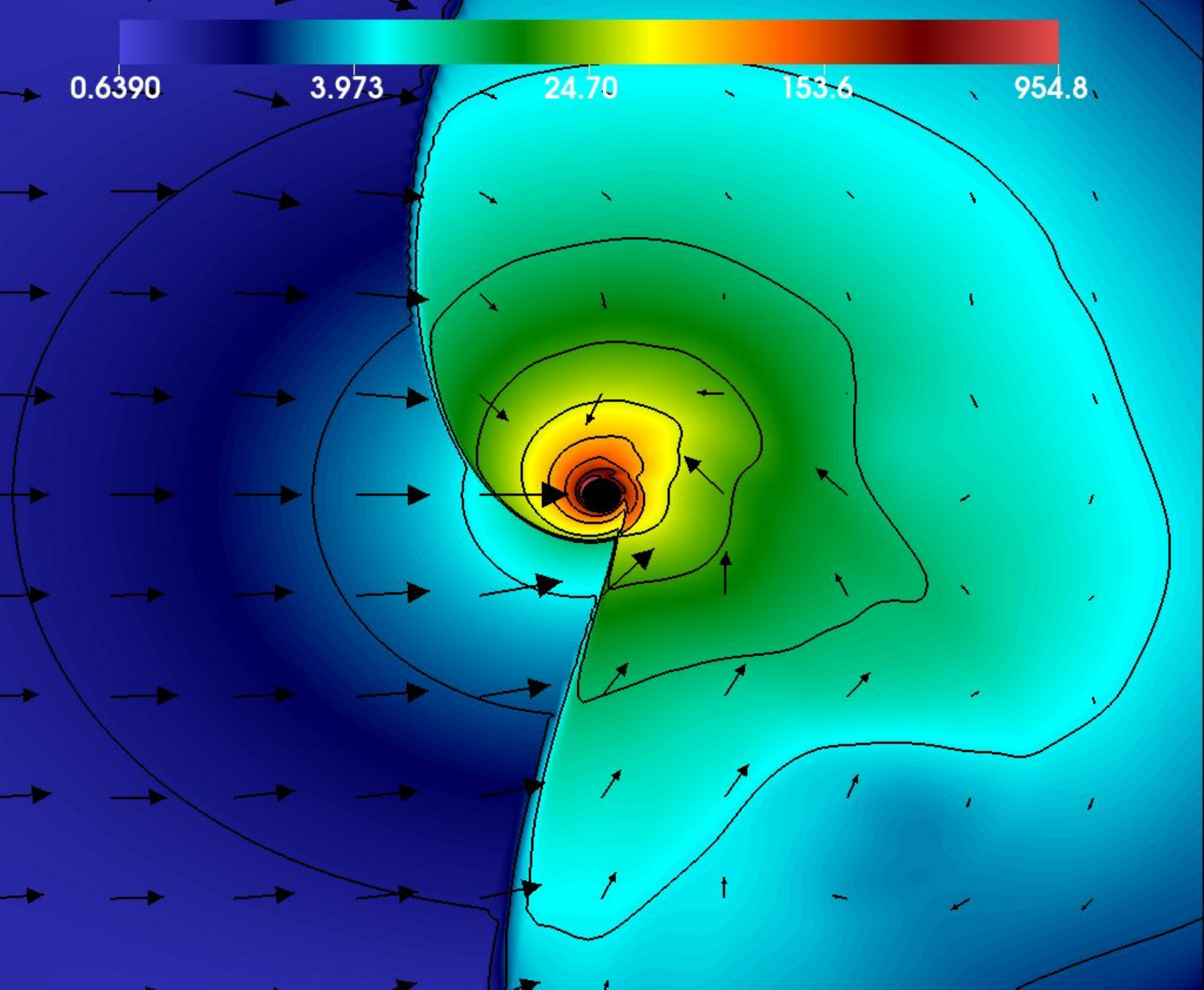,width=4.2cm,height=4.1cm}
  \psfig{file=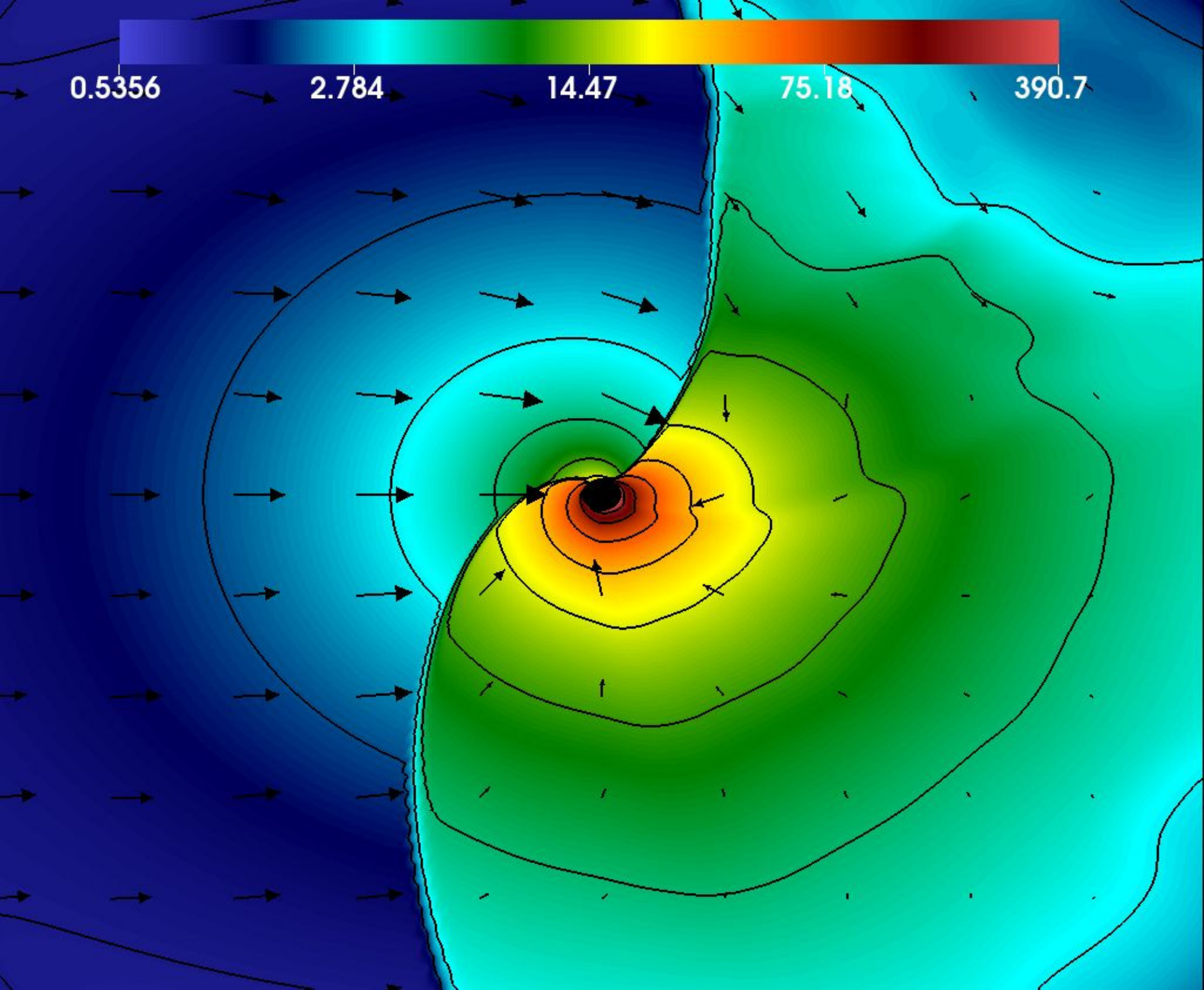,width=4.2cm,height=4.1cm}    
\caption{Same as Fig.~\ref{A1_animation}, but for Model $A_2$ ($a=0.9M$, $k\simeq-3\times10^{-6}\,M^{-2}$). A shock cone again forms at early times. As shocked gas accumulates downstream, the shock is pushed upstream, toward the incoming flow, and, as shown in the middle row, the opening angle of the cone grows until the shock becomes a single curved front just upstream of the black hole. This broadened shock then undergoes strong flip-flop oscillations in the downstream region, which persist until the end of the simulation at $t=3\times10^{4}M$ without forming a torus.}
\vspace{1cm}
\label{A2_animation}
\end{figure*}

Fig.~\ref{A2_animation} shows the evolution of Model $A_2$ ($a=0.9M$, $k\simeq-3.0\times10^{-6}\,M^{-2}$). Its $|k|$ is one tenth of that in Model $A_1$, so $r_k\simeq87M\simeq2.2R_{\rm acc}$ and $\eta\simeq0.19$. The curvature term therefore modifies the Newtonian pull at the accretion radius by about $20\%$ instead of dominating it. During the early stages, matter from the upstream region falls toward the black hole, and a downstream shock cone develops in a manner similar to that seen in Fig.~\ref{A1_animation}. The weaker additional pull still hinders the escape of the shocked gas downstream, but less than in $A_1$. The gas that accumulates behind the black hole raises the post-shock pressure and pushes the shock fronts outward and upstream, so the opening angle of the cone increases significantly over time. In the middle row of Fig.~\ref{A2_animation}, the two wings of the cone have merged into a single curved shock front standing just upstream of the black hole, with the whole downstream region filled with shocked gas. This broad, weakly confined configuration differs strongly from the narrow cone of the Kerr run with the same spin.

As seen in the later snapshots of Fig.~\ref{A2_animation}, the shock remains dynamically active. Its front swings from one side to the other, producing strong flip-flop oscillations and continuous deformation of the post-shock region, but the flow does not evolve into the dense toroidal configuration observed in Model $A_1$. Because $\eta<1$, the potential at large radii is deepened, but not enough to confine the shocked gas around the black hole, and the flow keeps a widened, oscillating shock throughout the simulation. The spin $a=0.9M$ introduces a mild asymmetry through frame dragging, which affects the inner part of the post-shock flow and the phase of the oscillations. Model $A_3$, whose $k$ differs from that of $A_2$ by only about $0.01\%$, although its $\gamma$ has the opposite sign and a different magnitude, shows the same flip-flop behavior, as the azimuthal density profiles below confirm. This is expected, since the metric depends on the Weyl coefficients only through $\widetilde{M}$ and $k$. The flip-flop regime therefore occupies an intermediate range of $|k|$: the curvature term is strong enough to widen and destabilize the cone, but too weak to trap the gas in a torus. For still smaller values (the models $B$ and $C$, $\eta\lesssim0.03$), the cone remains narrow and stable, as in the Kerr case.

\begin{figure*}[!ht]
  \vspace{1cm}
  \center
  \psfig{file=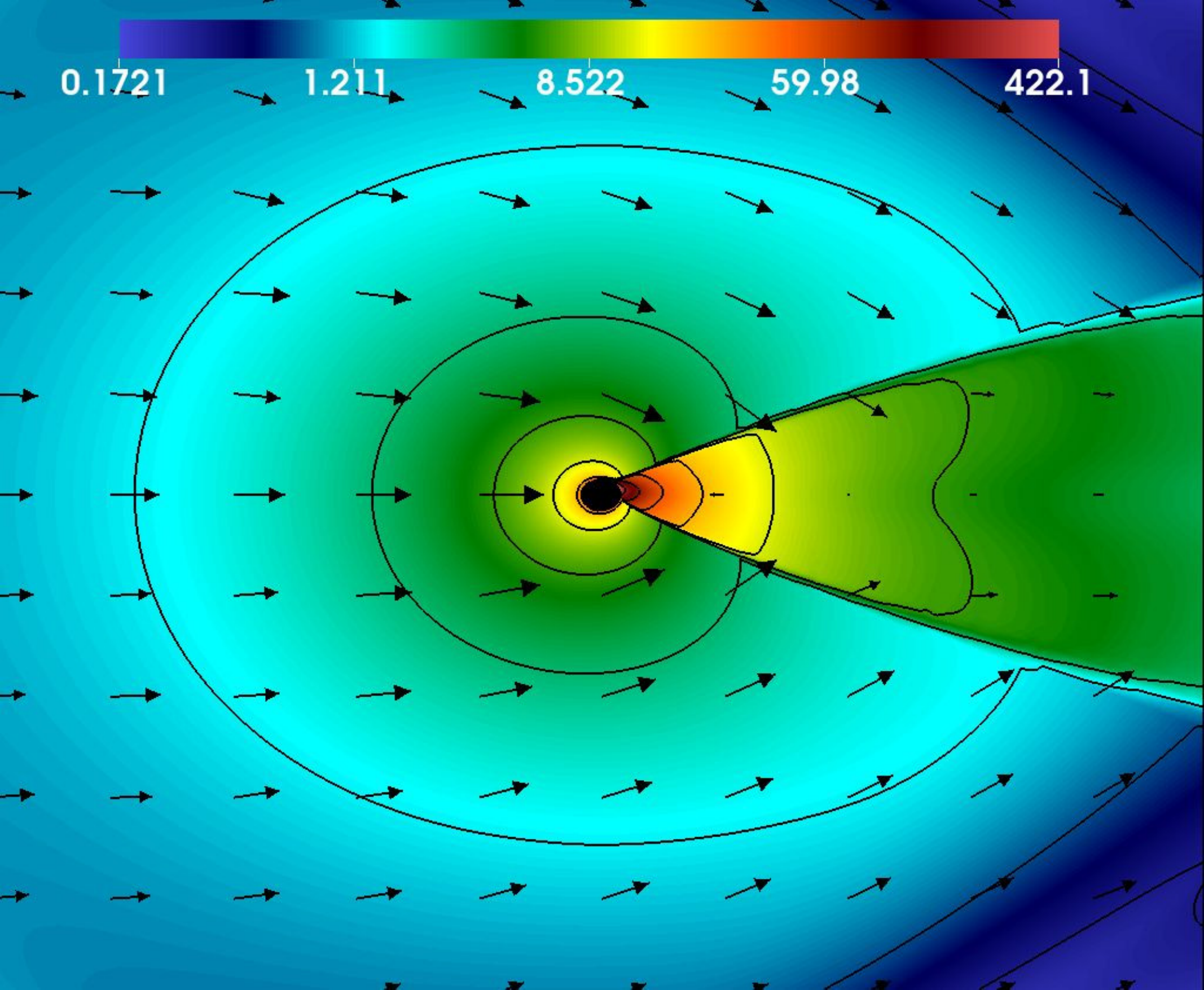,width=4.2cm,height=4.2cm}
  \psfig{file=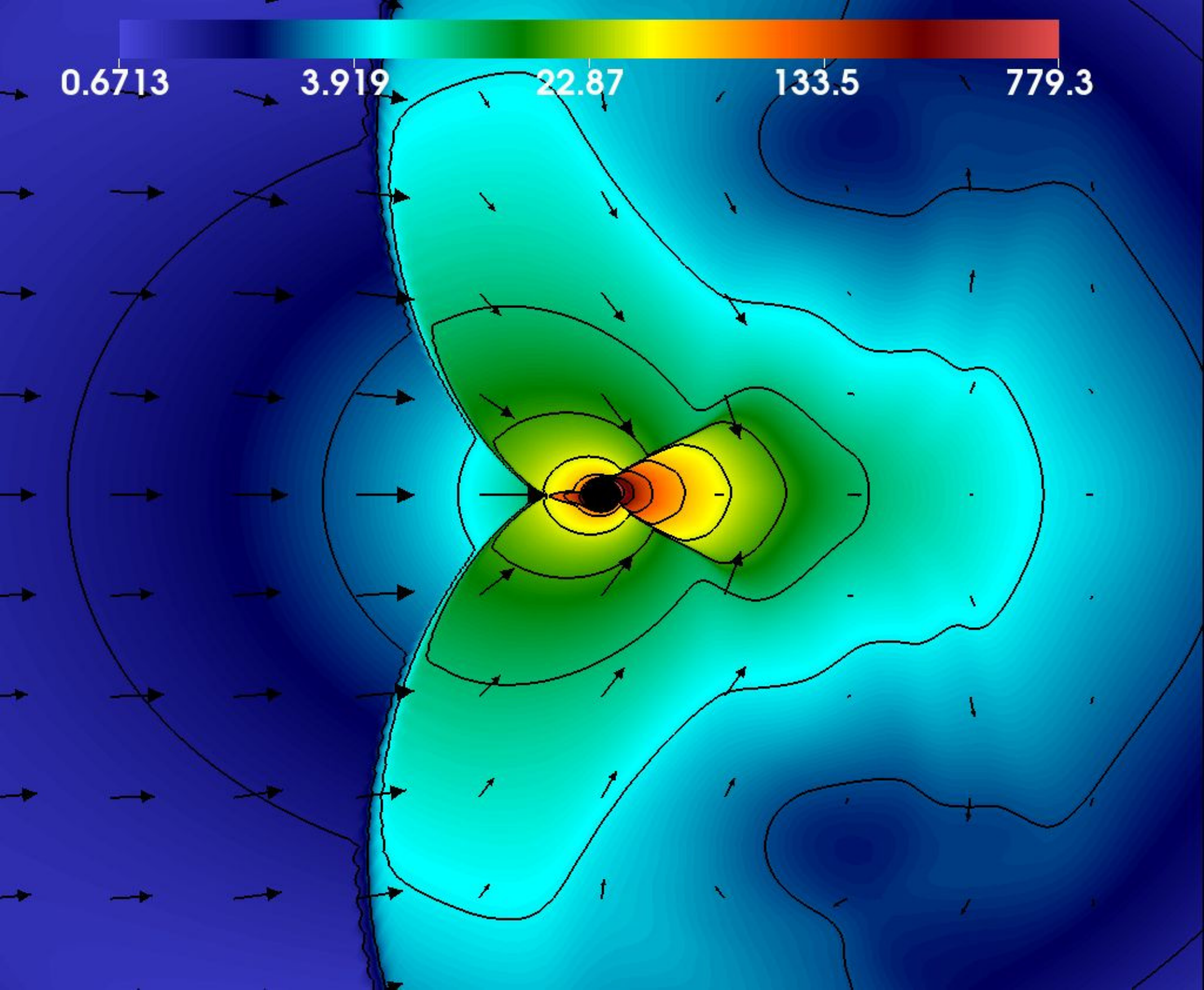,width=4.2cm,height=4.2cm}
  \psfig{file=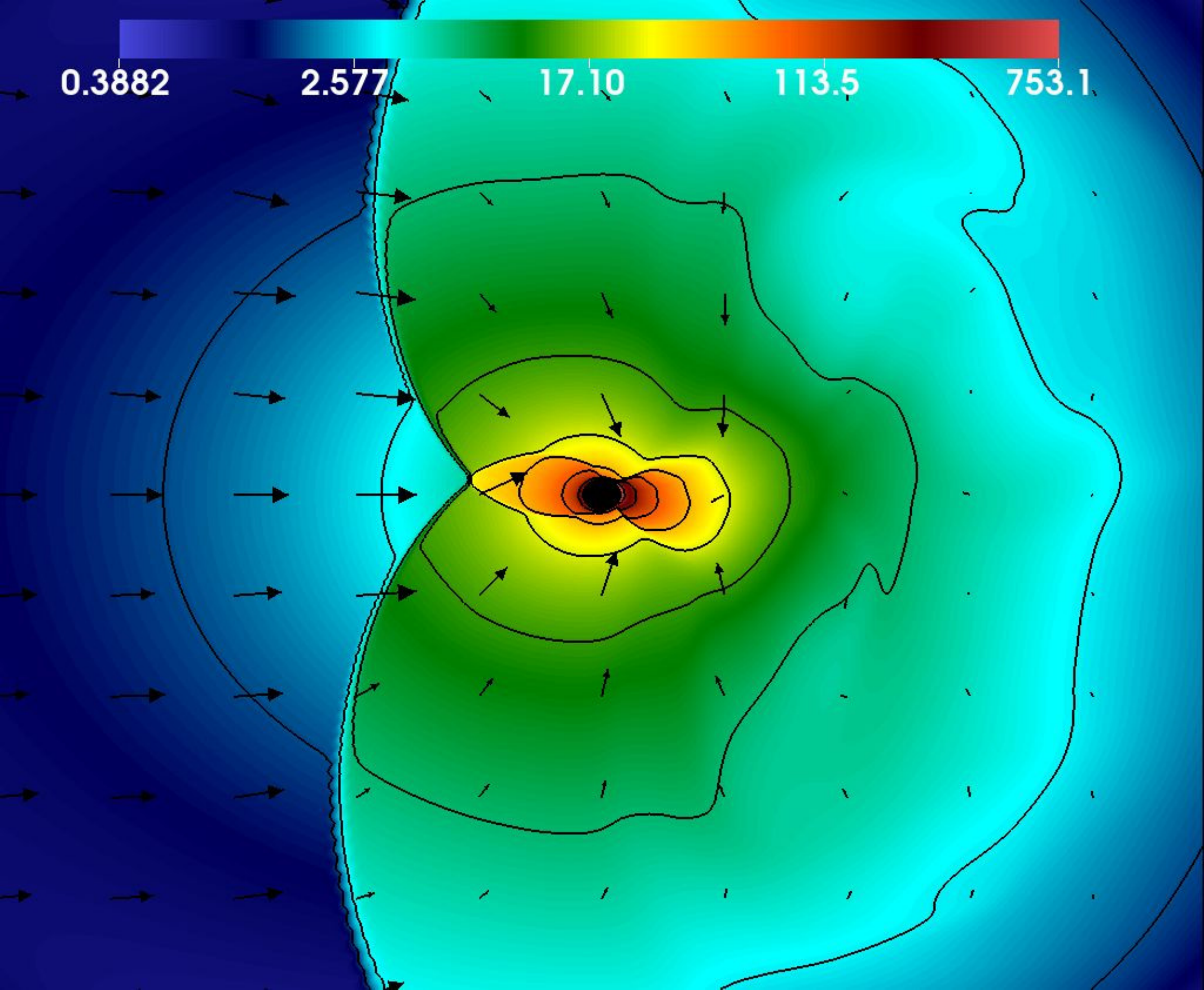,width=4.2cm,height=4.2cm}  
  \psfig{file=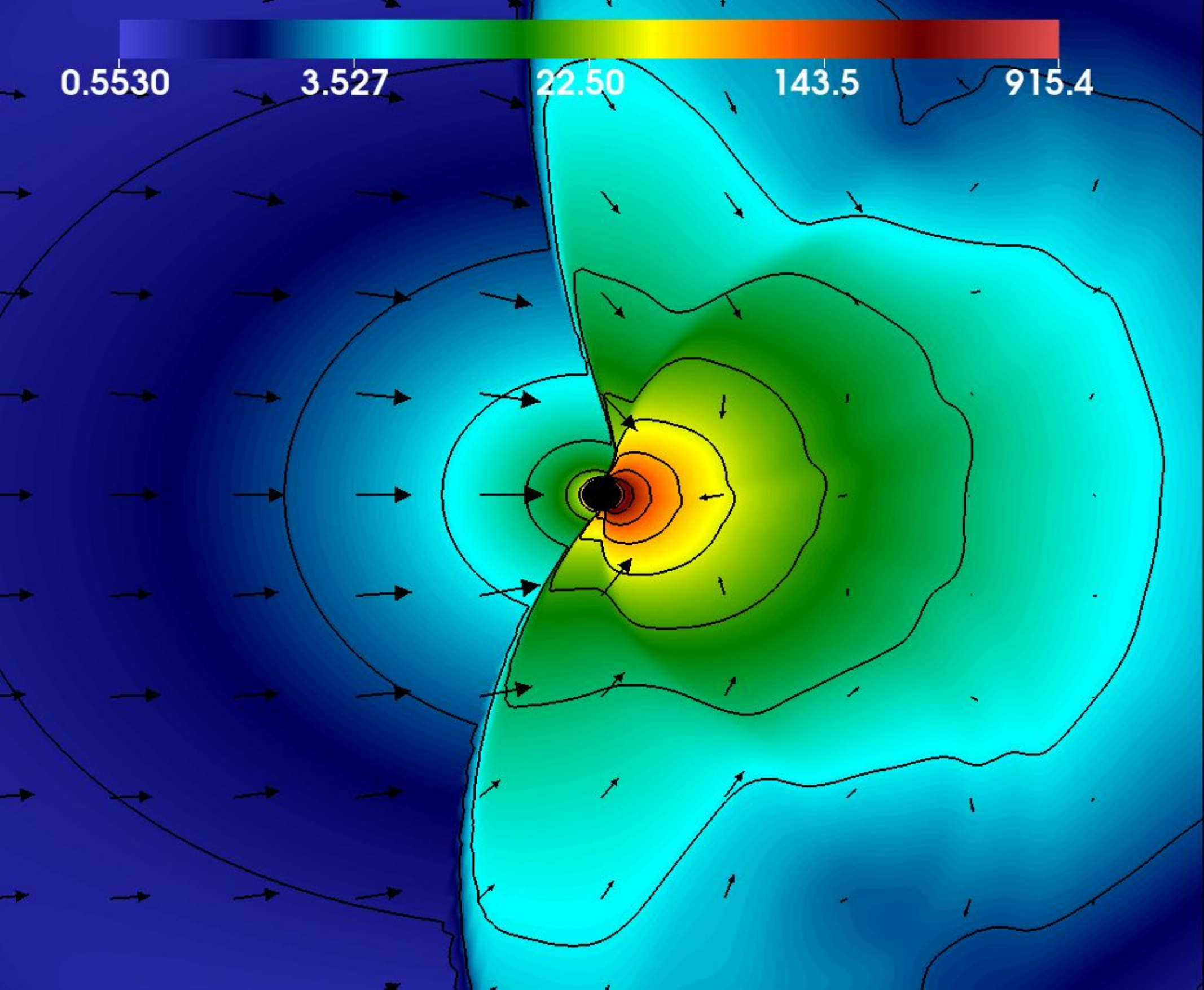,width=4.2cm,height=4.2cm}\\
  \psfig{file=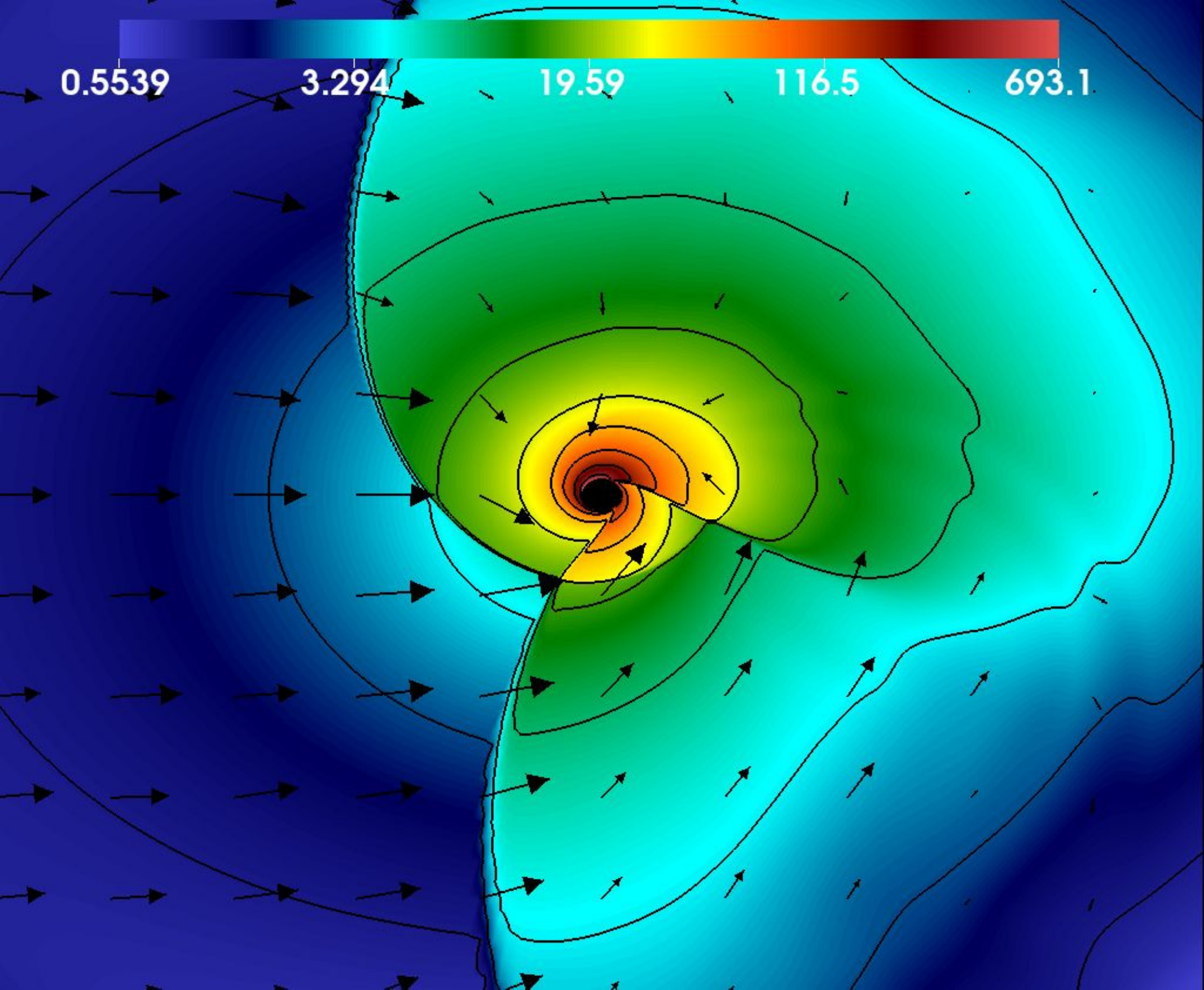,width=4.2cm,height=4.2cm}
  \psfig{file=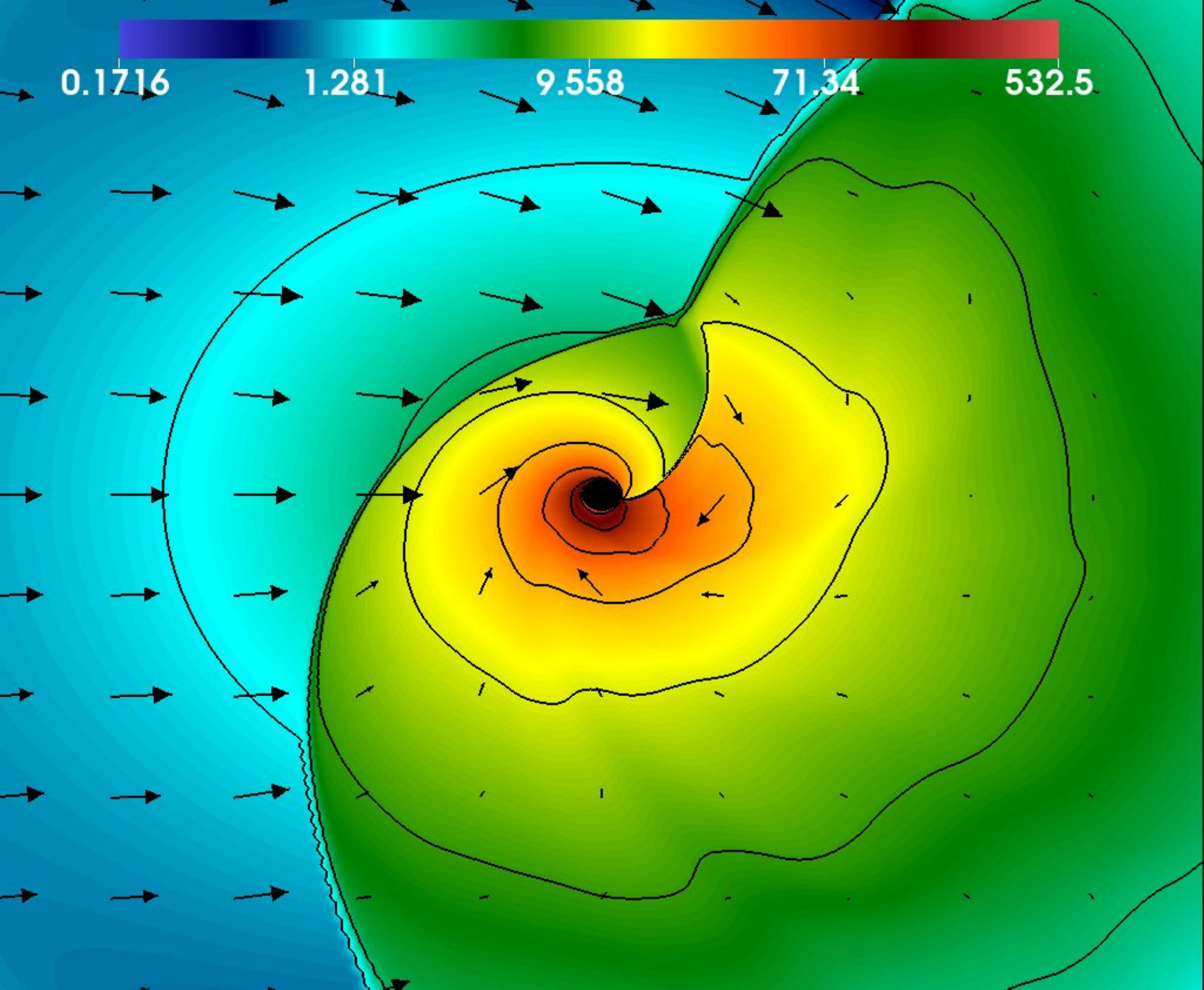,width=4.2cm,height=4.2cm} 
  \psfig{file=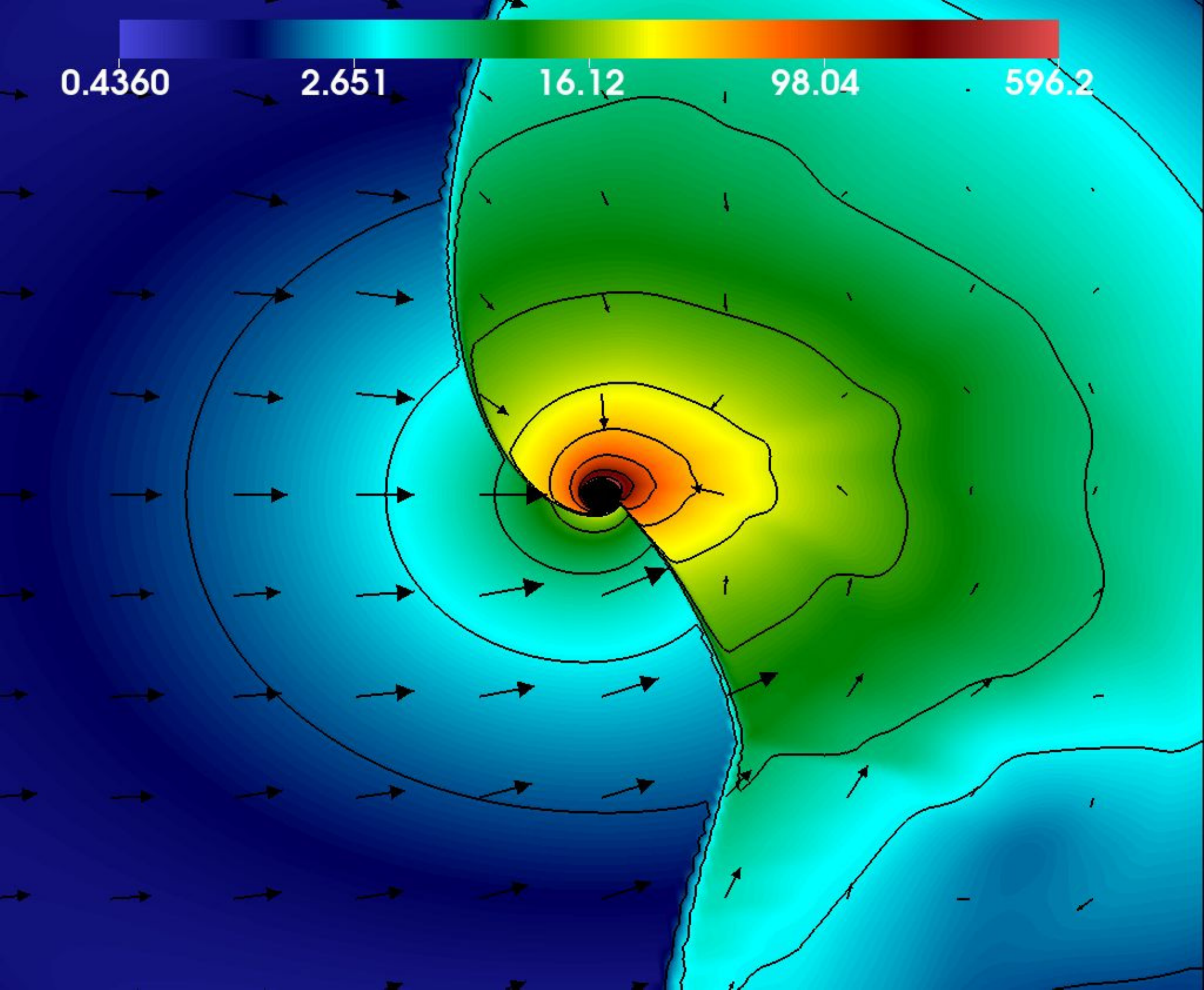,width=4.2cm,height=4.2cm}
  \psfig{file=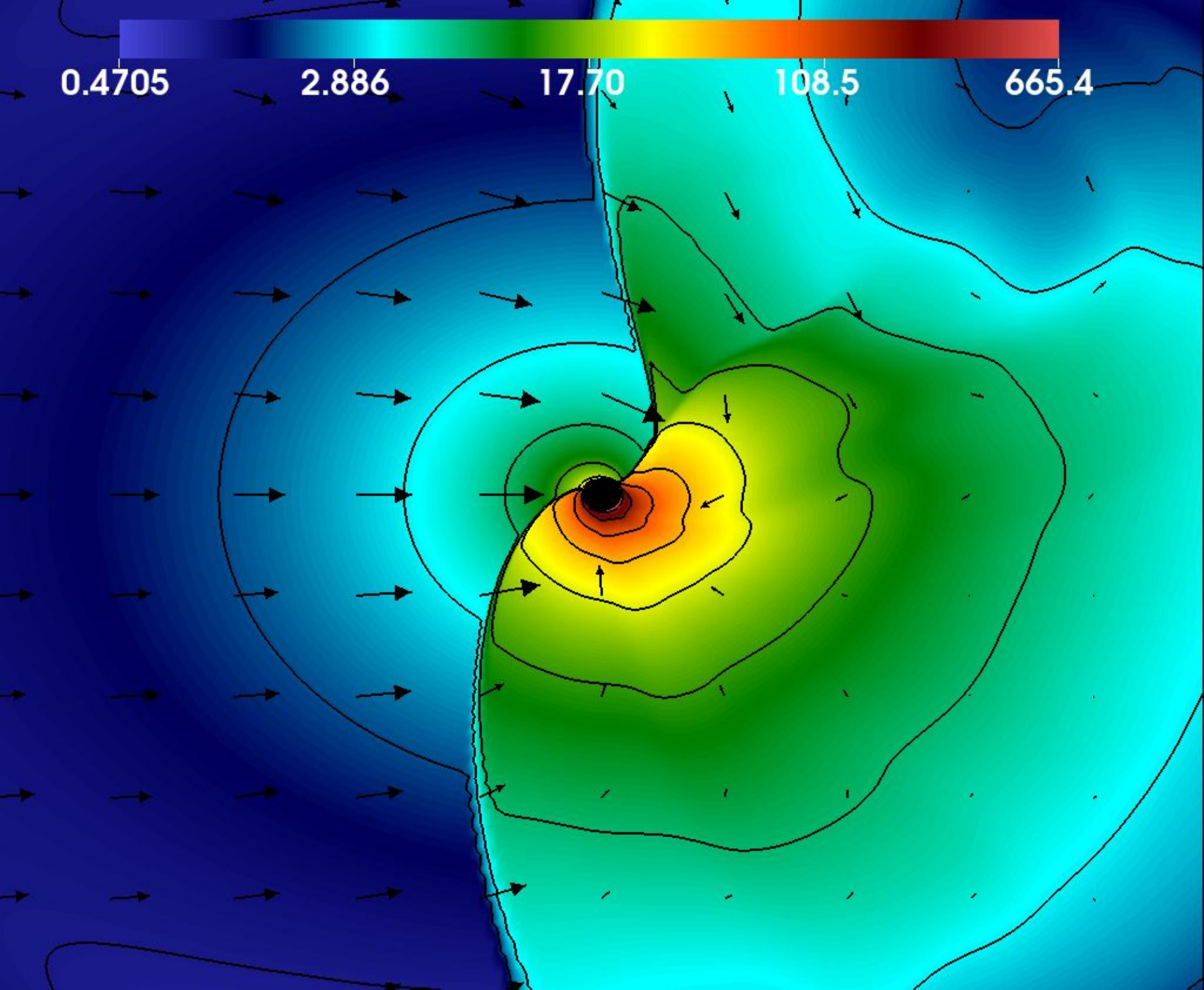,width=4.2cm,height=4.2cm}
\caption{Same as Figs.~\ref{A1_animation} and~\ref{A2_animation}, but for Model $A_4$ ($a=0.5M$, $k\simeq-4.7\times10^{-6}\,M^{-2}$), with snapshots from the beginning of the simulation up to $t=3\times10^{4}M$. The shock structure evolves very similarly to that of Model $A_2$ in Fig.~\ref{A2_animation}: the cone widens, the shock moves upstream, and the flow undergoes persistent flip-flop oscillations, during which the inner part of the shock wraps around the black hole into a spiral.
}
\vspace{1cm}
\label{A4_animation}
\end{figure*}

Fig.~\ref{A4_animation} shows Model $A_4$, in which the spin is lower, $a=0.5M$, while the curvature parameter, $k\simeq-4.7\times10^{-6}\,M^{-2}$ ($\eta\simeq0.30$), is comparable to that of Model $A_2$. In the early stages of the simulation, a shock cone develops in the downstream region under the influence of the black hole's gravitational field. Because $|k|$ is moderate, as in $A_2$, the additional gravitational focusing is much weaker than in Model $A_1$. As the evolution proceeds, the shock is pushed upstream, the opening angle of the cone widens, and the post-shock region remains comparatively diffuse. Toward the later snapshots, the dynamics of the shock around the black hole become dominated by flip-flop oscillations, and during the swings the inner part of the shock wraps around the black hole into a one-armed spiral (second row of Fig.~\ref{A4_animation}). Despite the continuous deformation of the post-shock region due to this instability, the system never transitions to the toroidal, quasi-steady configuration observed in Model $A_1$, and the overall morphology appears almost as a mirror image of the higher-spin configuration of Fig.~\ref{A2_animation}.

Together with the stable cones of the $B$, $C$ and Kerr runs at the same spins, the comparison between Figs.~\ref{A2_animation} and~\ref{A4_animation} shows that $k$, not the spin, plays the dominant role in determining whether a stable cone, an oscillating shock, or a toroidal structure develops. The curvature parameter governs the overall morphology of the flow, namely the confinement and opening angle of the cone and the possibility of forming a disk-like steady state. The spin parameter $a$, whose frame-dragging effect is confined to the inner few gravitational radii, mainly affects the inner dynamics, such as the near-horizon shear, the winding of the inner spiral, and the phase of the oscillations. Consequently, the low-spin Model $A_4$ and the high-spin Model $A_2$, which have similar values of $\eta$, display remarkably similar overall behavior, and the same holds for the pair $A_1$ and $A_5$ discussed below. These simulations thus indicate that the physical mechanisms operating around the black hole are primarily controlled by the curvature parameter of the spacetime. The large-scale motion of the shock and the inner, spin-dependent dynamics also operate on very different time scales. For small-amplitude oscillations of the accretion shock, the period is comparable to the Keplerian period at the accretion radius~\cite{Blondin2009}, which is of order $10^{3}M$ here, whereas the orbital period of test particles at $r\simeq6M$ is only of order $10^{2}M$~\cite{ZULQARNAIN2026117449}. The PSD analysis below shows which of these time scales actually appears in the variability of the flow.

\begin{figure*}[!ht]
  \vspace{1cm}
  \center
  \psfig{file=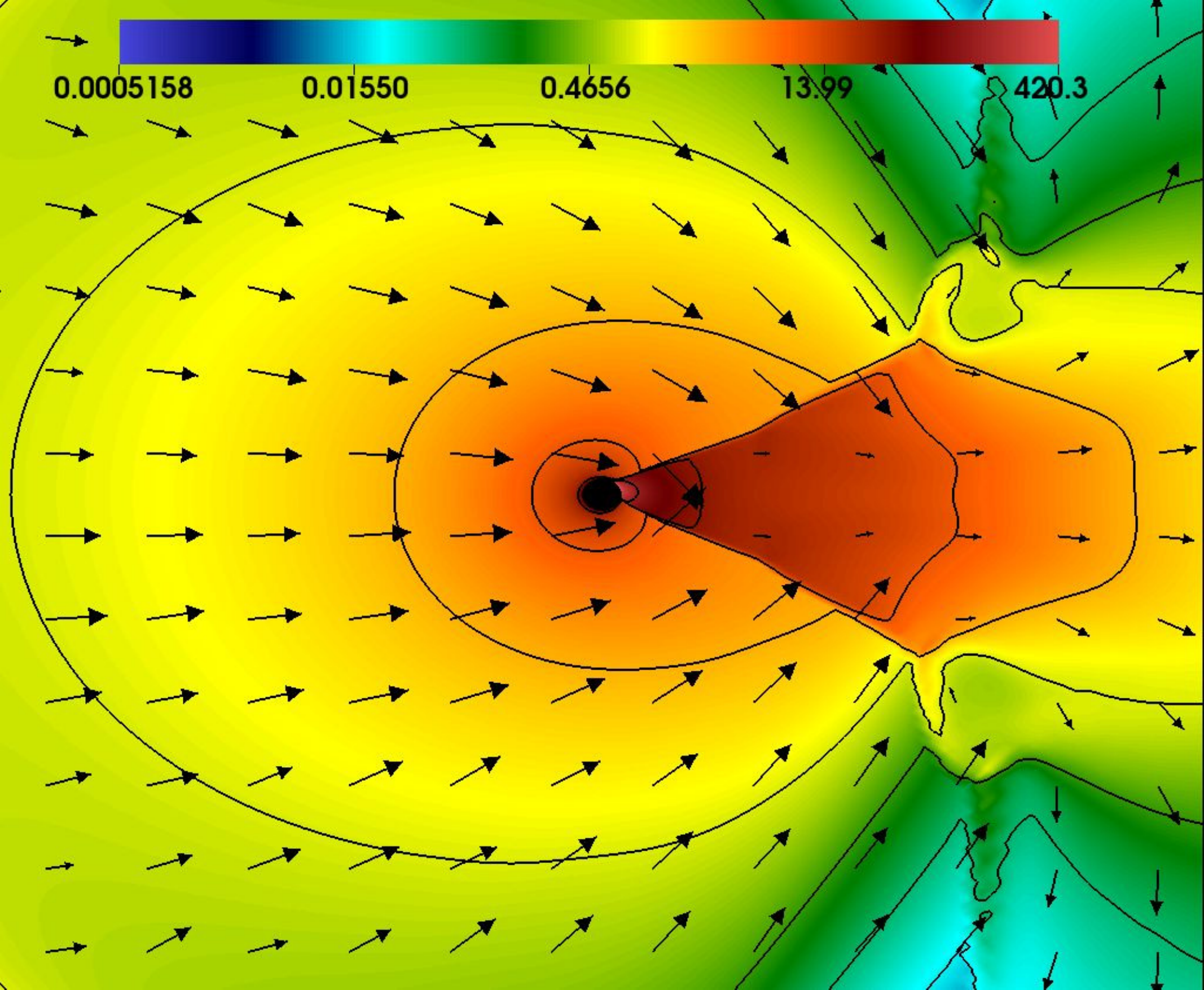,width=4.2cm,height=4.2cm} 
  \psfig{file=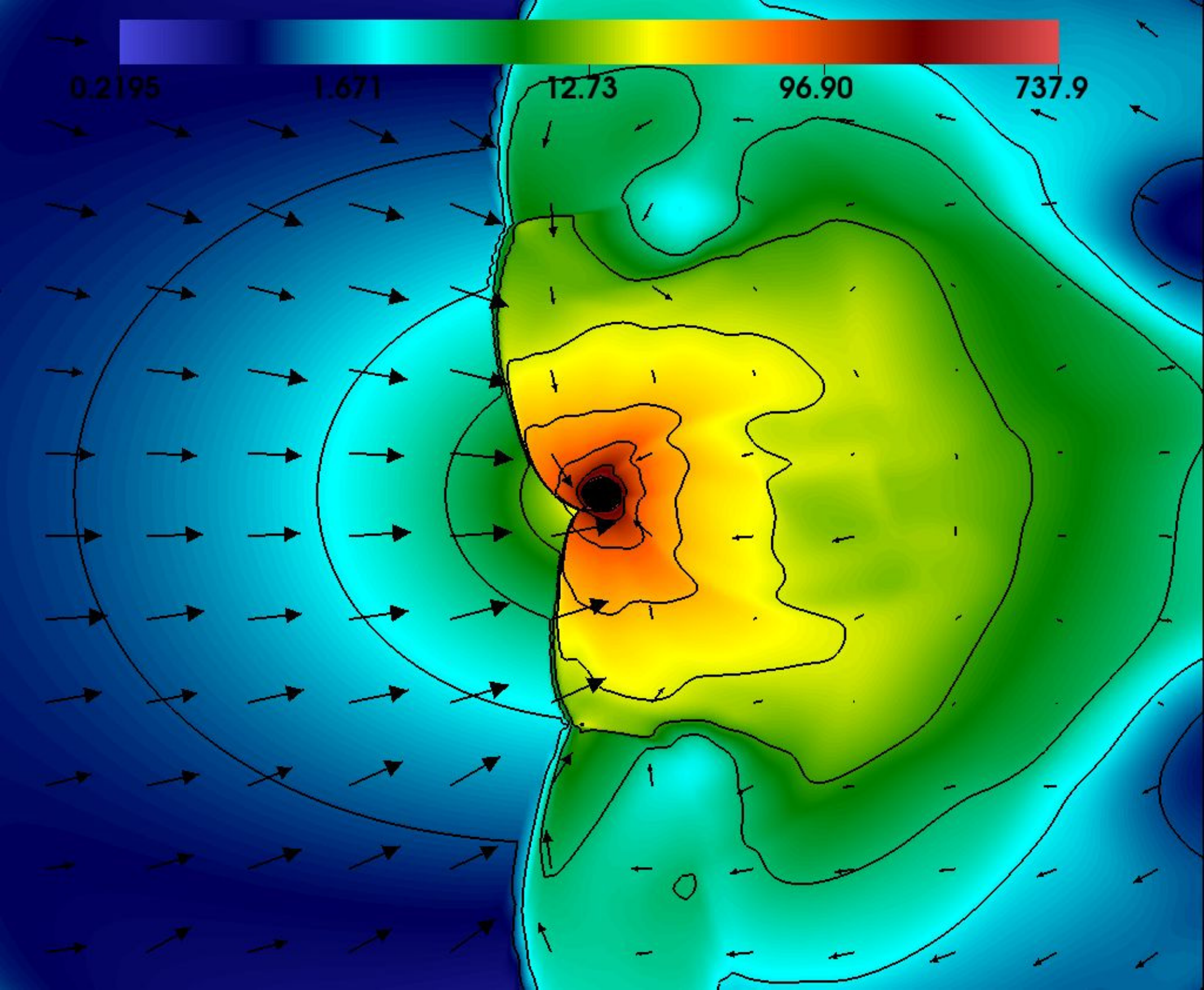,width=4.2cm,height=4.2cm}
  \psfig{file=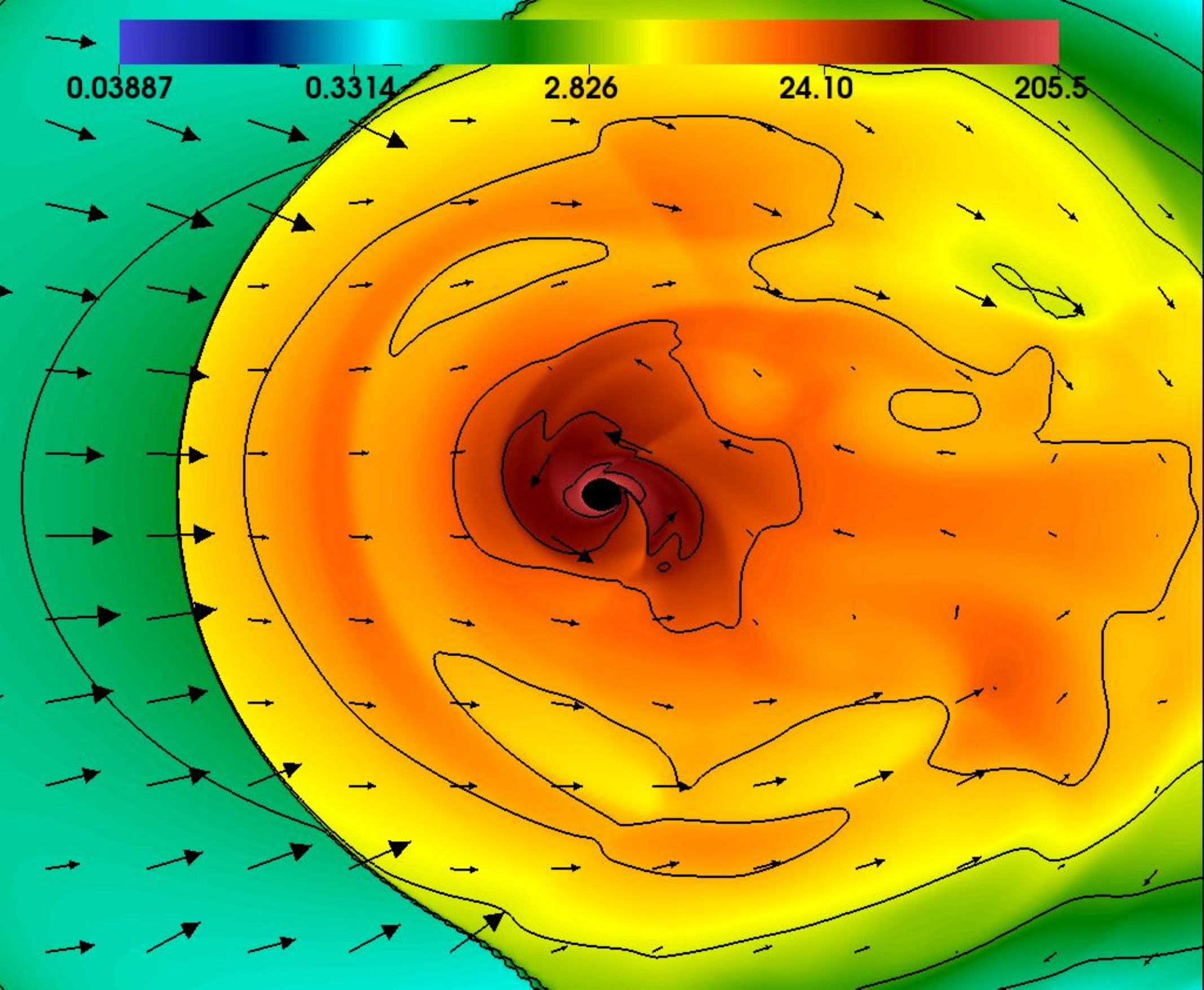,width=4.2cm,height=4.2cm} 
  \psfig{file=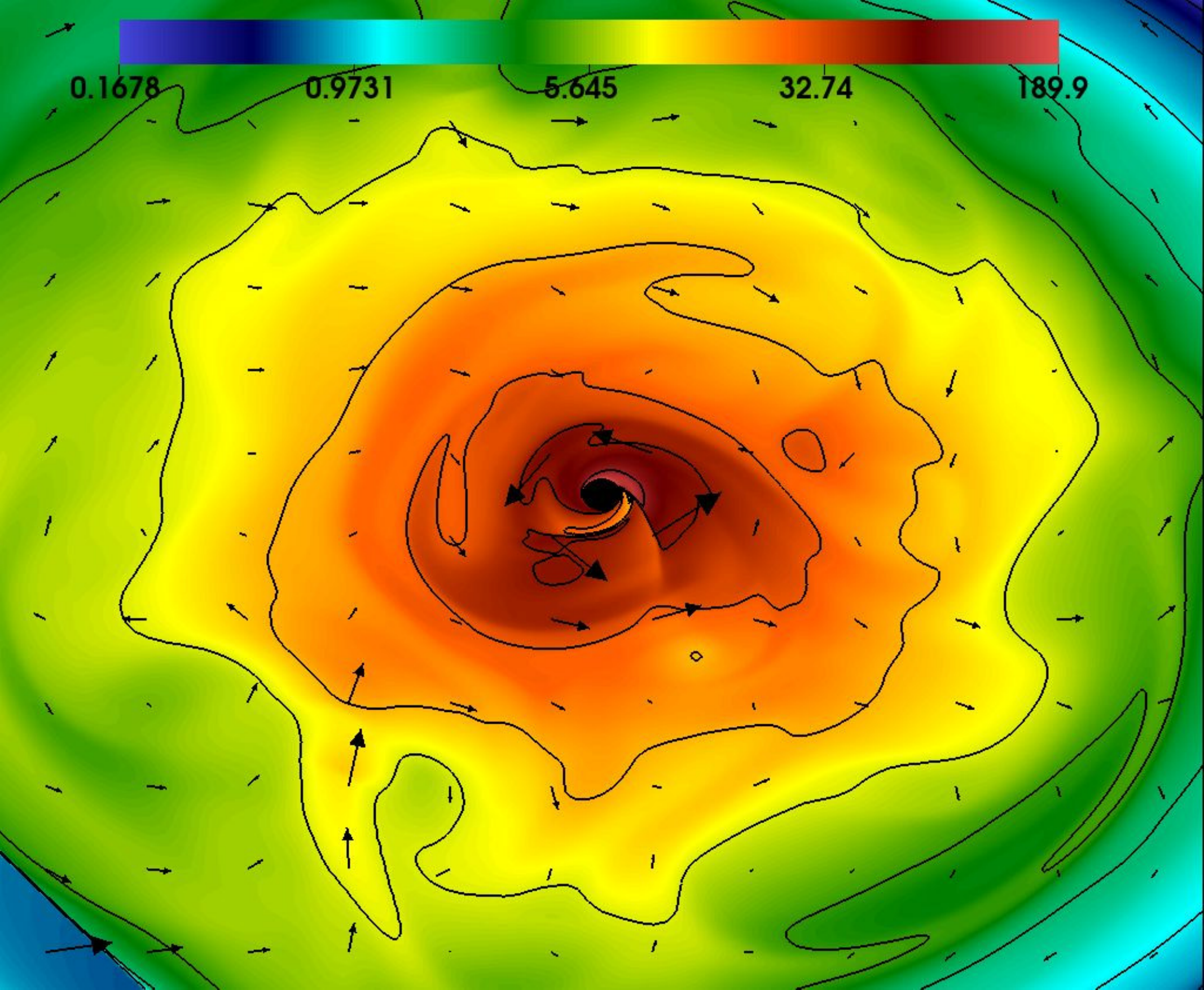,width=4.2cm,height=4.2cm}\\ 
  \psfig{file=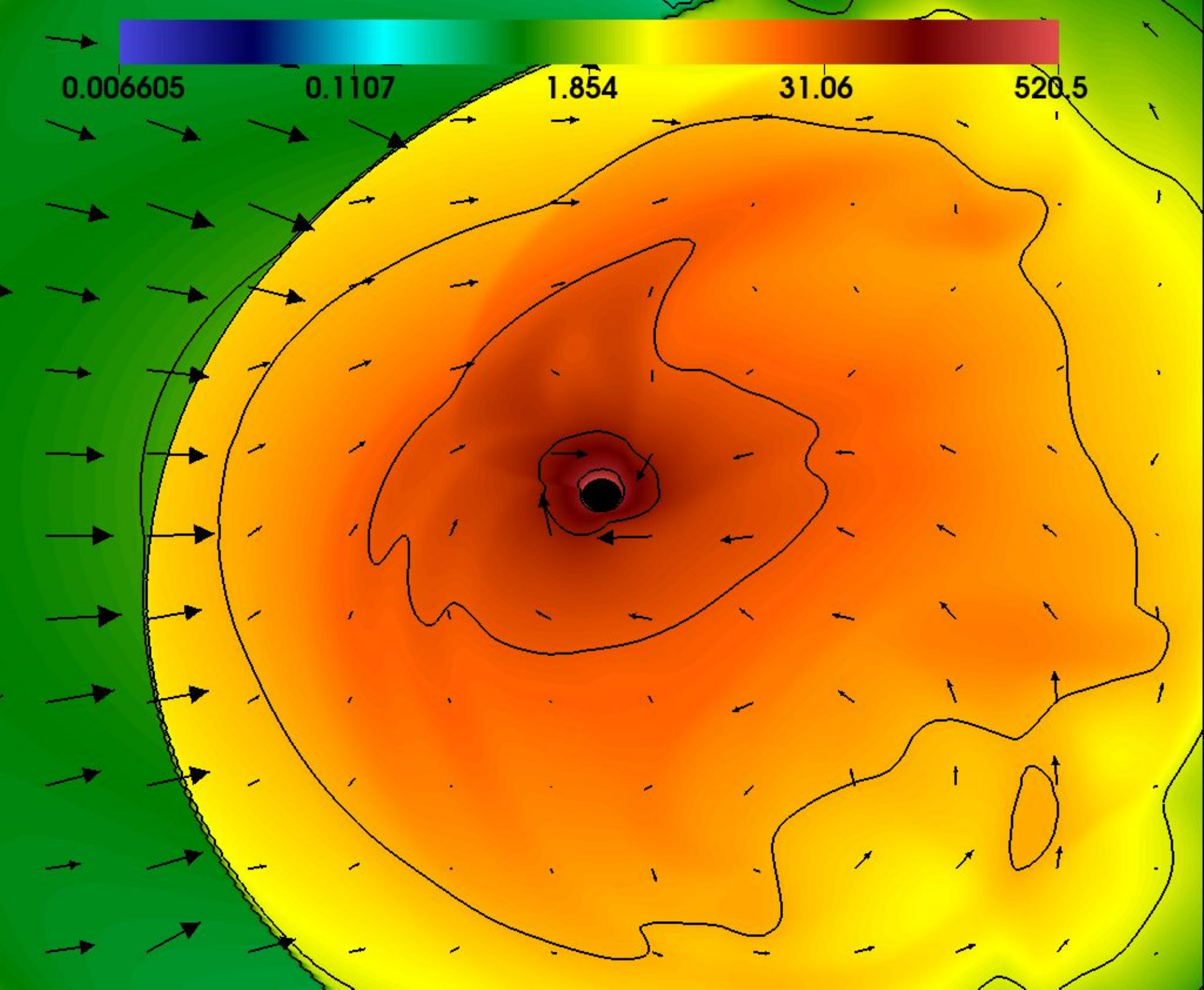,width=4.2cm,height=4.2cm}   
  \psfig{file=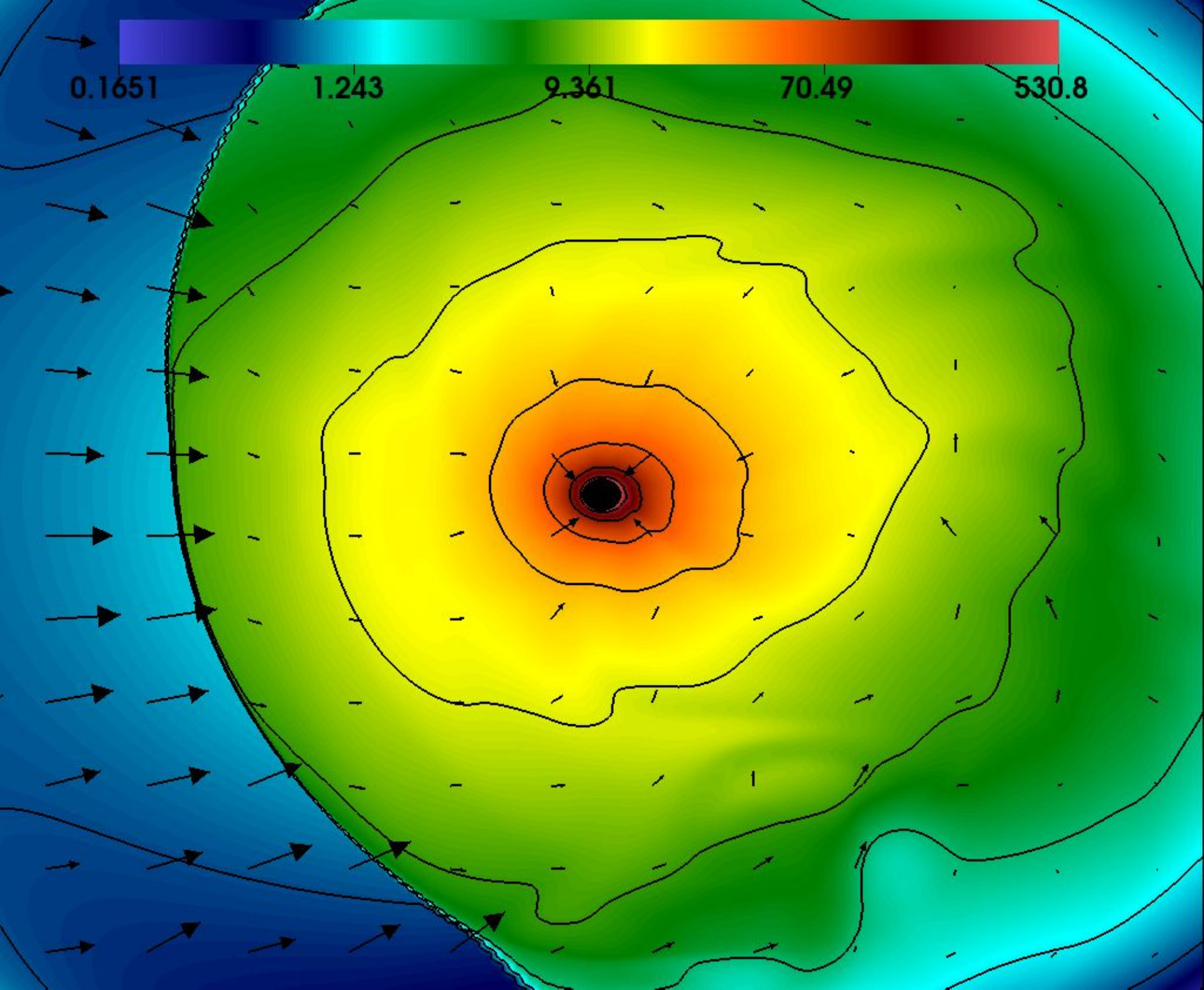,width=4.2cm,height=4.2cm} 
  \psfig{file=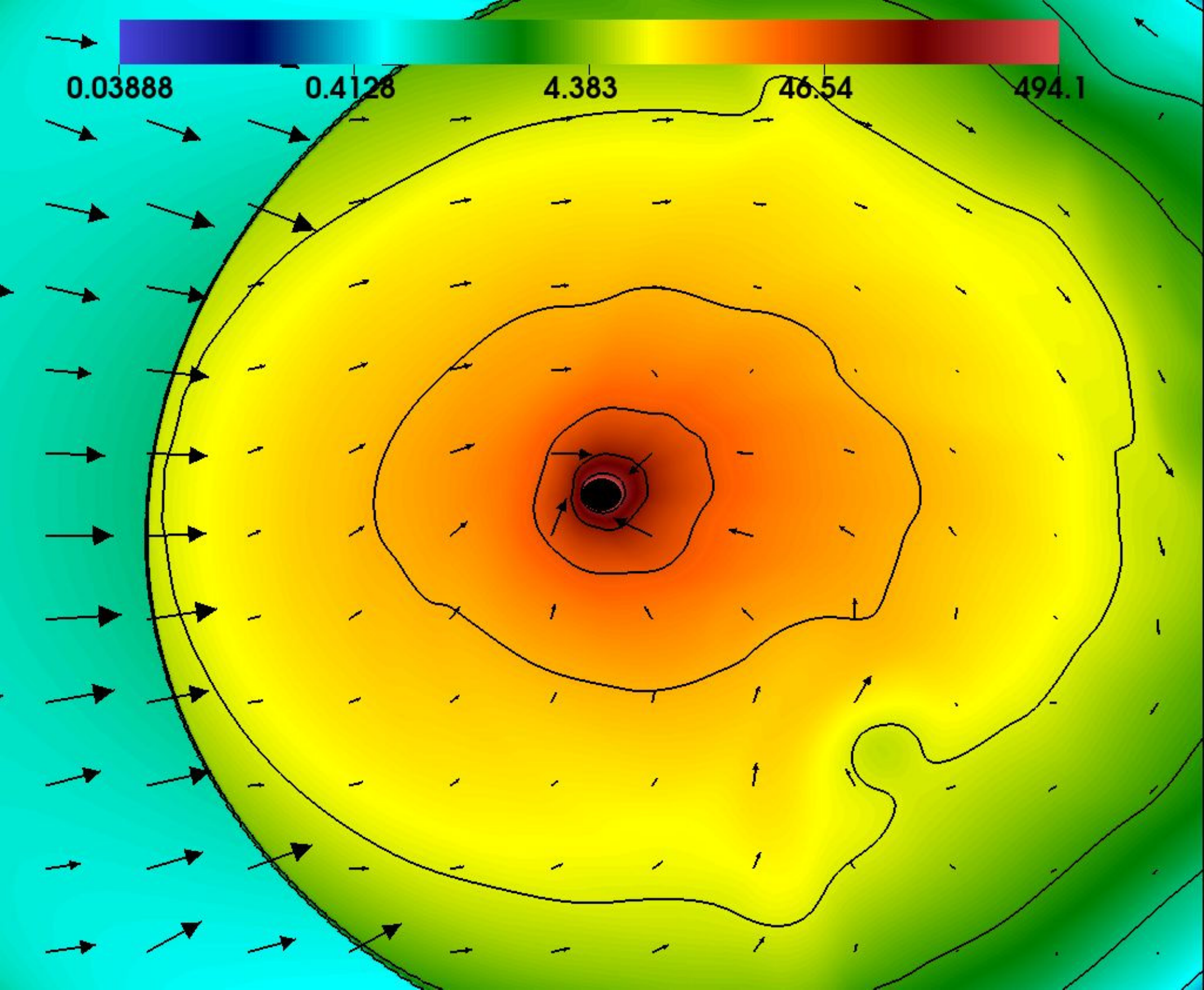,width=4.2cm,height=4.2cm}
  \psfig{file=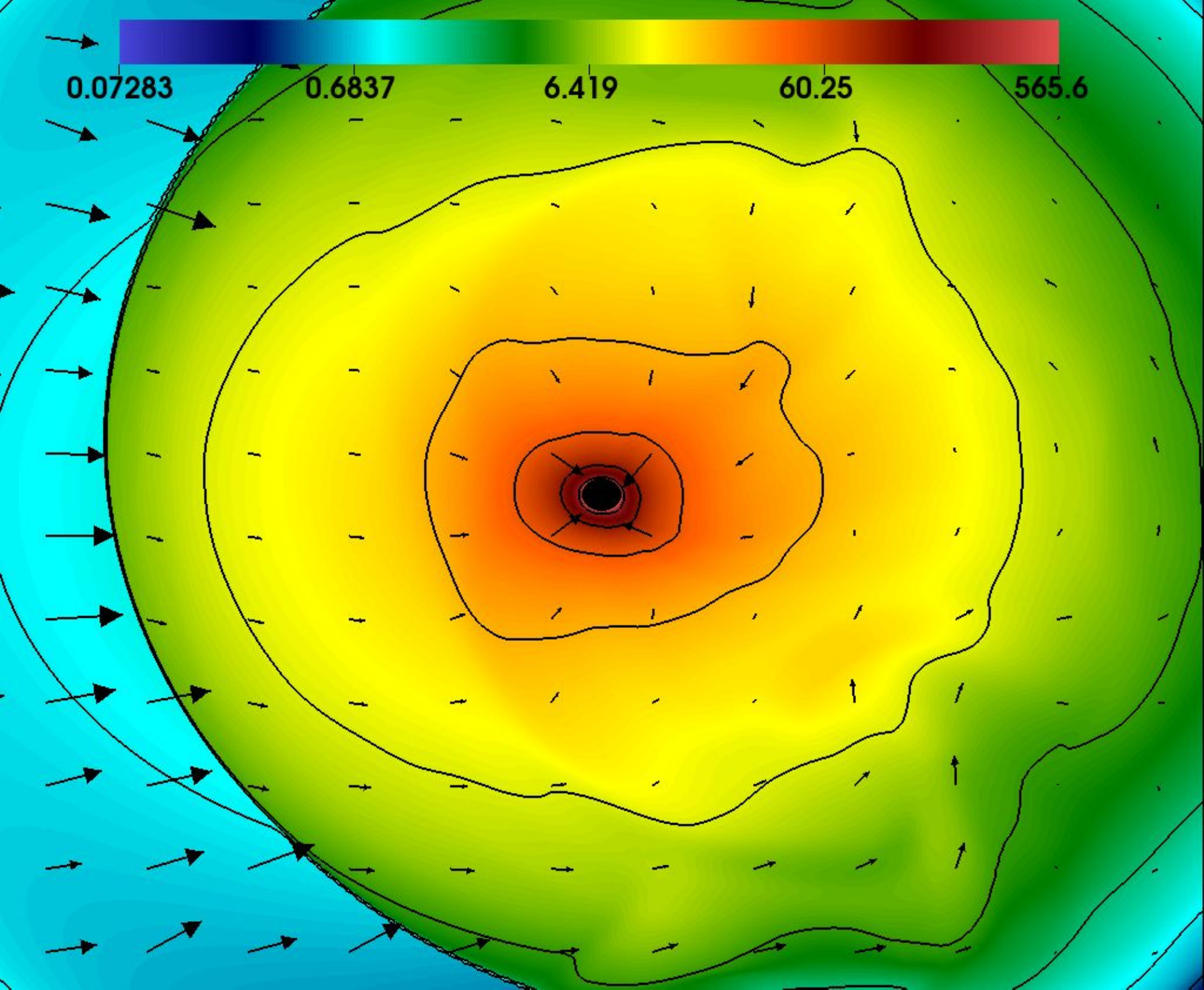,width=4.2cm,height=4.2cm}   
\caption{Same as Fig.~\ref{A1_animation}, but for Model $A_5$ ($a=0.5M$, $k\simeq-2.4\times10^{-5}\,M^{-2}$). After a short flip-flop phase, a plasma structure forms around the Kerr--Weyl black hole and settles into a quasi-steady toroidal configuration at approximately $t=3000M$, after which it oscillates around the black hole. The final snapshot corresponds to $t=3\times10^{4}M$.
}
\vspace{1cm}
\label{A5_animation}
\end{figure*}

For the moderately rotating black hole with $a=0.5M$, Fig.~\ref{A5_animation} shows the quasi-steady structure that emerges in Model $A_5$ ($k\simeq-2.4\times10^{-5}\,M^{-2}$, $\eta\simeq1.6$), whose curvature parameter is comparable to that of Model $A_1$. With a comparable value of $\eta$, the accretion morphology of $A_5$ is almost identical to that of $A_1$, independent of the black hole spin. The strongly negative value of $k$ makes the spacetime AdS-like, which enhances the gravitational attraction at large radii and deepens the effective potential. As in $A_1$, a well-defined shock cone first forms in the downstream region, on the side opposite to the incoming flow. As time progresses, the cone becomes unstable, the shock is pushed upstream, and the flow undergoes flip-flop oscillations. Around $t\simeq3000M$, the highly unstable shock evolves into a quasi-steady toroidal, or plasma-disk, structure surrounding the black hole. The newly formed configuration remains dynamically active, with strong continuous oscillations, up to the final snapshot at $t=3\times10^{4}M$.

As in Model $A_1$, the curvature-induced pull exceeds the Newtonian one inside the capture region ($\eta>1$), so the overall morphology of the accretion structure is primarily governed by $k$ rather than by the spin. The differences between $A_5$ and $A_1$ are confined to the inner region, where frame dragging is significant. They appear in the structure of the spiral shock waves within the strong-gravity region and in the time needed to reach the quasi-steady state, which is shorter in $A_5$ ($t\simeq3000M$) than in $A_1$ ($t\simeq1.5\times10^{4}M$). In all aggressive models, the variability is therefore produced by curvature-induced flip-flop and acoustic instabilities of the shocked flow, whose imprint on the accretion rate and its power spectrum is analyzed below.

\subsection{Temporal Evolution of the Mass Accretion Rate}
\label{mass_acc} 

The density maps show how the curvature parameter changes the global morphology of the flow. The mass accretion rate $dM/dt$ measures how these changes affect the supply of gas to the black hole, and its time dependence carries the variability that is later analyzed with the PSD. We compute $dM/dt$ at two radii. The inner radius, $r=2.3M$, lies just outside the outer horizon. It is practically at the prograde ISCO for $a=0.9M$ ($r_{\rm ISCO}\simeq2.32M$) and well inside it for $a=0.5M$ ($r_{\rm ISCO}\simeq4.23M$), so the gas there is plunging toward the horizon. The outer radius, $r=6.11M$, lies outside the ISCO for both spins, in the inner part of the flow where stable circular orbits exist. The ISCO of the Kerr--Weyl black hole and its dependence on the Weyl-gravity parameters are discussed in detail in Ref.~\cite{ZULQARNAIN2026117449}. For the values of $k$ used here, the curvature term shifts these ISCO radii by less than $0.01M$, so both radii have the same physical meaning in the Kerr and Kerr--Weyl runs. The rates are given in arbitrary units and are always compared between models at the same radius.

\begin{figure*}[!ht]
  \vspace{1cm}
  \center
  \psfig{file=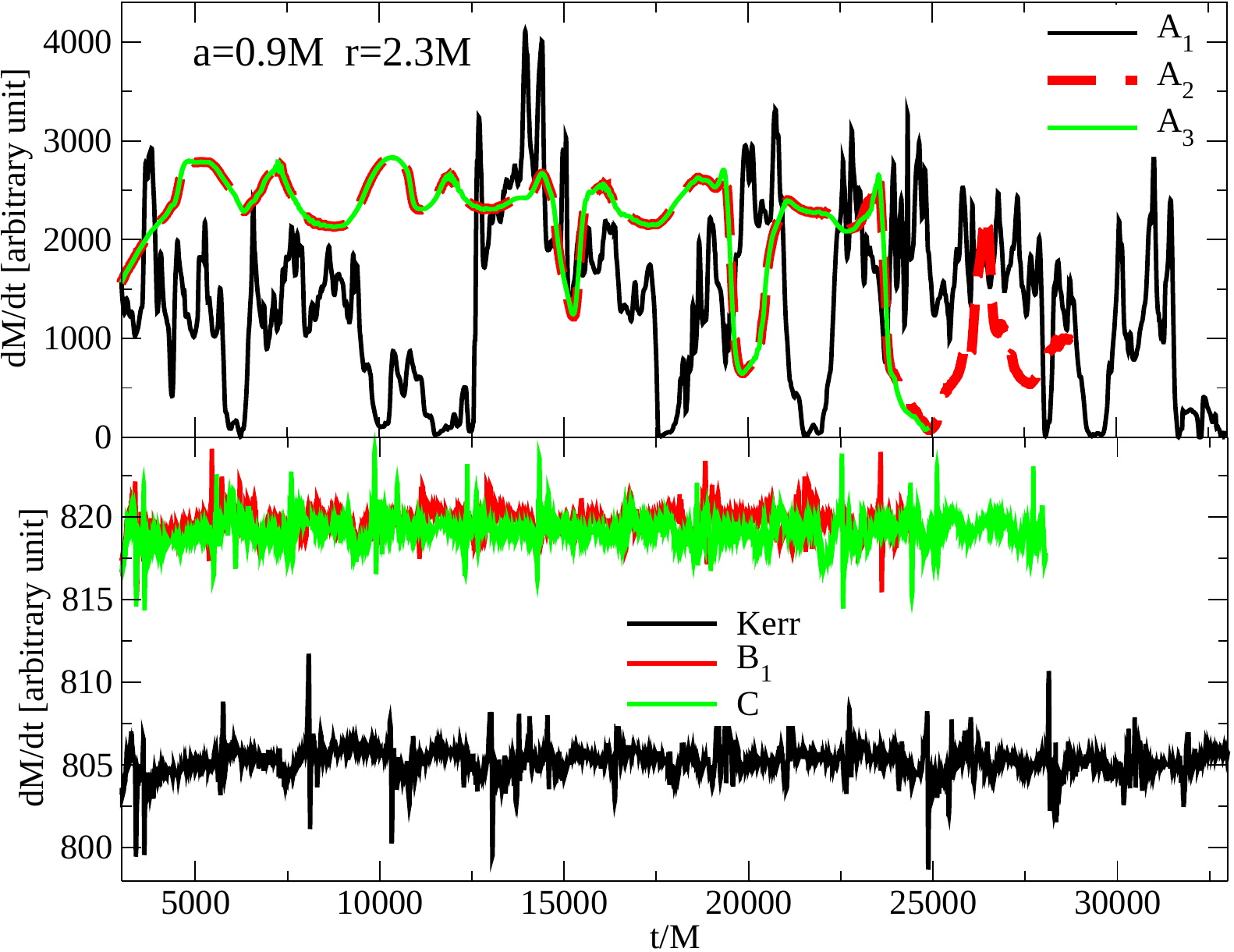,width=8.0cm,height=10cm}
  \psfig{file=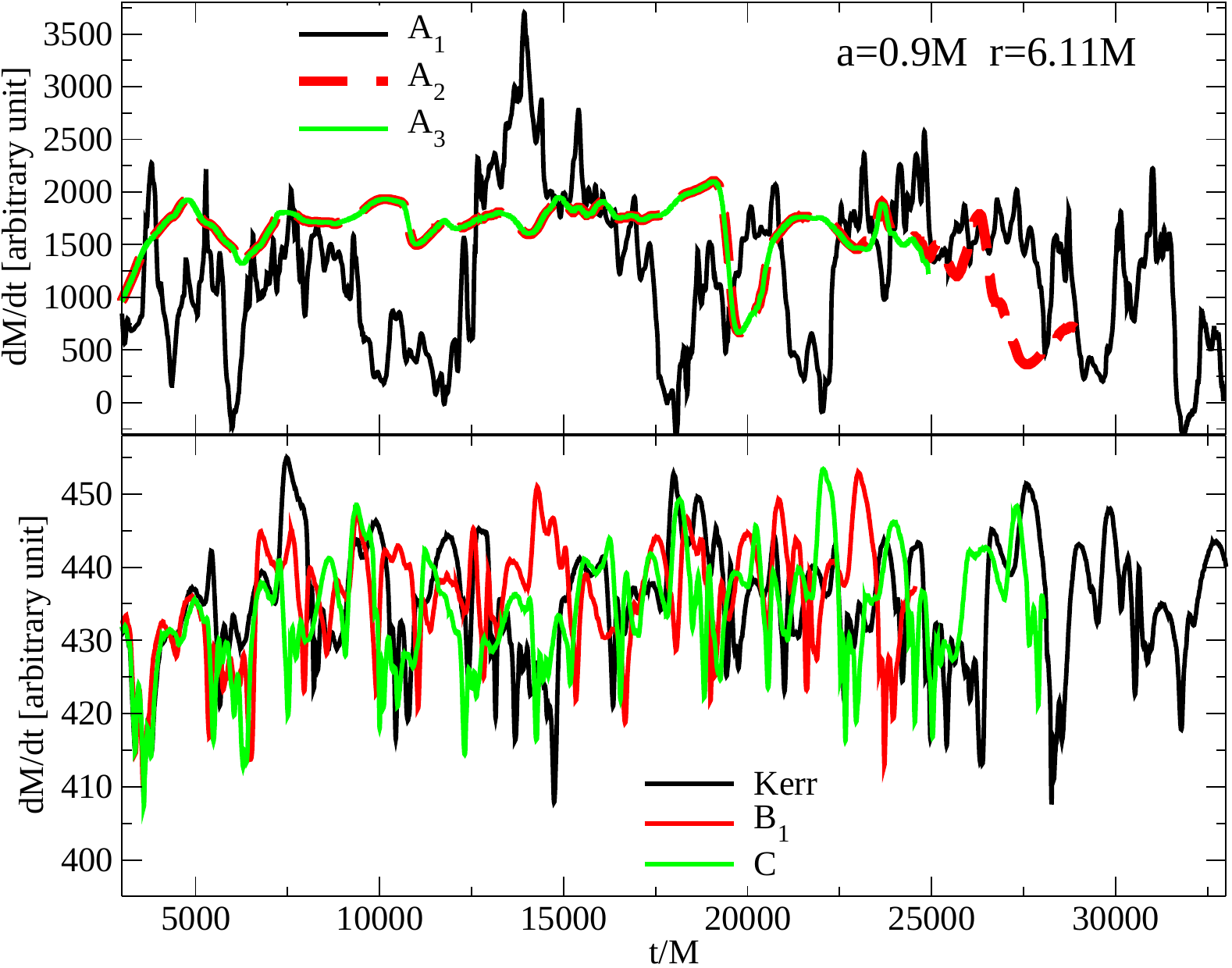,width=8.0cm,height=10cm}  
     \caption{Time evolution of the mass accretion rate for the Kerr--Weyl black hole with spin parameter $a=0.9M$. The left panels show $dM/dt$ at $r=2.3M$, close to the event horizon and practically at the ISCO, and the right panels at $r=6.11M$, outside the ISCO in the inner flow. In each set of plots, the upper panels show the aggressive models, illustrating the evolution of the accretion rate in strongly curved spacetimes, while the lower panels show the mild-deviation and near-Kerr models together with the Kerr solution. The rates are given in arbitrary units; note the different vertical scales of the upper and lower panels.
}
\vspace{1cm}
\label{mass_accc_a09_1}
\end{figure*}

Fig.~\ref{mass_accc_a09_1} shows $dM/dt$ for the rapidly rotating black hole with $a=0.9M$. It is evident that the curvature parameter $k$, and consequently the Kerr--Weyl coefficients, strongly  influences the accretion process. In the aggressive models ($A_1$, $A_2$ and $A_3$), the accretion rate near the horizon ($r=2.3M$) is strongly variable, and its time average is about $2$--$3$ times higher than in the Kerr run with the same spin. This confirms that the deeper potential and the suppressed downstream escape of the shocked gas, discussed above, increase the amount of gas reaching the black hole. Model $A_1$ ($\eta\simeq1.9$), with the most negative curvature parameter, shows the most irregular behavior. Its accretion rate varies between almost zero and about five times the Kerr value, and it repeatedly drops to nearly zero (e.g., at $t\simeq6\times10^{3}M$ and $1.8\times10^{4}M$). At the same times, the rate at $r=6.11M$ becomes slightly negative, which indicates episodes of net outflow. These rapid cycles of matter infall and ejection reflect the flip-flop phase and the subsequent oscillations of the torus. Because the torus temporarily stores part of the captured gas, the time-averaged rate of $A_1$ is lower than that of $A_2$ and $A_3$. In $A_2$ and $A_3$ ($\eta\simeq0.19$), the rate oscillates more regularly, by about $\pm15\%$ around its mean, with occasional deep minima associated with large swings of the shock (e.g., at $t\simeq1.5\times10^{4}M$ and $2\times10^{4}M$). The dominant modulation period is $\approx2\times10^{3}M$, of the same order as the Keplerian period at the accretion radius, $2\pi/\Omega_{\rm K}(R_{\rm acc})\simeq1.5\times10^{3}M$, where $\Omega_{\rm K}^{2}\simeq\widetilde{M}/R_{\rm acc}^{3}+|k|$ follows from Eq.~(\ref{g_r}). As the PSD analysis below shows, this modulation is the strongest harmonic of a slower fundamental oscillation with a period of $\approx4\times10^{3}M$, about three Keplerian periods. The curves of $A_2$ and $A_3$ are almost indistinguishable up to $t\simeq2.3\times10^{4}M$, even though their values of $\gamma$ have opposite signs. This is expected from the structure of the metric, and it shows that the flip-flop is a deterministic feature of the flow rather than a result of numerical noise.  At later times, small differences between the two runs grow, as expected for an unstable, nonlinear flow.

At $r=6.11M$, the behavior of the aggressive models remains strongly variable, and their deepest minima occur at nearly the same times as at $r=2.3M$. The variability is therefore imposed by the large-scale dynamics of the shock and transmitted through the inner flow, rather than generated locally. This is consistent with the appearance of the same PSD peaks at both radii, discussed below. In $A_2$ and $A_3$, the modulation amplitude at $r=6.11M$ is about half of that at $r=2.3M$ ($\approx\pm8\%$ around the mean), so the modulation grows as the gas converges toward the horizon. In contrast, the lower panels of Fig.~\ref{mass_accc_a09_1}, which correspond to the mild-deviation model ($B_1$) and the near-Kerr model ($C$), show that the accretion rates remain nearly steady, fluctuating only slightly around a constant mean value. These models have $|k|\lesssim10^{-8}\,M^{-2}$ ($\eta<10^{-3}$), so their shock cone remains stable and no strong nonlinear oscillations develop. In these models, the relative fluctuations are larger at $r=6.11M$ ($\approx\pm5\%$) than at $r=2.3M$ ($\lesssim\pm0.5\%$). The weak perturbations of the stable cone are therefore damped before the gas reaches the horizon. At $r=6.11M$, the $B_1$, $C$ and Kerr curves are statistically indistinguishable. At $r=2.3M$, the $B_1$ and $C$ rates lie about $2\%$ above the Kerr value. This offset is the same for $B_1$ ($k<0$) and $C$ ($k>0$), although the curvature term acts in opposite directions in the two cases, and $|k|r^{2}\lesssim10^{-7}$ at this radius. The offset therefore cannot be a direct effect of the curvature term and is most likely of numerical origin. We regard it as an estimate of the systematic uncertainty, about $2\%$, in the comparison between the Kerr and Kerr--Weyl accretion rates.

It is instructive to compare this variability with the characteristic frequencies of test-particle motion in the Kerr--Weyl metric. These frequencies, together with the periastron and Lense--Thirring precession frequencies, were derived in Ref.~\cite{ZULQARNAIN2026117449} and compared there with planar accretion onto the same black hole; here we extend this comparison to BHL accretion. For circular equatorial geodesics at $r=6.11M$, the orbital period is $\approx100M$ and the radial epicyclic period is $\approx153M$ for $a=0.9M$ ($\approx98M$ and $\approx191M$ for $a=0.5M$). Even in $A_1$ and $A_5$, the curvature term changes the orbital frequency by less than $0.3\%$ and the radial epicyclic frequency by about $2\%$, because $|k|r^{2}\sim10^{-3}$ at this radius. The strong variability of the aggressive models is therefore not caused by a modification of local test-particle motion. Its dominant time scale, of order $10^{3}M$, is set by the dynamics of the shock at $r\sim R_{\rm acc}$, where $k$ is important. This time scale is about an order of magnitude longer than the local orbital and epicyclic periods. It is also several times longer than the slowest in-plane geodesic period, that of periastron precession, $(\nu_\phi-\nu_r)^{-1}\approx290M$ for $a=0.9M$ and $\approx200M$ for $a=0.5M$. Local orbital and epicyclic motion of the inner flow can therefore contribute only at higher frequencies, and the PSD analysis below examines both frequency ranges. Vertical epicyclic oscillations and the associated nodal (Lense--Thirring) precession, although present for test particles in this spacetime~\cite{ZULQARNAIN2026117449}, are absent in the present equatorial simulations by construction.

\begin{figure*}[!ht]
  \vspace{1cm}
  \center
  \psfig{file=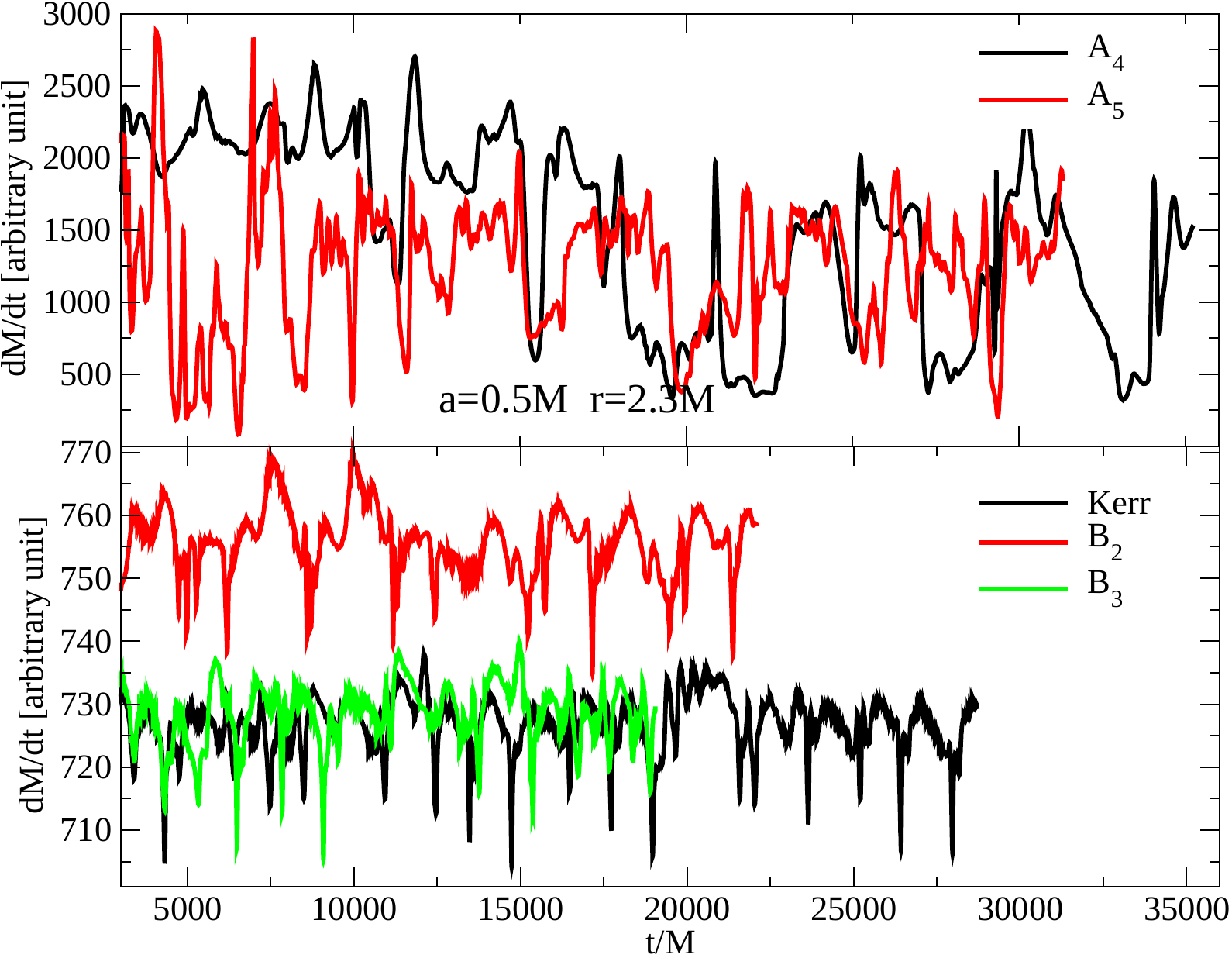,width=8.0cm,height=10cm}
  \psfig{file=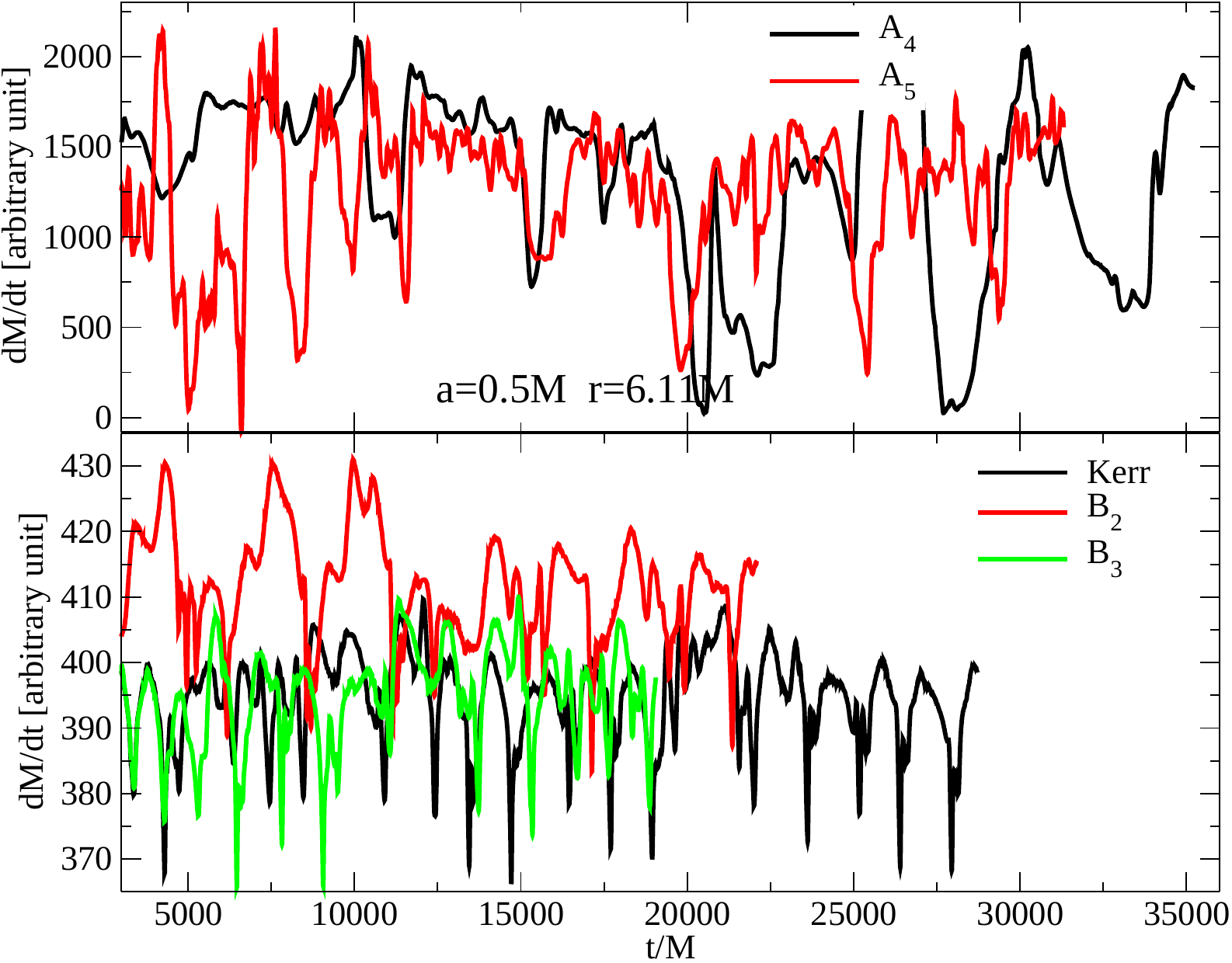,width=8.0cm,height=10cm}    
    \caption{Same as Fig.~\ref{mass_accc_a09_1}, but for the Kerr--Weyl black hole with spin parameter $a=0.5M$, for which $r=2.3M$ lies inside the ISCO. The upper panels show the aggressive models $A_4$ and $A_5$, and the lower panels the mild-deviation models $B_2$ and $B_3$ together with the Kerr solution.
}
\vspace{1cm}
\label{mass_accc_a05_1}
\end{figure*}

Like Fig.~\ref{mass_accc_a09_1}, Fig.~\ref{mass_accc_a05_1} shows the accretion rate for the moderately rotating black hole with $a=0.5M$. The aggressive models $A_4$ and $A_5$, given in Table~\ref{Inital_Con_1}, confirm that, regardless of the black hole spin, a larger $|k|$ deepens the effective potential at large radii and produces strongly variable accretion at both radii. Their time-averaged rates at $r=2.3M$ are about $2.2$ ($A_4$) and $1.7$ ($A_5$) times the Kerr value. The character of the variability follows the morphology found in the density maps. In $A_5$ ($\eta\simeq1.6$), where the cone is destroyed and a torus forms, the variability is fastest and most irregular while the newly formed torus settles ($t\lesssim10^{4}M$), with the rate swinging between about $0.3$ and $4$ times the Kerr value. Afterwards, the flow enters a phase of persistent torus oscillations. In $A_4$ ($\eta\simeq0.30$), the swings are slower and longer-lived, reflecting the widened and persistently oscillating shock cone associated with moderate values of $|k|$. As for $a=0.9M$, the time-averaged rate is lower in the torus-forming model than in the flip-flop model.

On the other hand, in the mild-deviation models $B_2$ and $B_3$, a stable shock cone forms and no flip-flop instability is observed. The accretion rate remains nearly steady but still shows weak quasi-periodic dips similar to those of the Kerr configuration. The two models differ in their mean accretion rate. In $B_3$ ($|k|\simeq7\times10^{-9}\,M^{-2}$, $\eta<10^{-3}$), the rate is indistinguishable from the Kerr value at both radii. In $B_2$ ($|k|\simeq5\times10^{-7}\,M^{-2}$, $\eta\simeq0.03$), it is about $4\%$ higher than in the Kerr run at both $r=2.3M$ and $r=6.11M$. This difference is consistent in magnitude with the $\approx3\%$ additional pull of the curvature term at the accretion radius, Eq.~(\ref{eta}). Model $B_2$ is therefore the clearest example of a curvature effect that changes the accretion rate without changing the morphology of the flow. As for $a=0.9M$, the relative fluctuations of the Kerr-like models are larger at $r=6.11M$ ($\approx\pm5\%$) than at $r=2.3M$ ($\approx\pm2\%$). In the aggressive models, by contrast, the strongest variations appear at both radii at nearly the same times.

Finally, comparing Figs.~\ref{mass_accc_a09_1} and~\ref{mass_accc_a05_1} for models with similar $\eta$ clarifies the role of the spin. The pairs $A_2$/$A_3$--$A_4$ (flip-flop) and $A_1$--$A_5$ (torus) show the same type of variability and similar ratios of the mean accretion rate to the Kerr value. The flow regime and the overall level of accretion are therefore set by $k$. The spin mainly acts in the inner region, where frame dragging is significant. It changes the phase and shape of the oscillations and the amplitude of the strongest excursions, which is greatest for the rapidly rotating Model $A_1$. The curvature parameter $k$ governs whether the system develops a disk-like configuration or a breathing shock cone, as well as the overall level and intensity of the resulting variability.

\subsection{Azimuthal Density Profiles and One-Dimensional Cuts}
\label{azimuthal_cut} 

The density maps and the accretion rates characterize the flow globally and at fixed radii. To follow the motion of the shock directly, we now record the rest-mass density along the azimuthal direction at a fixed radius, $r=6.14M$, as a function of time. This radius is practically the same as the outer radius $r=6.11M$ at which the accretion rate is measured, and it lies outside the ISCO for both spins. In these space--time ($\phi$--$t$) diagrams, $\phi=0$ ($\equiv2\pi$) corresponds to the downstream axis, where the shock cone forms, and $\phi=\pi$ to the upstream direction, from which the gas arrives. The diagrams separate the different kinds of motion identified in the density maps. A stationary cone appears as a dense band around $\phi=0$ whose edges are vertical lines. A shock that swings from side to side appears as edges that move back and forth in $\phi$. When the upstream part of the shock moves outward beyond $r=6.14M$, the low-density window around $\phi=\pi$, through which the unshocked gas falls in, closes. A torus fills the whole azimuthal range with dense gas. Since the orbital period at this radius is only $\approx100M$~\cite{ZULQARNAIN2026117449}, any structure that rotates with the gas crosses the full azimuthal range in a very small fraction of the time interval shown, and therefore appears as a nearly horizontal streak.

\begin{figure*}
  \vspace{1cm}
  \center
  \psfig{file=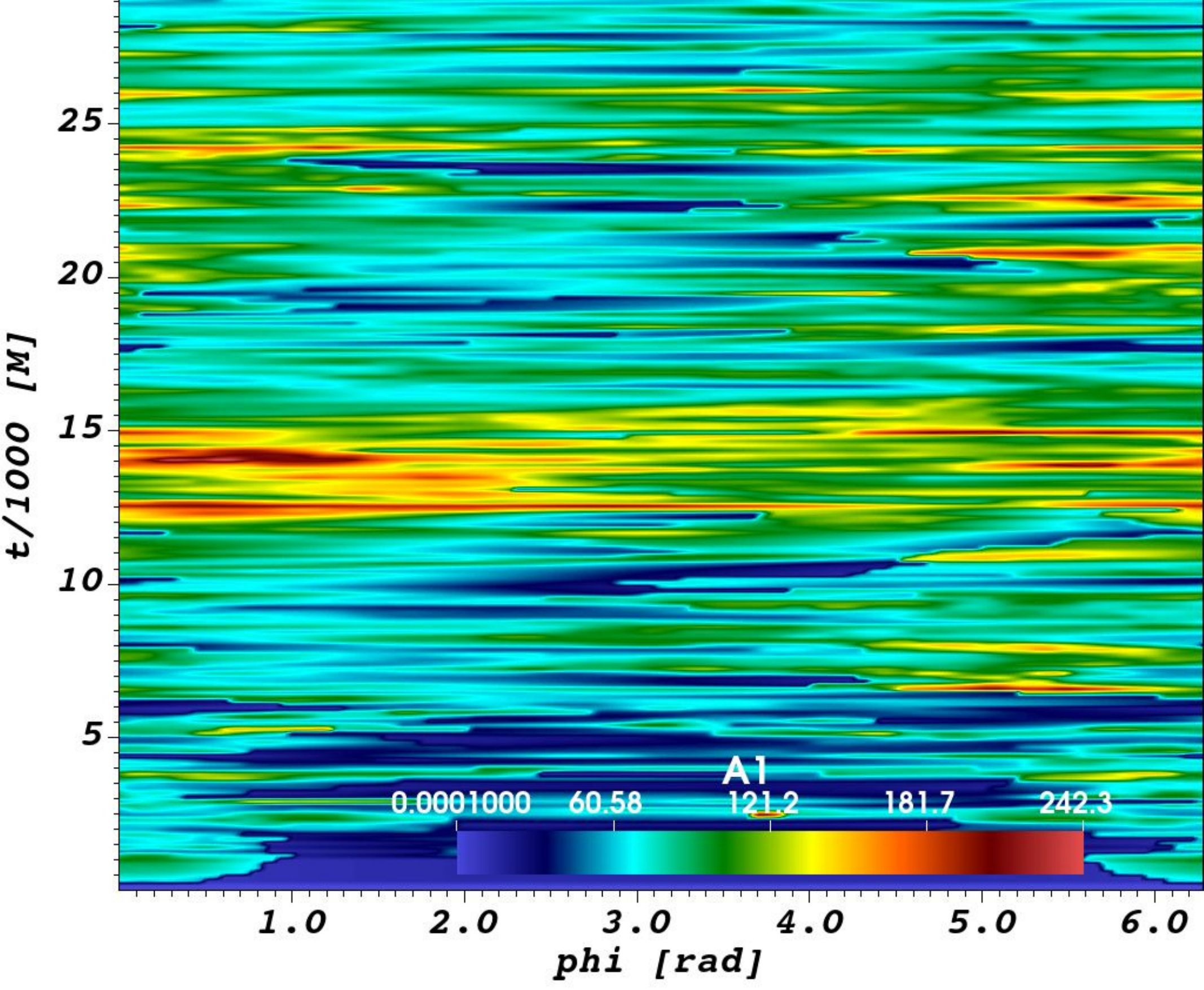,width=5.20cm,height=7cm} 
  \psfig{file=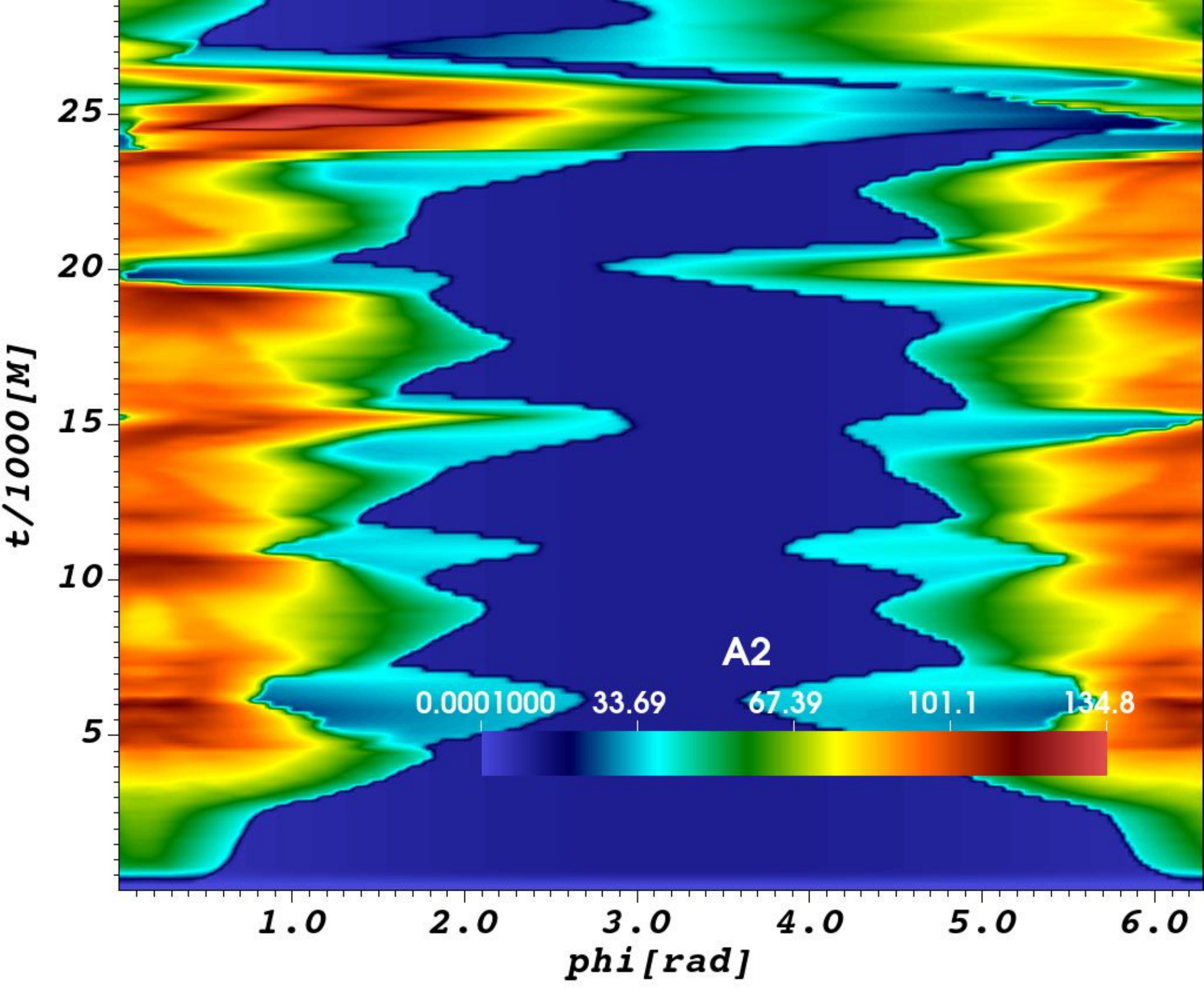,width=5.20cm,height=7cm} 
  \psfig{file=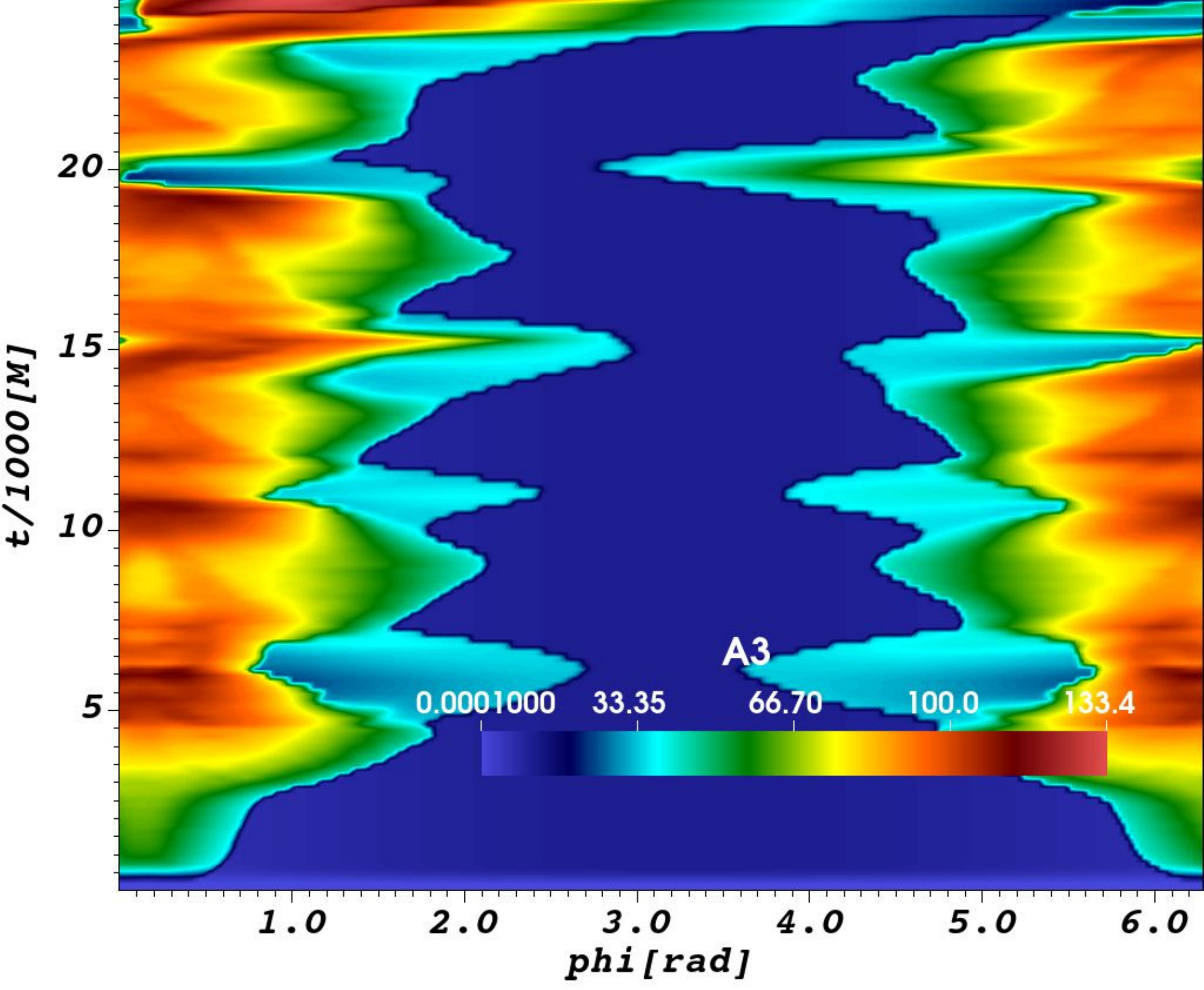,width=5.20cm,height=7cm}\\
  \psfig{file=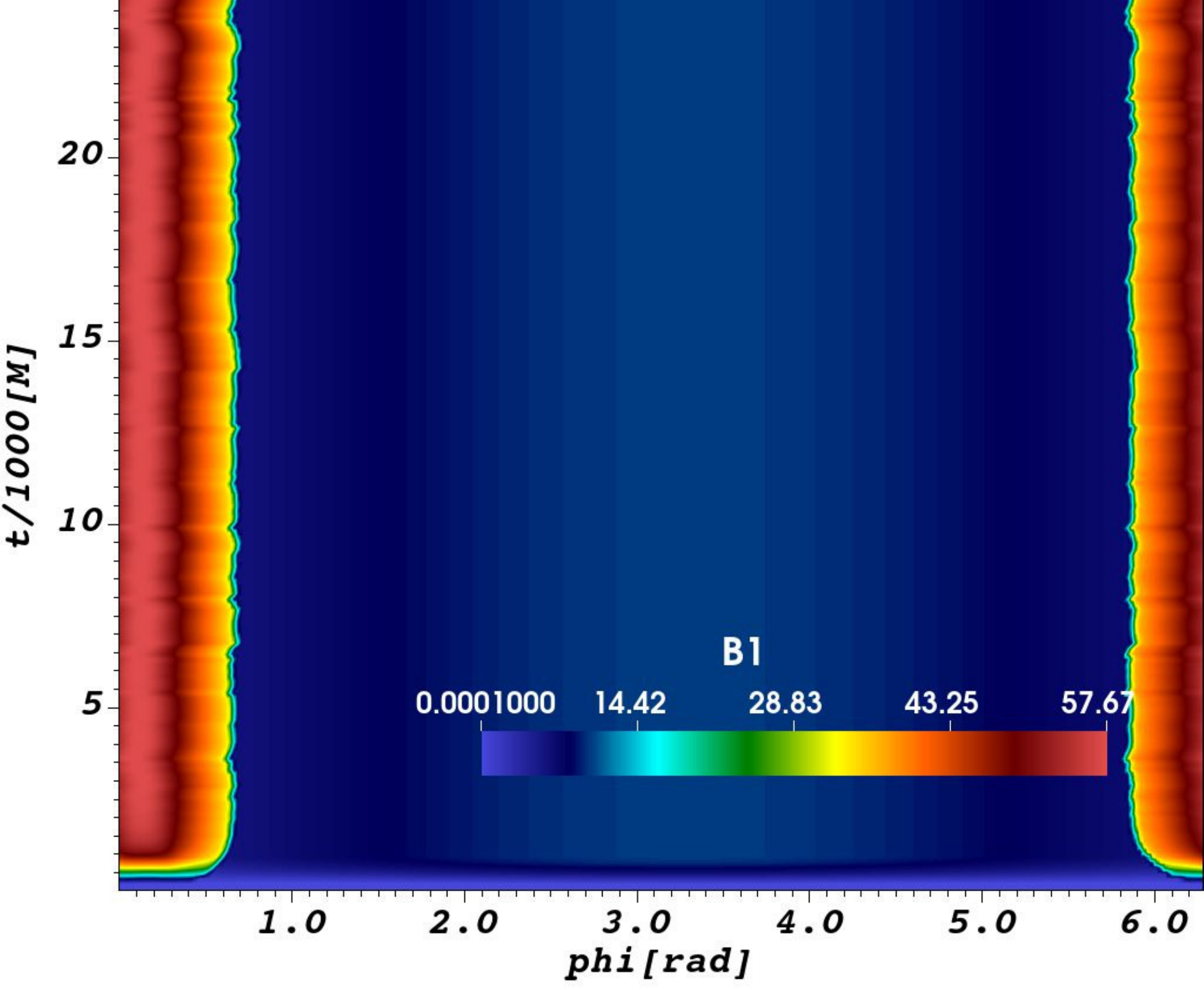,width=5.20cm,height=7cm}   
  \psfig{file=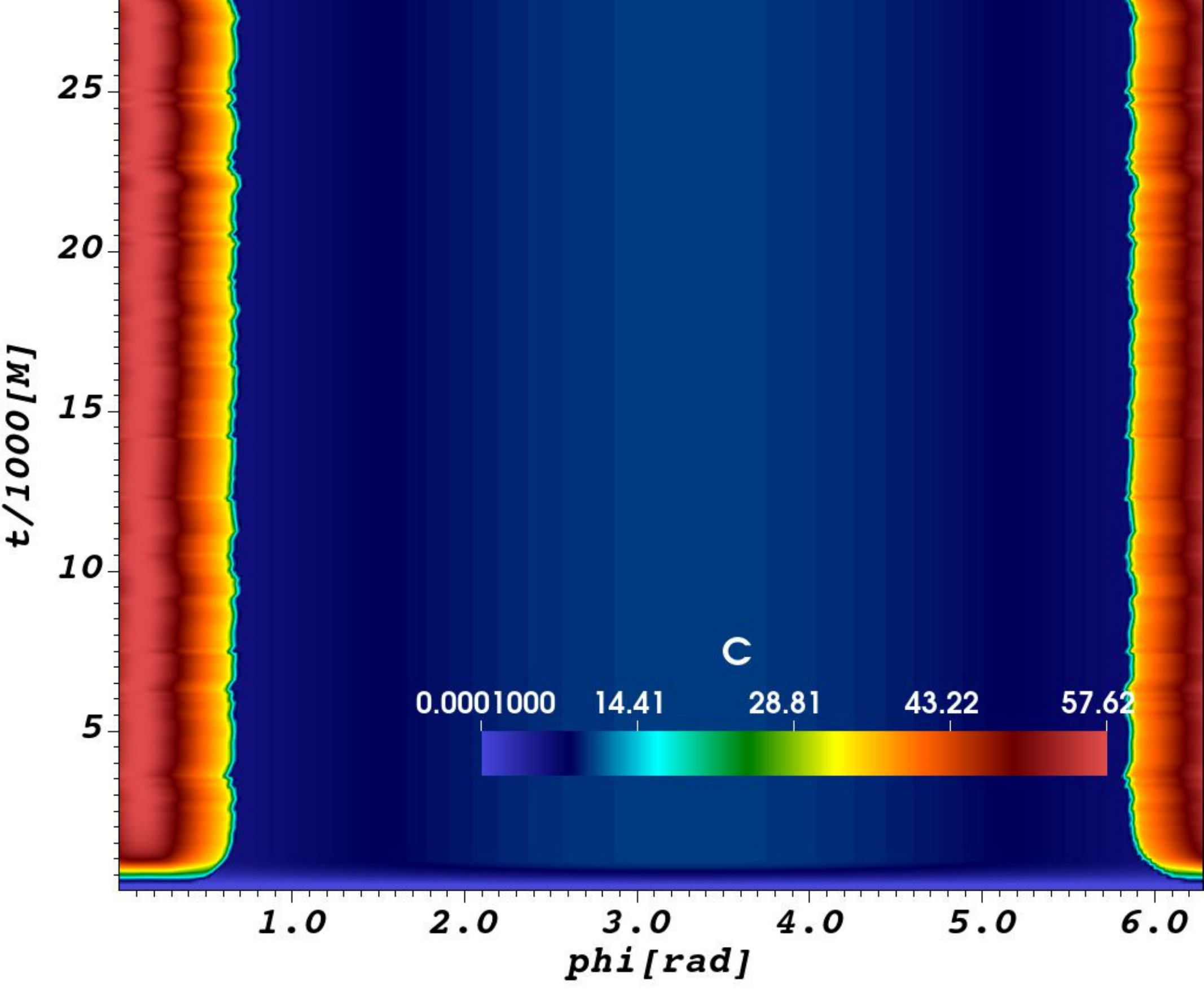,width=5.20cm,height=7cm}   
  \psfig{file=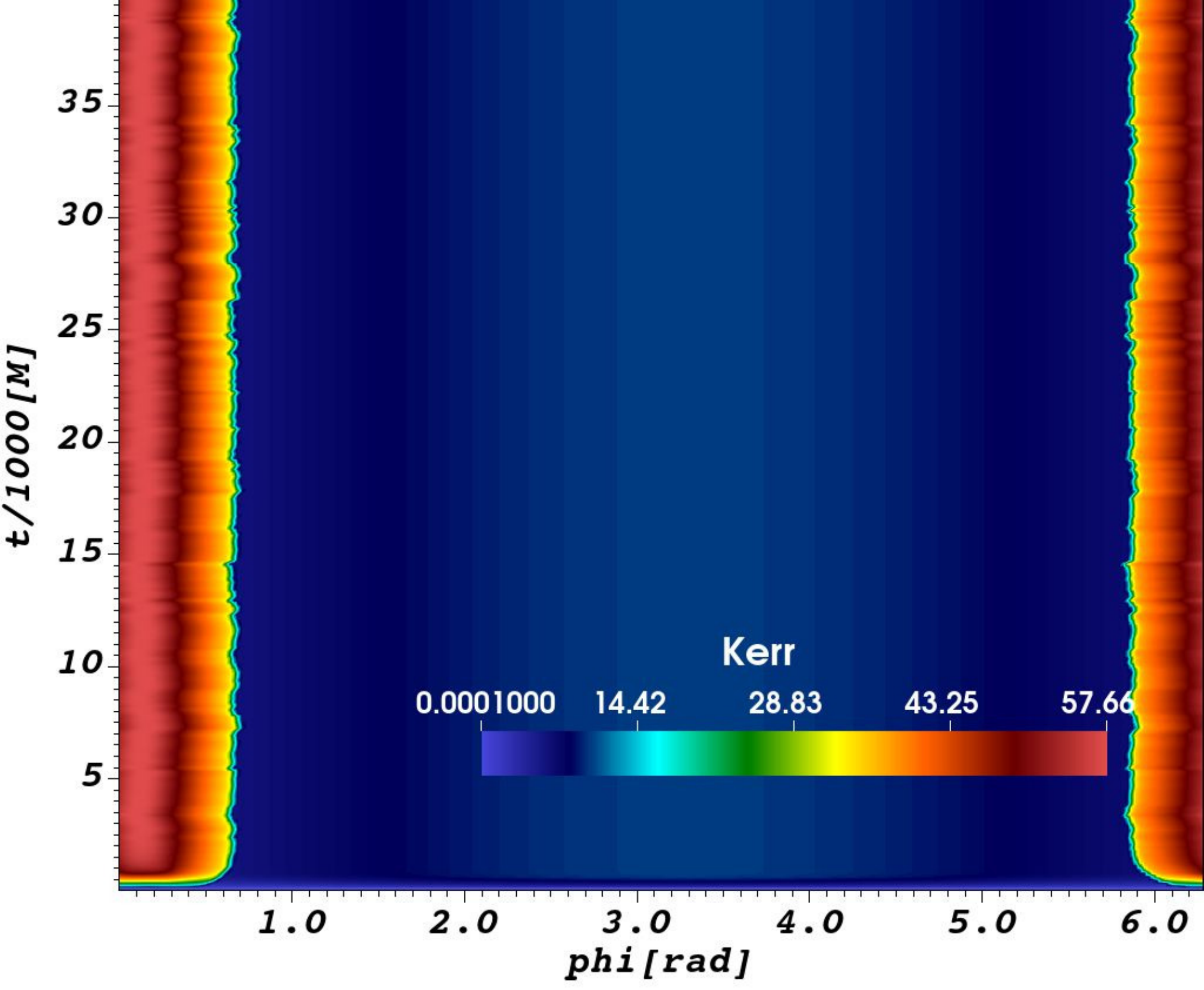,width=5.0cm,height=7cm}  
  \caption{Space--time ($\phi$--$t$) diagrams of the rest-mass density on the equatorial plane at $r=6.14M$, outside the ISCO, for the Kerr--Weyl black hole with spin parameter $a=0.9M$. The horizontal axis is the azimuthal angle, with $\phi=0$ on the downstream axis and $\phi=\pi$ in the upstream direction, and the vertical axis is the normalized time $t/1000\,[M]$. The upper panels show the aggressive models listed in Table~\ref{Inital_Con_1}. In $A_2$ and $A_3$, the low-density upstream window (dark blue) swings from side to side (flip-flop), while in $A_1$ it closes permanently and a torus surrounds the black hole. The lower panels show the $B_1$, $C$ and Kerr models, in which the stationary shock cone appears as two vertical bands around $\phi=0$. Note the different color scales and time ranges of the panels.
}
\vspace{1cm}
\label{color_azimut_a09_r614}
\end{figure*}

Fig.~\ref{color_azimut_a09_r614} shows the $\phi$--$t$ diagrams for the rapidly rotating black hole with $a=0.9M$. In the upper panels, corresponding to the aggressive models $A_1$, $A_2$ and $A_3$, the azimuthally averaged density at this radius, $\langle\rho\rangle_\phi$, is about $3$ ($A_2$, $A_3$) to $6$ ($A_1$) times higher than in the Kerr run. This is in line with the accumulation of shocked gas and the enhanced accretion rates discussed above. As anticipated from the values of $\eta$, a stronger curvature term leads to more extreme behavior. In Model $A_1$ ($\eta\simeq1.9$), the low-density upstream window closes at $t\simeq6\times10^{3}M$ and, apart from brief episodes, does not reopen, so dense gas surrounds the black hole at all azimuths from this time on. The azimuthal density contrast, $\sigma_\phi/\langle\rho\rangle_\phi$, drops from $\approx1$ in the Kerr run to $\approx0.2$, i.e., the flow at this radius becomes close to axisymmetric. The inner flow thus becomes nearly toroidal already during the flip-flop phase, while the shock at larger radii keeps swinging until $t\simeq1.5\times10^{4}M$ (Fig.~\ref{A1_animation}). The diagram is dominated by horizontal streaks, which reflect rotating, non-axisymmetric density structures and the time variation of the torus density. The most violent variations occur between $t\simeq6\times10^{3}M$ and $1.5\times10^{4}M$, when the temporal rms of $\langle\rho\rangle_\phi$ is about $30\%$; afterwards it decreases to about $18\%$ as the torus settles. These variations coincide with the features of the accretion rate. The lowest densities occur at $t\simeq(5$--$6)\times10^{3}M$, when the accretion rate drops almost to zero, and the highest at $t\simeq(1.25$--$1.5)\times10^{4}M$, around its largest peak.

Models $A_2$ and $A_3$ ($\eta\simeq0.19$) exhibit slower and more coherent flip-flop oscillations. After $t\simeq5\times10^{3}M$, the dense post-shock region covers on average $\approx4$\,rad ($\approx230^\circ$) of the circle, compared with $\approx1$\,rad in the Kerr run. The upstream window, of width $2.3\pm0.6$\,rad, swings from side to side around $\phi=\pi$. Its center moves by up to $\approx1$\,rad from the upstream axis, with an rms excursion of $\approx0.33$\,rad ($\approx19^\circ$). Successive turning points are separated by $\approx10^{3}M$ on average, which gives a swing period of $\approx2\times10^{3}M$. This is the same time scale as the dominant modulation of the accretion rate, which shows directly that the quasi-periodic bursts in $dM/dt$ are produced by the lateral motion of the shock. The swings are, however, not strictly periodic. The width of the window also varies on a longer time scale, $\approx(4$--$5)\times10^{3}M$, which sets the fundamental frequency identified in the PSD analysis below. The positions of the window in $A_2$ and $A_3$ agree to within $\approx0.02$\,rad up to $t\simeq2.2\times10^{4}M$. This agreement, expected because the metric depends on $\gamma$ and $\kappa$ only through $k$, shows that the swings of the shock are deterministic.  At $t\simeq(2.4$--$2.5)\times10^{4}M$, the window closes in both models and $\langle\rho\rangle_\phi$ rises by about $50\%$. At the same time, the accretion rate at $r=2.3M$ drops almost to zero, showing that gas temporarily accumulates around $r\approx6M$ instead of being accreted. These oscillations therefore correspond directly to the bursts and quasi-periodic modulations seen in the mass accretion rate (Fig.~\ref{mass_accc_a09_1}) and to the wide, swinging shock identified in the two-dimensional density maps (Figs.~\ref{A1_animation}--\ref{A5_animation}).

The lower panels of Fig.~\ref{color_azimut_a09_r614} show the results for the $B_1$, $C$ and Kerr models. Since $|k|\lesssim10^{-8}\,M^{-2}$ in these cases ($\eta<10^{-3}$), the three diagrams are practically identical. The dense post-shock region occupies a fixed band that extends $\approx0.65$\,rad on one side of the downstream axis and $\approx0.40$\,rad on the other. Its total angular extent at this radius is therefore $\approx1.05$\,rad ($\approx60^\circ$), and its edges fluctuate by only $\approx0.01$\,rad, at the resolution of the figure. The peak density in the cone agrees to within $0.2\%$ among the three models. These models therefore preserve a stationary shock cone without flip-flop oscillations, consistent with the nearly steady accretion rates in Fig.~\ref{mass_accc_a09_1}. The cone is not exactly centered on $\phi=0$. It is shifted by $\approx0.12$\,rad ($\approx7^\circ$) toward positive $\phi$, in the sense of the black hole rotation. The corresponding shift for $a=0.5M$ is about half as large (see below), so the shift is approximately proportional to $a$ and is produced by frame dragging. This agrees with the increasing wrapping of the tail shock around rapidly rotating black holes found by Font et al.~\cite{Font1999}. The Kerr-like models also provide a control for numerical effects. The amplitude of flip-flop oscillations in relativistic simulations has been shown to depend on the choice of coordinates and on the placement of the inner boundary outside the horizon~\cite{CruzOsorio2012}. Models $B_1$ and $C$ are evolved with the same numerical setup as $A_2$ and $A_3$, including the inner boundary at $r=2.3M$, and differ from them only in the value of $k$. The swings of the shock in $A_2$ and $A_3$ are about thirty times larger than the edge fluctuations in $B_1$ and $C$. They are therefore caused by the curvature term and not by the numerical treatment of the inner boundary.

\begin{figure*}
  \vspace{1cm}
  \center
   \psfig{file=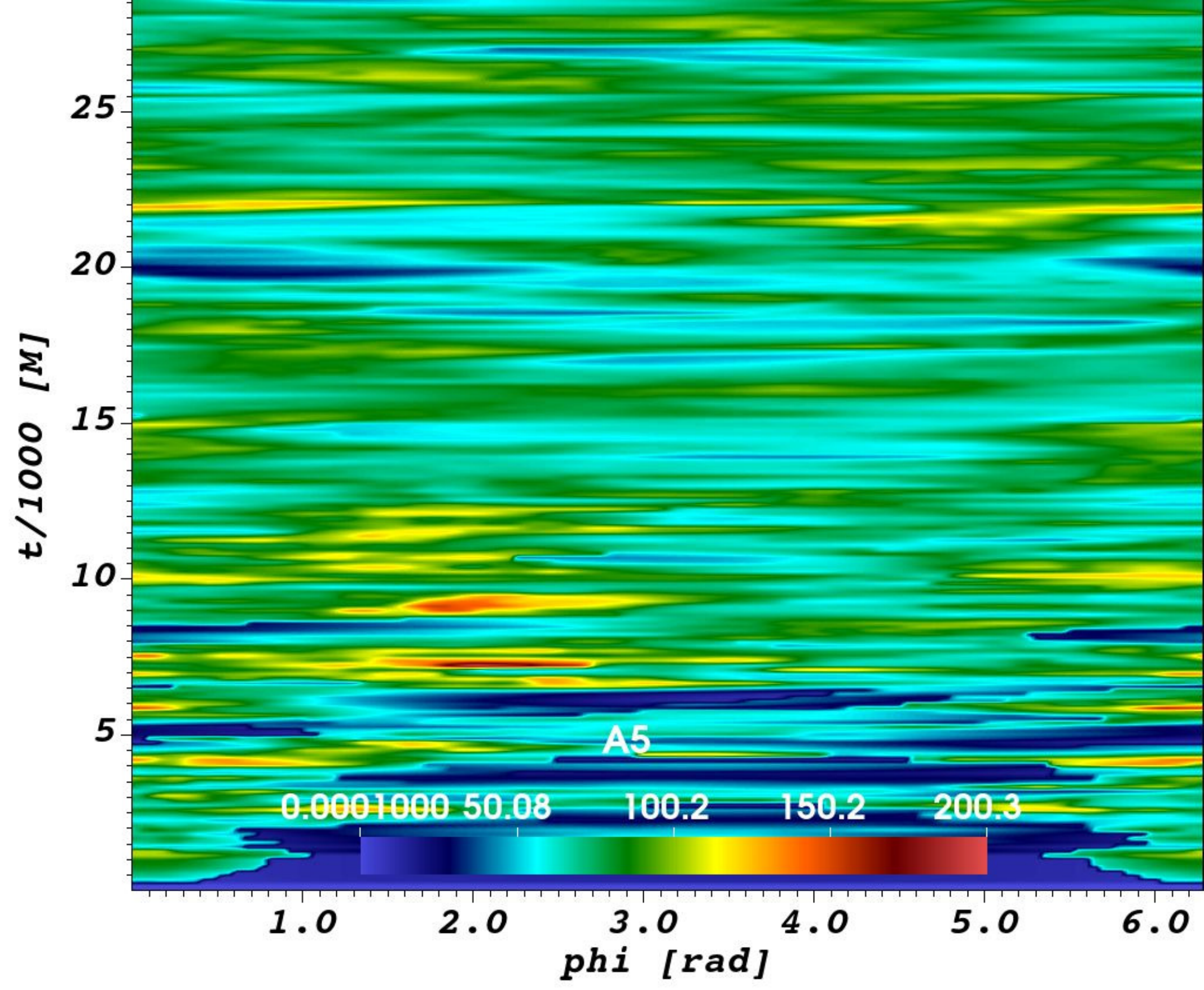,width=8.0cm,height=8cm}
  \psfig{file=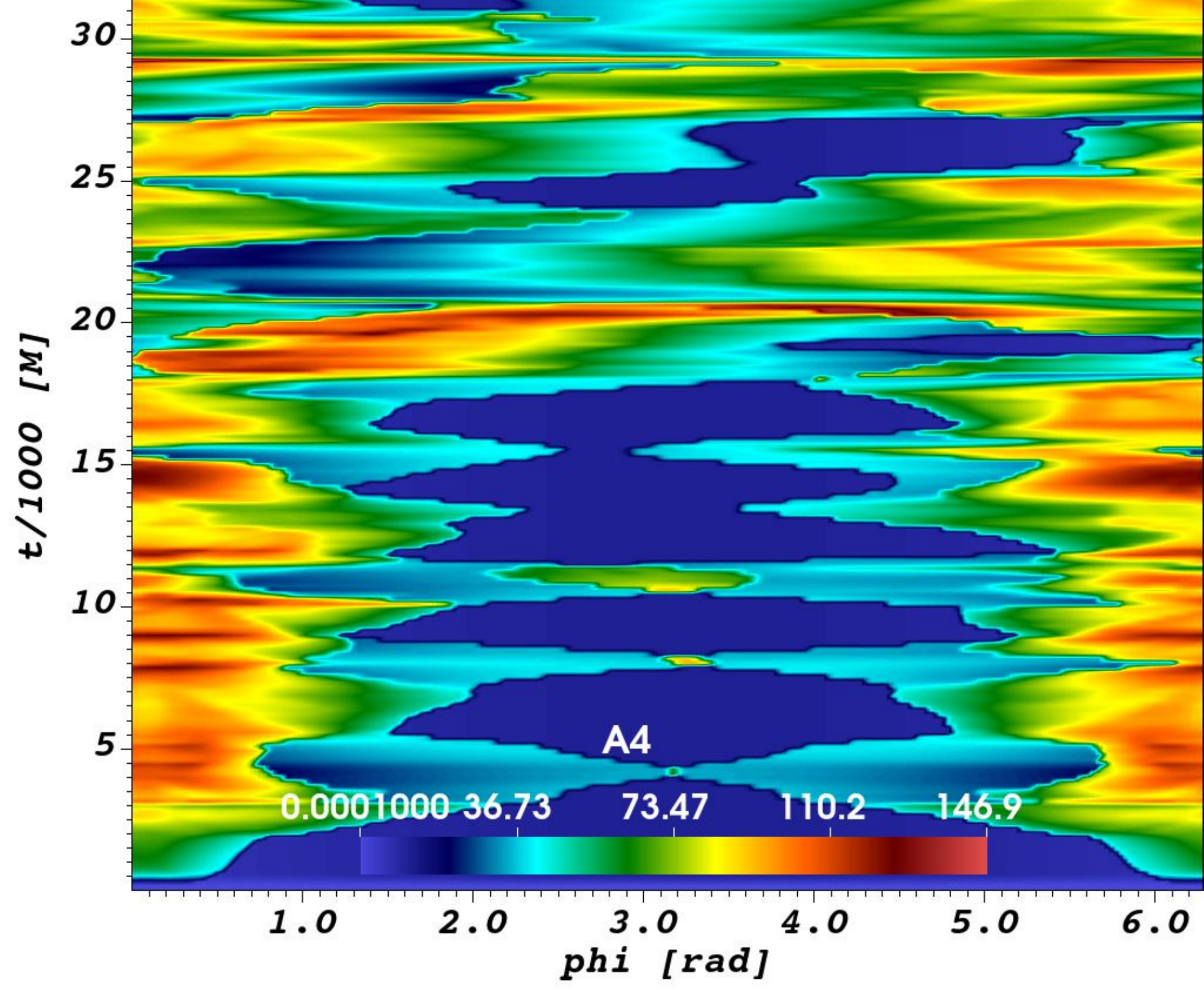,width=8.0cm,height=8cm} \\ 
  \psfig{file=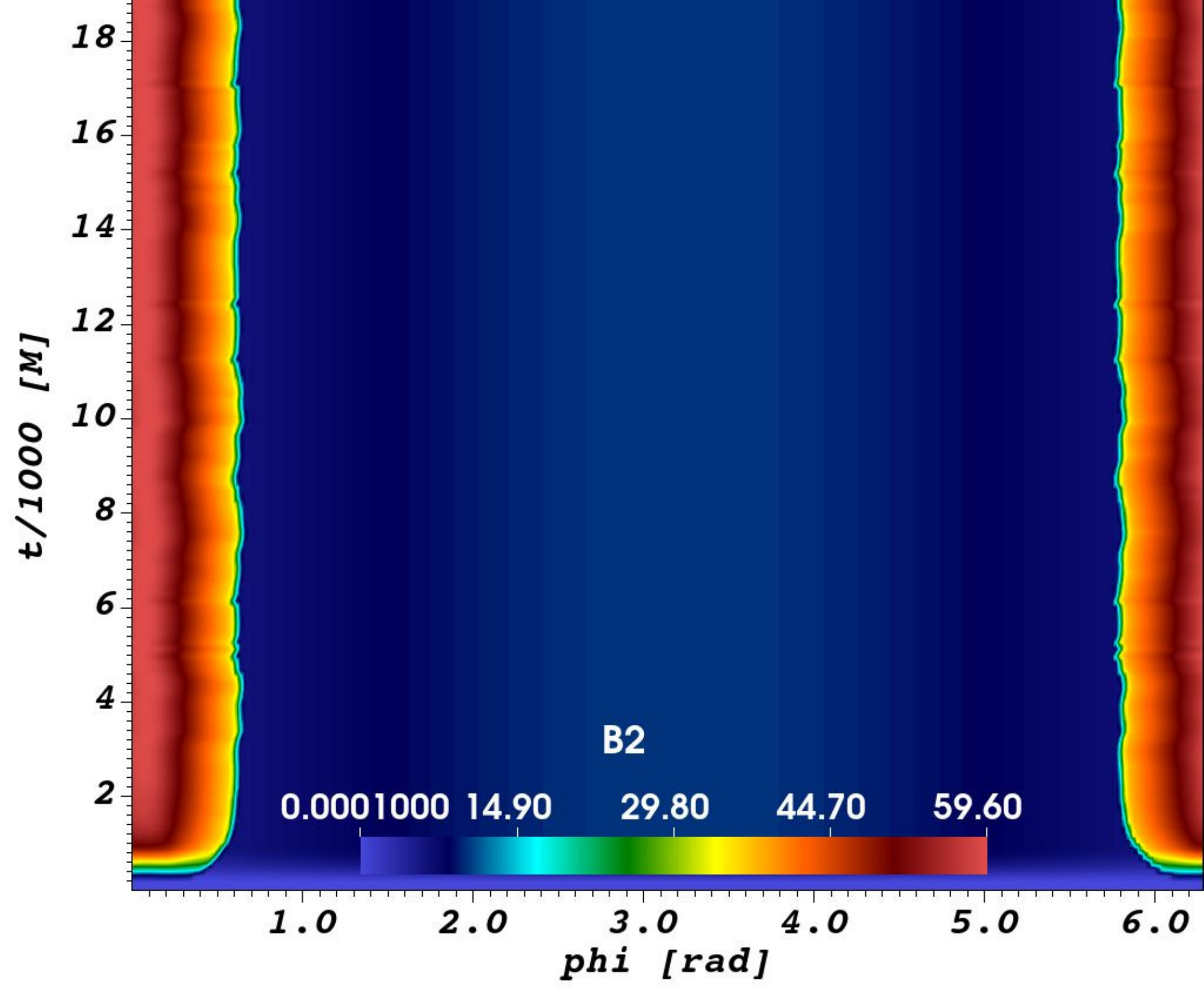,width=8.0cm,height=8cm}   
  \psfig{file=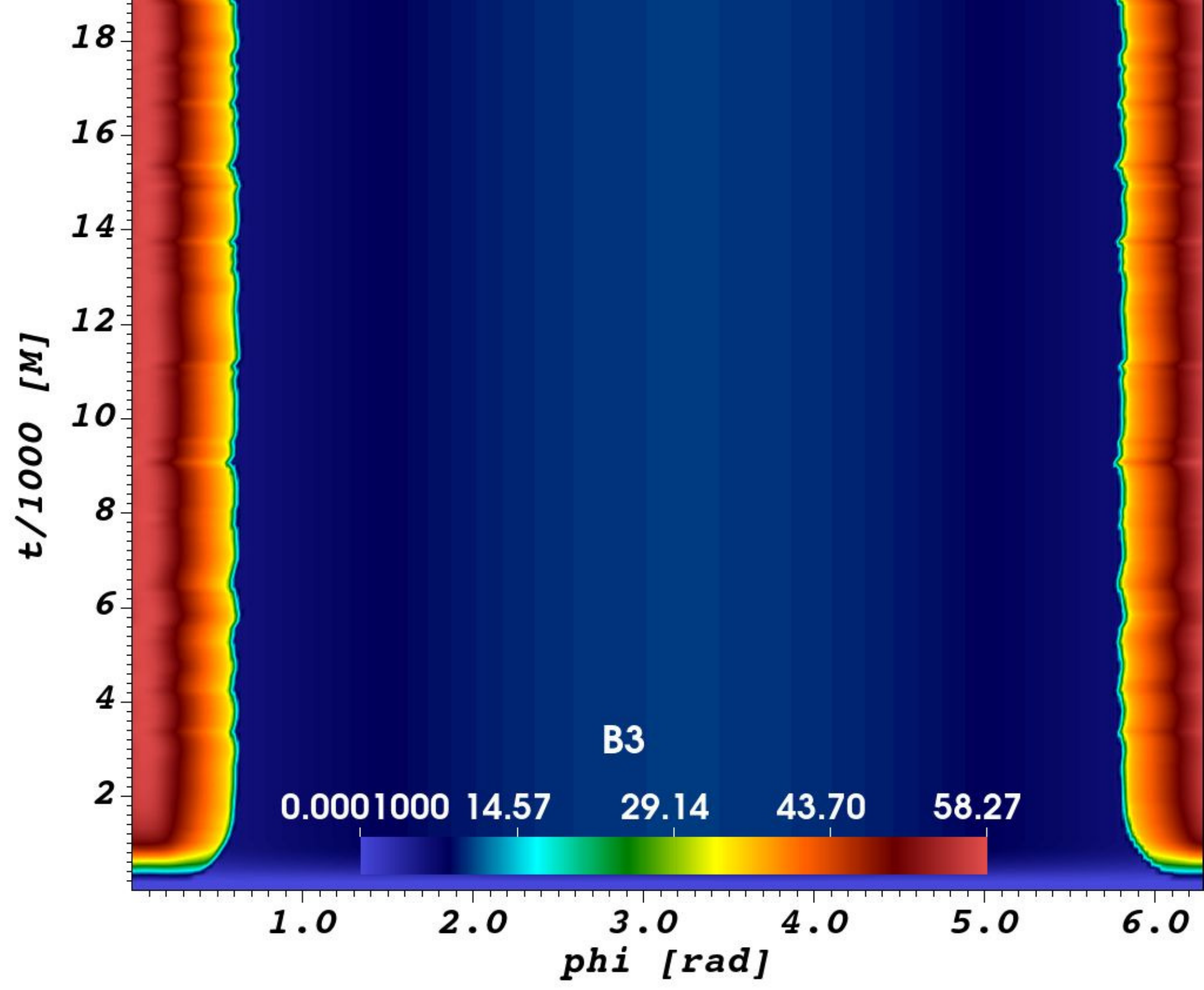,width=8.0cm,height=8cm}   
  \caption{Same as Fig.~\ref{color_azimut_a09_r614}, but for the rotating black hole with $a=0.5M$. The upper panels show the aggressive models $A_5$, in which a torus forms, and $A_4$, in which the shock undergoes flip-flop oscillations with intermittent closing of the upstream window. The lower panels show the mild-deviation models $B_2$ and $B_3$, in which a stable shock cone forms. The parameters of all models are listed in Table~\ref{Inital_Con_1}.
}
\vspace{1cm}
\label{color_azimut_a05_r614}
\end{figure*}

Fig.~\ref{color_azimut_a05_r614} illustrates the shock cone, the flip-flop instability and the toroidal structures around the Kerr--Weyl black hole with spin parameter $a=0.5M$. Like Fig.~\ref{color_azimut_a09_r614}, it presents the time evolution of the rest-mass density along the azimuthal direction for models with different values of the curvature parameter $k$ and of the Kerr--Weyl coefficients $(\gamma,\kappa,\beta)$, whose values are given in Table~\ref{Inital_Con_1}.

In the upper-left panel of Fig.~\ref{color_azimut_a05_r614}, Model $A_5$ ($\eta\simeq1.6$), the strongly curved AdS-like case for this spin, shows strong and irregular variations at all azimuths. From $t\simeq10^{3}M$ onward, the upstream window no longer appears, because the shock has been pushed outward beyond $r=6.14M$. A highly turbulent flow develops and, by $t\simeq3\times10^{3}M$, forms the toroidal structure seen in the density maps, which persists until the end of the simulation. After $t\simeq6\times10^{3}M$, the azimuthal contrast is $\approx0.15$--$0.2$, even lower than in $A_1$, and $\langle\rho\rangle_\phi$ is about five times the near-Kerr value. The flip-flop motion of the cone is therefore limited to the short transition phase. The horizontal streaks that dominate the diagram afterwards reflect rotating density structures and the oscillations of the torus, which continuously reconfigures itself between high- and low-density states. The temporal rms of $\langle\rho\rangle_\phi$ is $\approx11\%$, smaller than in $A_1$, in line with the smaller excursions of the accretion rate in $A_5$.

Although Model $A_4$ ($\eta\simeq0.30$) has a value of $|k|$ about one fifth of that of $A_5$, it still exhibits quasi-periodic flip-flop behavior. In this case, the oscillations are less frequent but longer-lived. The upstream window is open for about two-thirds of the time between $t\simeq6\times10^{3}M$ and $2.2\times10^{4}M$, and for less than half of the time afterwards. When open, it is up to $\approx3.8$\,rad wide and moves sideways, and it repeatedly narrows and closes at irregular intervals of a few $10^{3}M$. The shock in $A_4$ therefore oscillates both laterally, as in $A_2$ and $A_3$, and radially, with its upstream part moving back and forth across $r\approx6M$. This behavior is intermediate between that of $A_2$ and $A_3$ ($\eta\simeq0.19$), where the window remains open throughout the flip-flop phase, and that of $A_1$ and $A_5$ ($\eta>1$), where it closes permanently. The larger $\eta$ is, the more shocked gas accumulates and the farther upstream the shock is pushed. In contrast, the $B_2$ and $B_3$ models, in which $|k|$ is close to the Kerr limit $k=0$, show almost no temporal or azimuthal variation. The only exceptions are small displacements of the shock edges ($\approx0.01$\,rad) around their equilibrium positions, which correspond to the weak dips in the accretion rates of Fig.~\ref{mass_accc_a05_1}. No flip-flop instability is observed in these models, and the cone has the same angular extent, $\approx1.05$\,rad, as around the Kerr black hole. Its center is shifted by $\approx0.065$\,rad in the sense of rotation, about half of the shift found for $a=0.9M$. The only visible difference between $B_2$ and $B_3$ is that the cone of $B_2$ is about $2\%$ denser and $0.03$\,rad wider, and $\langle\rho\rangle_\phi$ is about $4\%$ higher. This is consistent with the $\approx4\%$ higher accretion rate of $B_2$ and with its small but non-negligible value $\eta\simeq0.03$.

The density oscillations found at this radius, namely the swings of the shock in $A_2$, $A_3$ and $A_4$ and the pulsations of the torus in $A_1$ and $A_5$, modulate the accretion rate quasi-periodically, as is known for oscillating tori~\cite{Zanotti2003,QPOs9}. Since $k$ changes the local epicyclic frequencies by only about $2\%$~\cite{ZULQARNAIN2026117449}, this modulation is set by the large-scale dynamics of the shock and the torus at $r\lesssim R_{\rm acc}$, while frame dragging only shifts the cone and the phase of the oscillations. Its frequencies, and their relation to the LFQPOs observed in black hole X-ray binaries~\cite{RemillardMcClintock2006,IngramMotta2019}, are determined next from the power spectra.

\subsection{Power Spectral Density Analysis of Quasi-Periodic Oscillations}
\label{PSD_analysis} 

The previous subsections identified three flow regimes around the Kerr--Weyl black hole, ordered by the parameter $\eta$ of Eq.~(\ref{eta}). These are a standing shock cone with a narrow opening angle ($\eta\lesssim0.03$), a flip-flop oscillating shock with a wide opening angle ($\eta\simeq0.2$--$0.3$), and a fully developed toroidal configuration that undergoes quasi-periodic oscillations around the black hole ($\eta\gtrsim1.6$). In this subsection, we analyze the PSD of the mass accretion rates shown in Figs.~\ref{mass_accc_a09_1} and~\ref{mass_accc_a05_1}. The aim is to identify the modes excited in each regime and to compare their frequencies and frequency ratios with the QPOs observed in black hole X-ray binaries. In these systems, LFQPOs occur at $\sim0.1$--$30$\,Hz and HFQPOs at $\sim40$--$450$\,Hz~\cite{RemillardMcClintock2006,IngramMotta2019}.

The simulations use geometrized units, in which time is measured in units of $GM/c^{3}$. A frequency $\hat f$, given in units of $M^{-1}$, is converted to physical units by
\begin{equation}
f=\frac{c^{3}}{GM}\,\hat f\simeq2.03\times10^{4}\left(\frac{10\,M_\odot}{M}\right)\hat f\ \ {\rm Hz}.
\label{f_Hz}
\end{equation}
We adopt $M=10\,M_\odot$, typical of stellar-mass black holes in X-ray binaries, so that a time interval of $1M$ corresponds to $4.93\times10^{-5}$\,s. For example, the $9.4$\,Hz peak discussed below corresponds to a period of $\approx2.2\times10^{3}M$, the dominant modulation period of the accretion rate in Models $A_2$ and $A_3$. Since $f\propto M^{-1}$, all frequencies quoted below scale inversely with the mass, whereas their ratios do not depend on it. For a supermassive black hole with $M=10^{6}$--$10^{9}\,M_\odot$, the same peaks would lie at $\sim10^{-4}$--$10^{-7}$\,Hz. The finite length $T$ of the time series sets the frequency resolution,
\begin{equation}
\Delta f=\frac{1}{T}\simeq0.68\left(\frac{3\times10^{4}M}{T}\right)\left(\frac{10\,M_\odot}{M}\right)\ {\rm Hz}.
\label{df_res}
\end{equation}
Peak frequencies are therefore uncertain by about $\pm\Delta f/2$, and the ratio $R=f_1/f_2$ of two peaks carries an uncertainty
\begin{equation}
\frac{\delta R}{R}\simeq\frac{\Delta f}{2}\left(\frac{1}{f_1^{2}}+\frac{1}{f_2^{2}}\right)^{1/2}.
\label{dR_ratio}
\end{equation}
In addition, a peak at frequency $f$ contains only $f/\Delta f$ oscillation cycles. Peaks below $\approx3.5$\,Hz therefore rely on fewer than about five cycles and should be regarded as tentative. A formal significance test against the red-noise continuum~\cite{Vaughan2005} would require longer time series or averaging over several independent segments~\cite{vanderKlis1989}. Here we regard a peak as robust only if it appears at both radii or in physically equivalent models. Finally, a periodic but strongly non-sinusoidal signal produces a ladder of harmonics,
\begin{equation}
f_n\simeq n f_0,\qquad f_0=\frac{\sum_n n f_n}{\sum_n n^{2}},
\label{harm_fit}
\end{equation}
in which neighboring members automatically stand in the ratios $2{:}1$, $3{:}2$, $4{:}3$, and so on. We use Eq.~(\ref{harm_fit}), with $f_0$ determined by least squares, to test whether the peaks of each model belong to such a ladder.

\begin{figure*}[!ht]
  \vspace{1cm}
  \centering
 \psfig{file=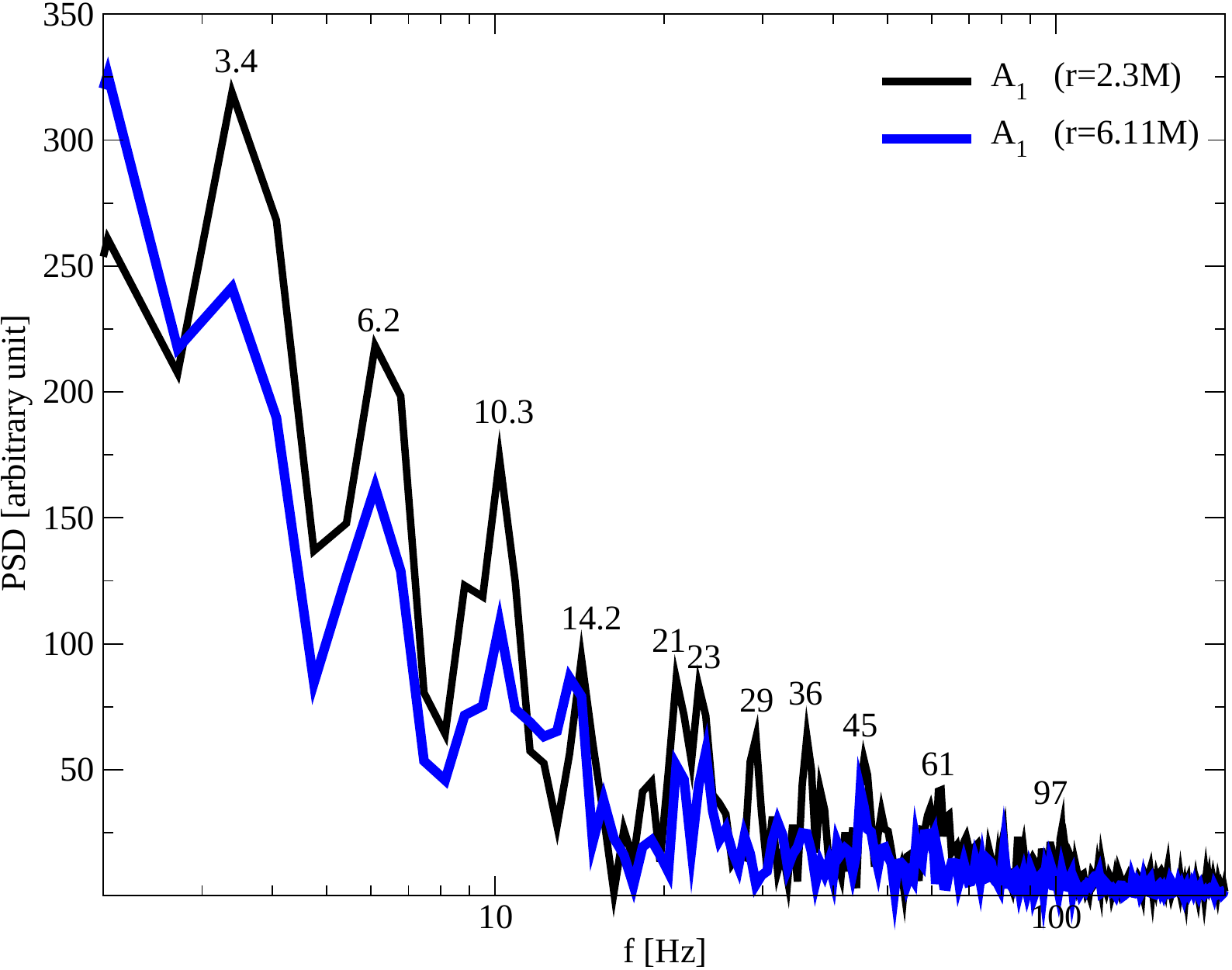,width=13.5cm,height=7.0cm}\\
 \vspace{0.3cm}
 \psfig{file=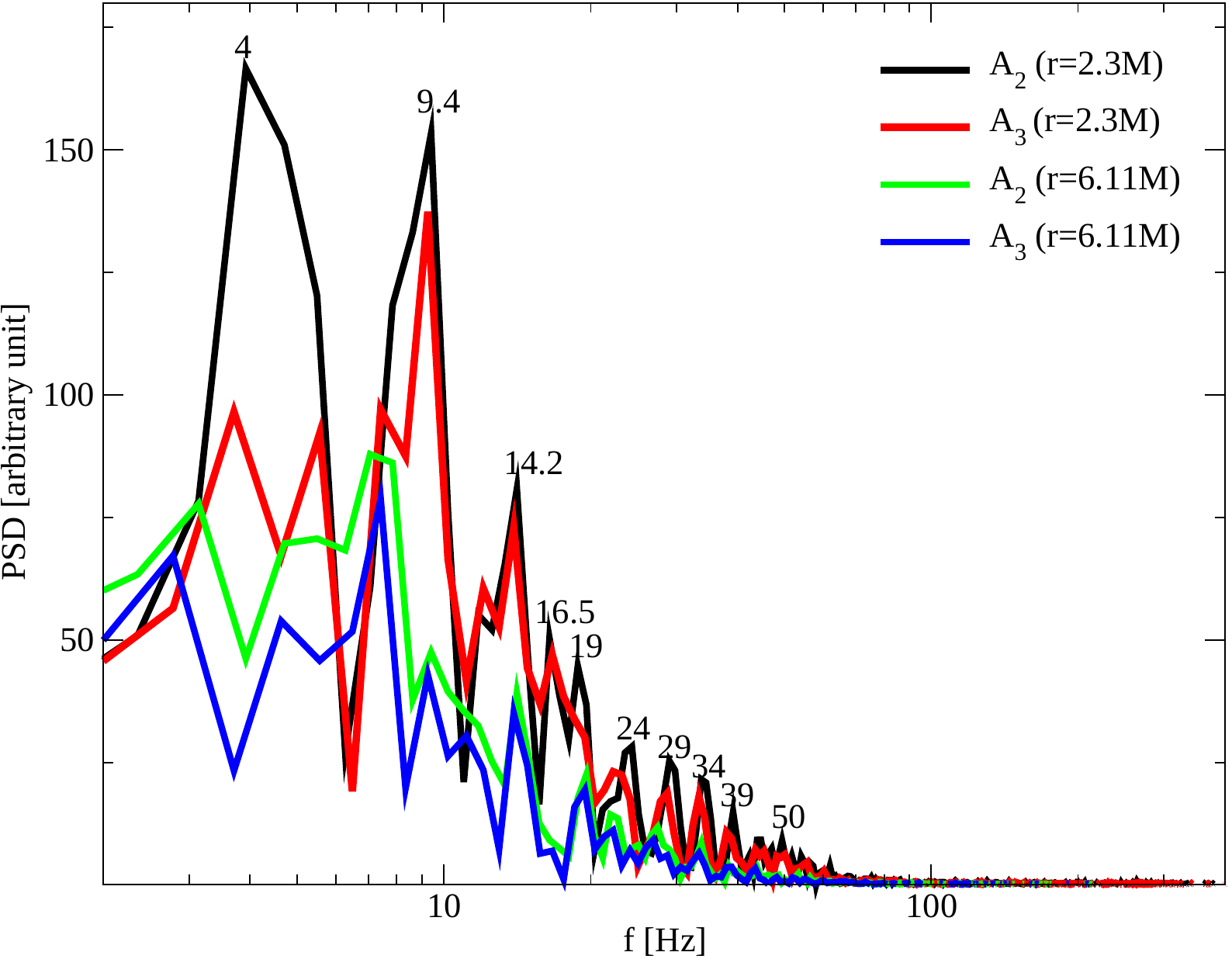,width=13.5cm,height=7.0cm}\\ 
 \vspace{0.3cm}
  \psfig{file=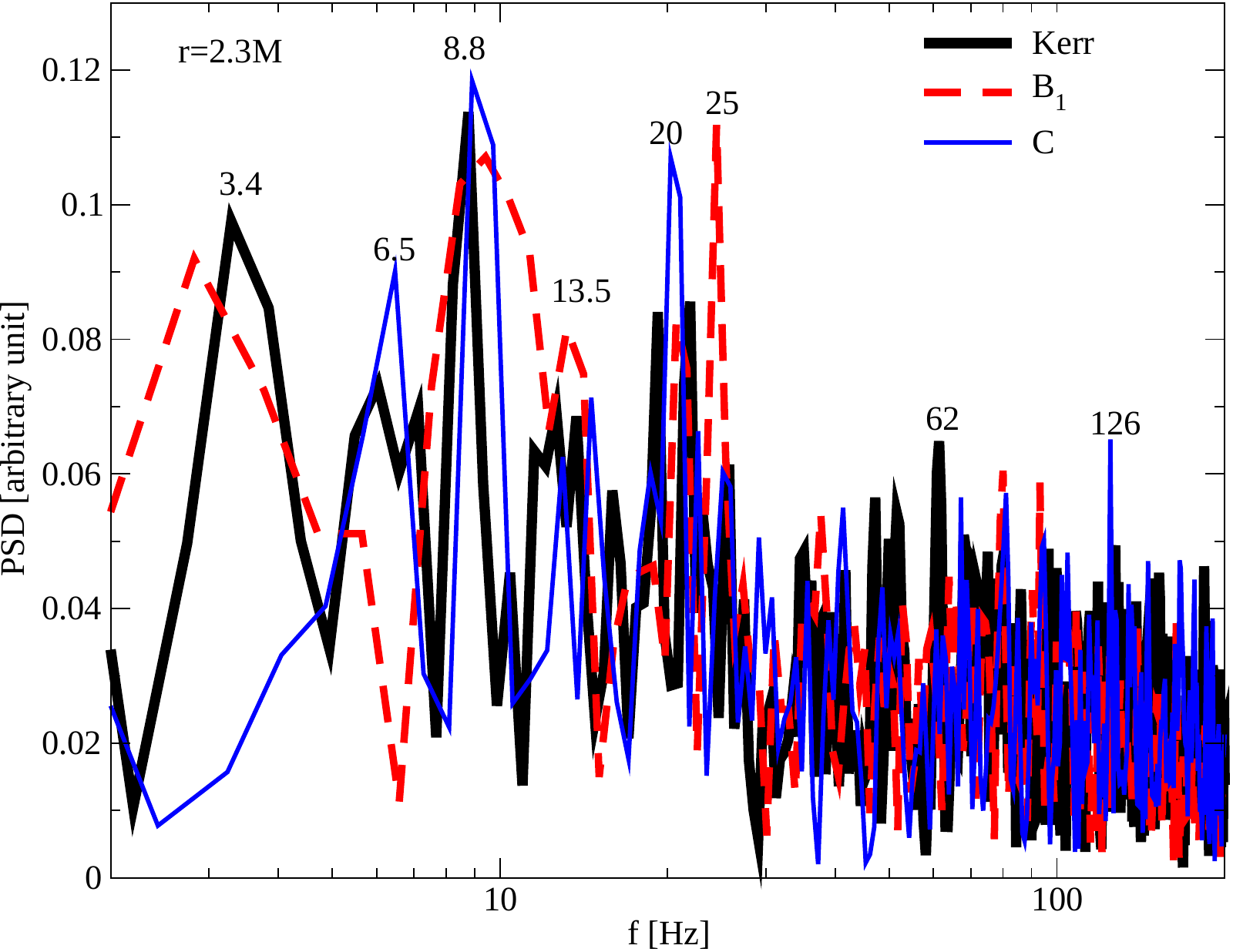,width=13.5cm,height=7.0cm}
  \caption{PSD of the mass accretion rate for the rapidly rotating black hole with $a=0.9M$, for the models given in Table~\ref{Inital_Con_1}. Frequencies are converted to Hz with Eq.~(\ref{f_Hz}) for $M=10\,M_\odot$; the frequency resolution is $\Delta f\approx0.7$\,Hz, and the labels give the frequencies of the main peaks. The upper panel shows the torus-forming Model $A_1$, and the middle panel the flip-flop Models $A_2$ and $A_3$, both at $r=2.3M$, close to the horizon, and at $r=6.11M$, outside the ISCO.
  The lower panel shows, at $r=2.3M$, the Kerr, $B_1$ and $C$ models in which a stable shock cone forms. The PSD is given in arbitrary units; note the different vertical scales of the panels.}
  \vspace{1cm}
  \label{QPO_a09}
\end{figure*}

Fig.~\ref{QPO_a09} shows the PSDs for $a=0.9M$. In Model $A_1$ ($\eta\simeq1.9$), the shock cone is replaced by an oscillating torus. Since the analyzed time series covers both the flip-flop phase and the torus phase, its spectrum combines the variability of the two. The upper panel shows a red-noise continuum that decreases with frequency, with a ladder of prominent peaks superposed on it at both $r=2.3M$ and $r=6.11M$. The four lowest peaks, at $3.4$, $6.2$, $10.3$ and $14.2$\,Hz, are present at both radii. They approximately follow the harmonic sequence $1{:}2{:}3{:}4$ of Eq.~(\ref{harm_fit}) with $f_0\simeq3.4$\,Hz (a period of $\approx5.9\times10^{3}M$), with deviations of at most one frequency bin. Although this fundamental contains only about five oscillation cycles,  it is robust by our criterion because it appears at both radii, and we identify it as the characteristic mode of the torus. At higher frequencies, peaks appear at $21$, $23$, $29$, $36$, $45$, $61$ and $97$\,Hz, but they exceed the local continuum by factors of only two to three. Among their ratios, $29/14.2=2.04\pm0.06$ is consistent with $2{:}1$. The ratios $61/29=2.10\pm0.03$ and $45/29=1.55\pm0.02$ lie close to $2{:}1$ and $3{:}2$, but formally differ from them by a few times the uncertainty of Eq.~(\ref{dR_ratio}). Multiple strong peaks are expected in $A_1$, because the transformation of the shock cone into a strongly oscillating torus excites global modes of the torus and their nonlinear couplings. As with the accretion rate itself, the peaks are stronger at $r=2.3M$ than at $r=6.11M$, while their frequencies coincide at the two radii. The QPOs therefore arise from global and persistent modes of the flow rather than from local instabilities.

In Models $A_2$ and $A_3$ ($\eta\simeq0.19$), the shock undergoes persistent flip-flop oscillations. The middle panel of Fig.~\ref{QPO_a09} show particularly strong peaks at $4$ and $9.4$\,Hz at $r=2.3M$. The peaks at $9.4$, $14.2$, $19$, $24$, $29$, $34$ and $39$\,Hz form an evenly spaced comb. A least-squares fit with Eq.~(\ref{harm_fit}) gives $f_0=4.83$\,Hz, and all seven peaks lie within $\Delta f/2$ of $nf_0$ with $n=2,\ldots,8$. The $4$\,Hz peak lies within about one bin of the fundamental itself. The spectrum therefore consists of a fundamental with a period of $\approx4.2\times10^{3}M$ and its harmonics. Its strongest component, $2f_0\simeq9.4$\,Hz, corresponds to $\approx2.2\times10^{3}M$, the dominant modulation period of the accretion rate. Both frequencies are also present in the azimuthal profiles at $r=6.14M$. There, the width of the upstream window varies with dominant periods of $\approx4.6\times10^{3}M$ and $\approx2.2\times10^{3}M$ ($\approx4.4$ and $9.4$\,Hz), so these peaks trace the breathing of the wide, swinging shock. The commensurate ratios $14.2/9.4=1.51\pm0.07\approx3{:}2$ and $29/14.2=2.04\pm0.06\approx2{:}1$ are the ratios of the third to the second and of the sixth to the third harmonic. They follow from the strongly non-sinusoidal shape of the modulation, in which smooth oscillations are interrupted by deep dips (Fig.~\ref{mass_accc_a09_1}). A resonance between independent modes is not required to explain them. The $9.4$\,Hz peak is equally strong in $A_2$ and $A_3$, whose evolutions are almost identical, whereas the $4$\,Hz peak is much stronger in $A_2$. The time series of $A_2$ extends beyond $t\simeq2.5\times10^{4}M$ and includes the episodes in which the upstream window closes and the accretion rate drops almost to zero. Part of the low-frequency power is therefore carried by a few large events. At $r=6.11M$, the power is reduced by about a factor of two and the strongest peaks shift to $\approx7.5$--$8$\,Hz and $\approx3$--$3.5$\,Hz, reflecting the weaker modulation of the flow farther from the horizon. Compared with Model $A_1$, the peaks of $A_2$ and $A_3$ are lower, consistent with the behavior of the accretion rate in Fig.~\ref{mass_accc_a09_1}. In the regime where the shock undergoes strong flip-flop oscillations, the dominant QPOs therefore correspond to a fundamental mode and its low-order harmonics, whose ratios are $2{:}1$ and $3{:}2$.

The physical origin of the QPOs in these three models can be summarized as follows. In $A_1$, the pressure-supported torus that forms after the destruction of the cone oscillates in its global modes. 

For small tori, the $p$-mode eigenfrequencies form the sequence $2{:}3{:}4{:}\ldots$ to within $5$--$10\%$~\cite{QPOs9}, and these oscillations modulate the accretion rate quasi-periodically~\cite{Zanotti2003}. Tori that are continuously perturbed by matter supplied from larger radii, as in the present BHL flow, also develop the Papaloizou--Pringle instability and oscillate around their equilibrium with frequencies that depend on the torus size and on the black hole spin~\cite{DonmezMNRAS2014}. The members $n=2$--$4$ of the ladder of $A_1$ (6.2, 10.3 and 14.2\,Hz) stand roughly in this ratio, although, with the present frequency resolution, a $p$-mode sequence cannot be distinguished from a pure harmonic ladder. Vertical (warp) modes cannot be excited in the present equatorial simulations. In $A_2$ and $A_3$, the flip-flop motion of the shock produces a fundamental oscillation, with a period of about three Keplerian periods at the accretion radius, together with its harmonics. In the near-Kerr and Kerr models, oscillations trapped in the stable shock cone~\cite{DonmezMNRAS2011} produce peaks of much smaller amplitude. Among them, a peak at $\approx9$\,Hz appears in all three models $B_1$, $C$ and Kerr, close to the $9.4$\,Hz harmonic of $A_2$ and $A_3$. This suggests that the flip-flop amplifies an oscillation already present in the stable cone. The modes in $A_2$ and $A_3$ and in $B_1$, $C$ and Kerr would then share the same physical origin, namely oscillations confined within the shock cone, but differ greatly in strength.

The lower panel of Fig.~\ref{QPO_a09} shows that the peaks of the Kerr-like, near-Kerr, and Kerr models ($B_1$, $C$, Kerr) are much weaker than those of the aggressive models. Their PSD is about three orders of magnitude lower in the units of the figure, in line with the very weak fluctuations of their accretion rates at $r=2.3M$ ($\lesssim\pm0.5\%$). These three models are physically equivalent ($\eta<10^{-3}$) and have identical azimuthal profiles, so the differences between their spectra measure the noise level of the PSD. Only the $\approx9$\,Hz peak is common to all three. The other peaks (e.g., at $3.4$, $6.5$, $13.5$, $20$ and $25$\,Hz) appear in only one or two of the models and exceed the high-frequency noise by factors of only two to three, so they cannot be identified as robust QPOs. The observability of QPOs is thus greatly enhanced by strong curvature. The same analysis that reveals a clear harmonic structure in the aggressive models finds only a weak feature in the Kerr-like cases. 
We now use Eq.~(\ref{f_Hz}) to convert into physical units the test-particle frequencies at $r=6.11M$ obtained in Sec.~\ref{mass_acc} from the Kerr–Weyl geodesics~\cite{ZULQARNAIN2026117449}. The orbital frequency is $\approx200$\,Hz for both spins. The radial epicyclic frequency is $\approx130$\,Hz for $a=0.9M$ and $\approx105$\,Hz for $a=0.5M$, and the periastron precession frequency is $\approx70$\,Hz and $\approx100$\,Hz, respectively. In no model does a peak rise clearly above the noise at these frequencies. The QPOs found here are therefore set by the large-scale dynamics of the shock and the torus rather than by local geodesic motion. This explains why their amplitudes and radial dependence differ systematically between models with different values of $\eta$.

\begin{figure*}[!ht]
  \vspace{1cm}
  \centering
\psfig{file=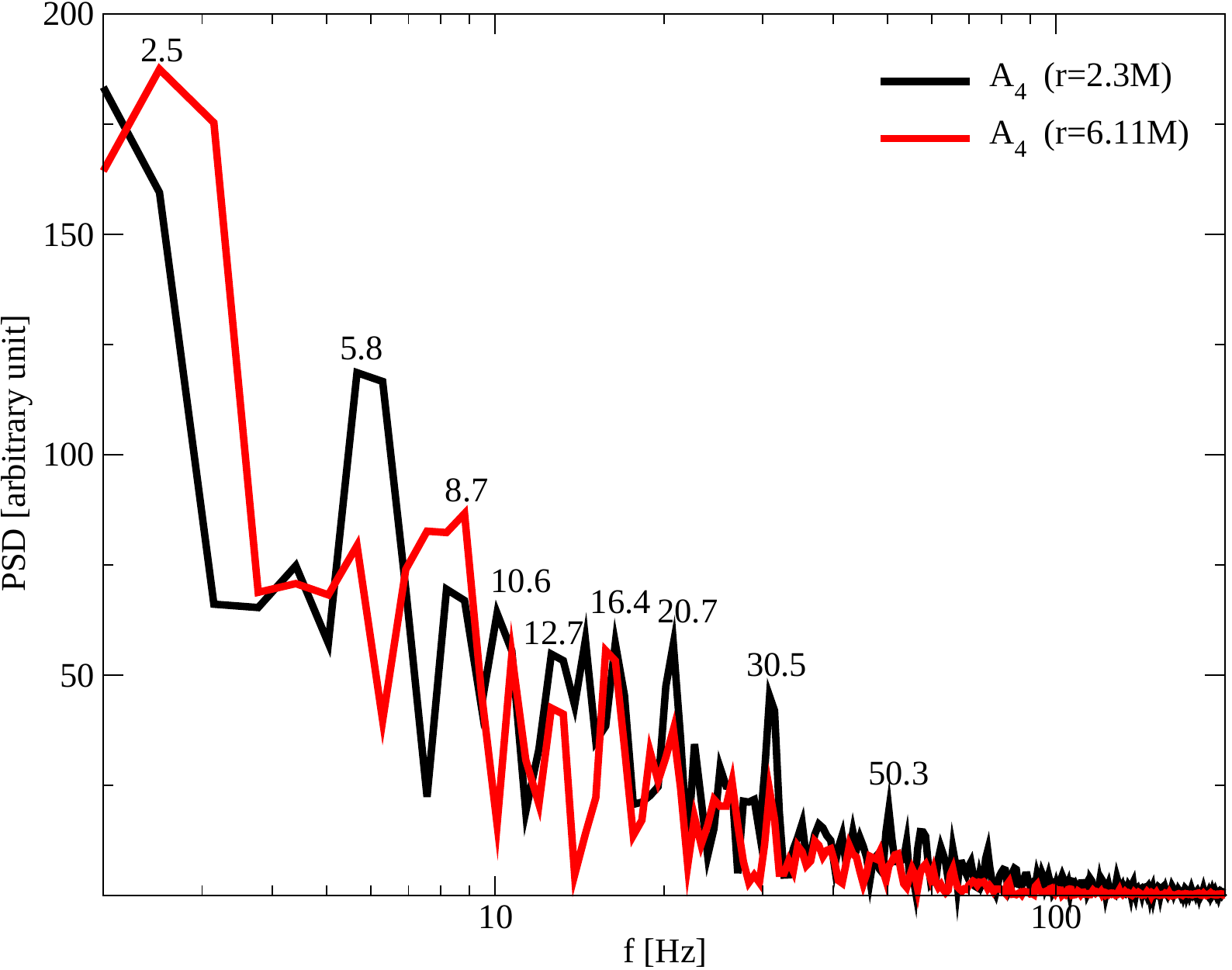,width=13.5cm,height=7.0cm}\\
\vspace{0.3cm}
\psfig{file=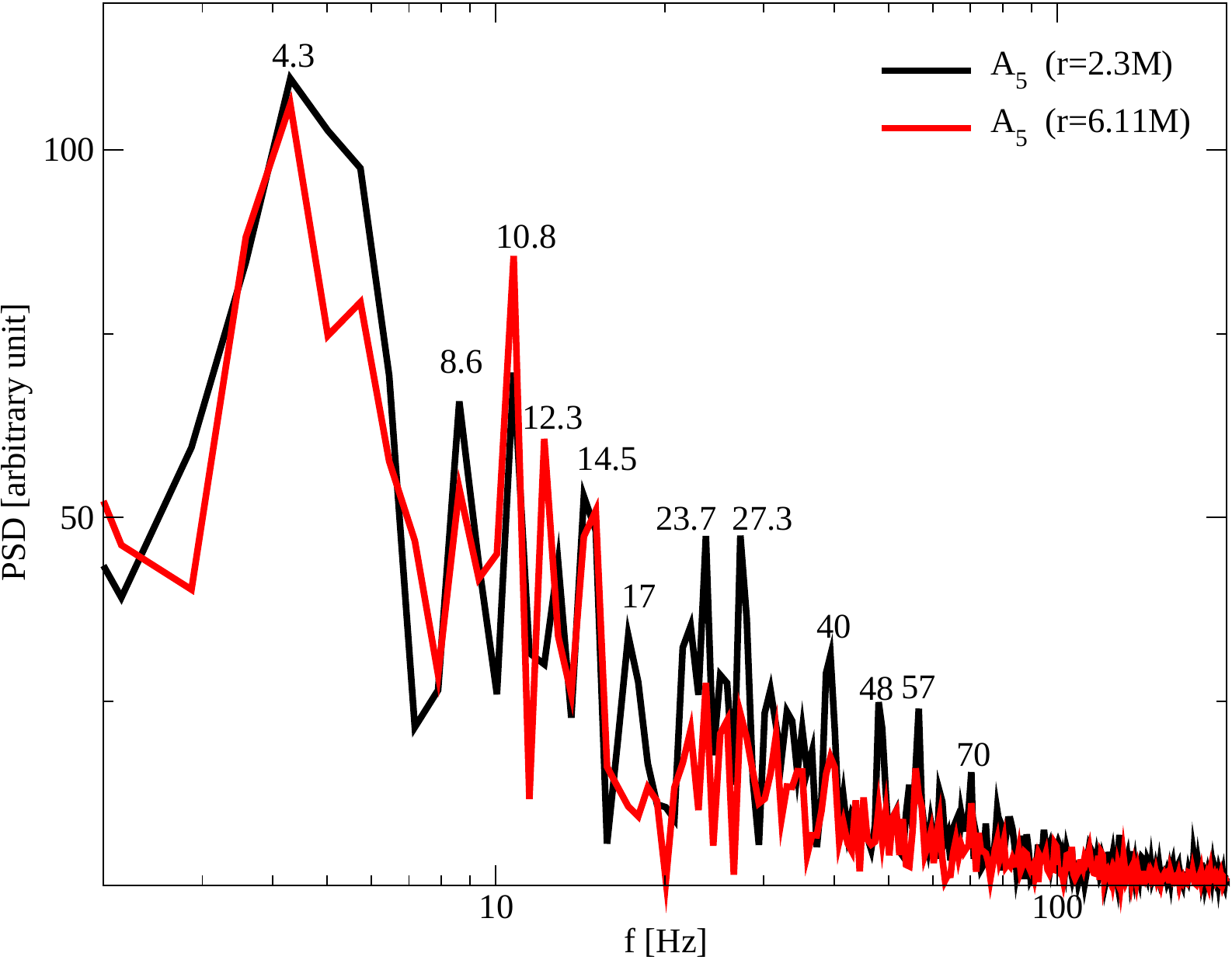,width=13.5cm,height=7.0cm}\\
\vspace{0.3cm}
\psfig{file=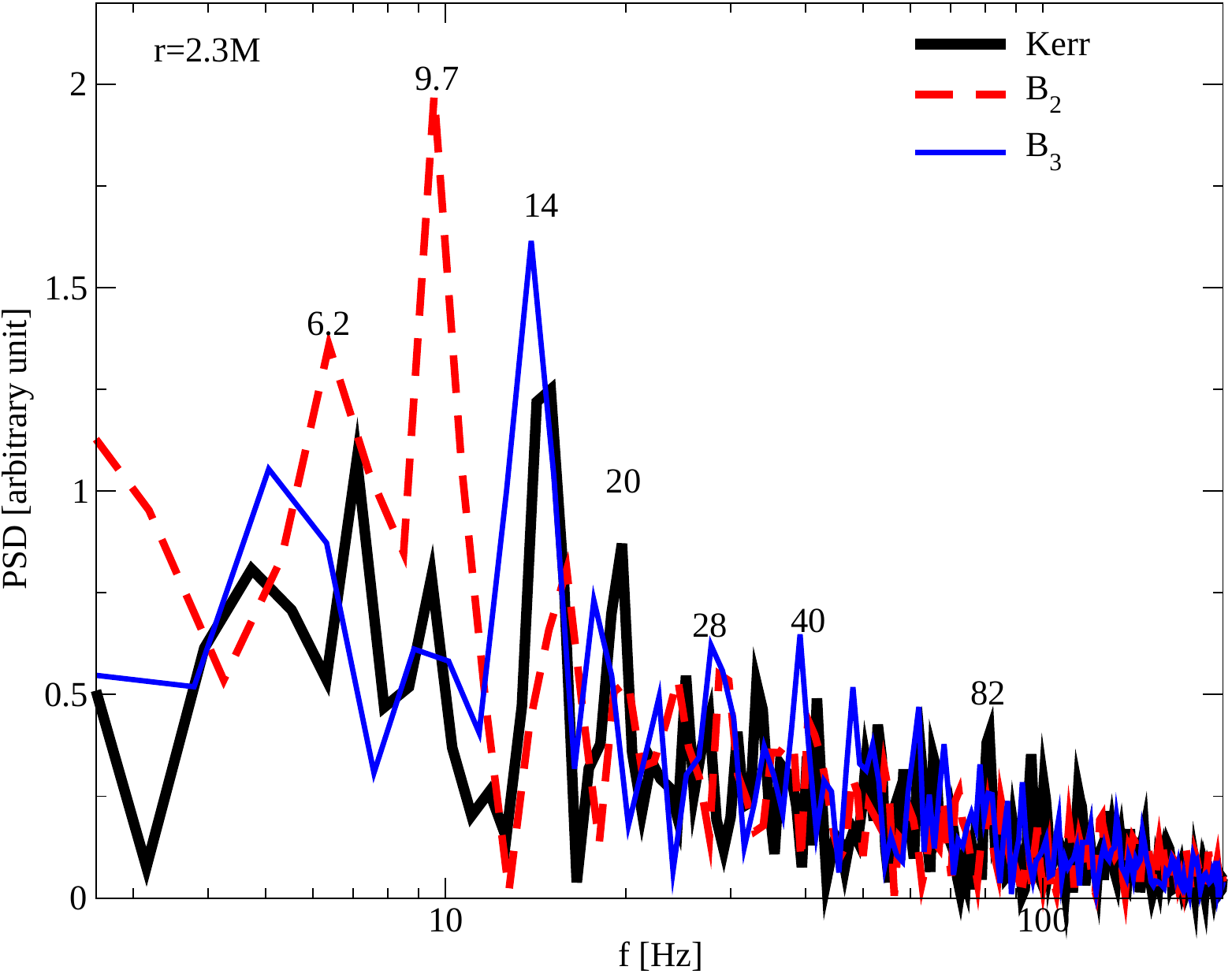,width=13.5cm,height=7.0cm} 
  \caption{Same as Fig.~\ref{QPO_a09}, but for the black hole with spin parameter $a=0.5M$. The upper panel shows the flip-flop Model $A_4$, and the middle panel the torus-forming Model $A_5$, both at $r=2.3M$ and $r=6.11M$.  The lower panel shows,  at $r=2.3M$, the Kerr, $B_2$ and $B_3$ models, in which a stable shock cone forms.
 The shorter $B_2$ and $B_3$ runs have a coarser frequency resolution, $\Delta f\gtrsim1$\,Hz.}
  \vspace{1cm}
  \label{QPO_a05}
\end{figure*}

For $a=0.5M$, Fig.~\ref{QPO_a05} shows the corresponding PSDs. In Model $A_4$ ($\eta\simeq0.30$), the shock undergoes flip-flop oscillations with intermittent closing of the upstream window. The upper panel again shows a red-noise continuum and a series of peaks at nearly identical frequencies at both radii, confirming that these modes are global. The peaks at $5.8$ and $8.7$\,Hz are the second and third harmonics of $f_0\simeq2.9$\,Hz (a period of $\approx7\times10^{3}M$). The fundamental appears as the $2.5$\,Hz peak at $r=6.11M$, but since it contains only about four cycles, its position is uncertain by about one bin. Further peaks appear at $10.6$, $12.7$, $16.4$, $20.7$, $30.5$ and $50.3$\,Hz, and above $\approx10$\,Hz they are stronger at $r=2.3M$ than at $r=6.11M$. Their ratios include $30.5/20.7=1.47\pm0.03\approx3{:}2$ and $16.4/8.7=1.89\pm0.09$, which is close to $2{:}1$.

In the middle panel, Model $A_5$ ($\eta\simeq1.6$), another torus-forming case, again shows peaks at the same frequencies at both radii, confirming that these are also global modes. The strongest peak, at $4.3$\,Hz (a period of $\approx4.7\times10^{3}M$), and the peaks at $8.6$, $12.3$ and $17$\,Hz form a harmonic sequence $1{:}2{:}3{:}4$ with $f_0=4.2$\,Hz, all within $\Delta f/2$ of $nf_0$. The corresponding ratios are $8.6/4.3=2.00\pm0.18$, $12.3/8.6=1.43\pm0.07$ and $17/8.6=1.98\pm0.09$. This pattern is consistent with a fundamental mode and its overtones in a thick, pressure-supported inner flow with toroidal or breathing dynamics~\cite{QPOs9,Zanotti2003}. The peak at $10.8$\,Hz, the strongest one at $r=6.11M$, does not belong to this sequence ($10.8/4.2\approx2.6$) and may correspond to an independent mode of the torus. The remaining peaks, at $14.5$, $23.7$, $27.3$, $40$, $48$, $57$ and $70$\,Hz, are weaker. In contrast, the peak amplitudes of the Kerr-like and Kerr models ($B_2$, $B_3$ and Kerr, lower panel) are much smaller, because the shock cone in these models is stable and exhibits only small-amplitude oscillations. The physically equivalent Kerr and $B_3$ runs ($\eta<10^{-3}$) share a dominant peak at $\approx14$\,Hz (a period of $\approx1.45\times10^{3}M$). $B_3$ also shows its second harmonic at $28$\,Hz, with $28/14=2.00\pm0.09$. Model $B_2$ ($\eta\simeq0.03$) instead has its strongest peaks at $6.2$ and $9.7$\,Hz, with $9.7/6.2=1.56\pm0.16\approx3{:}2$. The $B_2$ and $B_3$ runs are shorter, with $T\lesssim2\times10^{4}M$ and $\Delta f\gtrsim1$\,Hz, so longer simulations are needed to decide whether this shift is a physical effect of the small curvature term.

Comparing Figs.~\ref{QPO_a09} and~\ref{QPO_a05} shows that the effect of the black hole spin on the PSD depends on the flow regime. In the torus regime, the PSD of $A_1$ ($a=0.9M$) is about three times stronger than that of $A_5$ ($a=0.5M$) and contains more peaks, whose amplitudes vary rapidly. In the flip-flop regime, $A_2$ and $A_4$ have peak powers of the same order. In the Kerr-like regime, the $a=0.5M$ models show ten to twenty times more power at $r=2.3M$ than the $a=0.9M$ models, in line with the larger relative fluctuations of their accretion rates at that radius. In the aggressive models, the frequencies show no systematic dependence on the spin. Their fundamentals lie between $\approx2.9$ and $\approx4.8$\,Hz, corresponding to periods of $(4$--$7)\times10^{3}M$, and commensurate $3{:}2$ and $2{:}1$ ratios appear for both spins. In the stable cones, by contrast, the dominant peak lies at $\approx9$\,Hz for $a=0.9M$ and at $\approx14$\,Hz for $a=0.5M$ (Table~\ref{tab_QPO_obs}). The weak oscillations trapped in a stable cone are therefore sensitive to the spin, whereas the large-amplitude oscillations driven by the curvature term are not.

This follows from the separation of scales found above. Frame dragging is significant only within a few gravitational radii of the horizon, where it affects the shear of the inner flow and hence the amplitude of the oscillations measured at $r=2.3M$. The frequencies, by contrast, are set by the shock and the torus on scales $r\lesssim R_{\rm acc}$, which $k$ controls through $\eta$. In the flip-flop and torus regimes, the spin therefore mainly affects the observability of the QPOs, but not their fundamental frequencies or their commensurate ratios. The astrophysical interpretation of these frequencies is given in the next section.

 
\section{Astrophysical Signatures and Observational Prospects of Shock-Driven QPO Mechanisms}
\label{obser_signature} 

The accretion flow around the Kerr--Weyl black hole reveals a complex interaction among spacetime curvature, rotation and hydrodynamic instabilities, each of which leaves an imprint on the variability of the flow. The simulations show three distinct regimes: a standing shock cone, a flip-flop oscillating shock, and a quasi-steady oscillating torus. They are ordered by the ratio $\eta$ of the curvature-induced to the Newtonian acceleration at the accretion radius, Eq.~(\ref{eta}). Each regime corresponds to a different mechanism capable of trapping and exciting QPO modes. The PSD analysis showed that the resulting frequencies are set by the large-scale dynamics of the shock and the torus on scales $r\lesssim R_{\rm acc}$, rather than by local geodesic motion near the horizon~\cite{ZULQARNAIN2026117449}. In this section, we translate these results into observable quantities. We derive scaling relations for the regime boundaries and the QPO frequencies, convert the frequencies to physical units for representative black hole masses, estimate the detectability of the predicted signals, and discuss what observations can constrain. The main quantities are collected in Table~\ref{tab_QPO_obs}.

Since $\eta=|k|R_{\rm acc}^{3}/\widetilde{M}$ and $R_{\rm acc}=2\widetilde{M}/(v_\infty^{2}+c_s^{2})$, the curvature parameter at which a given regime appears depends on the asymptotic flow,
\begin{equation}
|k|=\frac{\eta\,(v_\infty^{2}+c_s^{2})^{3}}{8\widetilde{M}^{2}}.
\label{k_eta}
\end{equation}
For the flow simulated here, $|k|\simeq1.56\times10^{-5}\,\eta\,M^{-2}$, and the models bracket the boundaries between the regimes. The onset of the flip-flop instability lies between $\eta=0.03$ ($B_2$, stable) and $\eta=0.19$ ($A_2$, $A_3$), i.e., between $|k|\simeq5\times10^{-7}$ and $3\times10^{-6}\,M^{-2}$. The formation of a torus sets in between $\eta=0.30$ ($A_4$) and $\eta=1.6$ ($A_5$), i.e., between $|k|\simeq5\times10^{-6}$ and $2.4\times10^{-5}\,M^{-2}$. Because of the steep dependence on $v_\infty^{2}+c_s^{2}$ in Eq.~(\ref{k_eta}), these thresholds are not universal. For a wind with $(v_\infty^{2}+c_s^{2})^{1/2}=3\times10^{-3}c$ ($\approx900$\,km\,s$^{-1}$), $R_{\rm acc}\simeq2\times10^{5}M$, and the same values of $\eta$ are reached for $|k|\sim10^{-17}$--$10^{-16}\,M^{-2}$, provided that the regimes remain ordered by $\eta$ outside the simulated range.

The characteristic frequencies can be expressed in a similar way. The Keplerian frequency at the accretion radius is
\begin{equation}
f_{\rm K}(R_{\rm acc})=\frac{1}{2\pi}\left(\frac{GM}{R_{\rm acc}^{3}}\right)^{1/2}=\frac{(v_\infty^{2}+c_s^{2})^{3/2}}{2^{5/2}\pi}\,\frac{c^{3}}{GM},
\label{fK_Racc}
\end{equation}
which equals $12.8$\,Hz for the simulated flow and $M=10\,M_\odot$. Here $f_{\rm K}(R_{\rm acc})$ is evaluated without the curvature term, so that it depends on the asymptotic flow alone. Including the term $|k|$ in $\Omega_{\rm K}^2$, as in Sec.~\ref{mass_acc}, would raise it by a factor $(1+\eta)^{1/2}$, i.e., by $\approx9$--$14\%$ in the flip-flop models and by $\approx60$--$70\%$ in the torus models. Writing the fundamental frequency of each model as $f_0=\xi f_{\rm K}(R_{\rm acc})$, we find $\xi\simeq0.7$--$1.1$ for the dominant peaks of the stable cones and $\xi\simeq0.23$--$0.38$ for the flip-flop and torus regimes (Table~\ref{tab_QPO_obs}). The small-amplitude oscillations of a stable cone therefore occur close to the Keplerian frequency at the accretion radius, as found in Newtonian simulations~\cite{Blondin2009}. The large-amplitude flip-flop and torus oscillations are slower by a factor of three to four. 

Within each of these two regimes, $\xi$ is not constant but decreases with $\eta$. In the flip-flop regime, it falls from $0.38$ in $A_2$ and $A_3$ ($\eta\simeq0.19$) to $0.23$ in $A_4$ ($\eta\simeq0.30$), and in the torus regime from $0.33$ in $A_5$ ($\eta\simeq1.6$) to $0.27$ in $A_1$ ($\eta\simeq1.9$).  Writing $\xi\propto\eta^{-p}$ within each regime, we obtain from these pairs $p=1.1\pm0.3$ and $p=1.0\pm0.6$, respectively, where the uncertainties follow from the frequency resolution, $\delta f_0\simeq\Delta f/2$. The two values are consistent, and we adopt $p\simeq1.1$. This dependence is not an effect of the spin. The higher-$\eta$ member of each pair has the lower spin in the flip-flop regime ($A_4$, $a=0.5M$) but the higher spin in the torus regime ($A_1$, $a=0.9M$), and in both cases it has the lower frequency, whereas $A_2$ and $A_3$, which share the same $\eta$, have the same $f_0$. The trend instead reflects the behavior of the shock described in Sec.~\ref{azimuthal_cut}: the larger $\eta$, the more shocked gas accumulates and the farther upstream the shock is pushed, so the oscillating region becomes larger and its period longer. At the transition from the flip-flop to the torus regime, $\xi$ does not continue to decrease but rises from $0.23$ in $A_4$ to $0.33$ in $A_5$, both with $a=0.5M$, because the oscillating structure changes from a swinging shock to a torus. Since each regime is populated by only two models, this scaling should be regarded as provisional until it is confirmed by a finer scan in $\eta$.

If confirmed, this relation would turn the detection of a shock-driven QPO into a measurement of the curvature term. For a source with known mass and wind parameters, Eqs.~(\ref{k_eta}) and (\ref{fK_Racc}) together with $\xi=\xi_i(\eta/\eta_i)^{-p}$ give
\begin{equation}
|k|\simeq\frac{\eta_i\,(v_\infty^{2}+c_s^{2})^{3}}{8\widetilde{M}^{2}}\left[\frac{f_0}{\xi_i\,f_{\rm K}(R_{\rm acc})}\right]^{-1/p},
\label{k_from_f0}
\end{equation}
where $(\eta_i,\xi_i)=(0.19,0.38)$ in the flip-flop regime and $(1.6,0.33)$ in the torus regime, and each relation is applied only within the range of $\eta$ of its regime. The absence of shock-driven QPOs would bounds $|k|$ from above through Eq.~(\ref{k_eta}). Their detection, together with the frequency of the fundamental, would instead determine it.

Combined with the BHL accretion rate~\cite{Edgar2004},
\begin{equation}
\dot M_{\rm BHL}=\frac{4\pi G^{2}M^{2}\rho_\infty}{(v_\infty^{2}+c_s^{2})^{3/2}},
\label{Mdot_BHL}
\end{equation}
Eq.~(\ref{fK_Racc}) yields a testable prediction. For a given black hole, changes in the density of the supplied gas change the accretion rate but not the QPO frequency. 
Changes in its velocity at fixed density instead produce an anticorrelation between the QPO frequency and the accretion rate. For constant $\xi$, Eqs.~(\ref{fK_Racc}) and (\ref{Mdot_BHL}) give $f_0\propto\dot M_{\rm BHL}^{-1}$. Since $\eta\propto(v_\infty^{2}+c_s^{2})^{-3}$ at fixed $k$, the decrease of $\xi$ with $\eta$ makes this dependence much steeper, $f_0\propto(v_\infty^{2}+c_s^{2})^{3/2+3p}\propto\dot M_{\rm BHL}^{-(1+2p)}\approx\dot M_{\rm BHL}^{-3}$, as long as the flow remains in the same regime.
A shock-driven QPO in a wind-fed source should therefore not show the tight positive correlation between QPO frequency and accretion rate found in disk-fed systems such as GRS~1915+105~\cite{Liu:2020jig}.

The frequencies follow from Eq.~(\ref{f_Hz}) for any mass. Equivalently, a period $\hat P$ in units of $M$ corresponds to $P=\hat P\,GM/c^{3}\simeq4.93\times10^{-6}\,\hat P\,(M/M_\odot)$\,s. Table~\ref{tab_QPO_obs} lists the frequencies for $M=10\,M_\odot$, together with the corresponding periods for Sgr~A$^*$ ($M=4.30\times10^{6}M_\odot$~\cite{GRAVITY2022}) and M87$^*$ ($M=6.5\times10^{9}M_\odot$~\cite{EHT2019VI}). For stellar-mass black holes of $5$--$20\,M_\odot$, the fundamentals of the flip-flop and torus regimes lie at $\approx1.5$--$10$\,Hz, well inside the LFQPO band~\cite{RemillardMcClintock2006,IngramMotta2019}. For supermassive black holes, the same modes correspond to periods of about one to two days for Sgr~A$^*$ and four to seven years for M87$^*$. These values refer to the simulated flow with $R_{\rm acc}=40M$. For the slower winds expected in galactic nuclei, Eq.~(\ref{fK_Racc}) gives much longer periods.

For $\eta\lesssim0.03$, which corresponds here to $|k|\lesssim5\times10^{-7}\,M^{-2}$, BHL accretion produces a standing shock cone in the downstream region, behind the black hole. The cone is nearly stationary, its shape being set by the balance between the gravitational attraction of the black hole and the ram pressure of the incoming gas. It undergoes only small oscillations, driven by acoustic perturbations trapped in the post-shock region~\cite{DonmezMNRAS2011}. These oscillations produce weak peaks within the LFQPO band, the dominant ones at $\approx9$\,Hz for $a=0.9M$ and $\approx14$\,Hz for $a=0.5M$ (for $M=10\,M_\odot$). Their amplitude, however, is very small. The fractional rms of the accretion rate at $r=2.3M$ is only $\approx0.1\%$ for $a=0.9M$ and $\approx1\%$ for $a=0.5M$, whereas the type-C LFQPOs of black hole binaries reach fractional rms amplitudes of up to $\approx20\%$~\cite{Motta2016}. If the X-ray flux follows the accretion rate, the oscillations of a stable cone alone can therefore account only for weak LFQPOs. As an example, consider GRS~1915+105, whose black hole mass is $12.4^{+2.0}_{-1.8}\,M_\odot$~\cite{Reid2014}. For this mass, the predicted frequencies are $\approx7$--$11$\,Hz for the stable cone and $\approx2$--$4$\,Hz for the flip-flop and torus regimes, of the same order as the LFQPOs observed in this source~\cite{QPOs19,Liu:2020jig}. GRS~1915+105 is, however, a low-mass X-ray binary fed through Roche-lobe overflow~\cite{Greiner2001}, so this comparison concerns the frequency scale only. BHL accretion is most directly relevant to wind-fed X-ray binaries and to black holes moving through, or embedded in, ambient gas~\cite{Edgar2004}.

For intermediate values, $\eta\simeq0.19$--$0.30$ in our models ($|k|\simeq(3$--$5)\times10^{-6}\,M^{-2}$), the additional attraction suppresses the downstream escape of shocked gas, widens the cone and destabilizes it, leading to flip-flop oscillations. The shock front swings and breathes around the black hole, producing large-amplitude fluctuations of the density and of the accretion rate, whose fractional rms reaches $\approx20$--$40\%$. The PSD consists of a fundamental at $f_0\simeq2.9$--$4.8$\,Hz (for $M=10\,M_\odot$) and a ladder of its harmonics, so that $3{:}2$ and $2{:}1$ ratios appear between neighboring harmonics. These frequencies do not coincide with the local radial epicyclic frequency of test particles~\cite{ZULQARNAIN2026117449}, which is $\approx105$--$130$\,Hz at $r=6.11M$ for the same mass. Nor do they correspond to vertical modes, which cannot be excited in the present equatorial simulations. The $3{:}2$ pairs are therefore not the twin-peak HFQPOs observed in microquasars. For example, placing the $9.4$ and $14.2$\,Hz harmonics of $A_2$ at the $300$ and $450$\,Hz QPOs of GRO~J1655$-$40~\cite{AbramowiczKluzniak2001} would require a mass of only
\begin{equation}
M_{\rm req}=M_{\rm ref}\,\frac{f_{\rm sim}(M_{\rm ref})}{f_{\rm obs}}\simeq10\,M_\odot\times\frac{9.4\,{\rm Hz}}{300\,{\rm Hz}}\simeq0.3\,M_\odot ,
\label{M_req}
\end{equation}
far below that of any known black hole. Instead, the flip-flop regime produces strong LFQPOs with a harmonic structure, as is commonly observed for type-C QPOs~\cite{IngramMotta2019}.

For $\eta\gtrsim1.6$ ($|k|\gtrsim2.4\times10^{-5}\,M^{-2}$ here), the curvature-induced attraction exceeds the Newtonian one inside the capture region. The cone is destroyed, and the gas forms a torus around the black hole. The torus remains dynamically active and supports global oscillation modes, such as radial and azimuthal breathing modes. Periodic compression and rarefaction within the torus modulate the accretion rate~\cite{Zanotti2003}, which varies by $\approx30$--$60\%$ rms and, in $A_1$, is repeatedly interrupted when gas is stored in the torus. The PSD shows a red-noise continuum and a nearly harmonic sequence of peaks, with fundamentals at $3.4$ and 4.2\,Hz for $M=10\,M_\odot$. This is qualitatively consistent with the $2{:}3{:}4$ sequence of the $p$-mode eigenfrequencies of small tori~\cite{QPOs9}, while nonlinear couplings associated with the strong instabilities give rise to additional peaks. These frequencies are about twenty times lower than the periastron precession frequency of test particles at $r=6.11M$, $\approx70$--$100$\,Hz~\cite{ZULQARNAIN2026117449}. Moreover, Lense--Thirring precession of a tilted inner flow, the leading geometric model for type-C LFQPOs~\cite{IngramMotta2019}, whose test-particle frequency in this spacetime is given in Ref.~\cite{ZULQARNAIN2026117449}, cannot be captured by the present equatorial simulations. The torus oscillations therefore provide a hydrodynamic LFQPO mechanism that is distinct from, and potentially complementary to, the precession models.

The detectability of these signals can be estimated with the standard expression for the significance of a QPO of fractional rms amplitude $r$ and width $\Delta\nu$ in an observation of duration $T$~\cite{vanderKlis1989},
\begin{equation}
n_\sigma=\frac{1}{2}\,\frac{S^{2}}{S+B}\,r^{2}\left(\frac{T}{\Delta\nu}\right)^{1/2},
\label{n_sigma}
\end{equation}
where $S$ and $B$ are the source and background count rates. We assume that the X-ray flux follows the accretion rate, so that $r$ is approximately the fractional rms of $\dot M$. The exposure needed for a detection at the $n_\sigma$ level is then $T=\Delta\nu\,[2n_\sigma(S+B)/(S^{2}r^{2})]^{2}$. Consider a bright Galactic source with $S=10^{3}$\,counts\,s$^{-1}$, $B\ll S$, $\Delta\nu=1$\,Hz and $n_\sigma=5$. The weak oscillations of a stable cone then require $T\sim10^{4}$\,s for $a=0.5M$ and $T\sim10^{8}$\,s for $a=0.9M$, whereas for the flip-flop and torus regimes Eq.~(\ref{n_sigma}) gives $T<0.1$\,s. Since Eq.~(\ref{n_sigma}) assumes an observation covering many oscillation cycles, the exposure is in practice set not by photon statistics but by the need to sample many cycles, $T\gg f_0^{-1}\approx0.2$--$0.35$\,s, i.e., at least a few seconds. Because the rms amplitude of the QPO itself is smaller than the total rms of $\dot M$, these exposures are lower limits. They nevertheless show quantitatively that a strong curvature term makes the QPOs easily observable, whereas the cone oscillations of the Kerr-like models would remain practically undetectable.

Finally, the curvature parameter can be expressed in physical units, $k_{\rm phys}=\hat k\,(c^{2}/GM)^{2}$, where $\hat k$ is its value in units of $M^{-2}$:
\begin{equation}
k_{\rm phys}\simeq4.6\times10^{-18}\left(\frac{\hat k}{10^{-5}}\right)\left(\frac{10\,M_\odot}{M}\right)^{2}\ {\rm cm^{-2}}.
\label{k_phys}
\end{equation}
Fits of galactic rotation curves in conformal gravity require a linear coefficient $\gamma_0\simeq3.06\times10^{-30}$\,cm$^{-1}$ and a quadratic coefficient $\kappa\simeq9.54\times10^{-54}$\,cm$^{-2}$~\cite{MannheimOBrien2012}. Since $\gamma_0^{2}/4\sim10^{-60}$\,cm$^{-2}$ is negligible, they imply $|k|\sim10^{-53}$\,cm$^{-2}$, up to a factor of order unity that depends on the normalization of the potential. In units of $M^{-2}$, this corresponds to $\hat k\sim10^{-41}$ for $M=10\,M_\odot$, $\sim10^{-30}$ for Sgr~A$^*$ and $\sim10^{-23}$ for M87$^*$, i.e., $18$--$36$ orders of magnitude below the values used in our aggressive models. By Eq.~(\ref{k_eta}), a value $\eta\simeq0.2$ would then be reached only at a radius $R\simeq(\eta GM/c^{2}|k|)^{1/3}\approx10\,{\rm pc}\,(M/10\,M_\odot)^{1/3}$. This is far beyond the accretion radius of any X-ray binary, and for $M=10^{10}M_\odot$ it reaches $\approx10$\,kpc, the size of a galaxy, which is the scale on which the quadratic term was introduced. The regimes found here therefore probe local deviations from the Kerr geometry that are much larger than those implied by galactic dynamics. Observing the predicted signatures would point to such a local curvature term. Conversely, their absence in a source with known accretion parameters would constrain $|k|$ through Eq.~(\ref{k_eta}), with $\eta$ below the flip-flop threshold.

Compared with the Kerr geometry, a sufficiently strong curvature term thus produces strong LFQPOs with a harmonic structure, whose frequencies scale with the black hole mass and with the properties of the supplied gas, Eq.~(\ref{fK_Racc}). This distinguishes them from the orbital and precession frequencies of test particles, which depend only on $M$, $a$ and $k$~\cite{ZULQARNAIN2026117449}. Future high-throughput X-ray missions such as eXTP~\cite{Zhang:2025iae} and the Athena Wide Field Imager~\cite{Mayr:2024rbs} will be able to resolve such multi-peak QPO structures, building on current timing studies~\cite{Liu:2020jig}, and to test them against the predictions of Table~\ref{tab_QPO_obs}.

\begin{table*}[htbp]
\footnotesize
\setlength{\tabcolsep}{4pt}
\caption{Observational properties of the three flow regimes. For each group of models, we give the spin $a$; the parameter $\eta$ of Eq.~(\ref{eta}); the ratio $\xi=f_0/f_{\rm K}(R_{\rm acc})$ of the fundamental frequency to the Keplerian frequency at the accretion radius, Eq.~(\ref{fK_Racc}); the fundamental frequency $f_0$ for $M=10\,M_\odot$, converted with Eq.~(\ref{f_Hz}), together with other prominent peaks; the corresponding periods $P_0$ for Sgr~A$^*$ ($4.30\times10^{6}M_\odot$) and M87$^*$ ($6.5\times10^{9}M_\odot$); the fractional rms of the accretion rate at $r=2.3M$, measured from Figs.~\ref{mass_accc_a09_1} and~\ref{mass_accc_a05_1}; and the exposure $T_{5\sigma}$ for a $5\sigma$ detection from Eq.~(\ref{n_sigma}). The exposure assumes $S=10^{3}$\,counts\,s$^{-1}$, $B\ll S$, $\Delta\nu=1$\,Hz, and that the X-ray flux follows the accretion rate. For the stable-cone models, $f_0$ is the frequency of the dominant peak.}
\label{tab_QPO_obs}
\begin{center}
\begin{tabular}{llccccccccc}
\hline\hline
Regime & Models & $a\,(M)$ & $\eta$ & $\xi$ & $f_0$ (Hz) & Other peaks (Hz) & $P_0^{\rm Sgr\,A^*}$ (h) & $P_0^{\rm M87^*}$ (yr) & rms & $T_{5\sigma}$ (s)\\
\hline
Standing cone & $B_1$, $C$, Kerr & $0.9$ & $<10^{-3}$ & $0.69$ & $8.8$ & -- & $13.6$ & $2.3$ & $0.1\%$ & $\sim10^{8}$\\
 & $B_3$, Kerr & $0.5$ & $<10^{-3}$ & $1.10$ & $14$ & $28$ & $8.5$ & $1.5$ & $1\%$ & $\sim10^{4}$\\
 & $B_2$ & $0.5$ & $0.03$ & $0.76$ & $9.7$ & $6.2$ & $12.3$ & $2.1$ & $0.7\%$ & $\sim5\times10^{4}$\\
\hline
Flip-flop & $A_2$, $A_3$ & $0.9$ & $0.19$ & $0.38$ & $4.8$ & $9.4$, $14.2$, $29$ & $24.7$ & $4.3$ & $19\%$ & $<0.1$\\
 & $A_4$ & $0.5$ & $0.30$ & $0.23$ & $2.9$ & $5.8$, $8.7$, $30.5$ & $41.2$ & $7.1$ & $40\%$ & $<0.1$\\
\hline
Torus & $A_1$ & $0.9$ & $1.9$ & $0.27$ & $3.4$ & $6.2$, $10.3$, $14.2$ & $35.1$ & $6.1$ & $60\%$ & $<0.1$\\
 & $A_5$ & $0.5$ & $1.6$ & $0.33$ & $4.2$ & $8.6$, $10.8$, $12.3$ & $28.4$ & $4.9$ & $31\%$ & $<0.1$\\
\hline\hline
\end{tabular}
\end{center}
\end{table*}

\section{Conclusions}\label{S4} 
In this work, we have performed GRHD simulations of BHL accretion onto a rotating black hole in Weyl conformal gravity, described by the Kerr--Weyl metric~\cite{Asuncion:2025cxw}. This complements the test-particle and planar-accretion analysis of the same spacetime~\cite{ZULQARNAIN2026117449}. The flow was evolved on the equatorial plane for two spins, $a=0.9M$ and $a=0.5M$. All models used identical asymptotic conditions ($v_\infty=0.2c$, $c_s=0.1c$, $\Gamma=4/3$), so the differences between the runs are due to the spacetime alone. Nine Kerr--Weyl configurations, grouped into aggressive, mild-deviation and near-Kerr models (Table~\ref{Inital_Con_1}), were compared with Kerr runs of the same spin. The comparison used the rest-mass density, the mass accretion rate at $r=2.3M$ and $r=6.11M$, space--time diagrams of the azimuthal density at $r=6.14M$, and power spectra of the accretion rate.

The central result of this study is that the Kerr--Weyl correction controls the accretion flow through a single dimensionless quantity. This is the ratio $\eta=|k|R_{\rm acc}^{3}/\widetilde{M}$ of the curvature-induced to the Newtonian acceleration at the accretion radius, Eq.~(\ref{eta}). For the values of $k$ considered here, the horizon, the ISCO and the local test-particle frequencies are almost unchanged. The ISCO moves by less than $0.01M$, and the epicyclic frequencies change by at most $\approx2\%$. The $kr^{2}$ term, however, grows with radius and becomes comparable to the Newtonian term near $R_{\rm acc}$, where the gas is captured and the shock cone forms. As a result, all models fall into three regimes ordered by $\eta$. For $\eta\lesssim0.03$, a standing shock cone forms that is indistinguishable from the Kerr solution. At $r\approx6M$, its dense post-shock region spans $\approx60^\circ$, and its edges fluctuate by only $\approx0.01$\,rad. For $\eta\simeq0.2$--$0.3$, the cone widens until the dense post-shock region covers $\approx230^\circ$ of the circle, and the shock undergoes persistent flip-flop oscillations with rms excursions of $\approx0.3$\,rad. For $\eta\gtrsim1.6$, the cone is destroyed, and an oscillating, nearly axisymmetric torus forms around the black hole. For the simulated flow, the onset of the flip-flop instability lies between $|k|\simeq5\times10^{-7}$ and $3\times10^{-6}\,M^{-2}$, and the formation of a torus between $|k|\simeq5\times10^{-6}$ and $2.4\times10^{-5}\,M^{-2}$. These thresholds scale as $(v_\infty^{2}+c_s^{2})^{3}$, Eq.~(\ref{k_eta}), so they are properties of the flow as well as of the spacetime. This explains why no single critical value of $k$ separates the regimes.

The simulations also reveal the physical mechanism behind these regimes. For $k<0$, the AdS-like term adds an inward acceleration $|k|r$ that grows with distance. It deepens the potential at the scale of the cone, increases the inflow and, above all, prevents the shocked gas from escaping along the downstream axis. The gas accumulates behind the black hole, the post-shock pressure rises, and the shock is pushed upstream. The curvature term itself is smooth and does not oscillate. Instead, it drives the flow into the regime in which the flip-flop overstability of the BHL wake grows~\cite{Foglizzo2005,Blondin2009,DonmezMNRAS2012}, and, for $\eta>1$, it confines the shocked gas into a torus. Models $A_2$ and $A_3$, whose values of $\gamma$ differ in sign, evolve almost identically, with shock positions that agree to within $0.02$\,rad up to $t\simeq2.2\times10^{4}M$. As expected from the structure of the metric, the flow depends on the Weyl coefficients only through $k$, and this agreement shows that the flip-flop is deterministic rather than noise-driven.  The spin acts on a separate, smaller scale. Frame dragging is significant only within a few gravitational radii of the horizon. There, it shifts the stable cone in the sense of rotation, by an angle approximately proportional to $a$, and modulates the phase and amplitude of the oscillations, but it does not determine the flow regime. The Kerr-like models, evolved with the same numerical setup, show shock fluctuations at least thirty times smaller than the flip-flop swings. This rules out a numerical origin of the instability related to the treatment of the inner boundary~\cite{CruzOsorio2012}.

These regimes leave clear imprints on the accretion rate. In the aggressive models, the time-averaged rate near the horizon is about $1.7$--$3$ times the Kerr value. It is highest in the flip-flop models and lower in the torus-forming models, in which part of the captured gas is stored in the torus. In those models, accretion is repeatedly interrupted, at times with net outflow at $r=6.11M$. The fractional rms of the accretion rate reaches $\approx20$--$60\%$, compared with $0.1$--$1\%$ in the Kerr-like models. Even the mild-deviation model $B_2$, with $\eta\simeq0.03$, accretes about $4\%$ more than the Kerr black hole without any change in morphology. The accretion rate therefore responds to the curvature term before the structure of the flow does.

The power spectra show that the QPOs generated in these flows are global modes set by the large-scale dynamics of the shock and the torus. The same peaks appear at $r=2.3M$ and $r=6.11M$. Their frequencies are unrelated to the local orbital, epicyclic and precession frequencies of test particles~\cite{ZULQARNAIN2026117449}, which lie at $\approx70$--$200$\,Hz at $r=6.11M$ for $M=10\,M_\odot$. In the aggressive models, the spectra consist of a fundamental at $f_0\simeq2.9$--$4.8$\,Hz, for $M=10\,M_\odot$, and a ladder of its harmonics. In $A_2$ and $A_3$, for example, seven peaks lie within the frequency resolution of $nf_0$ with $f_0=4.83$\,Hz. The $3{:}2$ and $2{:}1$ ratios found in the spectra are therefore ratios between neighboring harmonics of a single, strongly non-sinusoidal oscillation, rather than the signature of a resonance between independent modes. The fundamental frequencies scale with the Keplerian frequency at the accretion radius, $f_0=\xi f_{\rm K}(R_{\rm acc})$, with $\xi\simeq0.7$--$1.1$ for the small-amplitude oscillations of a stable cone and $\xi\simeq0.23$--$0.38$  for the large-amplitude flip-flop and torus oscillations (Table~\ref{tab_QPO_obs}). Within the flip-flop and torus regimes, $\xi$ decreases with $\eta$ approximately as $\eta^{-1.1}$, so that a detected shock-driven QPO would measure, rather than only bound, the curvature term.

These results have direct observational consequences. For stellar-mass black holes of $5$--$20\,M_\odot$, the fundamentals fall at $\approx1.5$--$10$\,Hz, within the band of the type-C LFQPOs observed in black hole X-ray binaries~\cite{IngramMotta2019}. The $3{:}2$ pairs, however, lie at $\approx6$--$45$\,Hz, an order of magnitude below the HFQPO pairs of microquasars, and cannot be identified with them. For supermassive black holes and the simulated flow, the same modes correspond to periods of one to two days for Sgr~A$^*$ and four to seven years for M87$^*$. Under the assumptions of Table~\ref{tab_QPO_obs}, the large rms amplitudes of the flip-flop and torus regimes would make these QPOs detectable within a few seconds for a bright Galactic source. The cone oscillations of Kerr-like flows would instead require $10^{4}$--$10^{8}$\,s. The mechanism also makes a testable prediction. Since $f_0=\xi f_{\rm K}(R_{\rm acc})$ depends on the velocity of the supplied gas but not on its density, Eq.~(\ref{fK_Racc}), the frequency of a shock-driven QPO in a wind-fed source should be insensitive to changes in the density of the supplied gas. When the wind velocity changes, the frequency should instead be steeply anticorrelated with the accretion rate, approximately as $\dot M_{\rm BHL}^{-3}$ within a regime, unlike the positive correlation between frequency and accretion rate observed in disk-fed sources. Finally, the values of $|k|$ that produce the flip-flop and torus regimes exceed those inferred from galactic rotation curves~\cite{MannheimOBrien2012} by $18$--$36$ orders of magnitude. The signatures found here would therefore indicate a local curvature term much larger than the galactic one. Conversely, their absence in a source with known accretion parameters would constrain $|k|$ through Eq.~(\ref{k_eta}).

Several limitations of the present study define the next steps. The simulations are two-dimensional and restricted to the equatorial plane, so vertical modes and Lense--Thirring precession, which are central to geometric LFQPO models, cannot develop. 
Moreover, the flip-flop instability is known to be weaker in three dimensions. High-resolution Newtonian simulations found nearly axisymmetric three-dimensional Hoyle--Lyttleton flows~\cite{BlondinRaymer2012}, and three-dimensional relativistic simulations of supersonic winds onto black holes did not find the flip-flop instability in the parameter range explored~\cite{GraciaLinares2015}. The amplitude of the flip-flop oscillations found here should therefore be regarded as an upper limit. The mechanism that orders the regimes, the confinement of the shocked gas by the curvature term at $r\sim R_{\rm acc}$, follows from the radial force balance of Eq.~(\ref{eta}), which is the same in two and three dimensions. The ordering of the regimes by $\eta$ and the formation of a torus for $\eta>1$ are therefore expected to be more robust, but whether they persist in three dimensions remains to be tested. The flow is non-magnetized and non-radiative, so identifying the accretion rate with the X-ray flux remains an assumption. The asymptotic velocity $v_\infty=0.2c$, chosen to keep the accretion radius well inside the computational domain, is much higher than in most astrophysical winds. The extrapolation to realistic flows therefore relies on the scaling relations derived here. The runs cover $(2$--$3.5)\times10^{4}M$, which limits the frequency resolution to $\approx0.7$--$1$\,Hz for $M=10\,M_\odot$ and makes the lowest-frequency peaks tentative. In addition, the regime boundaries are bracketed rather than resolved. Several extensions are needed to test the universality of the $\eta$ classification and to turn the present predictions into quantitative observational tests. These include three-dimensional GRMHD simulations with radiative transfer, longer evolutions, a finer scan in $\eta$ and in the asymptotic flow, and an exploration of the de Sitter-like regime $k>0$ at comparable $|k|$. With the timing capabilities of current and future X-ray missions such as eXTP~\cite{Zhang:2025iae}, the combination of QPO frequencies, harmonic structure, amplitudes and their correlation with the accretion rate offers a promising route for probing conformal-gravity corrections in the strong-field regime around black holes.

\section*{Acknowledgments}
All numerical simulations were performed using the Phoenix High
Performance Computing facility at the American University of the Middle East (AUM), Kuwait.

\section*{Data Availability Statement}
The data generated and analyzed in this study are not publicly available because of their size and computational nature. They can, however, be obtained from the corresponding author upon reasonable request.

\bibliographystyle{apsrev4-2}  

\bibliography{Referances}

@article{Banyuls1997,
  author  = {Banyuls, F. and Font, J. A. and Ib{\'a}{\~n}ez, J. M. and Mart{\'\i}, J. M. and Miralles, J. A.},
  title   = {Numerical {3+1} general relativistic hydrodynamics: a local characteristic approach},
  journal = {Astrophys. J.},
  volume  = {476},
  pages   = {221},
  year    = {1997},
  doi     = {10.1086/303604}
}

@article{Font2008,
  author  = {Font, Jos{\'e} A.},
  title   = {Numerical hydrodynamics and magnetohydrodynamics in general relativity},
  journal = {Living Rev. Relativ.},
  volume  = {11},
  pages   = {7},
  year    = {2008},
  doi     = {10.12942/lrr-2008-7}
}

@article{DonatMarquina1996,
  author  = {Donat, R. and Marquina, A.},
  title   = {Capturing shock reflections: an improved flux formula},
  journal = {J. Comput. Phys.},
  volume  = {125},
  pages   = {42--58},
  year    = {1996},
  doi     = {10.1006/jcph.1996.0078}
}

@article{vanLeer1979,
  author  = {van Leer, B.},
  title   = {Towards the ultimate conservative difference scheme. {V}. {A} second-order sequel to {Godunov's} method},
  journal = {J. Comput. Phys.},
  volume  = {32},
  pages   = {101--136},
  year    = {1979},
  doi     = {10.1016/0021-9991(79)90145-1}
}

@article{MannheimOBrien2012,
  author  = {Mannheim, Philip D. and O'Brien, James G.},
  title   = {Fitting galactic rotation curves with conformal gravity and a global quadratic potential},
  journal = {Phys. Rev. D},
  volume  = {85},
  pages   = {124020},
  year    = {2012},
  doi     = {10.1103/PhysRevD.85.124020},
  eprint  = {1011.3495}
}

@article{GRAVITY2022,
  author  = {{GRAVITY Collaboration}},
  title   = {Mass distribution in the {Galactic Center} based on interferometric astrometry of multiple stellar orbits},
  journal = {Astron. Astrophys.},
  volume  = {657},
  pages   = {L12},
  year    = {2022},
  doi     = {10.1051/0004-6361/202142465}
}

@article{EHT2019VI,
  author  = {{Event Horizon Telescope Collaboration}},
  title   = {First {M87} {Event Horizon Telescope} Results. {VI}. {The} Shadow and Mass of the Central Black Hole},
  journal = {Astrophys. J. Lett.},
  volume  = {875},
  pages   = {L6},
  year    = {2019},
  doi     = {10.3847/2041-8213/ab1141},
  eprint  = {1906.11243}
}

@article{Reid2014,
  author  = {Reid, M. J. and McClintock, J. E. and Steiner, J. F. and others},
  title   = {A parallax distance to the microquasar {GRS 1915+105} and a revised estimate of its black hole mass},
  journal = {Astrophys. J.},
  volume  = {796},
  number  = {1},
  pages   = {2},
  year    = {2014},
  doi     = {10.1088/0004-637X/796/1/2},
  eprint  = {1409.2453}
}

@article{Motta2016,
  author  = {Motta, S. E.},
  title   = {Quasi periodic oscillations in black hole binaries},
  journal = {Astron. Nachr.},
  volume  = {337},
  pages   = {398},
  year    = {2016},
  doi     = {10.1002/asna.201612320},
  eprint  = {1603.07885}
}

@article{AbramowiczKluzniak2001,
  author  = {Abramowicz, M. A. and Klu{\'z}niak, W.},
  title   = {A precise determination of black hole spin in {GRO J1655-40}},
  journal = {Astron. Astrophys.},
  volume  = {374},
  pages   = {L19--L20},
  year    = {2001},
  doi     = {10.1051/0004-6361:20010791}
}

@article{Vaughan2005,
  author  = {Vaughan, S.},
  title   = {A simple test for periodic signals in red noise},
  journal = {Astron. Astrophys.},
  volume  = {431},
  pages   = {391--403},
  year    = {2005},
  doi     = {10.1051/0004-6361:20041453}
}

@incollection{vanderKlis1989,
  author    = {van der Klis, M.},
  title     = {Fourier techniques in {X}-ray timing},
  booktitle = {Timing Neutron Stars},
  editor    = {{\"O}gelman, H. and van den Heuvel, E. P. J.},
  series    = {NATO ASI Series C},
  volume    = {262},
  pages     = {27--69},
  publisher = {Springer},
  address   = {Dordrecht},
  year      = {1989},
  doi       = {10.1007/978-94-009-2273-0_3}
}

@article{CruzOsorio2012,
  author  = {Cruz-Osorio, A. and Lora-Clavijo, F. D. and Guzm{\'a}n, F. S.},
  title   = {Is the flip-flop behaviour of accretion shock cones on to black holes an effect of coordinates?},
  journal = {Mon. Not. Roy. Astron. Soc.},
  volume  = {426},
  number  = {1},
  pages   = {732--738},
  year    = {2012},
  doi     = {10.1111/j.1365-2966.2012.21794.x}
}

@article{Zanotti2003,
  author  = {Zanotti, O. and Rezzolla, L. and Font, J. A.},
  title   = {Quasi-periodic accretion and gravitational waves from oscillating `toroidal neutron stars' around a {Schwarzschild} black hole},
  journal = {Mon. Not. Roy. Astron. Soc.},
  volume  = {341},
  number  = {3},
  pages   = {832--848},
  year    = {2003},
  doi     = {10.1046/j.1365-8711.2003.06474.x},
  eprint  = {gr-qc/0210018}
}

@article{IngramMotta2019,
  author  = {Ingram, Adam R. and Motta, Sara E.},
  title   = {A review of quasi-periodic oscillations from black hole {X}-ray binaries: Observation and theory},
  journal = {New Astron. Rev.},
  volume  = {85},
  pages   = {101524},
  year    = {2019},
  doi     = {10.1016/j.newar.2020.101524},
  eprint  = {2001.08758}
}

@article{RemillardMcClintock2006,
  author  = {Remillard, Ronald A. and McClintock, Jeffrey E.},
  title   = {{X}-ray properties of black-hole binaries},
  journal = {Ann. Rev. Astron. Astrophys.},
  volume  = {44},
  pages   = {49--92},
  year    = {2006},
  doi     = {10.1146/annurev.astro.44.051905.092532},
  eprint  = {astro-ph/0606352}
}

@article{Foglizzo2005,
  author  = {Foglizzo, T. and Galletti, P. and Ruffert, M.},
  title   = {A fresh look at the unstable simulations of {Bondi--Hoyle--Lyttleton} accretion},
  journal = {Astron. Astrophys.},
  volume  = {435},
  pages   = {397--411},
  year    = {2005},
  doi     = {10.1051/0004-6361:20042201},
  eprint  = {astro-ph/0502168}
}

@book{RezzollaZanotti2013,
  author    = {Rezzolla, Luciano and Zanotti, Olindo},
  title     = {Relativistic Hydrodynamics},
  publisher = {Oxford University Press},
  address   = {Oxford},
  year      = {2013}
}

@article{Hoyle1939,
  author  = {Hoyle, F. and Lyttleton, R. A.},
  title   = {The effect of interstellar matter on climatic variation},
  journal = {Proc. Cambridge Phil. Soc.},
  volume  = {35},
  pages   = {405},
  year    = {1939}
}

@article{Bondi1944,
  author  = {Bondi, H. and Hoyle, F.},
  title   = {On the mechanism of accretion by stars},
  journal = {Mon. Not. Roy. Astron. Soc.},
  volume  = {104},
  pages   = {273},
  year    = {1944}
}

@article{Edgar2004,
  author  = {Edgar, Richard},
  title   = {A review of {Bondi--Hoyle--Lyttleton} accretion},
  journal = {New Astron. Rev.},
  volume  = {48},
  pages   = {843},
  year    = {2004},
  eprint  = {astro-ph/0406166}
}

@article{Mannheim1989,
  author  = {Mannheim, Philip D. and Kazanas, Demosthenes},
  title   = {Exact vacuum solution to conformal {Weyl} gravity and galactic rotation curves},
  journal = {Astrophys. J.},
  volume  = {342},
  pages   = {635},
  year    = {1989}
}

@article{FontIbanez1998,
  author  = {Font, J. A. and Ib{\'a}{\~n}ez, J. M.},
  title   = {Non-axisymmetric relativistic {Bondi--Hoyle} accretion on to a {Schwarzschild} black hole},
  journal = {Mon. Not. Roy. Astron. Soc.},
  volume  = {298},
  number  = {3},
  pages   = {835--846},
  year    = {1998},
  doi     = {10.1046/j.1365-8711.1998.01664.x},
  eprint  = {astro-ph/9804254}
}

@article{Font1999,
  author  = {Font, J. A. and Ib{\'a}{\~n}ez, J. M. and Papadopoulos, P.},
  title   = {Non-axisymmetric relativistic {Bondi--Hoyle} accretion on to a {Kerr} black hole},
  journal = {Mon. Not. Roy. Astron. Soc.},
  volume  = {305},
  number  = {4},
  pages   = {920--936},
  year    = {1999},
  doi     = {10.1046/j.1365-8711.1999.02459.x},
  eprint  = {astro-ph/9810344}
}

@article{Blondin2009,
  author  = {Blondin, John M. and Pope, T. Chris},
  title   = {Revisiting the ``flip-flop'' instability of {Hoyle--Lyttleton} accretion},
  journal = {Astrophys. J.},
  volume  = {700},
  number  = {1},
  pages   = {95--102},
  year    = {2009},
  doi     = {10.1088/0004-637X/700/1/95},
  eprint  = {0905.2769}
}

@article{DonmezMNRAS2012,
  author  = {D{\"o}nmez, Orhan},
  title   = {Relativistic simulation of flip-flop instabilities of {Bondi--Hoyle} accretion and quasi-periodic oscillations},
  journal = {Mon. Not. Roy. Astron. Soc.},
  volume  = {426},
  number  = {2},
  pages   = {1533--1545},
  year    = {2012},
  doi     = {10.1111/j.1365-2966.2012.21616.x}
}

@article{BHS1,
  author = {Einstein, A.},
  journal = {Ann. Phys.},
  volume = {49},
  pages = {769},
  year = {1916}
}

@article{BHS2,
  author = {Schwarzschild, K.},
  journal = {Sitzungsber. Preuss. Akad. Wiss. Berlin},
  volume = {1916},
  pages = {189},
  year = {1916}
}

@article{Asuncion:2025cxw,
    author = "Asuncion, Miguel Yulo and Horne, Keith and Kusano, Reinosuke and Dominik, Martin",
    title = "{Structure of Kerr black hole spacetimes in Weyl conformal gravity}",
    eprint = "2507.01379",
    archivePrefix = "arXiv",
    primaryClass = "gr-qc",
    reportNumber = "215005",
    doi = "10.1088/1361-6382/ae1109",
    journal = "Class. Quant. Grav.",
    volume = "42",
    number = "21",
    pages = "215005",
    year = "2025"
}

@article{BHS3,
  author = {Reissner, H.},
  journal = {Ann. Phys.},
  volume = {59},
  pages = {106},
  year = {1916}
}

@article{BHS4,
  author = {Nordström, G.},
  journal = {Proc. K. Ned. Akad. Wet.},
  volume = {20},
  pages = {1238},
  year = {1918}
}

@article{BHS5,
  author = {Finkelstein, D.},
  journal = {Phys. Rev.},
  volume = {110},
  pages = {965},
  year = {1958}
}

@article{BHS6,
  author = {Chandrasekhar, S.},
  journal = {Astrophys. J.},
  volume = {74},
  pages = {81},
  year = {1931}
}

@article{BHS7,
  author = {Oppenheimer, J. R. and Volkoff, G. M.},
  journal = {Phys. Rev.},
  volume = {55},
  pages = {374},
  year = {1939}
}

@article{BHS8,
  author = {Zeldovich, Y. B.},
  journal = {Dokl. Akad. Nauk SSSR},
  volume = {155},
  pages = {67},
  year = {1964}
}

@article{BHS9,
  author = {Salpeter, E. E.},
  journal = {Astrophys. J.},
  volume = {140},
  pages = {796},
  year = {1964}
}

@book{Ker1,
  author = {Chandrasekhar, S.},
  title = {The Mathematical Theory of Black Holes},
  publisher = {Oxford University Press},
  address = {Oxford},
  year = {1998}
}

@article{Ker2,
  author = {Carter, B.},
  journal = {Phys. Rev. Lett.},
  volume = {26},
  pages = {331},
  year = {1971}
}

@article{Ker3,
  author = {Robinson, D. C.},
  journal = {Phys. Rev. Lett.},
  volume = {34},
  pages = {905},
  year = {1975}
}

@article{Ker4,
  author = {Mars, M.},
  journal = {Class. Quantum Grav.},
  volume = {17},
  pages = {3353},
  year = {2000}
}

@article{Ker5,
  author       = {G{\"u}rlebeck, N.},
  title        = {},
  journal      = {Phys. Rev. Lett.},
  volume       = {114},
  pages        = {151102},
  year         = {2015},
}

@article{Ker7,
  author       = {Johannsen, T. and Psaltis, D.},
  title        = {},
  journal      = {Phys. Rev. D},
  volume       = {83},
  pages        = {124015},
  year         = {2011},
}

@article{Ker8,
  author       = {Yagi, K. and Stein, L. C.},
  title        = {},
  journal      = {Class. Quant. Grav.},
  volume       = {33},
  pages        = {054001},
  year         = {2016},
}

@article{Ker9,
  author       = {Bambi, C.},
  title        = {},
  journal      = {Mod. Phys. Lett. A},
  volume       = {26},
  pages        = {2453},
  year         = {2011},
}

@article{Ker10,
  author       = {Johannsen, T.},
  title        = {},
  journal      = {Class. Quant. Grav.},
  volume       = {33},
  pages        = {124001},
  year         = {2016},
}

@article{Ker11,
  author       = {Bambi, C. and C{\'a}rdenas-Avenda{\~n}o, A. and Dauser, T. and Garc{\'\i}a, J. A. and Nampalliwar, S.},
  title        = {},
  journal      = {Astrophys. J.},
  volume       = {842},
  pages        = {76},
  year         = {2017},
}

@article{Ker12,
  author       = {Bambi, C.},
  title        = {},
  journal      = {Phys. Rev. D},
  volume       = {87},
  pages        = {023007},
  year         = {2013},
}

@article{Ker13,
  author       = {Yunes, N. and Siemens, X.},
  title        = {},
  journal      = {Living Rev. Rel.},
  volume       = {16},
  pages        = {9},
  year         = {2013},
}

@article{Ker14,
  author       = {Gair, J. R. and Vallisneri, M. and Larson, S. L. and Baker, J. G.},
  title        = {},
  journal      = {Living Rev. Rel.},
  volume       = {16},
  pages        = {7},
  year         = {2013},
}

@article{Ker15,
  author       = {Abbott, B. P. and {et~al.}},
  title        = {},
  journal      = {Phys. Rev. Lett.},
  volume       = {118},
  pages        = {221101},
  year         = {2017},
}

@article{Ker16,
  author       = {Johannsen, T. and Psaltis, D.},
  title        = {},
  journal      = {Astrophys. J.},
  volume       = {718},
  pages        = {446},
  year         = {2010},
}

@article{Ker17,
  author       = {Bambi, C. and Freese, K. and Vagnozzi, S. and Visinelli, L.},
  title        = {},
  journal      = {Phys. Rev. D},
  volume       = {100},
  number       = {4},
  pages        = {044057},
  year         = {2019},
}

@article{Ker18,
  author       = {Akiyama, K. and {et~al.}},
  title        = {},
  journal      = {Astrophys. J. Lett.},
  volume       = {875},
  pages        = {L5},
  year         = {2019},
}

@article{Ker19,
  author       = {Konoplya, R. and Zhidenko, A.},
  title        = {},
  journal      = {Phys. Lett. B},
  volume       = {756},
  pages        = {350},
  year         = {2016},
}

@article{Ker20,
  author       = {Bamba, K. and Capozziello, S. and Nojiri, S. and Odintsov, S. D.},
  title        = {},
  journal      = {Astrophys. Space Sci.},
  volume       = {342},
  pages        = {155},
  year         = {2012},
}

@article{Ker21,
  author = {Nojiri, S. and Odintsov, S. D.},
  title = {},
  journal = {Int. J. Geom. Meth. Mod. Phys.},
  volume = {4},
  pages = {115},
  year = {2007}
}

@article{Ker22,
  author = {Clifton, T. and Ferreira, P. G. and Padilla, A. and Skordis, C.},
  title = {},
  journal = {Phys. Rep.},
  volume = {513},
  pages = {1},
  year = {2012}
}

@article{Ker23,
  author = {Will, C. M.},
  title = {},
  journal = {Living Rev. Rel.},
  volume = {9},
  pages = {3},
  year = {2006}
}

@article{Ker24,
  author = {Psaltis, D. and Perrodin, D. and Dienes, K. R. and Mocioiu, I.},
  title = {},
  journal = {Phys. Rev. Lett.},
  volume = {100},
  pages = {091101},
  year = {2008}
}

@article{Ker25,
  author = {de Felice, A. and Tsujikawa, S.},
  title = {},
  journal = {Living Rev. Rel.},
  volume = {13},
  pages = {3},
  year = {2010}
}

@article{Ker26,
  author = {Berti, E. and et al.},
  title = {},
  journal = {Gen. Rel. Grav.},
  volume = {50},
  pages = {49},
  year = {2018}
}

@article{QPOs1,
  author = {Rayimbaev, J. and Tadjimuratov, P. and Abdujabbarov, A. and Ahmedov, B. and Khudoyberdieva, M.},
  title = {},
  journal = {Galaxies},
  volume = {9},
  number = {4},
  pages = {75},
  year = {2021}
}

@article{QPOs2,
  author = {Jumaniyozov, S. and Khan, S. U. and Rayimbaev, J. and Abdujabbarov, A. and Urinbaev, S. and Murodov, S.},
  title = {},
  journal = {Eur. Phys. J. C},
  volume = {84},
  number = {9},
  pages = {964},
  year = {2024}
}

@article{QPOs3,
  author = {Faraji, S.},
  title = {},
  journal = {Astron. Rep.},
  volume = {67},
  pages = {S207--S213},
  year = {2023}
}

@article{QPOs4,
  author = {Stuchlík, Z. and Vrba, J.},
  title = {},
  journal = {JCAP},
  volume = {2021},
  number = {11},
  pages = {059},
  year = {2021}
}

@article{QPOs5,
  author = {Homan, J. and et al.},
  title = {},
  journal = {Astrophys. J.},
  volume = {586},
  pages = {1262},
  year = {2003}
}

@article{QPOs6,
  author = {Rayimbaev, J. and Ahmedov, B. and Bokhari, A. H.},
  title = {},
  journal = {Int. J. Mod. Phys. D},
  volume = {31},
  pages = {2240004},
  year = {2022}
}

@article{QPOs7,
  author = {Qi, M. and Rayimbaev, J. and Ahmedov, B.},
  title = {},
  journal = {Eur. Phys. J. C},
  volume = {83},
  pages = {730},
  year = {2023}
}

@article{QPOs9,
  author  = {Rezzolla, L. and Yoshida, S. and Maccarone, T. J. and Zanotti, O.},
  title   = {A new simple model for high-frequency quasi-periodic oscillations in black hole candidates},
  journal = {Mon. Not. R. Astron. Soc.},
  volume  = {344},
  pages   = {L37},
  year    = {2003},
  doi     = {10.1046/j.1365-8711.2003.07018.x},
  eprint  = {astro-ph/0307487}
}

@article{QPOs10,
  author = {Torok, G. and Kotrlova, A. and Sramkova, E. and Stuchlik, Z.},
  title = {},
  journal = {Astron. Astrophys.},
  volume = {531},
  pages = {A59},
  year = {2011}
}

@article{QPOs11,
  author = {Stuchlik, Z. and Kotrlova, A. and Torok, G.},
  title = {},
  journal = {Astron. Astrophys.},
  volume = {525},
  pages = {A82},
  year = {2011}
}

@article{QPOs12,
  author = {Stuchlík, Z. and Kotrlová, A. and Török, G.},
  journal = {Astron. Astrophys.},
  volume = {552},
  pages = {A10},
  year = {2013}
}

@article{QPOs13,
  author = {Mustafa, G. and Hussain, I. and Liu, W.-M.},
  journal = {Chin. J. Phys.},
  volume = {80},
  pages = {148},
  year = {2022}
}

@article{QPOs14,
  author = {Jiang, X. and Wang, P. and Yang, H. and Wu, H.},
  journal = {Eur. Phys. J. C},
  volume = {81},
  pages = {1043},
  year = {2021}
}

@article{QPOs15,
  author = {Amarilla, L. and Eiroa, E. F. and Giribet, G.},
  journal = {Phys. Rev. D},
  volume = {81},
  pages = {124045},
  year = {2010}
}

@article{QPOs16,
  author = {Liu, Y. and Mustafa, G. and Maurya, S. K. and Javed, F.},
  journal = {Eur. Phys. J. C},
  volume = {83},
  pages = {584},
  year = {2023}
}

@article{QPOs17,
  author = {Stuchlík, Z. and Kološ, M.},
  journal = {Astron. Astrophys.},
  volume = {586},
  pages = {A130},
  year = {2016}
}

@article{QPOs18,
  author = {Remillard, R. A. and et al.},
  journal = {Astrophys. J.},
  volume = {517},
  number = {2},
  pages = {L127},
  year = {1999}
}

@article{QPOs19,
  author  = {Morgan, E. H. and Remillard, R. A. and Greiner, J.},
  title   = {{RXTE} observations of {QPOs} in the black hole candidate {GRS 1915+105}},
  journal = {Astrophys. J.},
  volume  = {482},
  number  = {2},
  pages   = {993},
  year    = {1997}
}

@article{ZULQARNAIN2026117449,
  title   = {Probing rotating Weyl black hole through accretion-flow dynamics from planar spherical injection and frequencies analysis},
  journal = {Nuclear Physics B},
  volume  = {1027},
  pages   = {117449},
  year    = {2026},
  doi     = {10.1016/j.nuclphysb.2026.117449},
  author  = {Zulqarnain, Rana Muhammad and Donmez, Orhan and Bouzenada, Abdelmalek and Ditta, Allah and G{\"u}dekli, Ertan and Atamurotov, Farruh and Mardonov, Shukhrat}
}

@article{QPOs20,
  author = {Remillard, R. A. and et al.},
  journal = {Astrophys. J.},
  volume = {637},
  number = {2},
  pages = {1002},
  year = {2006}
}

@article{QPOs21,
  author = {Shahzadi, M. and Kološ, M. and Saleem, R. and Stuchlík, Z.},
  journal = {Class. Quantum Grav.},
  volume = {41},
  number = {7},
  pages = {075014},
  year = {2024}
}

@article{QPOs23,
  author = {Jumaniyozov, S. and et al.},
  journal = {Eur. Phys. J. C},
  volume = {85},
  number = {2},
  pages = {126},
  year = {2025}
}

@article{QPOs24,
  author = {Dönmez, O.},
  journal = {Eur. Phys. J. C},
  volume = {84},
  number = {5},
  pages = {524},
  year = {2024}
}

@article{QPOs25,
  author = {Garg, A. and Misra, R. and Sen, S.},
  journal = {Mon. Not. Roy. Astron. Soc.},
  volume = {514},
  number = {3},
  pages = {3285--3293},
  year = {2022}
}

@article{QPOs26,
  author = {Masterson, M. and et al.},
  title = {},
  journal = {Nature},
  volume = {638},
  pages = {370--375},
  year = {2025}
}

@article{QPOs27,
  author = {Bambi, C.},
  journal = {J. Cosmol. Astropart. Phys.},
  volume = {2012},
  number = {09},
  pages = {014},
  year = {2012}
}

@article{QPOs28,
  author = {Abdulkhamidov, F. and Narzilloev, B. and Hussain, I. and Abdujabbarov, A. and Ahmedov, B.},
  journal = {Eur. Phys. J. C},
  volume = {84},
  number = {4},
  pages = {420},
  year = {2024}
}

@article{QPOs29,
  author = {Török, G.},
  journal = {Astron. Nachr.},
  volume = {326},
  number = {9},
  pages = {856--860},
  year = {2005}
}

@article{QPOs31,
  author = {Vignarca, F. and Migliari, S. and Belloni, T. and Psaltis, D. and van der Klis, M.},
  title = {},
  journal = {Astron. Astrophys.},
  volume = {397},
  number = {2},
  pages = {729--738},
  year = {2003}
}

@article{QPOs33,
  author = {Wagoner, R. V.},
  title = {},
  journal = {Astrophys. J. Lett.},
  volume = {752},
  number = {2},
  pages = {L18},
  year = {2012}
}

@article{QPOs34,
  author = {Mustafa, G. and Demir, E. and Davlataliev, A. and Chaudhary, H. and Atamurotov, F. and Güdekli, E.},
  title = {},
  journal = {Phys. Dark Universe},
  volume = {46},
  pages = {101644},
  year = {2024}
}

@article{QPOs35,
  author = {Rayimbaev, J. and Murodov, S. and Shermatov, A. and Yusupov, A.},
  title = {},
  journal = {Eur. Phys. J. C},
  volume = {84},
  number = {10},
  pages = {1114},
  year = {2024}
}

@article{QPOs37,
  author = {Dönmez, O.},
  title = {},
  journal = {Astrophys. Space Sci.},
  volume = {293},
  pages = {323},
  year = {2004}
}

@article{QPOs38,
  author = {Montero, P. J. and Zanotti, O.},
  title = {},
  journal = {Mon. Not. Roy. Astron. Soc.},
  volume = {419},
  pages = {1507},
  year = {2011}
}

@article{QPOs41,
  author = {Boshkayev, K. and Luongo, O. and Muccino, M.},
  title = {},
  journal = {Phys. Rev. D},
  volume = {108},
  pages = {124034},
  year = {2023}
}

@article{QPOs42,
  author = {Boshkayev, K. and Konysbayev, T. and Kurmanov, Y. and Muccino, M. and Quevedo, H.},
  title = {},
  journal = {Mon. Not. Roy. Astron. Soc.},
  volume = {531},
  pages = {3876},
  year = {2024}
}

@article{QPOs43,
  author = {Boshkayev, K. and Idrissov, A. and Luongo, O. and Muccino, M.},
  title = {},
  journal = {Phys. Rev. D},
  volume = {108},
  pages = {044063},
  year = {2023}
}

@article{QPOs44,
  author = {Mitra, S. and Vrba, J. and Rayimbaev, J. and Stuchlik, Z. and Ahmedov, B.},
  title = {},
  journal = {Phys. Dark Univ.},
  volume = {46},
  pages = {101561},
  year = {2024}
}

@article{QPOs45,
  author = {Rayimbaev, J. and Eshimbetov, U. and Majeed, B. and Abdujabbarov, A. and Abduvokhidov, A. and Abdulazizov, B. and Xalmirzayev, A.},
  title = {},
  journal = {Chin. Phys. C},
  volume = {48},
  pages = {055104},
  year = {2024}
}

@article{QPOs46,
  author = {Abdulkhamidov, F. and Nedkova, P. and Rayimbaev, J. and Kunz, J. and Ahmedov, B.},
  title = {},
  journal = {Phys. Rev. D},
  volume = {109},
  pages = {104074},
  year = {2024}
}

@article{QPOs48,
  author = {Rayimbaev, J. and Abdujabbarov, A. and Abdulkhamidov, F. and Khamidov, V. and Djumanov, S. and Toshov, J. and Inoyatov, S.},
  title = {},
  journal = {Eur. Phys. J. C},
  volume = {82},
  pages = {1110},
  year = {2022}
}

@article{QPOs50,
  author = {Stuchlík, Z. and Kotrlová, A.},
  title = {},
  journal = {Gen. Relativ. Gravit.},
  volume = {41},
  pages = {1305},
  year = {2009}
}

@article{metric,
  author = {Kerr, R. P.},
  title = {},
  journal = {Phys. Rev. Lett.},
  volume = {11},
  pages = {237},
  year = {1963}
}

@article{B1,
  author  = {Ashraf, A. and et al.},
  title   = {},
  journal = {Phys. Dark Univ.},
  volume  = {47},
  pages   = {101787},
  year    = {2025}
}

@article{B2,
  author  = {Mustafa, G. and et al.},
  title   = {},
  journal = {Phys. Dark Univ.},
  volume  = {47},
  pages   = {101765},
  year    = {2025}
}

@article{B6,
  author  = {Ashraf, A. and et al.},
  title   = {},
  journal = {Phys. Dark Univ.},
  volume  = {47},
  pages   = {101823},
  year    = {2025}
}

@article{DonmezMNRAS2011,
    author = "Donmez, Orhan and Zanotti, Olindo and Rezzolla, Luciano",
    title = {},
    eprint = "1010.1739",
    archivePrefix = "arXiv",
    primaryClass = "astro-ph.HE",
    doi = "10.1111/j.1365-2966.2010.18003.x",
    journal = "Mon. Not. Roy. Astron. Soc.",
    volume = "412",
    pages = "1659--1668",
    year = "2011"
}

@article{DonmezEpjc2025,
    author = "Donmez, O.",
    title = {},
    doi = "10.1140/epjc/s10052-025-14779-6",
    journal = "Eur. Phys. J. C",
    volume = "85",
    number = "9",
    pages = "1019",
    year = "2025"
}

@article{DonmezJheap2024,
    author = "Donmez, Orhan",
    title = {},
    eprint = "2408.10102",
    archivePrefix = "arXiv",
    primaryClass = "astro-ph.HE",
    doi = "10.1016/j.jheap.2024.11.002",
    journal = "JHEAp",
    volume = "45",
    pages = "1--18",
    year = "2025"
}

@article{Donmezpdu2024,
    author = "Donmez, Orhan and Dogan, Fatih",
    title = {},
    eprint = "2407.01478",
    archivePrefix = "arXiv",
    primaryClass = "gr-qc",
    doi = "10.1016/j.dark.2024.101718",
    journal = "Phys. Dark Univ.",
    volume = "46",
    pages = "101718",
    year = "2024"
}

@article{KoyuncuMpla2014,
    author = {Koyuncu, Fahrettin and D{\"o}nmez, Orhan},
    title = {},
    doi = "10.1142/S0217732314501156",
    journal = "Mod. Phys. Lett. A",
    volume = "29",
    pages = "1450115",
    year = "2014"
}

@article{MustafaJcap2025,
    author = "Mustafa, G. and Ghosh, Sushant G. and Donmez, Orhan and Maurya, S. K. and Orzuev, Shakhzod and Atamurotov, Farruh",
    title = "{Testing quantum-corrected black holes with QPOs observations: a study of particle dynamics and accretion~flow}",
    eprint = "2506.16405",
    archivePrefix = "arXiv",
    primaryClass = "gr-qc",
    doi = "10.1088/1475-7516/2025/10/068",
    journal = "JCAP",
    volume = "10",
    pages = "068",
    year = "2025"
}

@article{Zhang:2025iae,
    author = "Zhang, Shuang-Nan and others",
    title = {},
    eprint = "2506.08101",
    archivePrefix = "arXiv",
    primaryClass = "astro-ph.HE",
    doi = "10.1007/s11433-025-2786-6",
    journal = "Sci. China Phys. Mech. Astron.",
    volume = "68",
    number = "11",
    pages = "119502",
    year = "2025"
}

@article{Mayr:2024rbs,
    author = {Mayr, Astrid and M{\"u}ller-Seidlitz, Johannes and Emberger, Valentin and Andritschke, Robert and Reiffers, Jonas and Schweingruber, Anna and Albrecht, Sebastian and Schnetler, Hermine},
    title = {Spectral performance budget for {ATHENA}'s Wide Field Imager},
    doi = "10.1117/12.3018737",
    journal = "Proc. SPIE Int. Soc. Opt. Eng.",
    volume = "13099",
    pages = "130992E",
    year = "2024"
}

@article{DonmezNPB2026,
  author  = {Donmez, Orhan},
  title   = {Evolution of shock structures and {QPOs} after halting {BHL} accretion onto {Kerr} black hole},
  journal = {Nucl. Phys. B},
  volume  = {1027},
  pages   = {117483},
  year    = {2026},
  doi     = {10.1016/j.nuclphysb.2026.117483}
}

@article{Greiner2001,
  author  = {Greiner, J. and Cuby, J. G. and McCaughrean, M. J.},
  title   = {An unusually massive stellar black hole in the {Galaxy}},
  journal = {Nature},
  volume  = {414},
  pages   = {522},
  year    = {2001},
  doi     = {10.1038/35107019},
  eprint  = {astro-ph/0111538}
}

@article{BlondinRaymer2012,
  author  = {Blondin, John M. and Raymer, Eric},
  title   = {Hoyle--Lyttleton accretion in three dimensions},
  journal = {Astrophys. J.},
  volume  = {752},
  pages   = {30},
  year    = {2012},
  doi     = {10.1088/0004-637X/752/1/30},
  eprint  = {1204.0717}
}

@article{DonmezMNRAS2014,
  author  = {D{\"o}nmez, Orhan},
  title   = {On the development of the {Papaloizou--Pringle} instability of the black hole--torus systems and quasi-periodic oscillations},
  journal = {Mon. Not. Roy. Astron. Soc.},
  volume  = {438},
  number  = {1},
  pages   = {846--858},
  year    = {2014},
  doi     = {10.1093/mnras/stt2255},
  eprint  = {1304.0584}
}

@article{Liu:2020jig,
    author = "Liu, Honghui and Ji, Long and Bambi, Cosimo and Jain, Pankaj and Misra, Ranjeev and Rawat, Divya and Yadav, J. S. and Zhang, Yuexin",
    title = "{Testing evolution of LFQPOs with mass accretion rate in GRS 1915+105 with Insight-HXMT}",
    eprint = "2012.01825",
    archivePrefix = "arXiv",
    doi = "10.3847/1538-4357/abdf65",
    journal = "Astrophys. J.",
    volume = "909",
    number = "1",
    pages = "63",
    year = "2021"
}

@article{Riess1998,
  author  = {Riess, A. G. and others},
  title   = {Observational evidence from supernovae for an accelerating universe and a cosmological constant},
  journal = {Astron. J.},
  volume  = {116},
  pages   = {1009},
  year    = {1998},
  doi     = {10.1086/300499},
  eprint  = {astro-ph/9805201}
}

@article{Perlmutter1999,
  author  = {Perlmutter, S. and others},
  title   = {Measurements of $\Omega$ and $\Lambda$ from 42 high-redshift supernovae},
  journal = {Astrophys. J.},
  volume  = {517},
  pages   = {565},
  year    = {1999},
  doi     = {10.1086/307221},
  eprint  = {astro-ph/9812133}
}

@article{Weinberg1989,
  author  = {Weinberg, S.},
  title   = {The cosmological constant problem},
  journal = {Rev. Mod. Phys.},
  volume  = {61},
  pages   = {1},
  year    = {1989},
  doi     = {10.1103/RevModPhys.61.1}
}

@article{Weyl1918,
  author  = {Weyl, H.},
  title   = {Reine {I}nfinitesimalgeometrie},
  journal = {Math. Z.},
  volume  = {2},
  pages   = {384},
  year    = {1918}
}

@article{Bach1921,
  author  = {Bach, R.},
  title   = {Zur {W}eylschen {R}elativit{\"a}tstheorie und der {W}eylschen {E}rweiterung des {K}r{\"u}mmungstensorbegriffs},
  journal = {Math. Z.},
  volume  = {9},
  pages   = {110},
  year    = {1921}
}

@article{Stelle1977,
  author  = {Stelle, K. S.},
  title   = {Renormalization of higher-derivative quantum gravity},
  journal = {Phys. Rev. D},
  volume  = {16},
  pages   = {953},
  year    = {1977},
  doi     = {10.1103/PhysRevD.16.953}
}

@article{Riegert1984,
  author  = {Riegert, R. J.},
  title   = {Birkhoff's theorem in conformal gravity},
  journal = {Phys. Rev. Lett.},
  volume  = {53},
  pages   = {315},
  year    = {1984},
  doi     = {10.1103/PhysRevLett.53.315}
}

@article{Mannheim2006,
  author  = {Mannheim, Philip D.},
  title   = {Alternatives to dark matter and dark energy},
  journal = {Prog. Part. Nucl. Phys.},
  volume  = {56},
  pages   = {340},
  year    = {2006},
  eprint  = {astro-ph/0505266}
}

@article{MannheimOBrien2011,
  author  = {Mannheim, Philip D. and O'Brien, James G.},
  title   = {Impact of a global quadratic potential on galactic rotation curves},
  journal = {Phys. Rev. Lett.},
  volume  = {106},
  pages   = {121101},
  year    = {2011},
  doi     = {10.1103/PhysRevLett.106.121101},
  eprint  = {1007.0970}
}

@article{BenderMannheim2008,
  author  = {Bender, Carl M. and Mannheim, Philip D.},
  title   = {No-ghost theorem for the fourth-order derivative {P}ais--{U}hlenbeck oscillator model},
  journal = {Phys. Rev. Lett.},
  volume  = {100},
  pages   = {110402},
  year    = {2008},
  doi     = {10.1103/PhysRevLett.100.110402},
  eprint  = {0706.0207}
}

@misc{Maldacena2011,
  author        = {Maldacena, Juan},
  title         = {Einstein gravity from conformal gravity},
  year          = {2011},
  eprint        = {1105.5632},
  archivePrefix = {arXiv},
  primaryClass  = {hep-th}
}

@article{EderyParanjape1998,
  author  = {Edery, A. and Paranjape, M. B.},
  title   = {Classical tests for {W}eyl gravity: deflection of light and time delay},
  journal = {Phys. Rev. D},
  volume  = {58},
  pages   = {024011},
  year    = {1998},
  doi     = {10.1103/PhysRevD.58.024011},
  eprint  = {astro-ph/9708233}
}

@article{Flanagan2006,
  author  = {Flanagan, {\'E}anna {\'E}.},
  title   = {Fourth order {W}eyl gravity},
  journal = {Phys. Rev. D},
  volume  = {74},
  pages   = {023002},
  year    = {2006},
  doi     = {10.1103/PhysRevD.74.023002},
  eprint  = {astro-ph/0605504}
}

@article{Carter1968,
  author  = {Carter, B.},
  title   = {Hamilton--{J}acobi and {S}chr{\"o}dinger separable solutions of {E}instein's equations},
  journal = {Commun. Math. Phys.},
  volume  = {10},
  pages   = {280},
  year    = {1968}
}

@article{GW150914,
  author  = {Abbott, B. P. and others},
  title   = {Observation of gravitational waves from a binary black hole merger},
  journal = {Phys. Rev. Lett.},
  volume  = {116},
  pages   = {061102},
  year    = {2016},
  doi     = {10.1103/PhysRevLett.116.061102},
  eprint  = {1602.03837}
}

@article{EHT2019I,
  author  = {{Event Horizon Telescope Collaboration}},
  title   = {First {M87} {Event Horizon Telescope} Results. {I}. {The} Shadow of the Supermassive Black Hole},
  journal = {Astrophys. J. Lett.},
  volume  = {875},
  pages   = {L1},
  year    = {2019},
  doi     = {10.3847/2041-8213/ab0ec7},
  eprint  = {1906.11238}
}

@article{EHT2022SgrA,
  author  = {{Event Horizon Telescope Collaboration}},
  title   = {First {Sagittarius A*} {Event Horizon Telescope} Results. {I}. {The} Shadow of the Supermassive Black Hole in the Center of the {Milky Way}},
  journal = {Astrophys. J. Lett.},
  volume  = {930},
  pages   = {L12},
  year    = {2022},
  doi     = {10.3847/2041-8213/ac6674}
}

@article{Psaltis2008,
  author  = {Psaltis, Dimitrios},
  title   = {Probes and tests of strong-field gravity with observations in the electromagnetic spectrum},
  journal = {Living Rev. Relativ.},
  volume  = {11},
  pages   = {9},
  year    = {2008},
  doi     = {10.12942/lrr-2008-9},
  eprint  = {0806.1531}
}

@article{Mizuno2018,
  author  = {Mizuno, Y. and Younsi, Z. and Fromm, C. M. and others},
  title   = {The current ability to test theories of gravity with black hole shadows},
  journal = {Nat. Astron.},
  volume  = {2},
  pages   = {585},
  year    = {2018},
  doi     = {10.1038/s41550-018-0449-5},
  eprint  = {1804.05812}
}

@article{JohannsenPsaltis2011,
  author  = {Johannsen, Tim and Psaltis, Dimitrios},
  title   = {Testing the no-hair theorem with observations in the electromagnetic spectrum. {III}. {Q}uasi-periodic variability},
  journal = {Astrophys. J.},
  volume  = {726},
  pages   = {11},
  year    = {2011},
  doi     = {10.1088/0004-637X/726/1/11},
  eprint  = {1010.1000}
}

@article{StellaVietri1998,
  author  = {Stella, L. and Vietri, M.},
  title   = {{L}ense--{T}hirring precession and quasi-periodic oscillations in low-mass {X}-ray binaries},
  journal = {Astrophys. J. Lett.},
  volume  = {492},
  pages   = {L59},
  year    = {1998},
  eprint  = {astro-ph/9709085}
}

@article{Stella1999,
  author  = {Stella, L. and Vietri, M. and Morsink, S. M.},
  title   = {Correlations in the quasi-periodic oscillation frequencies of low-mass {X}-ray binaries and the relativistic precession model},
  journal = {Astrophys. J. Lett.},
  volume  = {524},
  pages   = {L63},
  year    = {1999},
  eprint  = {astro-ph/9907346}
}

@article{Ingram2009,
  author  = {Ingram, A. and Done, C. and Fragile, P. C.},
  title   = {Low-frequency quasi-periodic oscillations spectra and {L}ense--{T}hirring precession},
  journal = {Mon. Not. Roy. Astron. Soc.},
  volume  = {397},
  pages   = {L101},
  year    = {2009}
}

@article{KluzniakAbramowicz2001,
  author  = {Klu{\'z}niak, W. and Abramowicz, M. A.},
  title   = {Strong-field gravity and orbital resonance in black holes and neutron stars --- k{H}z quasi-periodic oscillations ({QPO})},
  journal = {Acta Phys. Pol. B},
  volume  = {32},
  pages   = {3605},
  year    = {2001},
  eprint  = {astro-ph/0105057}
}

@article{Torok2005,
  author  = {T{\"o}r{\"o}k, G. and Abramowicz, M. A. and Klu{\'z}niak, W. and Stuchl{\'i}k, Z.},
  title   = {The orbital resonance model for twin peak k{H}z quasi periodic oscillations in microquasars},
  journal = {Astron. Astrophys.},
  volume  = {436},
  pages   = {1},
  year    = {2005},
  eprint  = {astro-ph/0401464}
}

@article{Strohmayer2001,
  author  = {Strohmayer, T. E.},
  title   = {Discovery of a 450 {H}z quasi-periodic oscillation from the microquasar {GRO J1655-40} with the {R}ossi {X}-ray {T}iming {E}xplorer},
  journal = {Astrophys. J. Lett.},
  volume  = {552},
  pages   = {L49},
  year    = {2001},
  eprint  = {astro-ph/0104487}
}

@article{Belloni2002,
  author  = {Belloni, T. and Psaltis, D. and van der Klis, M.},
  title   = {A unified description of the timing features of accreting {X}-ray binaries},
  journal = {Astrophys. J.},
  volume  = {572},
  pages   = {392},
  year    = {2002},
  eprint  = {astro-ph/0202213}
}

@article{Casella2005,
  author  = {Casella, P. and Belloni, T. and Stella, L.},
  title   = {The {ABC} of low-frequency quasi-periodic oscillations in black hole candidates: analogies with {Z} sources},
  journal = {Astrophys. J.},
  volume  = {629},
  pages   = {403},
  year    = {2005},
  eprint  = {astro-ph/0504318}
}

@article{Motta2014,
  author  = {Motta, S. E. and Belloni, T. M. and Stella, L. and Mu{\~n}oz-Darias, T. and Fender, R.},
  title   = {Precise mass and spin measurements for a stellar-mass black hole through {X}-ray timing: the case of {GRO J1655-40}},
  journal = {Mon. Not. Roy. Astron. Soc.},
  volume  = {437},
  pages   = {2554},
  year    = {2014},
  eprint  = {1309.3652}
}

@article{FryxellTaam1988,
  author  = {Fryxell, B. A. and Taam, R. E.},
  title   = {Numerical simulations of nonaxisymmetric adiabatic accretion flow},
  journal = {Astrophys. J.},
  volume  = {335},
  pages   = {862},
  year    = {1988}
}

@article{Petrich1989,
  author  = {Petrich, L. I. and Shapiro, S. L. and Stark, R. F. and Teukolsky, S. A.},
  title   = {Accretion onto a moving black hole: a fully relativistic treatment},
  journal = {Astrophys. J.},
  volume  = {336},
  pages   = {313},
  year    = {1989}
}

@article{Penner2011,
  author  = {Penner, A. J.},
  title   = {General relativistic magnetohydrodynamic {B}ondi--{H}oyle accretion},
  journal = {Mon. Not. Roy. Astron. Soc.},
  volume  = {414},
  pages   = {1467},
  year    = {2011},
  eprint  = {1011.2976}
}

@article{LoraClavijo2013,
  author  = {Lora-Clavijo, F. D. and Guzm{\'a}n, F. S.},
  title   = {Axisymmetric {B}ondi--{H}oyle accretion on to a {S}chwarzschild black hole: shock cone vibrations},
  journal = {Mon. Not. Roy. Astron. Soc.},
  volume  = {429},
  pages   = {3144},
  year    = {2013},
  eprint  = {1212.2139}
}

@article{GraciaLinares2015,
  author  = {Gracia-Linares, M. and Guzm{\'a}n, F. S.},
  title   = {Accretion of supersonic winds onto black holes in {3D}: stability of the shock cone},
  journal = {Astrophys. J.},
  volume  = {812},
  pages   = {23},
  year    = {2015},
  eprint  = {1510.05947}
}

\end{document}